%% file: doktoratMI.tex
\documentclass[12pt, a4paper, twoside, openany]{book}

\usepackage{calc}
\usepackage{rotating}
\usepackage[english]{babel}
\usepackage{hyperref}
\usepackage{graphicx}
\usepackage{subfig}
\usepackage{float}
\usepackage{amsmath,bm}
\usepackage{amssymb}
\usepackage{amsthm}
\usepackage{latexsym}
\usepackage{multirow}
\usepackage{cite}

\usepackage[toc,page]{appendix}

\usepackage{slashed}
\usepackage{cancel}
\usepackage{afterpage}
\usepackage{dcolumn}

\usepackage{verbatim}
\usepackage{graphicx}
\usepackage{slashed}
\usepackage{lscape}
\usepackage{rotating}

\usepackage{cite}
\usepackage{cancel}
\usepackage{caption}
\usepackage{epstopdf}
\usepackage{amsthm}
\usepackage{amsmath}
\usepackage{color}

\usepackage{hyperref}
\usepackage{braket}

\usepackage{booktabs}  
\usepackage{siunitx}
\usepackage{fancyhdr}

\DeclareGraphicsExtensions{.eps}

\usepackage{float}
\usepackage{numprint}

\usepackage{titlesec}
\titleformat{\part}[display]{\bf\LARGE\filcenter}{\titlerule[1.5pt] \vspace{1pc} \Huge\MakeUppercase{\partname} \thepart}{1pc}{\titlerule[1.5pt] \vspace{1pc} \Huge}
\titleformat{\chapter}[display]{\bf\Large\filcenter}{\titlerule[1pt] \vspace{1pc} \Huge\MakeUppercase{\chaptertitlename} \thechapter}{1pc}{\titlerule[1pt] \vspace{1pc} \LARGE}

\input macros.tex
\restylefloat{figure}

\titleclass{\chapter}{top}
\titleclass{\section}{straight}
\titleclass{\subsection}{straight}

\numberwithin{equation}{chapter}

\title{Yukawa matrix unification in the Minimal Supersymmetric Standard Model}
\author{Mateusz Kamil Iskrzy\'nski}
\date{August 2015}

\makeatletter  
\renewcommand\maketitle{%
  \begin{titlepage}%
    \let\footnotesize\small
    \let\footnoterule\relax
    \begin{center}%
     \par \vspace{1cm plus .8fill} 
         {\fontsize{20pt}{24pt}\selectfont\textbf{\@title}\par}
      \vspace{1cm}
      {\fontsize{17pt}{22pt}\selectfont\textsl{\@author}\par}
         \fontsize{12pt}{14pt}\selectfont
      \vspace{2cm}
      \begin{figure}[h!]
        \centering
        \includegraphics[scale=.4]{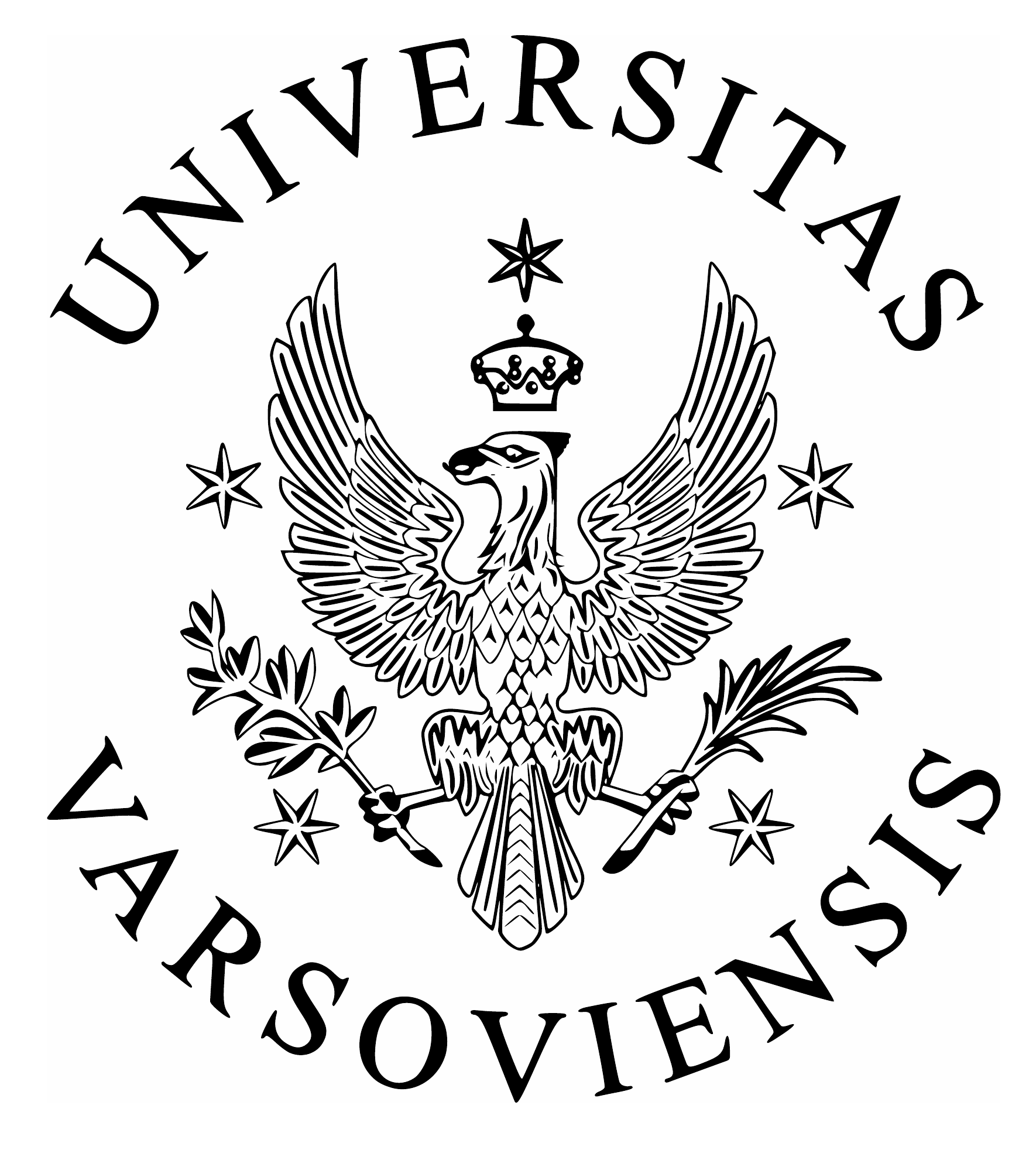}
       \end{figure}
       \vspace{2cm}
      \begin{flushright}
        \begin{tabular}{r}
          PhD dissertation under the supervision of\\[2pt]
          \bfseries prof. dr hab. Miko{\l}aj Krzysztof Misiak \\
          at the Institute of Theoretical Physics, Faculty of Physics, University of Warsaw\\[2pt]
        \end{tabular}
      \end{flushright}
      \vspace{20mm plus .1fill}
      {\fontsize{14pt}{18pt}\selectfont Warsaw, \space \@date\par}
    \end{center}
 \end{titlepage}%
}
\makeatother   

\begin{document}

\maketitle

  \cleardoublepage
  \chapter*{Abstract}
  \input{chapters/0_Abstract}

  \chapter*{Streszczenie}
  \input{chapters/0_Streszczenie}

  \chapter*{Acknowledgments}
  \input{chapters/Thx}

  \cleardoublepage
  \tableofcontents
  \markboth{Table of Contents}{Table of Contents}

  \listoffigures
  \markboth{List of Figures}{List of Figures}

  \listoftables
  \markboth{List of Tables}{List of Tables}
  \cleardoublepage
 
  \chapter*{Outline}
  \markboth{Outline}{Outline}
  \addcontentsline{toc}{chapter}{\protect Outline}
  \input{chapters/0_Outline}

  \chapter{Introduction}\label{TheoreticalOverviewChap}
   \input{chapters/1_0_intro}

  \section{Preliminaries}
   \input{chapters/1_1_Preliminaries}

  \section{The Standard Model}
    \input{chapters/1_2_StandardModel}

   \section{Grand Unified Theories}
     \input{chapters/1_3_GrandUnifiedTheories}

   \section{Supersymmetry}
     \input{chapters/1_4_Supersymmetry}

   \section{Yukawa coupling unification in the MSSM}
      \input{chapters/1_5_YukUni}

\chapter{Two MSSM scenarios for the SU(5) Yukawa matrix unification}\label{TwoScenarios}
 \input{chapters/2_1_2_Scenarios}

 \begin{section}{Threshold corrections in the diagonal soft term case}\label{TreshCorChap}
   \input{chapters/2_2_0_intro}
  \input{chapters/2_3_treshCor_SoftDiag}

 \end{section}
 \begin{section}{Threshold corrections with flavour mixing in the soft terms}\label{analysis}
  \input{chapters/2_4_treshCor_OffDiag}
 \end{section}    

\chapter{Phenomenological constraints}\label{PhenoIntro}

  In this chapter, some of the phenomenological low-energy constraints
  that matter for our analysis are described. They are selected
  according to their relevance for the discussion of the considered
  scenarios.

  \section{Flavour Changing Neutral Currents}\label{flavPheno}
    \input{chapters/3_1_Flavour}

  \begin{section}{Higgs boson mass measurement}\label{mhsect}
   \input{chapters/3_2_mh0}

  \end{section}

  \begin{section}{LHC SUSY searches}
   \input{chapters/3_3_SUSY_LHC}

  \end{section}

  \begin{section}{Electroweak vacuum stability}\label{EWSBstab}
   \input{chapters/3_4_vacStab}

  \end{section}
  
  \begin{section}{Dark matter}
    \input{chapters/3_5_DarkMatter}

  \end{section}

\chapter{Numerical Tools}\label{ToolsChap}
  \input{chapters/4_Tools}

\begin{chapter}{Large diagonal $A$ terms -- numerical results}\label{SoftDiagChap}
 
  \section{Regions with the successful $\mathbf{SU(5)}$ Yukawa matrix unification}\label{ExamplesSec}
    \input{chapters/5_1_RegionsSoftDiag}

   \section{Flavour observables}\label{flavSec}
     \input{chapters/5_2_FlavourSoftDiag}

   \section{Electroweak symmetry breaking}\label{vacSect}
     \input{chapters/5_3_EWSB_SoftDiag}

\end{chapter}

 \begin{chapter}{Numerical results for the $GFV_{23}$ scenario}\label{OffDiagChap}
    \input{chapters/6_00_introOffDiag}
    \section{Regions with the successful bottom-tau and strange-muon unification}
     \input{chapters/6_1_Regions_OffDiag}

   \section{Phenomenology of the $GFV_{23}$ scenario}
          
   \subsection{Dark matter}\label{sec:dm}
      \input{chapters/6_2_1_DM_OffDiag}

   \subsection{Higgs, flavour and electroweak observables}\label{sec:flav}
     \input{chapters/6_2_2_Flavour_OffDiag}

   \subsection{LHC direct SUSY searches}\label{sec:lhc} 
     \input{chapters/6_2_3_LHC_OffDiag}

   \subsection{Electroweak symmetry breaking}\label{OffDiagEWSBsec}
     \input{chapters/6_2_4_EWSB_Offdiag}

 \end{chapter}
 \begin{chapter}{Numerical results of the $GFV_{123}$ scenario}\label{GFV123Num}
	\input{chapters/7_0_introGFV123}

  \section{Regions with the $\mathbf{SU(5)}$ Yukawa matrix unification}\label{sec:unif}
    \input{chapters/7_1_Regions_OffDiag_123}
  \section{Phenomenology of the $GFV_{123}$ scenario}
     \input{chapters/7_2_phenoIntro}
   \subsection{Lepton Flavour Violating observables}
     \input{chapters/7_3_LFV}

   \subsection{EW vacuum stability}\label{vacSect.GFV}
      \input{chapters/7_4_EWSB_OffDiag}

 \end{chapter}
  
 \chapter{Conclusions}\label{concl}
  \input{chapters/8_conclusions}

    \chapter*{List of acronyms}
      \input{chapters/Acronyms}

 \input{chapters/bibl}

\end{document}

%% file: macros.tex
\newcommand{\newc}{\newcommand*}

\long\def\begincomment#1\endcomment{%
        \begingroup\sf\baselineskip12pt#1\endgroup}

\newc{\etal}{\textrm{et al.}} 
\newc{\eg}{\textrm{e.g.}} 
\newc{\ie}{\textrm{i.e.}}
\newc{\etc}{\textrm{etc}}
\newc\vs{\textrm{vs.}}
\newc{\cl}{\rm {C.L.}}

\newcommand{\nc}{\newcommand}
\nc{\be}{\begin{equation}}
\nc{\ee}{\end{equation}}
\nc{\munu}{^{\mu}_{\phantom{\mu}\nu}}
\nc{\p}{\phantom{\mu}}
\nc{\ph}{\varphi}
\nc{\tph}{\tilde{\varphi}}
\nc{\eps}{\varepsilon}
\nc{\bgt}{\begin{gather}}
\nc{\egt}{\end{gather}}

\def\sla#1{\not\!\!#1}
\nc{\Gmna}{G_{\mu\nu}^{A}}
\nc{\jd}{\tfrac{1}{2}}
\nc{\psb}{\bar{\psi}}
\nc{\pss}{\psi^{*}}
\nc{\pht}{\tilde{\varphi}}

\newc{\ev}{\ensuremath{\,\mathrm{eV}}}
\newc{\kev}{\ensuremath{\,\mathrm{keV}}}
\newc{\mev}{\ensuremath{\,\mathrm{MeV}}}
\newc{\gev}{\ensuremath{\,\mathrm{GeV}}}
\newc{\tev}{\ensuremath{\,\mathrm{TeV}}}
\newc{\MeV}{\mev} 
\newc{\TeV}{\tev}
\newc{\invpb}{\ensuremath{/\text{pb}}}
\newc{\invfb}{\ensuremath{/\text{fb}}}
\newc\nb{\ensuremath{\,\mathrm{nb}}} \newc\pb{\ensuremath{\,\mathrm{pb}}} \newc\fb{\ensuremath{\,\mathrm{fb}}}
\newc\pc{\ensuremath{\,\mathrm{pc}}}
\newc\kpc{\ensuremath{\,\mathrm{kpc}}}
\newc\mpc{\ensuremath{\,\mathrm{Mpc}}}
\newc\ps{\ensuremath{\,\mathrm{ps}}} 
\newc\cmeter{\ensuremath{\,\mathrm{cm}}} 
\newc\meter{\ensuremath{\,\mathrm{m}}} 
\newc\kmeter{\ensuremath{\,\mathrm{km}}}
\newc\second{\ensuremath{\,\mathrm{s}}}
\newc\msecond{\ensuremath{\,\mathrm{ms}}}
\newc\nsecond{\ensuremath{\,\mathrm{ns}}}
\newc\psecond{\ensuremath{\,\mathrm{ps}}}

\newc{\chisqmin}{\ensuremath{\chi^2_{\mathrm{min}}}}
\newc{\Delchisq}{\ensuremath{\Delta\chi^2}}
\newc{\chisq}{\ensuremath{\chi^2}}
\newc{\like}{\ensuremath{\mathcal{L}}}

\newc\lsim{\ensuremath{\mathrel{\rlap{\lower4pt\hbox{\hskip1pt$\sim$}}\raise1pt\hbox{$<$}}}}
\newc\gsim{\ensuremath{\mathrel{\rlap{\lower4pt\hbox{\hskip1pt$\sim$}}\raise1pt\hbox{$>$}}}}
\newc{\VEV}[1]{\ensuremath{\langle #1 \rangle}}
\newc{\dl}{\ensuremath{\stackrel{\leftarrow}{D}}}
\newc{\dr}{\ensuremath{\stackrel{\rightarrow}{D}}}

\newc{\bcenter}{\begin{center}}    \newc{\ecenter}{\end{center}}
\newc{\bfl}{\begin{flushleft}}    \newc{\efl}{\end{flushleft}}
\newc{\bfr}{\begin{flushright}}    \newc{\efr}{\end{flushright}}

\newc{\bi}{\begin{itemize}}
\newc{\ei}{\end{itemize}}
\newc{\bed}{\begin{description}}
\newc{\eed}{\end{description}}
\newc{\ben}{\begin{enumerate}}
\newc{\een}{\end{enumerate}}

\newc{\bea}{\begin{eqnarray}}
\newc{\eea}{\end{eqnarray}}
\newc{\bfle}{\begin{flalign}}
\newc{\efle}{\end{flalign}}
\newc{\ra}{\rightarrow}

\newc{\ymu}{$SU(5)$ Yukawa matrix unification}

\newc{\vsm}{\tfrac{v}{\sqrt{2}}}
\newc{\smgg}{$SU(3)\times SU(2)\times U(1)$}
\newc{\yuku}{\mathbf{Y^u}}
\newc{\yuke}{\mathbf{Y^e}}
\newc{\yukd}{\mathbf{Y^d}}
\newc{\alphas}{\ensuremath{\alpha_s}}
\newc{\alphatwo}{\ensuremath{\alpha_2}}
\newc{\alphaone}{\ensuremath{\alpha_1}}
\newc{\alphai}[1]{\ensuremath{\alpha_{#1}}}
\newc{\alphaem}{\ensuremath{\alpha_{\mathrm{em}}}}
\newc{\alphaeff}{\ensuremath{\alpha_{\mathrm{eff}}}}
\newc{\sineff}{\ensuremath{\sin^2 \theta_{\mathrm{eff}}}}
\newc{\sinsqeff}{\ensuremath{\sin^2 \theta_{\mathrm{eff}}}}
\newc{\dalphahad}{\ensuremath{\Delta \alpha_{\mathrm{had}}}}
\newc{\yt}{\ensuremath{h_t}} \newc{\yb}{\ensuremath{h_b}} \newc{\ytau}{\ensuremath{h_{\tau}}}
\newc\mz{\ensuremath{m_Z}} 
\newc\mw{\ensuremath{m_W}}
\newc\mZ{\mz}        \newc\mW{\mw}
\newc\mhsm{\ensuremath{ m_{H_{\mathrm{SM}}}}}
\newc{\mtop}{\ensuremath{ m_t}}               
\newc{\mbottom}{\ensuremath{ m_b}} 
\newc{\mtau}{\ensuremath{ m_{\tau}}}
\newc{\mt}{\mtpole}
\newc{\mb}{\mbottom} 
\newc{\rtwogg}{\ensuremath{R_{h_2}(\gamma\gamma)}}
\newc{\rtwozz}{\ensuremath{R_{h_2}(ZZ)}}
\newc{\ronegg}{\ensuremath{R_{h_1}(\gamma\gamma)}}
\newc{\ronezz}{\ensuremath{R_{h_1}(ZZ)}}
\newc{\rsiggg}{\ensuremath{R_{h_\textrm{sig}}(\gamma\gamma)}}
\newc{\rsigzz}{\ensuremath{R_{h_\textrm{sig}}(ZZ)}}
\newc{\llbar}{\ensuremath{\ell\bar{\ell}}}
\newc{\tauptaum}{\ensuremath{ \tau^+\tau^-}}
\newc{\qqbar}{\ensuremath{ q\bar{q}}} \newc{\ppbar}{\ensuremath{ p\bar{p}}}
\newc{\bbbar}{\ensuremath{ b\bar{b}}} \newc{\ttbar}{\ensuremath{ t\bar{t}}}
\newc{\ffbar}{\ensuremath{ f\bar{f}}} \newc{\tautaubar}{\ensuremath{ \tau\bar{\tau}}}

\newc{\mchi}{\ensuremath{m_\neutone}}
\newc{\squark}{\ensuremath{\tilde{q}}}
\newc{\slepton}{\ensuremath{\tilde{l}}}
\newc{\gluino}{\ensuremath{\tilde{g}}} 
\newc{\mgluino}{\ensuremath{{m_{\gluino}}}}
\newc{\wino}{\ensuremath{\tilde{W}}} 
\newc{\mwino}{\ensuremath{{m_{\wino}}}}
\newc{\tone}{\ensuremath{{\tilde{t}_1}}}
\newc{\Hone}{\ensuremath{{\tilde{H}_{1}}}}
\newc{\Htwo}{\ensuremath{{\tilde{H}_{2}}}}
\newc{\Hhtwo}{\ensuremath{{H_{2}}}}
\newc{\qli}{\ensuremath{{\tilde{Q}_{i}}}}
\newc{\uri}{\ensuremath{{\tilde{u}_{i}}}}
\newc{\dri}{\ensuremath{{\tilde{d}_{i}}}}
\newc{\lli}{\ensuremath{{\tilde{L}_{i}}}}
\newc{\eri}{\ensuremath{{\tilde{e}_{i}}}}

\newc{\sthw}{\ensuremath{ \sin\theta_W}}              \newc{\cthw}{\ensuremath{\cos\theta_W}}
\newc{\tanthw}{\ensuremath{ \tan\theta_W}}              \newc{\cotthw}{\ensuremath{\cot\theta_W}}
\newc{\ssqthw}{\ensuremath{\sin^2 \theta_W}}
\newc{\msbar}{\ensuremath{\overline{MS}}} \newc{\drbar}{\ensuremath{\overline{DR}}}
\newc{\mtmtsmmsbar}{\ensuremath{ m_t(m_t)^{\msbar}_{{\mathrm{SM}}}}}
\newc{\mtmtsmdrbar}{\ensuremath{ m_t(m_t)^{\drbar}_{{\mathrm{SM}}}}}
\newc{\mtmtmssmdrbar}{\ensuremath{ m_t(m_t)^{\drbar}_{{\mathrm{SUSY}}}}}
\newc{\mbmbmsbar}{\ensuremath{ m_b^{\msbar}(m_b)}}
\newc{\mcmbmsbar}{\ensuremath{ m_c^{\msbar}(m_c)}}
\newc{\msmbmsbar}{\ensuremath{ m_s^{\msbar}}}
\newc{\mdmbmsbar}{\ensuremath{ m_d^{\msbar}}}
\newc{\mumbmsbar}{\ensuremath{ m_u^{\msbar}}}
\newc{\mtaupole}{\ensuremath{m_\tau^{\rm pole}}}
\newc{\mmupole}{\ensuremath{m_\mu^{\rm pole}}}
\newc{\mepole}{\ensuremath{m_e^{\rm pole}}}
\newc{\mzpole}{\ensuremath{M_Z^{\rm pole}}}
\newc{\mbmbsmmsbar}{\ensuremath{ m_b(m_b)^{\msbar}_{{\mathrm{SM}}}}}
\newc{\mbmzsmmsbar}{\ensuremath{ m_b(\mz)^{\msbar}_{{\mathrm{SM}}}}}
\newc{\mbmzsmdrbar}{\ensuremath{ m_b(\mz)^{\drbar}_{{\mathrm{SM}}}}}
\newc{\mbmzmssmdrbar}{\ensuremath{ m_b(\mz)^{\drbar}_{{\mathrm{SUSY}}}}}
\newc{\mtaumzsmmsbar}{\ensuremath{ m_{\tau}(\mz)^{\msbar}_{{\mathrm{SM}}}}}
\newc{\mtaumzsmdrbar}{\ensuremath{ m_{\tau}(\mz)^{\drbar}_{{\mathrm{SM}}}}}
\newc{\mtaumzmssmdrbar}{\ensuremath{ m_{\tau}(\mz)^{\drbar}_{{\mathrm{SUSY}}}}}
\newc{\alphasmzms}{\ensuremath{\alpha_s^{\overline{MS}}(M_Z)}}
\newc{\alphaimzms}[1]{\ensuremath{\alpha_{#1}(M_Z)^{\overline{MS}}}}
\newc{\alphaemmz}{\ensuremath{\alpha_{\mathrm{em}}^{-1}(M_Z)}}

\newc{\mzero}{\ensuremath{{m_0}}}
\newc{\mhalf}{\ensuremath{ M_{1/2}}}
\newc{\tanb}{\ensuremath{\tan\beta}}
\newc{\azero}{\ensuremath{ A_0}}
\newc{\signmu}{\ensuremath{\rm{sgn}\,\mu}}
\newc{\atau}{\ensuremath{{A_{\tau}}}}
\newc{\mueff}{\ensuremath{\mu_{\rm{eff}}}}
\newc{\lam}{\ensuremath{{\lambda}}}
\newc{\kap}{\ensuremath{{\kappa}}}
\newc{\alam}{\ensuremath{{A_{\lambda}}}}
\newc{\akap}{\ensuremath{{A_{\kappa}}}}
\newc{\hs}{\ensuremath{ H_s}}      
\newc{\mhs}{\ensuremath{ m_{H_s}}} 
\newc{\mgut}{\ensuremath{M_{\textrm{GUT}}}}
\newc{\gut}{\ensuremath{{\rm GUT}}}
\newc{\mplanck}{\ensuremath{ M_{\rm P}}}      \newc{\mpl}{\ensuremath{ M_{\rm Pl}}}
\newc{\msusy}{\ensuremath{ M_{\rm SUSY}}}      \newc{\ms}{\ensuremath{ M_{\rm S}}}
\newc{\mew}{\ensuremath{ M_{\rm EW}}}  
 \newc{\hu}{\ensuremath{ H_u}}       \newc{\hd}{\ensuremath{ H_d}}
 \newc{\mhu}{\ensuremath{ m_{H_u}}}       \newc{\mhd}{\ensuremath{ m_{H_d}}}
 \newc{\mhuew}{\ensuremath{ m^{\ast}_{H_u}}}       \newc{\mhdew}{\ensuremath{ m^{\ast}_{H_d}}}
 \newc{\mhuewsq}{\ensuremath{ m^{\ast\, 2}_{H_u}}}       \newc{\mhdewsq}{\ensuremath{ m^{\ast\, 2}_{H_d}}}
 \newc{\mhl}{\ensuremath{m_\hl}} 
 \newc{\mhone}{\ensuremath{m_{h_1}}} 
 \newc{\mhtwo}{\ensuremath{m_{h_2}}} 
 \newc{\mhi}{\ensuremath{m_{\tilde{h}}}} 
 \newc{\mul}{\ensuremath{m_{\tilde{u}_L}}} 
 \newc{\mtone}{\ensuremath{m_{\tilde{t}_1}}} 
 \newc{\ma}{\ensuremath{m_A}} 
 \newc{\mH}{\ensuremath{m_H}} 
 \newc{\maone}{\ensuremath{m_{a_1}}} 
 \newc{\matwo}{\ensuremath{m_{a_2}}}
 \newc{\hone}{\ensuremath{h_1}}
 \newc{\htwo}{\ensuremath{h_2}}
 \newc{\aone}{\ensuremath{a_1}}
 \newc{\atwo}{\ensuremath{a_2}}
 \newc{\mqthree}{\ensuremath{m_{\tilde{q}_3}^2}}
 \newc{\muthree}{\ensuremath{m_{\tilde{u}_3}^2}}
 \newc{\mql}{\ensuremath{m_{\tilde{q}}}}
 \newc{\mqlij}{\ensuremath{(m^2_{\tilde{q}})_{ij}}}
 \newc{\mur}{\ensuremath{m_{\tilde{u}}}}
 \newc{\mdr}{\ensuremath{m_{\tilde{d}}}}
 \newc{\murij}{\ensuremath{(m^2_{\tilde{u}})_{ij}}}
 \newc{\md}{\ensuremath{m_{\tilde{D}}}}
 \newc{\me}{\ensuremath{m_{\tilde{E}}}}
 \newc{\muu}{\ensuremath{m_{\tilde{U}}}}
 \newc{\mdrij}{\ensuremath{(m^2_{\tilde{d}})_{ij}}}
 \newc{\mll}{\ensuremath{m_{\tilde{l}}}}
 \newc{\mllij}{\ensuremath{(m^2_{\tilde{l}})_{ij}}}
 \newc{\mdlij}{\ensuremath{(m^2_{dl})_{ij}}}
 \newc{\mer}{\ensuremath{m_{\tilde{e}}}}
 \newc{\merij}{\ensuremath{(m^2_{\tilde{e}})_{ij}}}
 \newc{\ts}{\ensuremath{T_{SUSY}}}

\newc{\sigsip}{\ensuremath{\sigma^{\rm SI}_{p}}}	\newc{\sigsin}{\ensuremath{\sigma^{\rm SI}_{n}}}
\newc{\sigsdp}{\ensuremath{\sigma^{\rm SD}_{p}}}	\newc{\sigsdn}{\ensuremath{\sigma^{\rm SD}_{n}}}
\newc{\sigsi}{\ensuremath{\sigma^{\rm SI}}}	\newc{\sigsd}{\ensuremath{\sigma^{\rm SD}}}
\newc{\abund}{\ensuremath{ \Omega h^2}}
\newc{\omegadm}{\ensuremath{ \Omega_{{\rm DM}}}}     \newc{\abunddm}{\ensuremath{ \Omega_{{\rm DM}} h^2}} 
\newc{\omegam}{\ensuremath{ \Omega_{{\rm m}}}}       \newc{\abundm}{\ensuremath{ \Omega_{{\rm m}} h^2}}
\newc{\omegab}{\ensuremath{ \Omega_{{\rm b}}}}	\newc{\abundb}{\ensuremath{ \Omega_{{\rm b}} h^2}}
\newc{\omegatot}{\ensuremath{ \Omega_{{\rm TOT}}}}
\newc{\omegacdm}{\ensuremath{ \Omega_{{\rm CDM}}}}   \newc{\abundcdm}{\ensuremath{ \Omega_{{\rm CDM}} h^2}}
\newc{\omegalambda}{\ensuremath{ \Omega_{\Lambda}}} \newc{\abundlambda}{\ensuremath{ \Omega_{\Lambda} h^2}}
\newc{\omegarad}{\ensuremath{ \Omega_{{\rm rad}}}}  \newc{\abundrad}{\ensuremath{ \Omega_{{\rm rad}} h^2}}
\newc{\rhocrit}{\ensuremath{ \rho_{\rm crit}}}
\newc{\rhochi}{\ensuremath{ \rho_{\chi}}}
\newc{\abunchi}{\ensuremath{\Omega_\chi h^2}}
\newc{\abundlsp}{\ensuremath{\Omega_{\rm LSP}h^2}}
\newcommand*{\abundchi}{\ensuremath{\Omega_\chi h^2}}

\newc{\amu}{\ensuremath{ a_{\mu}}}        \newc{\amususy}{\ensuremath{ a_{\mu}^{\mathrm{SUSY}}}}
\newc{\amuexpt}{\ensuremath{ a_{\mu}^{\mathrm{expt}}}}        \newc{\amusm}{\ensuremath{ a_{\mu}^{\mathrm{SM}}}}
\newc\deltaamu{\ensuremath{\Delta a_{\mu}}} \newc{\deltaamususy}{\ensuremath{\delta a_{\mu}^{\mathrm{SUSY}}}}
\newc\gmtwo{\ensuremath{ (g-2)_{\mu}}} 
\newc{\deltagmtwomususy}{\ensuremath{\delta\left(g-2\right)_{\mu}^{\mathrm{SUSY}}}}
\newc{\deltagmtwomu}{\ensuremath{\delta\left(g-2\right)_{\mu}}}
\newc\bsgamma{\ensuremath{ b\rightarrow s \gamma }}
\newc\bxsgamma{\ensuremath{\overline{B}\rightarrow X_{s}\gamma}}
\newc\brbsgamma{\ensuremath{{\mathcal B}\left(\bsgamma\right)}}
\newc\brbxsgamma{\ensuremath{{\mathcal B}\left(\bxsgamma\right)}}
\newc\brmuegamma{\ensuremath{{\mathcal B}\left( \mu^+ \rightarrow e^+ \gamma  \right)}}
\newc\brmueee{\ensuremath{{\mathcal B}\left( \mu^+ \rightarrow e^+ e^+ e^-  \right)}}
\newc\brtaummm{\ensuremath{{\mathcal B}\left( \tau^+ \rightarrow \mu^+ \mu^+ \mu^-  \right)}}
\newc\brtmg{\ensuremath{{\mathcal B}\left( \tau^+ \rightarrow \mu^+ \gamma  \right)}}
\newc\bsmumu{\ensuremath{B_s\to\mu^+\mu^-}}
\newc\bdmumu{\ensuremath{B_d\to\mu^+\mu^-}}
\newc\brbsmumu{\ensuremath{{\mathcal B}\left(B_s\to\mu^+\mu^-\right)}}
\newc\brbdmumu{\ensuremath{{\mathcal B}\left(B_d\to\mu^+\mu^-\right)}}
\newc\bdmmumu{\ensuremath{\overline{B}_d\to\mu^+\mu^-}}
\newc\bbbarmix{\ensuremath{\overline{B}_s\mbox{-}B_s}}      
\newc\delmbs{\ensuremath{\Delta M_{B_s}}}
\newc\thc{\ensuremath{t\to h c}}
\newc\thu{\ensuremath{t\to h u}}
\newc{\butaunu}{\ensuremath{B_u \rightarrow \tau \nu}}
\newc{\brbutaunu}{\ensuremath{{\mathcal B}\left(B_u \rightarrow \tau \nu\right)}}

\newcommand*{\reffig}[1]{Fig.~\ref{#1}}
 
        \newcommand*{\refeq}[1]{Eq.~(\ref{#1})}
     \newcommand*{\refsec}[1]{Sec.~\ref{#1}}

\newcommand*{\neutone}{\ensuremath{\tilde{\chi}^0_1}}
\newcommand*{\neuttwo}{\ensuremath{\tilde{{\chi}}^0_2}}

\newcommand*{\charone}{\ensuremath{\tilde{{\chi}}^{\pm}_1}}

\newcommand*{\eight}{\ensuremath{\sqrt{s}=8\tev}}
\newcommand*{\four}{\ensuremath{\sqrt{s}=14\tev}}

\newcommand*{\dsusy}{DarkSUSY}
\newcommand*{\higgsbounds}{H{\scriptsize IGGS}B{\scriptsize OUNDS}}
\newcommand*{\higgssignals}{H{\scriptsize IGGS}S{\scriptsize IGNALS}}
\newcommand*{\feynhiggs}{\texttt{FeynHiggs}}
\newcommand*{\spheno}{\texttt{SPheno}}

\newcommand*{\multinest}{MultiNest}

\newcommand*{\susyflav}{\texttt{SUSY\textunderscore FLAVOR}}

\let\oldcite\cite
\renewcommand*{\cite}{~\oldcite}

\newcommand*{\hl}{\ensuremath{h}}

\def\sla#1{\not\!\!#1}

\newc{\Yb}{\ensuremath{Y_b}}
\newc{\Ys}{\ensuremath{Y_s}}
\newc{\Ym}{\ensuremath{Y_{\mu}}} 
\newc{\mtpole}{\ensuremath{m_t^{\rm pole}}}

\newcommand{\SidePlotsTwoMM}[5]
{
\begin{figure}[t]
\begin{center}
\includegraphics[width=0.45\textwidth]{#1}
\hspace{5mm}
\includegraphics[width=0.45\textwidth]{#2}
\end{center}
\caption[#5]{#3}
\end{figure}
}

\newcommand{\SidePlotsTwoMMx}[5]
{
\begin{figure}[t]
\centering
\subfloat[]{\includegraphics[width=0.5\textwidth]{#1}
}
\subfloat[]{\includegraphics[width=0.5\textwidth]{#2}
}
\caption[#5]{#3}
\end{figure}
}

\newcommand{\SidePlotsTwo}[5]
{
\begin{figure}[!ht]
\centering
\subfloat[]{\includegraphics[width=0.5\textwidth]{#1}
}
\subfloat[]{\includegraphics[width=0.5\textwidth]{#2}
}
\caption[#5]{#3}
\end{figure}
}

\newcommand{\SidePlotsTwoW}[4]
{
\begin{figure}[!ht]
\centering
\subfloat[]{
  \includegraphics[width=\textwidth]{#1}
}

\subfloat[]{
  \includegraphics[width=\textwidth]{#2}
  }
\caption[#4]{#3}
\end{figure}
}


%% file: chapters/0_Abstract.tex

In this dissertation, the Minimal Supersymmetric Standard Model (MSSM) is
    studied as a low-energy theory stemming from the $SU(5)$ Grand Unified
    Theory (GUT).  The well-known gauge coupling unification in the MSSM is
    considered to be one of the main virtues of the model. Similarly, Yukawa
    couplings of the bottom quark and tau lepton become relatively close to
    each other at the GUT scale \mgut~$\simeq 2\times 10^{16}\,$GeV. However,
    it is not the case for the first- and second-generation Yukawa couplings,
    unless large threshold corrections arise at the scale $\mu_{\rm sp}$ at
    which the superpartners are being decoupled.  Here, we investigate a
    possibility of satisfying the minimal $SU(5)$ boundary condition
    $\mathbf{Y}^d=\mathbf{Y}^{e\,T}$ for the full $3\!\times\!3$ down-quark
    and lepton Yukawa matrices at the GUT scale within the $R$-parity
    conserving MSSM.
    
    We give numerical evidence in favour
    of the statement: 
    
    \emph{There exist regions in the parameter space of the R-parity
    conserving MSSM for which the unification of the down-quark and lepton
    Yukawa matrices takes place, while the predicted values of flavour,
    electroweak and other collider observables are consistent with
    experimental constraints.}
    
    Furthermore, we find evidence that the bottom-tau and strange-muon Yukawa
    unification is possible with a stable MSSM vacuum in the standard form, where only 
    the neutral Higgs fields acquire non-vanishing vacuum expectation values.
     However, if the equality of the electron and down-quark Yukawa couplings at
    \mgut{} is demanded, only such cases remain, for which the standard MSSM vacuum is
    metastable, though sufficiently long-lived.

We investigate two separate scenarios of the soft supersymmetry breaking terms
at \mgut. In the first one, it is assumed that the soft terms are
non-universal but flavour-diagonal in the super-CKM basis.  In such a case,
the trilinear Higgs-squark-squark $A$-terms can generate large threshold
corrections to the Yukawa matrix $\mathbf{Y}^d$ at the superpartner decoupling
scale $\mu_{\rm sp}$, while no significant new contributions to the Flavour
Changing Neutral Current processes are generated. In effect, the $SU(5)$
boundary condition $\mathbf{Y}^d=\mathbf{Y}^{e\,T}$ at the GUT scale can be
satisfied. However, the large trilinear terms make the usual Higgs vacuum
metastable (though sufficiently long-lived). We broaden the previous studies
of such a scenario by including results from the first LHC phase, notably the
measurement of the Higgs particle mass, as well as a quantitative
investigation of the relevant flavour observables.

In the second scenario, we consider non-vanishing flavour off-diagonal entries
in the soft SUSY-breaking mass matrices. As we aim to alleviate the
metastability problem, the diagonal $A$-terms are assumed to be proportional
to the respective Yukawa couplings.  We show that a non-trivial flavour
structure of the soft SUSY-breaking sector can allow a precise bottom-tau and
strange-muon Yukawa coupling unification, while satisfying all
phenomenological constraints.

%% file: chapters/0_Streszczenie.tex
W niniejszej rozprawie rozwa\.zany jest Minimalny Supersymetryczny Model
Standardowy (MSSM) jako niskoenergetyczna teoria efektywna wynikaj\c{a}ca z
Teorii Wielkiej Unifikacji (GUT) opartej o symetri\c{e} $SU(5)$. Dobrze znana
unifikacja sprz\c{e}\.ze\'n cechowania w MSSM jest uwa\.zana za jedn\c{a} z
g{\l}\'ownych zalet tego modelu. Podobnie, tak\.ze sta{\l}e Yukawy kwarku
bottom i leptonu tau maj\c{a} zbli\.zone warto\'sci przy skali GUT,
\mgut~$\simeq 2\times 10^{16}\,$GeV. Nie ma to jednak miejsca w przypadku
pierwszej i drugiej generacji, o ile du\.ze poprawki progowe nie
wyst\c{a}pi\c{a} przy skali odprz\c{e}gania superpartner\'ow $\mu_{\rm sp}$. W
niniejszej pracy analizowana jest mo\.zliwo\'s\'c spe{\l}nienia w ten spos\'ob
minimalnych warunk\'ow brzegowych $SU(5)$ przy skali GUT
$\mathbf{Y}^d=\mathbf{Y}^{e\,T}$ dla pe{\l}nej ($3\!\times\!3$) macierzy
Yukawy kwark\'ow dolnych i lepton\'ow w ramach MSSM zachowuj\c{a}cego
R-parzysto\'s\'c.

W pracy zaprezentowane s\c{a} numeryczne argumenty przemawiaj\c{a}ce za
nast\c{e}puj\c{a}c\c{a} tez\c{a}: \emph{Istniej\c{a} obszary przestrzeni
parametr\'ow zachowuj\c{a}cego R-parzysto\'s\'c MSSM, w kt\'orych zachodzi
unifikacja macierzy Yukawy kwark\'ow dolnych i lepton\'ow, a przewidywane
przez model warto\'sci obserwabli zapachowych, elektros{\l}abych oraz innych
kluczowych pomiar\'ow dokonanych w zderzaczach cz\c{a}stek s\c{a} zgodne z
ograniczeniami do\'swiadczalnymi.}

Pokazujemy r\'ownie\.z, \.ze unifikacja sta{\l}ych Yukawy kwarku bottom i
leptonu tau oraz kwarku dziwnego i mionu jest mo\.zliwa bez naruszenia warunku
stabilno\'sci standardowej pr\'o\.zni MSSM, w kt\'orej tylko niena{\l}adowane
pola Higgsa przyjmuj\c{a} pr\'o\.zniowe warto\'sci oczekiwane. Je\'sli
\.z\c{a}da si\c{e} tak\.ze r\'owno\'sci sta{\l}ych Yukawy kwarku dolnego i
elektronu przy \mgut{}, pozostaj\c{a} tylko przypadki, w kt\'orych standardowa
pr\'o\.znia MSSM jest metastabilna, ale dostatecznie d{\l}ugo \.zyj\c{a}ca.

Rozwa\.zamy dwa odr\c{e}bne scenariusze dla cz{\l}on\'ow mi\c{e}kko
{\l}ami\c{a}cych supersymetri\c{e} przy skali \mgut. W pierwszym z nich
zak{\l}adamy, \.ze cz{\l}ony mi\c{e}kkie s\c{a} nieuniwersalne, ale diagonalne
w przestrzeni zapachu w bazie super-CKM. W takim wypadku, tr\'ojliniowe
sprz\c{e}\.zenia Higgs-skwark-skwark (cz{\l}ony $A$) mog\c{a} wygenerowa\'c
du\.ze poprawki progowe do macierzy Yukawy $\mathbf{Y}^d$ przy skali
odprz\c{e}gania superpartner\'ow $\mu_{\rm sp}$. Jednocze\'snie nie
pojawiaj\c{a} si\c{e} \.zadne nowe znacz\c{a}ce wk{\l}ady do 
proces\'ow z pr\c{a}dami neutralnymi zmieniaj\c{a}cymi zapach. W rezultacie,
warunek brzegowy $\mathbf{Y}^d=\mathbf{Y}^{e\,T}$ mo\.ze by\'c spe{\l}niony
przy skali GUT. Jednak\.ze, du\.ze cz{\l}ony tr\'ojliniowe czyni\c{a}
zwyczajn\c{a} pr\'o\.zni\c{e} Higgsa metastabiln\c{a} (cho\'c o czasie \.zycia
d{\l}u\.zszym ni\.z historia Wszech\'swiata). Niniejsza praca poszerza
uprzednie wyniki tego scenariusza poprzez uwzgl\c{e}dnienie wynik\'ow
z pierwszej fazy dzia{\l}ania LHC, zw{\l}aszcza pomiaru masy bozonu Higgsa, jak
te\.z ilo\'sciowo rozwa\.zaj\c{a}c istotne obserwable zapachowe.

W drugim scenariuszu rozwa\.zamy nieznikaj\c{a}ce pozadiagonalne w zapachu
elementy ma-cierzy mas mi\c{e}kko {\l}ami\c{a}cych supersymetri\c{e}. W celu
unikni\c{e}cia problemu metastabilno\'sci za{\l}o\.zone diagonalne cz{\l}ony
$A$ s\c{a} proporcjonalne do odpowiednich sprz\c{e}\.ze\'n Yukawy. Wykazujemy,
i\.z nietrywialna struktura zapachowa cz{\l}on\'ow mi\c{e}kko {\l}ami\c{a}cych
superymetri\c{e} pozwala na precyzyjn\c{a} unifikacj\c{e} sta{\l}ych Yukawy
kwarku bottom i leptonu tau oraz kwarku dziwnego i mionu, przy spe{\l}nieniu
wszystkich ogranicze\'n fenomenologicznych.

%% file: chapters/Thx.tex
First, this work would not have been undertaken, pursued and
completed without the supervision of Professor Miko{\l}aj Misiak. His intellect,
scientific passion, optimism, best intentions and kindness have often taken as
conversational input a sceptical, wavering PhD student and returned a
refreshed wanderer ready to focus on the next mountain to climb.

The second part of the work which has led to results that are
presented here owes much to Dr.\ Kamila Kowalska, her
proficiency, energy and experience in handling all the technical issues in
supersymmetric phenomenology. Her understanding what is
feasible, and when to stop researching a problem chiefly contributed to the
successful completion of this thesis. I am truly grateful for the
spirit-raising experience of our collaboration.

The TTP Institute of the Karlsruhe Institute of Technology will remain in my
memory as a perfectly organised and welcoming place, teeming with extremely
specialised, energetic research. Professor Ulrich Nierste and Dr.\ Andreas Crivellin
have largely inspired the research projects I have pursued. I am indebted as
well to the hospitality of CERN, by far the most motivating physics
environment I have been working in, thanks to the invitation and personal
kindness of Professor Christophe Grojean. The helpful remarks of Professor Gian
Giudice brought the vacuum metastability issue to my attention.

Everything I did owes much to the kind support of my Parents and the rest of
the family. It is their early formation that cradled my inquisitive curiosity
and ambition. Later, it was dr El\.zbieta Zawistowska, the exceptional teacher
and the spirit of the XIV Stanisław Staszic High School in Warsaw, that
inspired me to choose the field of physics. I would not have pursued in particle
physics if Professor Jan Kalinowski's lecture had not shown us the
logic in it.

I would like to thank all the people whose friendship, kindness and love
accompanied me in life, in its ups and downs. My sisters and brothers in arms
Ola, Micha{\l}, Radek, Marcin and especially Maciek -- our long evening
conversations at Nowolipki street were helpful way beyond Mathematica tips. Those who
shared our cramped nest at Ho\.za street, notably Ola, Wojtek and Lis. The
homemade-lunch lovers Bogusia, J\c{e}drek and Arek -- our discussions were an
important daily break. Rysiek for his uniqueness and our underground seminars.  The most
reliable friends not only from the high peaks, Klara, Jola, Janek, Karol and
Tomek. The whole Epifania Choir, especially our director Wiesław Jele\'n, who
left us the memory and testimony of his great open heart, patience and love to
music.

I would like to thank the kind people I met during my international exchanges. Otto for his lasting friendship. The whole Karlsruhe community, Ricarda, Guillaume, Jens, Jens, Nikolai. Remembering Yannan, who is not among us anymore. The happy CERN company, especially Christine, Robb, Juan, Mindaugas.

I owe very special gratitude to all those who were close to me during those
years, they profoundly changed me and made this time truly genuine, even if it
did not last. There would have been no final success if not the love and happiness
I enjoyed. Thank you.

%% file: chapters/0_Outline.tex
 
In the years when the research resulting in this thesis was performed,
a major breakthrough in particle physics took place. In
2012, the experiments ATLAS and CMS at the Large Hadron Collider
(LHC)\footnote{
Definitions of all the abbreviations used in this work are collected in
the List of Acronyms on page~\pageref{acronymsRef}.}
at CERN announced the discovery of a particle whose properties 
overlap with those of the Standard Model (SM) Higgs boson. Thus, all
the particles postulated by the SM have already been observed, and,
at the time of writing, all its parameters are known with an
arguably good precision. On the other hand, despite decades of
dedicated studies, no clear contradiction between the SM predictions and
experimental data has been observed, provided the SM is supplemented with
dimension-five operators that generate the neutrino masses. Also, the
Dark Matter (DM) in the Universe is not explained within the SM, which can
be resolved by adding an extra (super)weakly interacting particle.
Admittedly, there are a few earth-based measurements in which tensions
with the SM predictions occur, but they either seem to be debatable or at
least statistically allowed given the large number of observables considered.
At present, there is no principle that would clearly show a direction for
phenomenology beyond the SM. In this work, we make a step on a path that
follows a very traditional, popular, and still uncontested direction, namely
unification of all the known gauge interactions within what might be its
simplest realisation possible.

Supersymmetric Grand Unified Theories (SUSY GUTs) have been the topic of
numerous studies since the original formulation of the $SU(5)$
model\cite{Dimopoulos:1981zb}. A successful unification of the gauge couplings
at \mgut~$\simeq 2\times 10^{16}\,$GeV in the Minimal Supersymmetric Standard
Model (MSSM) is a phenomenological triumph of this programme. GUT symmetries
are decisively helpful, as they provide boundary conditions at the high energy
scale. In effect, the dimensionality of the huge MSSM
parameter space gets reduced.

A not yet completely solved quantitative issue of SUSY GUTs is a consistent
treatment of their Yukawa sector. In typical MSSM scenarios, an
equality of the down-quark and lepton Yukawa matrices at \mgut{} would
contradict the measured masses of the first- and second-generation
fermions. To reconcile those two constraints without introducing
non-minimal Higgs sectors above \mgut{}, one needs to consider large
threshold corrections either at \mgut{}, or at the
superpartner decoupling scale $\mu_{\rm sp}$. In this work,
we focus on adjusting the threshold corrections at the matching scale 
$\mu_{\rm sp}$ between the SM and the MSSM, while assuming that those
at \mgut{} are small. We analyse two MSSM scenarios, an
essential difference between which lies in the role of flavour.  The
first one assumes the soft supersymmetry breaking terms
(soft terms) to be flavour-diagonal in the super-CKM basis,
and can be nicknamed ``large diagonal $A$-term scenario''.
The second one makes use of the soft-term flavour-off-diagonal entries,
which implies considering the General Flavour Violating (GFV) MSSM. 
We first analyzed the large $A$-term scenario, and the obtained
results served as a motivation for our second attempt to resolve the 
Yukawa matrix unification problem.

Our analysis of the first scenario updates that of Ref.\cite{Enkhbat:2009jt}
with a broader range of $\tan\beta$ (reaching 40) and a universal scalar mass
$m_0$. We have taken into account new experimental data, and performed a
quantitative study of flavour observables. Results from the LHC experiments at
7 and 8 TeV have constrained the superpartner masses both via direct searches
and by fixing the lightest Higgs boson mass. This calls for an up-to-date
analysis of the Yukawa unification. We confirm that the $SU(5)$ Yukawa
coupling unification of all the three families is phenomenologically viable,
and attainable for a wide range of \tanb. However, its price in this case is
that the standard MSSM vacuum (in which only the neutral Higgs fields acquire
non-vanishing Vacuum Expectation Values (VEVs)) becomes a metastable
one. Fortunately, it remains sufficiently long-lived, i.e. its estimated
lifetime is significantly longer than the present age of the Universe.  The
outcome of this research was published in Ref.\cite{Iskrzynski:2014zla}.

The second framework assumes that the diagonal entries of the trilinear terms
have the same hierarchy as the Yukawa couplings. However, non-zero
off-diagonal entries in the sfermion mass matrices and trilinear
terms are allowed.

First, we restrict ourselves to the cases of bottom-tau and strange-muon
Yukawa couplings. In the strange-muon case, the necessary threshold
corrections are the largest and most troublesome. We begin with introducing
flavour violation into the GUT-scale soft terms solely via the
$(m^2_{\tilde{d}})_{23}$ and $(m^2_{\tilde{l}})_{23}$ elements of
the down-type squark and lepton-doublet soft mass matrices,
respectively. The role of $(m^2_{\tilde{d}})_{23}$ is to
``transmit'' the large threshold correction that affects the bottom quark
Yukawa coupling to the strange quark one. We call this scenario
\emph{$GFV_{23}$}, as only the second and third generations are involved.

  We perform a full phenomenological analysis of this scenario,
taking into account such observables as the mass and decay rates of the
lightest Higgs boson, electroweak (EW) precision tests, flavour
observables, relic density of the neutralino DM, limits on the
spin-independent proton-neutralino scattering cross-section, as well as the
8\tev\ LHC exclusion bounds from the direct SUSY searches. We find that the
$SU(5)$ Yukawa unification condition for the third and second generations is
consistent at $3\sigma$ with all the considered experimental constraints, and
it does \emph{not} imply metastability of the standard MSSM vacuum, contrary to 
the large diagonal $A$-term scenario.

The phenomenological features of the $GFV_{23}$ scenario make it an
alternative to models that assume Minimal Flavour Violation (MFV).  A crucial
role in the successful Yukawa coupling unification in the GFV case is
played by the chirality-preserving mixing term $(m^2_{\tilde
d})_{23}$ between the second and third generations of down-type
squarks. This particular mixing is less constrained by the Flavour Changing
Neutral Current (FCNC) processes, as compared to other off-diagonal entries in
the soft terms.  Moreover, low values of \tanb\ and heaviness of the squarks
make supersymmetric contributions to the flavour observables relatively small
in the considered scenario.


In the GFV MSSM, one may also attempt to unify the electron
and down-quark Yukawa couplings. However, it requires relatively large values
of the flavour-violating soft terms that involve the first generation. 
%
%
We analyse this framework as the so-called \emph{$GFV_{123}$}
scenario. Investigating the Lepton Flavour Violating (LFV) observables,
we find that for the points with a full $SU(5)$ Yukawa matrix
unification, the predicted values of \brmuegamma{} and
\brmueee{} exceed the experimental 90\% \cl\ upper limits.

Our analysis of the GFV scenarios has been published\footnote{
See the current (v2) arXiv version of the article where both the $GFV_{23}$
and $GFV_{123}$ scenarios are described. The journal version is restricted to
the $GFV_{123}$ case, while the LFV constraint, indicated near Eq.~(3.2) there, is
discussed further in the ``note added in the proof''.}
in Ref.\cite{Iskrzynski:2014sba}, prepared in collaboration with dr
K.~Kowalska.  She was the main contributor to the derivation of constraints
from dark matter, as well as from the direct SUSY searches at the LHC.

In our discussion of the $SU(5)$ unification of the Yukawa couplings, we
restrict ourselves solely to the minimal case in which one demands
equality of the lepton and down-quark Yukawa matrices at the GUT scale. We
treat the up-quark Yukawa matrix as unrelated, and use a basis in which the SM
quark flavour mixing originates from this matrix only. Moreover, we assume
that the neutrino masses and mixings originate from decoupling of superheavy
right-handed neutrinos whose Majorana mass matrix has an arbitrary flavour
structure. Thus, no constraints from the neutrino experiments affect our
analysis.

The thesis is organised as follows. In Chapter ~\ref{TheoreticalOverviewChap},
the theoretical background and reasoning introducing the problem of Yukawa
unification in the MSSM is described. In Chapter~\ref{TwoScenarios}, the two
scenarios for the soft terms are defined, and an illustrative analysis of
supersymmetric threshold corrections to the Yukawa couplings is
performed. Chapter~\ref{PhenoIntro} introduces the phenomenological
constraints imposed on the model. In Chapter~\ref{ToolsChap}, numerical
tools employed in the analysis are described. Chapter~\ref{SoftDiagChap}
scrutinises the results obtained in the first scenario -- with large diagonal 
$A$-terms. Chapter~\ref{OffDiagChap} shows the outcome of our
investigation of the $GFV_{23}$ scenario, namely the successful bottom-tau and
strange-muon Yukawa coupling unification. Then,
Chapter~\ref{GFV123Num} describes how the $GFV_{123}$ scenario is capable of
the full $SU(5)$ Yukawa matrix unification, but comes short of satisfying the 
LFV constraints. We conclude in Chapter~\ref{concl}.

%% file: chapters/1_0_intro.tex
The problems addressed in this work can be described by a perturbative 
Quantum Field Theory (QFT). Throughout 
the whole introduction, we aim at recalling the definitions 
that are necessary for a qualitative understanding of the presented
results, rather than at introducing quantitative
techniques of the perturbative QFT from the outset. These are
covered in an extensive collection of textbooks, see
e.g. Refs.\cite{QFT1,QFT2,QFT3,QFT4}.

%% file: chapters/1_1_Preliminaries.tex
Before introducing the models we deal with in this work, let us briefly review
a few of the basic concepts that matter for our purpose. In
particular, one should mention the idea of \emph{effective theories} 
that is crucial for the phenomenology of fundamental interactions. It
provides a language for describing the same phenomena in terms of several QFTs
defined at different scales, and rules for correlating their parameters.

We shall deal with the simplest version of an effective QFT that originates
from heavy particle decoupling. In the framework of perturbative
QFT, it is based on the following observation.  If a model
contains particles with masses heavier than the center-of-mass energy of a
given process, all the diagrams influencing the process can contain
those heavy fields only as intermediate states. They can never appear at the
external legs. Therefore, in an effective theory to which the considered
model can be reduced, those heavy fields are not explicitly present as
degrees of freedom. They only modify interactions between the remaining
particles.

For example, the SM can be viewed as an effective field theory derived from a
model containing additional particles. This, in turn, can also be an effective
theory that has a more fundamental source. In this way, our description of the
fundamental interactions can consist of layers of theories, unravelling when
higher energies are reached. In this dissertation, we will consider the
SM as originating from the MSSM at the scale $\mu_{\rm sp}$, further
described by a supersymmetric GUT, valid above \mgut.

\begin{subsection}{Renormalisation Group Equations}

Values of the renormalised constants of a given effective QFT are
determined at specified energy scales. Each scale corresponds to the
energy scale of the process which is used for the considered parameter
determination. Once the parameters are fixed, we often need to use them in
processes that take place at very different energy scales. This requires
modifying the renormalisation scheme for the parameters to avoid introducing
spurious perturbative uncertainties. Throughout the present thesis, unless explicitly
specified, we shall assume that our parameters are renormalised in the
modified Minimal Subtraction ($\overline{MS}$) scheme at various renormalisation
scales $Q$. Values of the renormalised (running) couplings $c_i(Q)$
at any scale $Q$ can be determined by solving the Renormalisation Group
Equations (RGEs) of a form
\begin{equation}
 Q\frac{d}{dQ}\begin{pmatrix}
              c_1 (Q) \\ c_2(Q) \\ ...
             \end{pmatrix}
             =f(c_1(Q), c_2(Q), ...).
\end{equation}
The function $f$ can be calculated order-by-order in perturbation
theory from the $\overline{MS}$ renormalisation constants of the considered
QFT. The RGEs are systems of ordinary differential equations. As they might
be coupled and non-linear, they are often unsolvable analytically. 
In such cases, testing a high-energy theory with the help of low-energy
observables requires scanning the high-energy theory parameter space, and
numerically solving the RGEs point-by-point in this space.

\end{subsection}
\begin{subsection}{Threshold corrections}
  \begin{figure}[t]\begin{center}
    \includegraphics[width=0.7\textwidth]{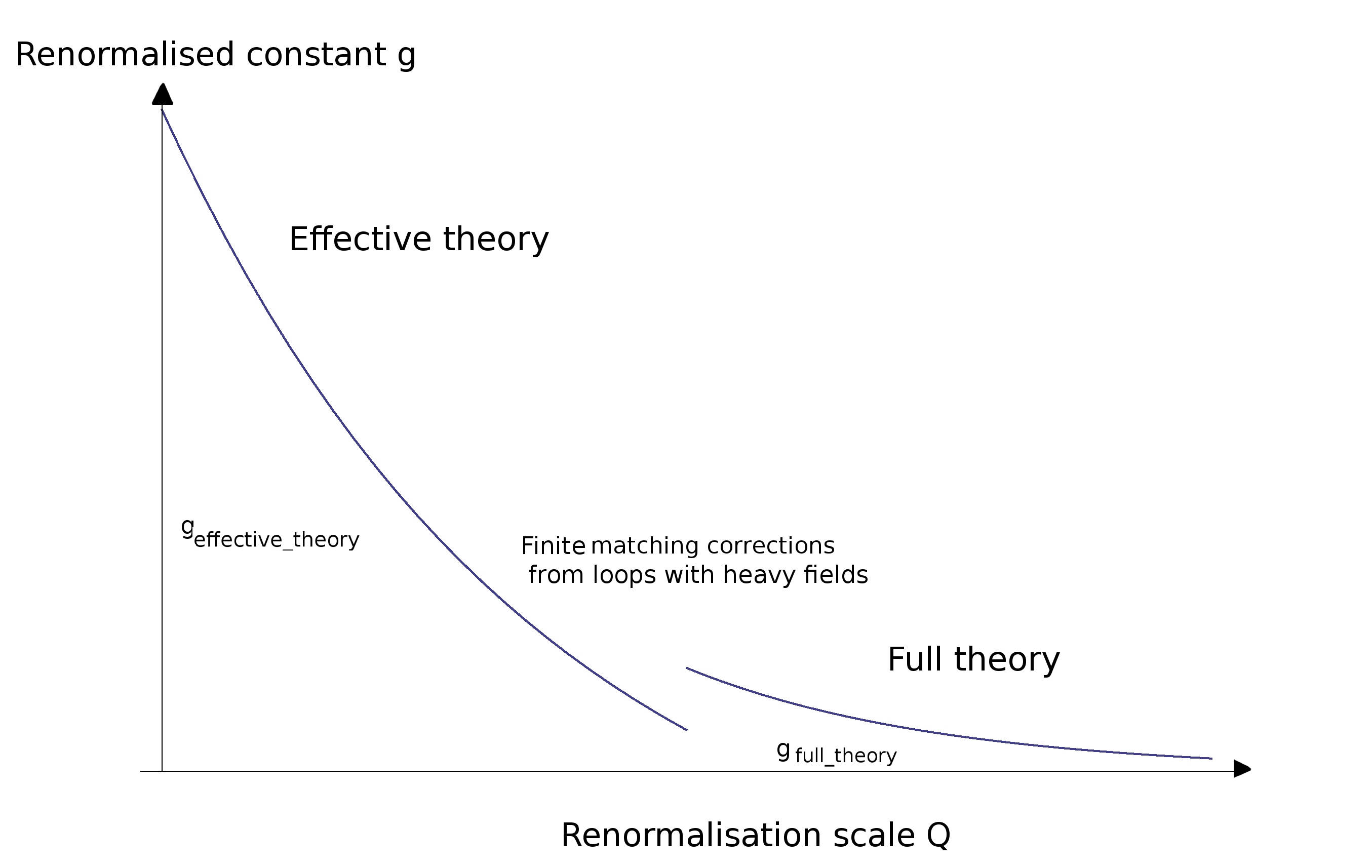}
    \caption[Explanation of threshold corrections.]{An illustration of the
    matching for a parameter $g$ between the full and effective theories.
    The discontinuity is given by the threshold correction (see the text).\label{TCexpl}}
  \end{center}\end{figure}

Having renormalised a given effective QFT, and having solved its RGEs,
we know the values of its renormalisation-scale-dependent couplings at
all scales. However, when one works with several layers of
effective theories, a new question arises: how to translate parameters of one
theory to the parameters of another? In other words, what are the 
relations between renormalised couplings of two different effective
theories that describe the same physical system?  We illustrate this in
Fig.~\ref{TCexpl}, where $g$ stands for a particular parameter that is
present both theories (e.g., the QCD gauge coupling in the MSSM and SM).
In the plot, the two theories are nicknamed ``full theory'' and ``effective
theory'', but they can actually both be effective. In each of them, the
coupling $g$ depends on the renormalisation scale $Q$. The dependence is
different, because the spectra of the two theories are different.  We want to
use the full (effective) theory above (below) a certain scale that we shall
call $\mu_{\rm sp}$, to maintain the analogy with the case of passing from the
MSSM to the SM. The scale $\mu_{\rm sp}$ is called the matching scale. 
The relation between the couplings at this scale takes the form
\begin{equation}
 g_{\text{effective theory}}(\mu_{\rm sp})=g_{\text{full theory}} (\mu_{\rm sp}) + \Delta g (\mu_{\rm sp}),
\end{equation}
where $\Delta g (\mu_{\rm sp})$ is called the \emph{threshold
correction}. It is responsible for the discontinuity in the curve
presented in Fig.~\ref{TCexpl}.  Its presence means that the ``same''
renormalised coupling $g$ means different things in both models. The
consistency requirement from which the actual values of threshold
corrections are derived is that both theories must give the same predictions
for physical processes involving only light particles (i.e. the low-energy
theory degrees of freedom). Instead of matching the physical amplitudes, it
is often more convenient to match several light-particle Green's functions
with arbitrary off-shell momenta. If, in addition, an expansion in these
momenta is applied prior to the loop momentum integration, evaluation of the
threshold corrections becomes particularly simple. In such a case, they are
given by finite parts of the loop diagrams involving heavy fields on the full
theory side, as indicated in Fig.~\ref{TCexpl}. 

\end{subsection}

%% file: chapters/1_2_StandardModel.tex
In this section we introduce the foundation of our present understanding of
particle physics - the Standard Model. It is an experimentally unchallenged
starting point of any high energy speculations.

\subsection{Field content and Lagrangian density \label{subsec:SMfields}}

The Standard Model (SM) is a renormalisable gauge theory, based on three Lie
groups $U(1)$, $SU(2)$ and $SU(3)$. Starting from the successful
unified theory of electroweak interactions by Sheldon Glashow\cite{Glashow},
it was conceived independently by Steven Weinberg and Abdus
Salam\cite{SM}. They applied the mechanism worked out in Refs.\cite{EB,H1,H2,GHK}, 
further referred to as the \emph{Higgs mechanism}, to resolve the
electroweak theory problems with unitarity and weak boson masses. The SM
contains the following fields.\footnote{
The indices $i=1,2$ and $a=1,2,3$ -- weak isospin and
colour -- correspond to the fundamental representations of $SU(2)$ and
$SU(3)$, respectively, $I=1,2,3$ and $A=1,\ldots,8$ -- to
their adjoint representations, whereas $p=1,2,3$ enumerates the
fermion generations called flavours.}

\ \\[-1mm]
\setlength{\parskip}{1ex plus 0.5ex minus 0.2ex}
Gauge fields:
\begin{itemize}
 \item gluons:~~~~~ $G_{\mu}^{A}$, \par
 \item W bosons:~   $W_{\mu}^{I}$, \par
 \item B boson:~~~  $B_{\mu}$, 
\end{itemize}
Matter fermion fields: 
\begin{itemize}
 \item left-handed leptons: $l_i^p =\begin{pmatrix}
 \nu^{p} \\ 
 e_L^{p}
\end{pmatrix}_i$~~, \par
 \item right-handed leptons: $e_R^p$~~,\par
 \item left-handed quarks: $q_i^{ap} =\begin{pmatrix}
 u_L^{ap} \\ 
 d_L^{ap}
\end{pmatrix}_i$~~,  \par
 \item right-handed quarks: $u_R^{a p}, d_R^{a p}$,
\end{itemize}
Higgs scalar fields: 
\begin{itemize}
 \item The Higgs doublet: $\ph_i$.
\end{itemize}
Using
$$\eps=\left( \begin{array}{cc} 0& 1\\-1 & 0 \end{array}\right)$$
we also define $\p \tilde{\ph}=\eps\ph^{*}$, which implies that $\tilde{\ph}^{\dag}=-\ph^{T}\eps $. 
\begin{table}
\begin{center}
\begin{tabular}{| c | c | c | c |}
\hline
\multirow{2}{*}{field} 
& \multicolumn{2}{c|}{representation}  & hypercharge \\\cline{2-4}
& $SU(3)$ & $SU(2)$ & $U(1)_{Y}$ \\
\hline
$G_{\mu}$  & $\mathbf{8}$ & 1 & 0 \\[0.6ex]
$W_{\mu}$ & 1 & $\mathbf{3}$ & 0 \\[0.6ex]
$B_{\mu}$ & 1 & 1 & 0 \\[0.6ex]
$q$ & $\mathbf{3}$ & $\mathbf{2}$ & $\frac{1}{6}$ \\[0.6ex]
$u_R$ & $\mathbf{3}$ & 1 & $\frac{2}{3}$ \\[0.6ex]
$d_R$ & $\mathbf{3}$ & 1 & $-\frac{1}{3} $\\[0.6ex]
$l$ & 1 & $\mathbf{2}$ & $-\frac{1}{2}$ \\[0.6ex]
$e_R$ & 1 & 1 & $-1$ \\[0.6ex]
$\ph$ & 1 & $\mathbf{2}$ & $\frac{1}{2}$ \\[0.6ex]
\hline 
\end{tabular}
\end{center}
\caption{Field content of the Standard Model. \label{ms}}
\end{table}

To concisely define the SM Lagrangian density in terms of its fields, we
need to introduce a few more objects. A covariant derivative
of the left-handed quark field has the following form:
\be
D_{\mu} q  =
\left( \partial_{\mu} +ig_{s}G^{A}_{\mu} T^{A} +ig W^{I}_{\mu} S^{I}+ ig' Y_q \right) q,
\ee
or, with explicit isospin and colour indices
\be
\left( D_{\mu} q \right)^a_j  =
\left( \partial_{\mu} + ig' Y_q \right) q^a_j  
+ ig_{s}G^{A}_{\mu} T^{A}_{ab} q^b_j + ig W^{I}_{\mu} S^{I}_{jk} q^a_k. 
\ee
Here, $T^A$ and $S^I$ are the fundamental representation
generators of, respectively, the SU(2) and SU(3) groups,
while $Y_q = \frac16$ denotes the hypercharge of the field
$q$. Covariant derivatives of other matter, gauge and Higgs fields are
defined in an analogous manner, following the field assignment to various
representations of $SU(2)$ and $SU(3)$, as given in Tab.~\ref{ms}.  Generators
of singlet representations are equal to zero.

The field-strength tensors of the SM gauge fields are given by
\begin{itemize}
\item $SU(3)$
\be
G_{\mu\nu}^{A}=\partial_{\mu}G_{\nu}^{A}-\partial_{\nu}G_{\mu}^{A}-g_{s}f^{ABC}G_{\mu}^{B}G_{\nu}^{C},
\ee 
\item $SU(2)$
\be
W^{I}_{\mu\nu}=\partial_{\mu}W_{\nu}^{I}-\partial_{\nu}W_{\mu}^{I}-g\eps^{IJK}W_{\mu}^{J}W_{\nu}^{K},
\ee
\item $U(1)_{Y}$
\be
B_{\mu\nu}=\partial_{\mu}B_{\nu}-\partial_{\nu}B_{\mu}.
\ee
\end{itemize}

In the following, we will suppress the isospin, colour and flavour indices, 
as contractions of them will always be straightforward, e.g.,
$$ (\bar{q}\,\yuku^\star u_R)\tilde{\ph} ~~\equiv~~ 
\bar{q}_{i}^{a p}\, \yuku^\star_{\!\!\!\!\! pp'}\, u_R^{a p'}\, \tilde{\ph}_i. $$

With the above-listed objects at hand, the Lagrangian density of the Standard
Model can be written as follows:\footnote{
We take the Yukawa coupling matrices complex conjugated here to make the notation
    consistent with the standard MSSM expression~\eqref{MSSMsup} for the superpotential.}
\be \label{L4}
\begin{split}
\mathcal{L}_{\rm SM}& =-\tfrac{1}{4}G_{\mu\nu}^{A}G^{A\mu\nu}-\tfrac{1}{4}W_{\mu\nu}^{I}W^{I\mu\nu}
-\tfrac{1}{4}B_{\mu\nu}B^{\mu\nu}\\
&  + (D_{\mu}\varphi)^{\dag}(D^{\mu}\varphi)+m^{2} \varphi^{\dag}\varphi-\tfrac{1}{2}\lambda(\varphi^{\dag}\varphi)^{2}\\
& +i \left[ \bar{l}\sla{D}l +\bar{e}_R\sla{D}e_R +\bar{q}\sla{D}q +\bar{u}_R\sla{D}u_R +\bar{d}_R\sla{D}d_R \right]\\
& -\left[(\bar{l}\yuke^\star e_R)\varphi+ 
         (\bar{q}\yuku^\star u_R)\tilde{\ph}+ 
         (\bar{q}\yukd^\star d_R)\varphi + {\rm h.c.}\right].\\
\end{split}\end{equation}

To describe the observed neutrino masses and mixings without introducing
new degrees of freedom, the SM needs to be thought about as an effective
theory, the so-called Standard Model Effective Field Theory (SMEFT). Its
Lagrangian density has the following general form:
\be \label{Lsmeft}
\mathcal{L}_{\rm SMEFT} ~=~ \mathcal{L}_{\rm SM} ~+~ \frac{1}{\Lambda } \sum_{k} C_k^{(5)} Q_k^{(5)}
~+~ \frac{1}{\Lambda^2} \sum_{k} C_k^{(6)} Q_k^{(6)} ~+~ \ldots~,
\ee
where $Q_k^{(n)}$ stand for dimension-$n$ operators built out of the SM
fields, $C_k^{(n)}$ are the corresponding coupling constants (Wilson
coefficients), and $\Lambda$ is a certain scale that defines the validity
region of our effective theory. All the phenomenological predictions are given
as series in powers of $1/\Lambda$, with masses and kinematical invariants
standing in the numerators. Thus, the quantities in the numerators must be
much smaller than $\Lambda$.

Imposing the SM gauge symmetry on $Q_n^{(5)}$ leaves out just a single
operator\cite{Weinberg:1979sa}, up to a Hermitian conjugation and flavour
assignments. It reads
\be \label{qnunu}
Q_{\nu\nu}^{pr} = \eps_{jk} \eps_{mn} \ph_j \ph_m (l^p_k)^T C l^r_n 
~\equiv~ (\tilde{\ph}^\dag l^p)^T C (\tilde{\ph}^\dag l^r),
\ee
where $C$ is the charge conjugation matrix.  After the Electroweak Symmetry
Breaking (EWSB), such operators generate the neutrino masses and mixings. Since the
neutrino masses are very small, it is conceivable that the scale $\Lambda$
standing in front of $Q_{\nu\nu}^{pr}$ in Eq.~\eqref{Lsmeft} is very large,
many orders of magnitude above the electroweak scale. If this is the case, the
neutrinos can be treated as massless for our purpose. We are going to assume
this is the case, and ignore the presence of $Q_{\nu\nu}^{pr}$ in what
follows.

As far as the Dark Matter is concerned, we shall not introduce any candidate
for it in the SM but rather assume that it will come with the MSSM degrees of
freedom, above the $\mu_{\rm sp}$ matching scale.

\subsection{Spontaneous Electroweak Symmetry Breaking and fermion masses}

In the Standard Model with unbroken gauge symmetry, all the gauge
bosons are massless. On the other hand, we know that in Nature,
to the contrary, they are among the heaviest known elementary particles.
Giving masses to the gauge bosons was the purpose for introducing the Higgs
field into the SM Lagrangian, and then invoking the Higgs
mechanism. It is a spontaneous breaking of the $SU(2)\times U(1)_Y$ symmetry
to the residual electromagnetic gauge symmetry $U(1)_{em}$. The Higgs
field $\ph$ (being a complex-valued $SU(2)$ doublet) acquires a vacuum
expectation value (VEV). Without loss of generality (by using the gauge symmetries), 
one can write the VEV as
\be
\left \langle 0 | \ph |  0 \right \rangle = \frac{1}{\sqrt{2}} \begin{pmatrix}
 0 \\ 
 v
\end{pmatrix}.
\ee
Substituting the above VEV into $\mathcal{L}_{\rm SM}$, one finds the tree-level
masses of the $W$ and $Z$ bosons, namely $M_W = \frac12 vg$ and $M_Z = \frac12
v\sqrt{g^2 + {g'}^2}$. The measured values of these masses and gauge couplings
imply that $v \approx 246$ GeV.

In fact, not only do the $W$ and $Z$ bosons acquire masses from their interactions
with the Higgs doublet $\ph$. Also the fermion mass terms originate from their
Yukawa interactions with $\ph$:
\begin{equation}
\mathcal{L}_{\rm f.\,mass} ~=~ -\frac{v}{\sqrt{2}} \left[
  \bar{u}_L \yuku^\star u_R + 
  \bar{d}_L \yukd^\star d_R + 
  \bar{e}_L \yuke^\star e_R + \textrm{h.c.}\right].
\end{equation}

The physical Higgs field $h^0$ is a real scalar field that
parameterises the departure of $\ph$ from the VEV. In the so-called
unitary gauge, we have
\begin{equation}
 \ph=\frac{1}{\sqrt{2}} \begin{pmatrix}
 0 \\ 
 v + h^0
\end{pmatrix}.
\end{equation}

The fermion mass sector is the main focus of the analysis performed in this
work. The basis in which we have defined the SM in Eq.~\eqref{L4} is
known as the \emph{weak-eigenstate basis} or the
\emph{interaction basis}. The Yukawa couplings, being matrices in flavour
space, are not diagonal in that basis. Their singular value
decomposition leads to the \emph{fermion mass-eigenstate basis}
$\left\{ \hat{u}_L, \hat{u}_R, \hat{d}_L, \hat{d}_R, \hat{e}_L,
\hat{e}_R \right\}$, in which the mass matrices $\mathbf{m}^u,
\mathbf{m}^d, \mathbf{m}^e$ are real and diagonal:
\begin{align}
 \bar{u}_L \yuku^\star u_R \vsm + \textrm{h.c.}~ 
&=~ \overline{\hat{u}}_L (V_{uL}^\dagger \yuku^\star V_{uR})\hat{u}_R \vsm + \textrm{h.c.}~ 
=~ \bar{\hat{u}}_L \mathbf{m}^u \hat{u}_R ~+~ \textrm{h.c.}~, \\
\bar{d}_L \yukd^\star d_R \tfrac{v}{\sqrt{2}} + \textrm{h.c.}~ 
&=~ \bar{\hat{d}}_L (V_{dL}^\dagger \yukd^\star V_{dR}) \hat{d}_R \vsm + \textrm{h.c.}~\,
=~ \bar{\hat{d}}_L \mathbf{m}^d \hat{d}_R ~+~ \textrm{h.c.}~, \\
\bar{e}_L \yuke^\star e_R \tfrac{v}{\sqrt{2}} + \textrm{h.c.}~ 
&=~ \bar{\hat{e}}_L (V_{eL}^\dagger \yuke^\star V_{eR}) \hat{e}_R \vsm + \textrm{h.c.}~~ 
=~ \bar{\hat{e}}_L \mathbf{m}^e \hat{e}_R ~+~ \textrm{h.c.}~.
\end{align}
Passing to the mass-eigenstate basis undergoes via the following
unitary redefinitions of the fields in the flavour space:
\begin{equation} \label{flav.transf}
u_L = V_{uL} \hat{u}_L\,, \hspace*{15pt}
u_R = V_{uR} \hat{u}_R\,, \hspace*{15pt}
d_L = V_{dL} \hat{d}_L\,, \hspace*{15pt}
d_R = V_{dR} \hat{d}_R\,, \hspace*{15pt}
e_L = V_{eL} \hat{e}_L\,, \hspace*{15pt}
e_R = V_{eR} \hat{e}_R\,. 
\end{equation}
Such a transformation is neither a symmetry of $\mathcal{L}_{\rm
SM}$, nor it can be compensated by a redefinition of its parameters. This
fact is a consequence of that different flavour transformations are applied
separately to particular components of the $SU(2)$ doublets. However, apart
from the Yukawa terms, the only other part of the Lagrangian density
that gets affected by the considered basis change are the
interactions of quarks with charged $W$ bosons
\begin{equation}
 W^{\pm}=\tfrac{1}{\sqrt{2}}(W^1 \mp i W^2).
\end{equation}
Rewriting them in the mass-eigenstate basis, we get 
\begin{equation}
 \mathcal{L}_{\rm SM} \ni -g  \bar{u}_L \gamma^{\mu} d_L W_{\mu}^+ + \textrm{h.c.} = 
-g  \overline{\hat{u}}_L (V_{uL}^\dagger  V_{dL}) \gamma^{\mu} \hat{d}_L   W_{\mu}^+ + \textrm{h.c.}~.
\end{equation}
Here,
\begin{equation}
 (V_{uL}^\dagger  V_{dL}) \equiv V_{CKM}
\end{equation}
stands for the Cabibbo-Kobayashi-Maskawa\cite{Cabibbo,KobayashiMaskawa} matrix
(CKM matrix). This matrix is the only source of flavour- and CP-violation in the 
SM, once the fields are expressed in the mass-eigenstate basis.

It is important to stress that the flavour- and CP-violation are
basis-dependent concepts. Before we pass to the mass-eigenstate basis, we are
not even able to define the discrete CP symmetry that is being tested in
experiments.\footnote{
The transformation $\psi \to C \bar\psi^T$ applied to each fermionic field
does not commute with the linear transformations~\eqref{flav.transf} that involve
\emph{complex} unitary matrices.}
Flavour in the interaction basis is violated by the Yukawa couplings but
conserved by the gauge interactions. On the other hand, in the mass-eigenstate
basis, the Yukawa interactions with $h^0$ conserve flavour, but the gauge interactions do
not.

Although all the SM constituents correspond to distinct experimentally 
established particles, its underlying structure remains complicated.  
We have three independent gauge groups and three generations, each
involving five different matter representations. Thus, there seems to remain
some space for a further simplification of our description of fundamental interactions.

%% file: chapters/1_3_GrandUnifiedTheories.tex
In this section, the grand unification idea is motivated, and the $SU(5)$ GUT,
the simplest model realising this idea, is introduced. Gauge coupling unification is
discussed as the historical quantitative argument for the GUTs. In the last
part, we observe how the hierarchy problem arises in the considered framework.
 
 \begin{subsection}{Algebraical considerations}

 The involved algebraical structure of the SM symmetry group \smgg{} and its
 representations have motivated efforts to find a simpler gauge
 symmetry group. It would get spontaneously broken to the SM group, by 
 analogy to the breaking of the SM group to $SU(3)\times U(1)_{em}$.
 This idea becomes even more natural at present, once the SM Higgs mechanism 
 has already been experimentally confirmed by the discovery of
 the Higgs boson at the LHC.
 
 The smallest simple Lie group that can embed the SM group is $SU(5)$.\footnote{
 Both are rank-4 groups.} 
 It was also the first one proposed for the role of a
 GUT group\cite{GeorgiGlashow}. The lowest-dimensional irreducible
 representations of $SU(5)$ are two complex 5-dimensional ones
 ($\mathbf{5}$, $\mathbf{\bar{5}}$), and two complex 10-dimensional ones
 ($\mathbf{10}$, $\mathbf{\overline{10}}$).  The crucial issue that
 governs the assignments of matter fields to their $SU(5)$ 
 representations are the branching rules that describe how a $SU(5)$
 representation is decomposed into a direct sum of the
 $SU(3)\times SU(2)\times U(1)$ representations. Below, we specify the
 representations by their dimensions in the cases of $SU(3)$ and
 $SU(2)$, and by the corresponding $U(1)$ generator
 eigenvalue.
%
 
 To discuss the representation embedding, it is convenient to deal with
     fermions of fixed chirality. By convention, one chooses the left-handed ones.
     The fermionic $SU(2)$ doublets of the SM are taken as they stand, while
     the SU(2) singlets are described in terms of charge-conjugates of the original
     right-handed fields. Thus, we deal with a set of five left-handed fermionic 
     representations of the SM group, namely $q$, $u_R^{c}$, $d_R^{c}$, 
     $l$, $e_R^{c}$. It appears that those fields can be embedded 
     into the $\boldmath{\bar{5}}$ and $\boldmath{10}$ representations of 
     $SU(5)$ as follows:
\begin{align} 
 \underbrace{(\mathbf{\bar 3},\mathbf{1},\tfrac{1}{3})}_{d_R^{c}} \oplus \underbrace{(\mathbf{1},\mathbf{2}, 
  -\tfrac{1}{2})}_l &= \underbrace{\mathbf{\bar 5}}_{\Psi_{\bar 5}}, \notag \\ 
\underbrace{(\mathbf{3},\mathbf{2},\tfrac{1}{6})}_q \oplus \underbrace{(\mathbf{\bar 3},
   \mathbf{1},-\tfrac{2}{3})}_{u_R^{c}} \oplus \underbrace{(\mathbf{1},\mathbf{1},1)}_{e_R^{c}} &= 
\underbrace{\mathbf{10}}_{\Psi_{10}}, \label{embeddings}
\end{align}
where we have used the conventional SM normalisation for the
hypercharges $Y$. The proper GUT normalisation $Y_1=\sqrt{\tfrac{3}{5}}Y$ is
needed to ensure a uniform normalisation condition for all the $SU(5)$
generators: ${\rm Tr}(a_{SU(5)}^i a_{SU(5)}^j)=\jd \delta^{ij}$. It is 
achieved by rescaling the coupling constant $g'$. The choice of
$\mathbf{\bar 5}$ and $\mathbf{10}$ is necessary to fit the SM
hypercharges. Remarkably, it also guarantees that the considered GUT gauge
symmetry is anomaly-free.

In fact, it is easy to understand why the embedding of down-quarks and leptons
into a single $\mathbf{\bar 5}$ representation means that their respective
hypercharges must follow the proportion of $-2\!:\!3$.  The reason lies in the
dimensions of the corresponding representations of $SU(3)$ and $SU(2)$ -- the
$SU(5)$ generator corresponding to $U(1)_Y$ must be traceless. A similar
argument holds in the case of $\mathbf{10}$.  In this way, the $SU(5)$ model
solves the puzzle of charge quantization in the SM, explaining why they are
commensurable at all, and it does so without resort to anomalies.
   
\end{subsection}
  
\begin{subsection}{Gauge interactions}

   In addition to the matter fields $\Psi_{\bar 5}$ and $\Psi_{10}$, the $SU(5)$
GUT includes the gauge fields $\Sigma$, belonging to the adjoint
24-dimensional representation. Under the action of the \smgg{}
subgroup, the gauge bosons decompose as:
\begin{equation}
 \underbrace{(\mathbf{8},\mathbf{1},0)}_{G}
+\underbrace{(\mathbf{1},\mathbf{3},0)}_{W}
+\underbrace{(\mathbf{1},\mathbf{1},0)}_{B}
+\underbrace{(\mathbf{3},\mathbf{2},-\tfrac{5}{6})}_{X}
+\underbrace{(\mathbf{\bar{3}},\mathbf{2},\tfrac{5}{6})}_{X^*}
~=~\underbrace{\mathbf{24}}_{\Sigma}
   \end{equation}
   The gauge multiplet contains all the SM gauge fields -- gluons $G$,
   electroweak bosons $W$ and $B$, as well as additional gauge fields $X$ that
   carry both the $SU(2)$ and $SU(3)$ indices. The latter fields, if present
   in Nature, would mediate the yet-unobserved proton decay.  The current
   lower bounds on the proton lifetime\cite{Agashe:2014kda} imply that the masses of
   the $X$ bosons must be very large (often called \emph{superheavy}). These
   masses are by construction of the same order as \mgut, i.e. the
   scale at which the GUT gauge symmetry is broken to the SM one. Thus,
   experimental bounds on the proton lifetime provide important constraints on
   \mgut{} (see below).

   The scalar sector has to serve a twofold purpose. The SM Higgs mechanism of
   weak-scale mass generation is realised by a field belonging to the
   fundamental representation $H_{\mathbf{5}}$. Other scalar fields,
   e.g., $\Phi_{\mathbf{24}}$ are responsible for breaking the
   $SU(5)$ symmetry to \smgg. Particular scalar sectors of the
   non-supersymmetric GUTs have been discussed in detail in
   Ref.\cite{Buras:1977yy}.

   In the following, we shall review two features of the model: the gauge
   coupling unification and the \emph{hierarchy problem}. To formulate them,
   let us write explicitly the kinetic terms for the matter and
   $H_{\mathbf{5}}$ Higgs fields. They read
   \begin{equation}\label{lagrgaugesu5}
   \mathcal{L}_{SU(5)}\ni i\overline{\Psi}_{\bar 5}\sla{D}\Psi_{\bar 5}
   +i\overline{\Psi}_{10}\sla{D}\Psi_{10} + (D_\mu H_5)^\dagger (D^\mu H_5),
   \end{equation}
with the covariant derivatives acting in the following way
  \begin{equation}\label{gaugeIntsu5}
   \begin{split}
    (D_{\mu} H_5 )^a & = \partial_{\mu} H_5^a +i g_5 (\Sigma_{\mu})^a_{~b} H_5^b, \\ 
    (D_{\mu}\Psi_{\bar 5})^a &= \partial_{\mu}\Psi_{\bar 5}^a 
    +i g_5 (-\Sigma_{\mu}^{\star})^a_{~b}\Psi_{\bar 5}^b, \\ 
    (D_{\mu}\Psi_{10})^{ab} &=\partial_{\mu}(\Psi_{10})^{ab}
    + ig_5 (\Sigma_{\mu})^a_{~c}(\Psi_{10})^{cb}
    + ig_5 (\Sigma_{\mu})^{b}_{~d} (\Psi_{10})^{ad}.
   \end{split}
  \end{equation}
   \end{subsection}

  \begin{subsection}{Gauge coupling unification}

  The simplicity of the $SU(5)$ GUT, apparent in Eq.~\eqref{gaugeIntsu5}, relies on
  existence of only a single gauge coupling $g_5$. The strong and electroweak
  interactions are unified at the scale $\mgut \gg M_Z$. However, they must attain
  their significantly hierarchical coupling strengths at the low scales. At
  the time the model was conceived, the quantitative unification of gauge couplings 
  based on the SM RGEs worked quite well within experimental errors of the 1970's. 
  All the three properly normalised gauge couplings appeared to meet at a scale 
  $\mgut^{1970s} \sim 10^{14\div 15}\,$GeV. The current knowledge is described further in the text.
  
  The big hierarchy between the scales $M_Z$ and \mgut{} leaves a natural question of what
  happens at the intermediate scales. An assumption that no new particles
  exist at intermediate scales might be motivated by the successful gauge
  coupling unification. The unification could be spoiled by possible new physics 
  that does not fit into the GUT framework.

  \end{subsection}
  
  \begin{subsection}{Hierarchy problem}

  \begin{figure}
   \begin{center}
    \includegraphics[width=0.9 \textwidth]{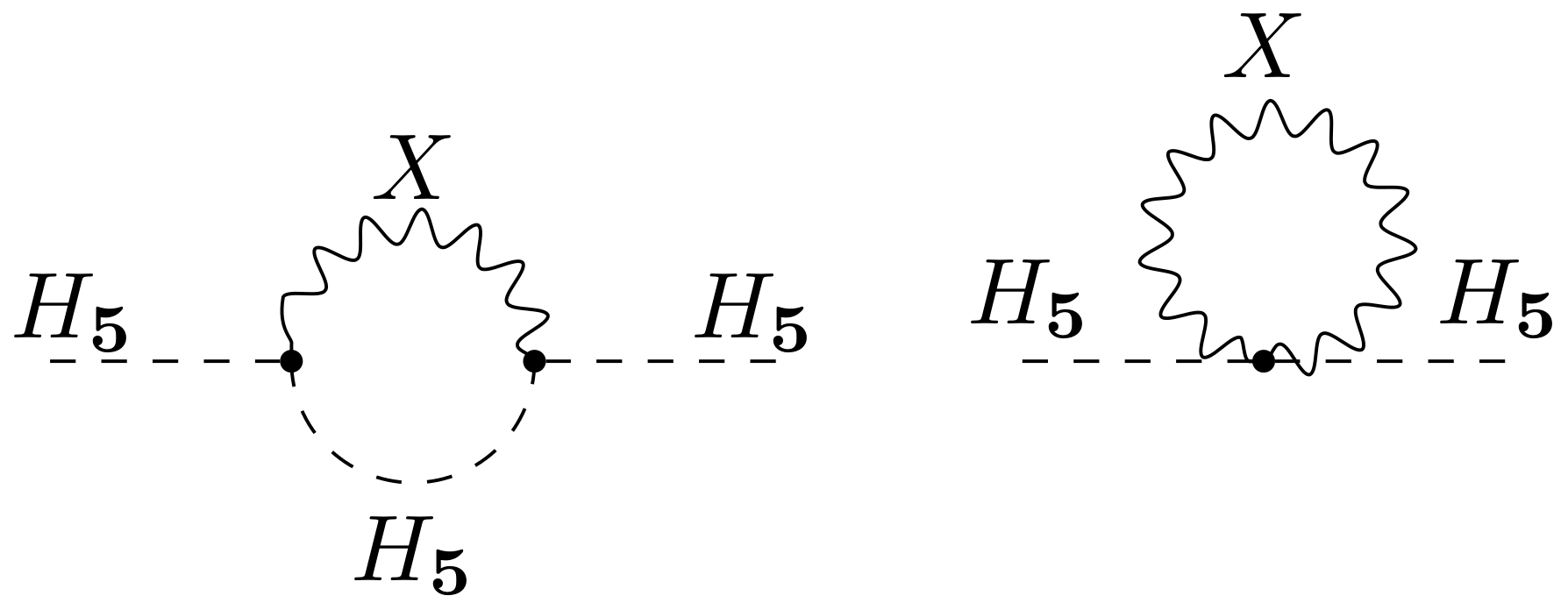}
    \caption[Diagrams illustrating the hierarchy problem in the non-SUSY $SU(5)$ GUT.]{
    One-loop self-energy diagrams for the Higgs multiplet $H_{\mathbf{5}}$ 
    that involve the gauge bosons $X$. They renormalise the scalar mass
    squared by contributions the order of $\mgut^2$.
\label{higgsXdiags}}
   \end{center}
  \end{figure}
  
  One of the features of Eqs.~\eqref{lagrgaugesu5} 
  and~\eqref{gaugeIntsu5} is an interaction between the SM Higgs 
  doublet $\ph$ (a part of the $H_5$ multiplet) and the
  superheavy gauge bosons $X$. The corresponding one-loop
  radiative corrections to the Higgs propagator are given by the
  diagrams in Fig.~\ref{higgsXdiags}. They generate threshold
  corrections of the order of $\mgut^{2}$ in the GUT-SM matching
  condition for the Higgs mass parameter $m$ in the SM
  Lagrangian~\eqref{L4}:
  \begin{equation}\label{HierarchyEq}
   m^2 = m^2_{\rm GUT} + {\mathcal O}(\mgut^2),
  \end{equation}
  where $m^2_{\rm GUT}$ is the corresponding mass parameter in the GUT Lagrangian.
  The value of $m$ of the order of ${\mathcal O}(100\,{\rm GeV})$ could only be a result of an
  ``accidental'' cancellation of contributions that are greater by many orders
  of magnitude. Such a fine-tuning problem occurs in all the ``particle desert''
  scenarios that assume the SM at the weak scale, a GUT at the high scale, and
  nothing in between. It is often referred to as the \emph{hierarchy problem}.
 
 There are two main phenomenological routes of addressing the hierarchy
 problem, where the ``particle desert'' assumption is relaxed above
 a certain scale $\mu_0$. They rely either on compositeness of the Higgs
 boson or on supersymmetry.\footnote{
 In the case of supersymmetry, $\mu_0$ is given by the superpartner decoupling scale $\mu_{\rm sp}$.}
 Both of them lead to the Higgs mass parameter being naturally of the
 order of ${\mathcal O}\left(\mu_0^2\right)$ instead of ${\mathcal
 O}(\mgut^2)$. The first strategy treats the Higgs doublet as a bound
 state analogous to QCD mesons with a confinement scale $\mu_0 \sim
 1\,$TeV. Supersymmetric extensions of the SM are introduced in the next
 section. Interestingly, the physical Higgs mass is an
 observable where both scenarios yield different predictions. Supersymmetry
 generically predicts a smaller mass, whereas the composite Higgs
 models -- a bigger one. The LHC measurement has posed a
 challenge for both frameworks, yet not excluding definitely any of them. 
 A vast literature on this subject is reviewed, e.g., in
 Refs.\cite{Arbey:2015exa, Barducci:2014oha, Djouadi:2013lra}.
 
\end{subsection}

%% file: chapters/1_4_Supersymmetry.tex
In this section, supersymmetry (SUSY) is introduced as a framework in which both the
hierarchy problem can be solved, and a precise gauge coupling
unification can be achieved. First, it is defined as an algebraical
concept. Next, the absence of quadratic radiative corrections to the
Higgs boson mass is discussed. Subsequently, the Minimal Supersymmetric
Standard Model (MSSM) is described, including the gauge coupling
unification issue. At this point, the stage for considering the
Yukawa coupling unification is set. For a detailed introduction
to supersymmetry, see, e.g., Refs.\cite{WestSUSY,MartinPrimer}.

\begin{subsection}{The SUSY algebra}

As we have already noted, the SM has significant shortcomings if treated as a direct
foundation for Grand Unified Theories. One of the possible directions for
extending the model is to ask whether new principles of symmetry
could be introduced simultaneously with extending the particle spectrum.

The complete symmetry of the SM as a relativistic QFT is given
by its gauge group \smgg, and the Poincar\'e group ${\mathcal
P}$ of the spacetime symmetries. The latter is generated by
$P_{\mu}$ and $M_{\mu\nu}$ (with $\mu, \nu=0,1,2,3$) obeying the
following commutation rules:
\begin{equation}\label{PoincareAlg}
\begin{split}
 [P_{\mu}, P_{\nu}]&=0,  \\
 [M_{\mu\nu}, P_{\rho}]&=-i(\eta_{\mu\rho}P_\nu - \eta_{\nu\rho}P_{\mu}), \\ 
 [M_{\mu\nu}, M_{\rho\xi}]&=i (
   \eta_{\nu\rho} M_{\mu\xi}
 - \eta_{\nu\xi}  M_{\mu\rho}
 - \eta_{\mu\rho} M_{\nu\xi}
 + \eta_{\mu\xi}  M_{\nu\rho}).
 \end{split}
\end{equation}
Coleman and Mandula showed in Ref.\cite{ColemanMandula} that for
any local relativistic QFT in four dimensions, generators
$T_s$ of any Lie group that describes an internal symmetry must
commute with the Poincar\'e algebra ones
\begin{equation}\label{PoincareGauge}
 [P_{\mu}, T_s]=0=[M_{\mu\nu}, T_s].
\end{equation}
This is equivalent to saying that fields of different spin must
transform independently under internal symmetries that are
described by Lie groups. It seemed therefore, that the only way of changing
the symmetries governing the model was to introduce different gauge 
groups, or different global symmetries, either continuous or
discrete ones.

However, it turns out that there exist possible symmetries which are
not described by Lie groups, and which nontrivially extend the
Poincar\'e transformations. It was shown a few years later by
Golfand and Likhtman\cite{GolfandLikhtman}, as well as by Wess and
Zumino\cite{WessZumino}. The mathematical framework governing such
symmetries is based on Lie superalgebras, which differ from Lie algebras by
being defined through both commutation and anticommutation relations
among their elements.

As first derived in Ref.\cite{HaagLopuszanskiSohnius}, a supersymmetric
algebra contains the above-mentioned Poincar\'e generators $P_{\mu}$ and
$M_{\mu\nu}$, together with a set of anticommuting supersymmetry
generators in the $(\jd,0)$ and $(0,\jd)$ representations of the Lorentz
group. The latter are denoted by $Q_{\alpha}$ and $\bar{Q}_{\dot{\beta}}$ with
$\alpha, \dot{\beta}=1,2$ enumerating their spinorial indices. They are
Hermitian conjugates of each other. In general, there could be more
generations of $Q$'s, but their number, called $N$, is restricted to 1 in 
most of the phenomenological applications.

The algebra of $N=1$ supersymmetry has the following structure in addition
to~\refeq{PoincareAlg}
\begin{equation}\label{QPcom} 
 \begin{split}
 \{ Q_{\alpha}, \bar{Q}_{\dot{\beta}} \} &= 2 (\sigma^{\mu})_{\alpha\dot{\beta}}P_{\mu}, \\
 \{ Q_{\alpha}, Q_{\beta} \} &= 0 = \{ \bar{Q}_{\dot{\alpha}}, \bar{Q}_{\dot{\beta}} \}, \\
 [Q_{\alpha}, P_{\mu}] &= 0 = [\bar{Q}_{\dot{\alpha}}, P_{\mu}],   \\ 
 [Q_{\alpha}, M_{\mu\nu}] &= -\frac12 (\sigma_{\mu\nu})_{\alpha}^{~\,\beta}\;Q_{\beta}, \\
 [\bar{Q}_{\dot{\alpha}}, M_{\mu\nu}] &= \frac12 
 (\bar{\sigma}_{\mu\nu})_{~\,\dot{\alpha}}^{\dot{\beta}}\;\bar{Q}_{\dot{\beta}}. 
 \end{split}
\end{equation}
The latter two commutation rules above are consistent with the fact
that $Q$ and $\bar{Q}$ transform respectively as the $(\jd,0)$ and
$(0,\jd)$ Weyl spinors under the spacetime transformations.  In
the $N=1$ case, the SUSY generators commute with other internal symmetry
generators, including the gauge group ones.

\end{subsection}

\begin{subsection}{Softening the hierarchy problem}

  \begin{figure}[t]
   \begin{center}
    \includegraphics[width=0.7 \textwidth]{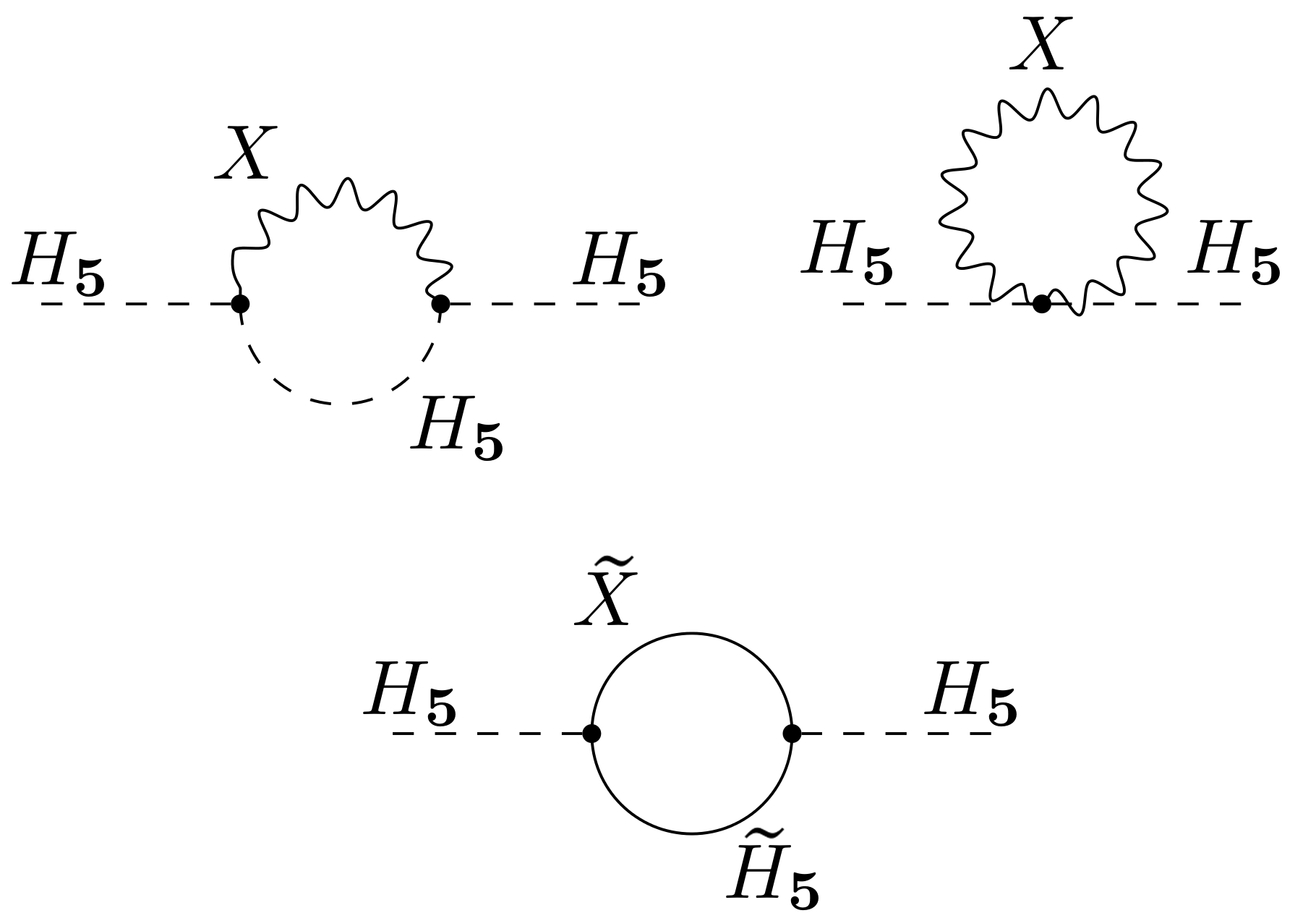}
    \caption[Diagrams illustrating the SUSY solution to the hierarchy problem.]{
    One-loop self-energy diagrams for the Higgs multiplet
    $H_{\mathbf{5}}$ that involve the gauge bosons $X$ 
    and their superpartners -- the gauginos $\tilde{X}$.\label{higgsXsusydiags}}
   \end{center}
  \end{figure}

  Supersymmetry imposes relations between fermionic and
  bosonic states. The action of $Q$ or $\bar Q$ changes the spin of a
  state by $\pm\jd$, while preserving its energy and momentum,
  as follows from~\refeq{QPcom}. Physical fields of different
  spin that are related by supersymmetry transformations are
  called \emph{superpartners}. The same coupling constants parameterise
  interactions of both superpartners with other fields, which allows for
  cancellations among otherwise unrelated Feynman diagrams. This fact 
  turns out to give a solution to the hierarchy problem, which
  is among the main phenomenological arguments in favour of SUSY. In a
  supersymmetric GUT, radiative corrections to the Higgs boson
  mass coming from loops with superheavy gauge bosons $X$ from
  Fig.~\ref{higgsXdiags} are accompanied by the ones involving their
  superpartners, as illustrated in Fig.~\ref{higgsXsusydiags}.  
  Contributions of the order of ${\mathcal O}(\mgut^2)$ from these three
  diagrams to the Higgs boson mass squared cancel in an exact manner thanks to
  supersymmetric relations between the gauge boson and gaugino couplings, and to 
  equality of their masses in the unbroken SUSY limit.

  Similar cancellations occur also in higher-loop diagrams, which follows from
  the so-called non-renormalisation theorem\cite{nonreno} that holds for any
  SUSY model. The theorem states that certain terms in the
  Lagrangian\footnote{
specifically, the terms that originate from differentiating the superpotential - see below.}
  do not get renormalised, i.e., all the relevant Feynman diagrams
  cancel out.  In our context it means that loop corrections can affect only
  the gauge-kinetic terms in the Lagrangian, but not the mass terms or Yukawa
  interactions. An immediate consequence is the absence of quadratic
  ultraviolet divergences because the scalar mass terms are the only ones for
  which quadratic divergences would be allowed by power counting.\footnote{
We are not considering quantum corrections to the vacuum energy here, and we
assume absence pure singlets, i.e. fields being in singlet representations of all the symmetries.}

  The non-renormalisation theorem implies that in a SUSY GUT, contrary to
  Eq.~\eqref{HierarchyEq}, the Higgs mass squared does not receive any
  threshold corrections scaling as $\mgut^2$. This statement
  holds to all orders in perturbation theory.
  
\end{subsection}
  
\begin{subsection}{The Minimal Supersymmetric Standard Model}

  As we have seen, supersymmetry possesses certain phenomenological
  advantages. Let us now describe a theory that
  acts as a supersymmetric extension of the SM. It
  is known as the Minimal Supersymmetric Standard Model (MSSM). 
\begin{table}[t]
\begin{center}
\begin{tabular}{|c|c|c|}
\hline 
Superfields & Left-handed & Complex \\ 
            & fermions           &     scalars \\ \hline && \\[-3mm]
    $Q=  \begin{pmatrix}  U_L\\   D_L \end{pmatrix} $     &  $q=\begin{pmatrix}  u_L \\
 d_L     \end{pmatrix}$    &    $ \tilde{q}=\begin{pmatrix}  \tilde{u}_L \\ 
 \tilde{d}_L     \end{pmatrix}$    \\[3.2ex]
      $U_R^{c}$  &  $u_R^{c}$ & $\tilde{u}_R^{c}$ \\[1.2ex]
      $D_R^{c}$  &  $d_R^{c}$ & $\tilde{d}_R^{c}$ \\[1.2ex]
      $L=  \begin{pmatrix}  N \\   
                            E_L \end{pmatrix} $      &   
      $l=  \begin{pmatrix}  \nu \\   e_L \end{pmatrix} $        &     
      $\tilde{l}=  \begin{pmatrix}  \tilde{\nu} \\   
                                    \tilde{e}_L \end{pmatrix} $     \\[3.2ex]
   $E_R^{c}$ &  $e_R^{c}$ &  $\tilde{e}_R^{c}$   \\[1.2ex]
   $H_d~ = \begin{pmatrix}  H_d^0 \\ H_d^-     \end{pmatrix} $  &  
                      $ \tilde{h}_d= \begin{pmatrix}  \tilde{h}_d^0 \\  
                                                       \tilde{h}_d^-     \end{pmatrix}$     &          
                       $  h_d       = \begin{pmatrix}  h_d^0 \\  
                                                       h_d^-     \end{pmatrix}$ \\[3.2ex]
    $  H_u ~ = \begin{pmatrix}  H_u^+ \\ H_u^0     \end{pmatrix} $  &  
                    $ \tilde{h}_u= \begin{pmatrix}  \tilde{h}_u^+ \\  
                                                    \tilde{h}_u^0     \end{pmatrix}$     &          
                    $ h_u= \begin{pmatrix}  h_u^+ \\  h_u^0     \end{pmatrix}$    \\[2ex] 
    \hline
\end{tabular}
\end{center}
\caption[MSSM matter fields.]{The MSSM matter field content. 
For brevity, the colour and flavour indices have been omitted, as
they are identical to those in Sec.~\ref{subsec:SMfields}. To introduce the necessary
notation, the isospin structure has been made explicit. \label{MSSMcontent}}
\end{table}

In the MSSM, we find each of the SM fields from
Tab.~\ref{ms} embedded in a \emph{superfield} that contains 
also another physical field -- its scalar or fermionic
partner.\footnote{
We restrict to describing the MSSM explicitly in terms of the
  physical component fields.}
In this way, one extends the field content to get partners of
quarks, leptons and Higgs scalars, called \emph{squarks}, \emph{sleptons}
and \emph{higgsinos}, respectively. The SM vector gauge fields
have their corresponding spin-$\jd$ partners, called
\emph{gauginos}:
\begin{itemize}
 \item gluinos:~~ $\tilde{g}_{\mu}^{A}$ \hspace{4ex} ($A=1,\ldots,8$), \par
 \item winos:~~~~ $\tilde{W}_{\mu}^{I}$ \hspace{3ex} ($I=1,\ldots,3$), \par
 \item bino:~~~~~ $\tilde{B}_{\mu}$. 
\end{itemize}
To write a symbol for each superpartner, a tilde is added
over the corresponding SM symbol. A summary of the MSSM
matter field content is given in Tab.~\ref{MSSMcontent}. The
squarks and sleptons will commonly be called \emph{sfermions}. Let us note the
appearance of two distinct Higgs doublets $H_d$ and $H_u$. It is necessary due
to holomorphy constraints on the SUSY interactions, which forces us to replace
$\ph$ and $\tilde{\ph}$ in the Yukawa terms by two independent
fields. Moreover, making the gauge symmetries free of chiral anomalies
requires having two higgsino fields with opposite hypercharges. The $H_d$ and
$H_u$ hypercharges are equal to $-\jd$ and $\jd$, respectively. 

A generic formula for the supersymmetric part of the MSSM Lagrangian
density is a sum of kinetic terms $\mathcal{L}_{kin}$, interactions
$\mathcal{L}_f$ involving fermions, and a scalar potential term $(-V_{\phi})$
  \begin{equation}
   \mathcal{L}_{\rm MSSM}^{\rm SUSY}=\mathcal{L}_{kin} + \mathcal{L}_f - V_{\phi}.
  \end{equation}
The kinetic terms are given by
  \begin{equation}
   \mathcal{L}_{kin} ~=~ \sum_k (D_{\mu}\phi_k)^\dag (D^{\mu}\phi_k) ~+~
                     i\sum_{j}\bar{\psi}_j\sla{D}\psi_j ~+~ \mathcal{L}_{kin}^{gauge},
  \end{equation}
where the indices $i$ and $j$ enumerate all the MSSM scalars and
left-handed fermions (including the gauginos),
respectively. The last term stands for the gauge boson kinetic terms
which are identical to those in the SM
\begin{equation}
 \mathcal{L}_{kin}^{gauge}=-\tfrac{1}{4}G_{\mu\nu}^{A}G^{A\mu\nu}
-\tfrac{1}{4}W_{\mu\nu}^{I}W^{I\mu\nu}-\tfrac{1}{4}B_{\mu\nu}B^{\mu\nu}.
\end{equation}

For specifying the remaining parts of the Lagrangian, it is
convenient to introduce a functional $W$ called the
\emph{superpotential}. It defines the non-gauge interactions between
the MSSM superfields as
\begin{equation}\label{MSSMsup}
  W_{\rm MSSM}=  Q \mathbf{Y^u} U_R^{c} H_u 
                 +  Q \mathbf{Y^d} D_R^{c} H_d 
                 +  L \mathbf{Y^e} E_R^{c} H_d + \mu H_d H_u,
\end{equation}
where $\mu$ is a dimensionful parameter. For all practical purposes
within the MSSM phenomenology, it is enough to view the superpotential
just as a step in an algorithm of defining the Lagrangian
density. The $\mathcal{L}_f$ terms have the following form:
\begin{equation}
 \mathcal{L}_f=
-\jd \sum_{i,j} \psi^T_i C \psi_j 
\left( \frac{\partial^2 W}{\partial \Phi_i \partial \Phi_j}\right)_{\Phi \rightarrow \phi} 
 - \sqrt{2} \sum_{i,G,a} g_G \phi_i^{\dag} T^a_{G} (\lambda_G^a)^T C \psi_{i} 
~~+~~ \text{h.c.}\;,
\end{equation}
where $\Phi \rightarrow \phi$ means replacing the superfield by its scalar
component. The indices $i,j$ run over all the (scalar, fermion)
multiplets.  In the second term, $\lambda_G^a$ denotes a gaugino
corresponding to the group $G$, with $a$ being its adjoint representation index.

The scalar potential $V_{\phi}$ can be split into parts usually called
D- and F-terms,
\begin{equation}
V_{\phi} ~=~ \sum_i |F_i|^2 + \jd \sum_{G,a} g^2_{G} D^a_{G} D^a_{G}
\end{equation}
with
\begin{equation}
 F_i=\left(\frac{\partial W}{\partial \Phi_i}\right)_{\Phi \rightarrow \phi} 
\hspace{30pt} \mbox{and} \hspace{30pt} 
D^a_{G} = \sum_{i}(\phi_i^{\dag} T^a_{G} \phi_i).
\end{equation}

If $\mathcal{L}_{\rm MSSM}^{\rm SUSY}$  defined as above was the
complete Lagrangian density, the superpartners would have the same
mass as the corresponding SM fields. This would contradict the
experimental fact that no superpartner has been observed so far. One needs to
include additional mass and interaction terms that explicitly break SUSY but,
at the same time, do not violate the key property of non-renormalisation of
the superpotential terms (and thus cancellation of the ultraviolet quadratic
divergences). It can be done by adding the following bi- and
trilinear terms:
\begin{equation}
\begin{split}
 \mathcal{L}^{\rm soft}_{\rm MSSM}&= 
-\jd [m_{\tilde{g}} (\tilde{G}^a)^T C\tilde{G}^a 
    + m_{\tilde{W}} (\tilde{W}^{I})^T C \tilde{W}^I 
    + m_{\tilde{B}}\tilde{B}^T C \tilde{B} +\text{\emph{h.c.}}] 
- m_{h_d}^2 h_d^{\dag} h_d - m_{h_u}^2 h_u^{\dag} h_u \\
 & - \tilde{q}^{\dag} (\mathbf{\mathbf{M}}_{\tilde{q}}^2) \tilde{q} 
   - (\tilde{u}_R^{c})^{\dag} (\mathbf{M}_{\tilde{u}}^2) (\tilde{u}_R^{c}) 
   - (\tilde{d}_R^{c})^{\dag} (\mathbf{M}_{\tilde{d}}^2) (\tilde{d}_R^{c}) 
   - \tilde{l}^{\dag} (\mathbf{M}_{\tilde{l}}^2) \tilde{l} 
   - (\tilde{e}_R^{c})^{\dag} (\mathbf{M}_{\tilde{e}}^2) (\tilde{e}_R^{c})\\
 & + \tilde{q} \mathbf{A^u} \tilde{u}_R^{c} h_u +  \tilde{q} \mathbf{A^d} \tilde{d}_R^{c} h_d  
 +  \tilde{l} \mathbf{A^e} \tilde{e}_R^{c} h_d + B\mu h_d h_u ~~+~~ {\rm h.c.}\;,
\end{split} \label{Lsoft}
\end{equation}
where the sfermion masses $\mathbf{M}^{2}_{\tilde{f}}$ and 
the trilinear Higgs-sfermion-sfermion couplings $\mathbf{A}^f$ are matrices in
the flavour space. Thus, the complete Lagrangian density reads
\be \label{LMSSMfull}
\mathcal{L}_{\rm MSSM} = \mathcal{L}_{\rm MSSM}^{\rm SUSY} + \mathcal{L}^{\rm soft}_{\rm MSSM}.
\ee

The SUSY-breaking terms introduced in $\mathcal{L}^{\rm soft}_{\rm MSSM}$
are all proportional to parameters of positive dimension in the units of
mass. Thus, by power counting, they cannot introduce quadratic ultraviolet
divergences. For this reason, the considered SUSY breaking is called
\emph{soft}.  The soft-SUSY-breaking terms in $\mathcal{L}^{\rm soft}_{\rm
MSSM}$ are called \emph{soft terms}, and the corresponding parameters
-- \emph{soft parameters}.

All the soft parameters are assumed to be of the order of $\mu_{\rm sp}
    \sim {\mathcal O}(1\,{\rm TeV})$ which is much lower than \mgut.  Thus,
    they do not reintroduce the hierarchy problem in the (SUSY GUT)-MSSM
    matching conditions despite breaking SUSY explicitly. One may ask whether
    keeping the soft terms much smaller than \mgut{} is natural in any sense.
    In 1979, Gerardus 't~Hooft defined the concept of \emph{naturalness}, used
    as a way of selecting theories among those equally plausible from the
    experimental point of view. According to him, a theory can be called
    natural, if ``at an energy scale $Q$, a physical parameter or a set of
    physical parameters $c_i(Q)$ is allowed to be very small only if the
    replacement $c_i(Q)\rightarrow 0$ would increase the symmetry of the
    system''\cite{'tHooft:1979bh}. In our case, setting the soft terms
    to zero increases the symmetry, so the considered solution to the 
    hierarchy problem is natural in the sense of 't~Hooft's criterion.

The complete MSSM Lagrangian~\eqref{LMSSMfull} is closed under
renormalisation. However, contrary to the SM case, it does not contain all
the terms that are allowed for the given field content by the gauge symmetries
and renormalisability alone. In the SM, there is a non-anomalous global $U(1)$
symmetry that comes as accidental, i.e., it is not imposed in the theory
definition but still turns out to be a global symmetry of the Lagrangian. This
symmetry is generated by $B-L$, where $B$ is the baryon number and $L$ is the
lepton number. For all the quarks (leptons) in Tab.~\ref{ms}, we have
$B=\frac13$ and $L=0$ ($B=0$ and $L=1$). Thus, for the multiplets in
Tab.~\ref{MSSMcontent}, we have $B-L =
\{\frac13,-\frac13,-\frac13,-1,1,0,0\}$.  In the MSSM, if we did not impose
the global $B-L$ symmetry in the theory definition, we could write the
following additional terms in the superpotential:
\be
W_{\begin{picture}(20,10) \put(0,0){\line(3,1){17}} \end{picture} 
\hspace{-8mm} B-L} = a U_R^c D_R^c D_R^c + b QL D_R^c +  c LL E_R^c + d L H_u,
\ee
where the appropriate (unique) colour and isospin contractions are understood,
while the (arbitrary) flavour structure is encoded in the coefficients $a,b,c,d$.

Since violation of $B$ and/or $L$ is strongly constrained by experimental
data (including the nucleon lifetime bounds), one usually assumes that the
above terms are absent.  It can be achieved by either imposing the very $B-L$
global symmetry or some discrete symmetry.  The simplest example of a
sufficient discrete symmetry is the so-called \emph{matter parity} under which
all the matter superfields change sign, while the Higgs ones do not. Another
example is the \emph{R-parity} defined for each component field by
$(-1)^{3B+L+2S}$, with $B$, $L$ and $S$ being the baryon, lepton number and
spin, respectively. The R-parity seems to be the least straightforward but is
nevertheless the most popular for forbidding the unwanted superpotential
terms. In the following, we shall always assume that the MSSM we consider is
R-parity conserving.

It is easy to observe in Tab.~\ref{MSSMcontent}, that all the SM particles are
R-even, while their superpartners (called \emph{sparticles}) are R-odd. This
means that all physical processes have to involve even numbers of 
the superpartners, and the Lightest Supersymmetric Particle (LSP)
is stable. The LSP is thus a natural candidate for a Dark Matter (DM)
particle. Moreover, the R-parity conserving MSSM is able to successfully
explain the observed DM relic density.

As far as the soft terms are concerned, the gauge symmetries, R-parity and
the requirement that all the coupling constants are dimensionful are still not
sufficient to restrict $\mathcal{L}^{\rm soft}_{\rm MSSM}$ to the form given
in Eq.~\eqref{Lsoft}. In principle, one might add to $\mathcal{L}^{\rm
soft}_{\rm MSSM}$ all the terms with dimensionful couplings that are present
in $\mathcal{L}_{\rm MSSM}^{\rm SUSY}$, but ignoring the SUSY-induced
correlations between the couplings, namely the fact that the triple-scalar
terms in $\mathcal{L}_{\rm MSSM}^{\rm SUSY}$ are uniquely determined by the
Yukawa matrices and the higgsino mass $\mu$. Such non-holomorphic
triple-scalar soft terms would not introduce quadratic ultraviolet
divergences, as the MSSM contains no pure singlets. However, loop corrections
from them would renormalise the extra soft contribution to the higgsino mass
term, so the total higgsino mass would no longer be protected by the SUSY
non-renormalisation theorem. Although the presence of such extra soft terms is
conceivable (see, e.g., Refs.\cite{Rosiek:1989rs,Jack:1999fa,Cakir:2005hd}),
they are not necessary, as the theory closes under renormalisation without
them. In fact, they are not included in most of the MSSM analyses in the
literature. In the present work we shall assume their absence, too, taking the
soft Lagrangian as it stands in Eq.~\eqref{Lsoft}.

\end{subsection}
  
\begin{subsection}{The MSSM spectrum and field bases} \label{sec:MSSMspectr}

Similarly to the SM case, one needs to pass from the interaction basis to
the mass eigenstate basis after the EWSB takes place. We assume that the MSSM
parameters are such that only the two Higgs doublets acquire non-vanishing
VEVs. Then we use the $SU(2)$ gauge symmetry to simplify the scalar potential. 
Finally, we find that in the minimum the two VEVs can be written as
\begin{equation}
\left\langle 0| h_u |0 \right\rangle = \frac{1}{\sqrt{2}} \begin{pmatrix} v_u \\ 0 \end{pmatrix}, 
\hspace{2cm} 
\left\langle 0| h_d |0 \right\rangle = \frac{1}{\sqrt{2}} \begin{pmatrix} 0 \\ v_d  \end{pmatrix}. 
\end{equation}
They spontaneously break the electroweak symmetry $SU(2) \times
U(1)_Y$ down to $U(1)_{em}$. Once we define $v = \sqrt{v_d^2 + v_u^2}$, the
gauge boson masses are given by the same expressions as in the SM in terms of
$v$ and the gauge couplings. A useful parameter is the ratio of the
two VEVs:
\begin{equation}
   \tanb = \frac{v_u}{v_d}.
\end{equation}
  
  Most of the MSSM mass eigenstates are mixtures of some of the
  interaction eigenstate fields. The electrically neutral
  gauginos (the bino $\tilde{B}$ and the neutral wino $\tilde{W}_3$) mix with
  the neutral higgsinos $\tilde{h}_d^0$ and $\tilde{h}_u^0$ to form four
  \emph{neutralinos} $\chi_i^0, i=1,\ldots,4$. Those are especially relevant
  for phenomenology, as they provide the best candidate for a dark
  matter particle. Analogously, the charged winos $\tilde{W}_{1,2}$ mix with
  the charged higgsinos $\tilde{h}_u^+$ and $\tilde{h}_d^-$ to form
  Dirac fermions called \emph{charginos} $\chi^{\pm}_j, j=1,2$. The chiral matter
  fermions combine to the Dirac ones in the same way as in the SM, except for
  that \tanb{} affects relations between their Yukawa couplings and
  masses. The CKM matrix arises after the Yukawa matrix diagonalization, again
  in the same way as in the SM. The sfermion mass matrices require, in
  general, an additional diagonalization, which gives three distinct
  complex-scalar mass eigenstates for the sneutrinos, six for sleptons, and
  six for each type of squarks (up- and down-type). As compared to the SM, the
  number of particles with (potentially) different masses is more than
  tripled,
%
%
  and the number of free parameters increases roughly by a factor of 6.
\begin{figure}[t]
\begin{center}
\includegraphics[width=0.9\textwidth]{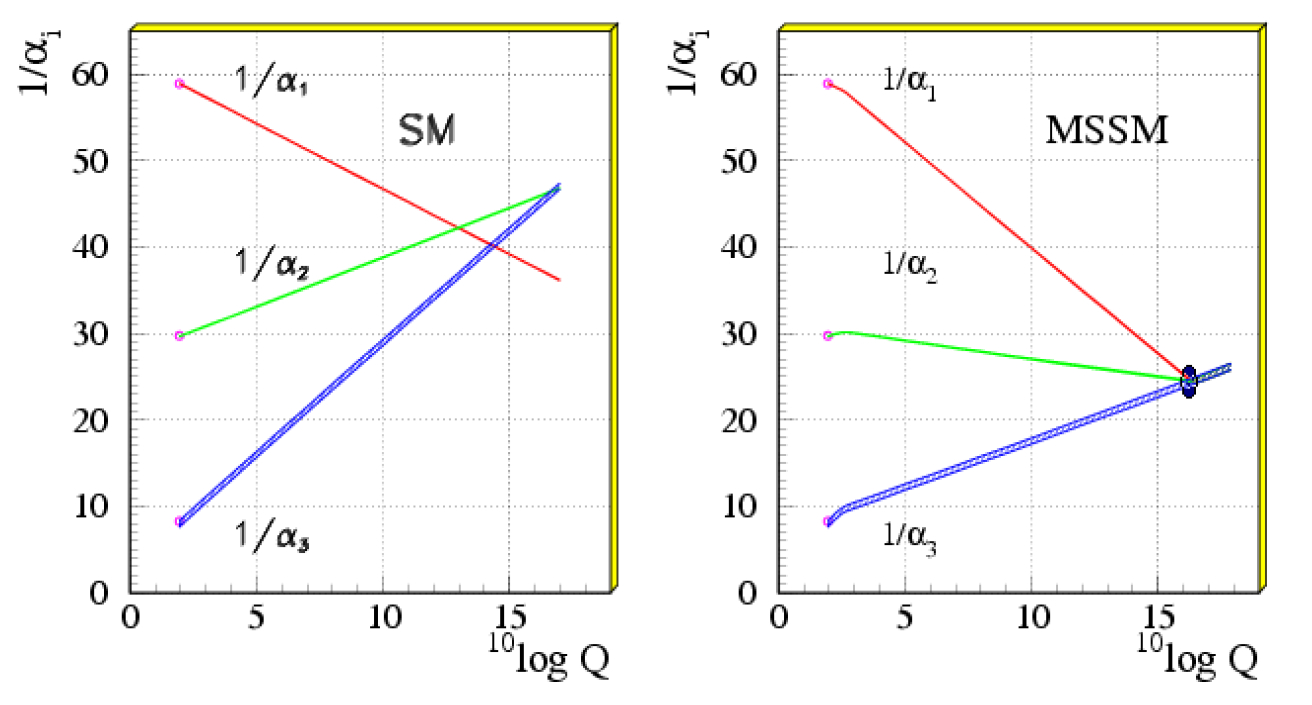}
\caption[Renormalisation scale dependence of the SM and MSSM gauge couplings.]{Illustration 
of the gauge coupling unification in the SM (left) and MSSM (right), adopted from
Fig.~16.1 of Ref.\cite{Agashe:2014kda}. \label{pdgRun}}
\end{center}
\end{figure}

  As far as the choice of the flavour basis is concerned, one of the
  most popular bases to express the MSSM fields is the \emph{super-CKM}
  basis. It is defined as a basis in which the $Q$, $U$ and $D$
  superfields are assumed to have been rotated from the interaction
  basis, in order to diagonalise the fermion masses, exactly as in
  the SM. It means that the super-CKM basis Yukawa matrices are diagonal,
  whereas the soft terms have been rotated in the same way as 
  the fermions from their interaction basis. A detailed description of the
  transformation between those two bases for all the soft terms can be found
  in Ref.\cite{SLHA2}.
\begin{figure}[t]
\begin{center}
\includegraphics[width=0.7\textwidth]{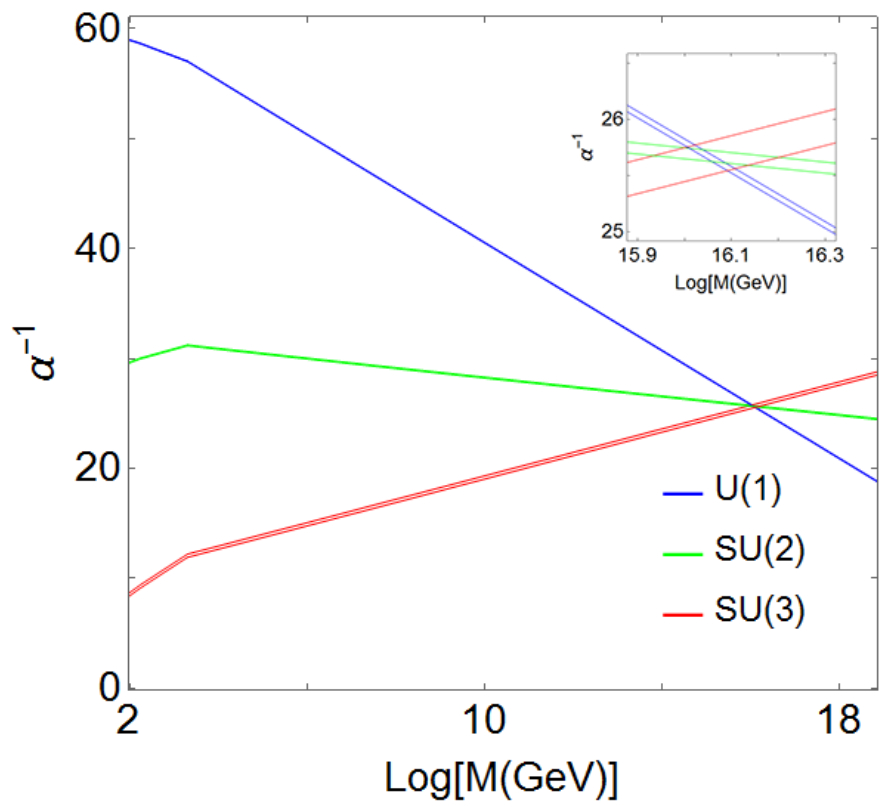}
\caption[Renormalisation scale dependence of the MSSM gauge couplings.]{Illustration 
of the gauge coupling unification in the MSSM, adopted from
Fig.~1 of Ref.\cite{Berezhiani:2015vea}. The plot has been obtained using 2-loop
RGEs, and for the sparticle masses around $2\,$TeV (see the
text).\label{1505.04950Run}}
\end{center}
\end{figure}

  \end{subsection}
  
  \begin{subsection}{Gauge coupling unification}

When the GUTs were postulated, the uncertainty in estimation of
the strong interaction coupling $\alpha_s \equiv \alpha_3$
and the ratio $\alpha_1/\alpha_2$ were so large that all 
the three couplings seemed to attain the same value at the
scale $\mgut^{1970s} \sim 10^{14\div15}\,$GeV. Later however, as the
measurements and lattice calculations became more accurate, it turned out that
at this scale they miss each other by around twelve standard
deviations\cite{Agashe:2014kda}.  This is illustrated in the left plot of
Fig.~\ref{pdgRun} where the inverse gauge couplings are plotted as functions
of the renormalisation scale.

In the MSSM, the RGEs governing the gauge coupling
evolution are different. At the one-loop level, they depend on
nothing but the particle spectrum and the gauge couplings themselves. The
evolution of the three couplings in the MSSM is illustrated in the right plot
of Fig.~\ref{pdgRun}, and, more precisely, in Fig.~\ref{1505.04950Run} that is
adopted from Ref.\cite{Berezhiani:2015vea}.  The inset in the latter figure
shows a focus of the crossing point at $\mgut \sim 10^{16}\,$GeV. The bands
correspond to the experimental $3\sigma$ ranges. Although the applied two-loop
RGEs depend only on the gauge and Yukawa couplings, the precise outlook of the
crossing point at \mgut{} depends on the remaining MSSM parameters that affect
the threshold corrections at $\mu_{\rm sp}$.  In the considered example, the
gauge couplings cross at a single point with an accuracy better than the
experimental $1\sigma$.

However, with the current high-precision experimental determination of the
gauge couplings,\footnote{
The least precisely determined $\alpha_3(M_Z) = 0.1184 \pm 0.0007$\cite{Agashe:2014kda}
is known with a 0.6\% accuracy.}
measuring the accuracy of their apparent coincidence at \mgut{} in terms of
experimental errors is an academic issue. Much more important effects come
from the GUT-scale threshold corrections which depend on the particle spectra
above \mgut. For each of the three couplings $\alpha_i$ we have
\be
\alpha_i(\mgut) ~=~ \alpha_{\rm GUT}(\mgut) ~+~ c_i\, \alpha_{\rm GUT}^2(\mgut) ~+~ \ldots\;,
\ee
with the numbers $c_i$ being of order unity, and the ellipses standing for
higher order corrections. Thus,
\be
\frac{1}{\alpha_i(\mgut)} ~=~ \frac{1}{\alpha_{\rm GUT}(\mgut)} ~-~ c_i ~+~ \ldots\;.
\ee
By convention, we define \mgut{} in such a way that $c_1=c_2$, and then test
the unification quality by calculating $\Delta_c \equiv c_3-c_2$ in a
particular GUT model.  If the physics above \mgut{} is not specified, we can
only verify whether $\Delta_c$ is not too large as for an order-unity
number. The example in Fig.~\ref{1505.04950Run} implies that $\Delta_c$ in the
MSSM can be arbitrarily small. However, in generic points of the MSSM
parameter space, one usually finds $\Delta_c \sim -1$. On the other hand, the
left plot of Fig.~\ref{pdgRun} implies that in the SM one would need $\Delta_c
\sim +5$, which might be difficult to obtain. Thus, the gauge coupling
unification issue speaks in favour of the MSSM (as compared to the SM) as a
candidate for the correct effective theory below $\mgut$. However, it is not a
decisive argument. The main motivation for SUSY in the GUT context comes from
the hierarchy problem.

The scale \mgut{} in the MSSM turns out to be significantly higher than
$\mgut^{1970s}$, which is easily seen by comparing the two plots in
Fig.~\ref{pdgRun}. This is a welcome feature because the superheavy gauge
boson contributions to the proton decay rate scale like $m_p^5/\mgut^4$,
where $m_p$ is the proton mass. With $\mgut \sim 10^{16}\,$GeV, such
contributions are not in conflict with the experimental lower bounds on the
proton lifetime. However, SUSY implies existence of potentially dangerous
contributions to the proton decay rate that scale like $m_p^5/(\mgut^2 \mu_{\rm
sp}^2)$. They may originate from the $SU(3)$-triplet higgsinos that are
present in the $H_5$ and $H_{\bar 5}$ Higgs superfields of the $SU(5)$ SUSY
GUT (see, e.g., Ref.\cite{Nilles:1983ge}). The role of such contributions
depends on the way how the so-called doublet-triplet splitting problem is
solved in a given GUT model, i.e. how the superheavy masses are generated for
the triplets without being simultaneously generated for the usual MSSM Higgs
doublets. After the GUT symmetry is broken, the triplets might receive masses
from trilinear superpotential terms that do \emph{not} depend simultaneously
on \emph{both} $H_5$ and $H_{\bar 5}$, at the cost of introducing extra $\mathbf{5}$
and $\mathbf{\bar 5}$ superfields non-interacting with the MSSM matter. In
such a case, no dangerous contributions to the proton decay would be generated
by the heavy triplet higgsinos. There are also alternative methods of
suppressing the unwanted contributions to the proton decay, the simplest of
which is making the triplet higgsinos significantly heavier than \mgut{}.

In the present work, we shall analyse the GUT-constrained MSSM without
specifying the details of the theory above \mgut{}. The doublet-triplet
splitting problem will not be addressed. Apart from what has been mentioned
above, we shall not consider any specific mechanism of avoiding the triplet
higgsino contributions to the proton decay. We shall work under the assumption
that the problem gets resolved in some way, and that the proton lifetime 
satisfies the current experimental bounds.
   
\end{subsection}

%% file: chapters/1_5_YukUni.tex
Motivated by the successful gauge coupling unification, we assume
the MSSM to be a correct effective theory below \mgut. Moreover, we
assume that the SUSY theory above \mgut{} can be described in terms of the
$SU(5)$ gauge symmetry, and that the Yukawa interactions of the MSSM matter
and Higgs fields are generated in a minimal manner, i.e. they originate from
the following part of the $SU(5)$ GUT superpotential
\begin{equation} \label{yukGUT}
{\cal W} \ni \Psi_{10} \mathbf{Y^{de}} \Psi_{\bar 5} H_{\bar 5} 
+  \Psi_{10} \mathbf{Y^u} \Psi_{10} H_5.
\end{equation}
Here, $\Psi_{\bar 5}$ and $\Psi_{10}$ are the GUT superfields that contain the
MSSM matter ones, according to the embeddings given in Eq.~\eqref{embeddings}.
The  Higgs superfields $H_{\bar 5}$ and $H_5$ contain the 
MSSM ones $H_d$ and $H_u$, respectively.

All the SM fermion masses are thus given by only two independent
$3\times 3$ matrices: $\mathbf{Y^{de}}$ and $\mathbf{Y^u}$.  In other words, the considered SUSY $SU(5)$ GUT 
induces a boundary condition for the Yukawa couplings of the
MSSM. Thus, we should address a question whether 
the Yukawa sector under such a constraint is compatible with the
low-energy data.  Our boundary condition for the MSSM RGEs is given by
the equality of the matrices $\mathbf{Y^d}$ and $\mathbf{Y^e}^T$ at
\mgut. This requirement is true up to possible one-loop threshold
corrections at this scale. In our numerical analysis, we are
going to allow for moderate threshold corrections at the GUT scale without
investigating their origin, as we do not specify details of the model
above \mgut. Once the minimal form~\eqref{yukGUT} of the GUT Yukawa terms is
assumed, the GUT-scale threshold corrections to the corresponding MSSM Yukawa
couplings are protected by the non-renormalization theorem. It means that they
can only originate from threshold effects in the field renormalization. In
consequence, they are naturally of moderate size, as long as one assumes that
Eq.~\eqref{yukGUT} is the only source of flavour violation above the GUT
scale.

The unification conditions for $\mathbf{Y^d}$ and 
$\mathbf{Y^e}$ take the simplest form in a basis where the superpotential
flavour mixing is entirely included in $\mathbf{Y^u}$, while $\mathbf{Y^d}$
and $\mathbf{Y^e}$ are real and diagonal. In such a case, it is enough to
require (approximate) equality of the diagonal entries at the GUT
scale,
\begin{equation}\label{yukunif}
Y^d_{ii}(\mgut) ~~\simeq~~ Y^e_{ii}(\mgut), \hspace{20pt} i=1,2,3.
\end{equation}

\begin{figure}[t]
  \centering
  \includegraphics[width=0.7\textwidth]{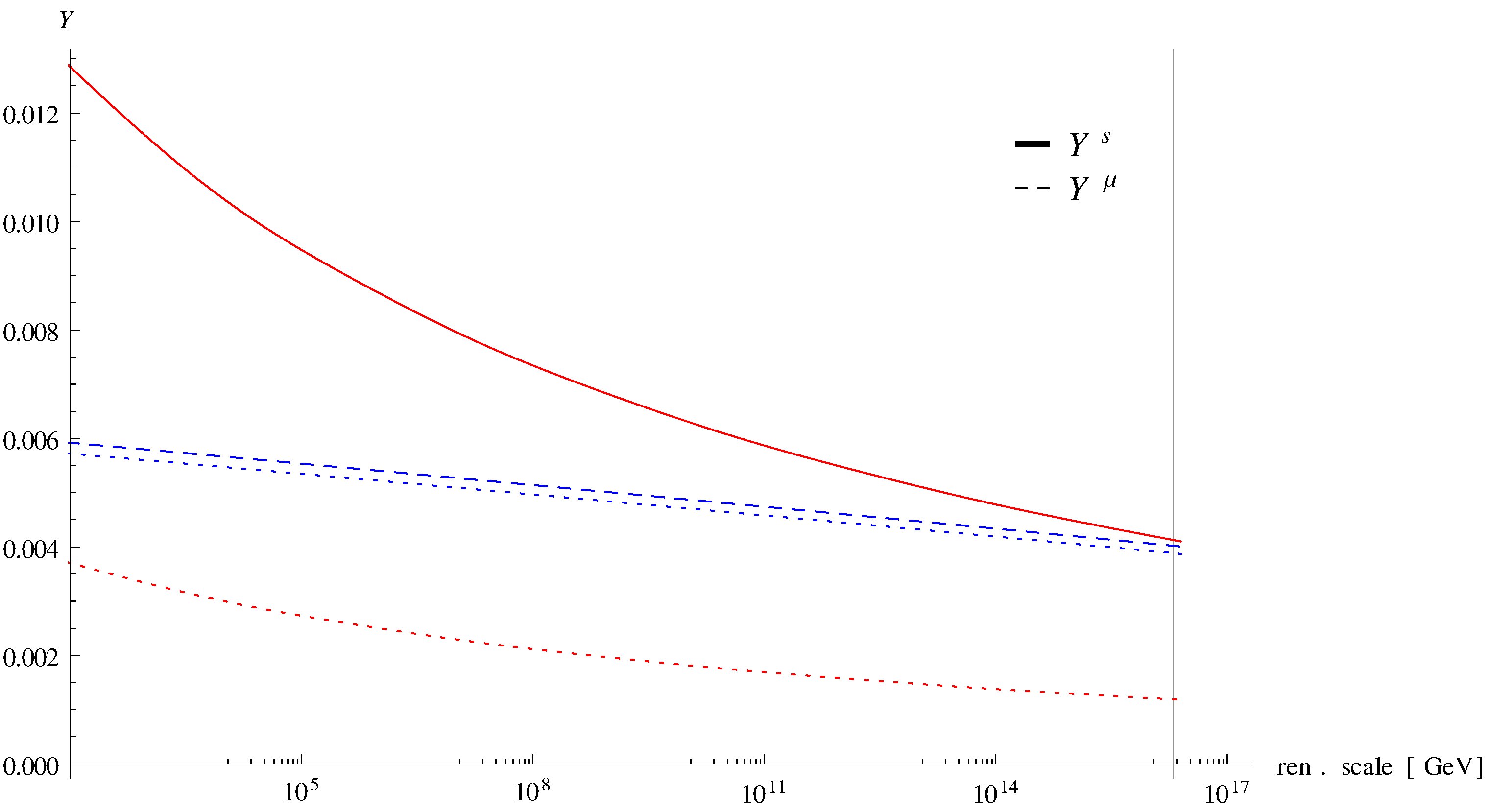}
\caption[RG-running of \Ys{} and \Ym{} in the MSSM with an
adjusted lower-end boundary.]{RG-running of \Ys{} 
(red lines) and \Ym{} (blue lines) between $\mu_{\rm sp}=M_Z$ and
\mgut{} for two sample points in the MSSM parameter space. Dotted lines
describe a situation with vanishing threshold corrections at $\mu_{\rm sp}$.
For the solid (red) and dashed (blue) lines, the threshold corrections at
$\mu_{\rm sp}$ have been adjusted to achieve unification at the GUT
scale.\label{YukUniShift}}
\end{figure}

Let us illustrate the problem with the case of the strange quark and
muon Yukawa couplings, as they require the largest relative threshold
corrections at $\mu_{\rm sp}$. As shown by the dotted lines in
Fig.~\ref{YukUniShift}, in a situation when the SUSY-scale threshold
corrections are set to zero, the considered Yukawa couplings differ
by a factor of around three at \mgut.

An important fact is that the RGEs for the superpotential
couplings are independent of the soft terms.\footnote{
For a reference, the complete two-loop MSSM RGEs can be found in
Ref.\cite{Martin:1993zk}.}
In general, no dimensionful couplings can influence 
the RG-evolution of dimensionless couplings. Thus, for the gauge and
Yukawa couplings of the MSSM we have
\begin{equation}
  Q\frac{d}{dQ} \begin{pmatrix}
              g_i (Q) \\ Y_i (Q) 
             \end{pmatrix} = f (g_i(Q), Y_i(Q) ).
\end{equation}
The MSSM RGEs have therefore a subsystem that can be solved independently of
the remaining equations, as long as the boundary 
conditions for $g_i(\mu_{\rm sp})$ and $Y_i(\mu_{\rm
sp})$ are fixed. Because of that, the gauge coupling running
cannot be modified by manipulating the soft terms, which
implies that these terms can have only a minor influence on the gauge
coupling unification issue. Analogously, neither the Yukawa coupling
running can be affected by the soft terms. However, in the case of 
the first two generation Yukawa couplings, huge effects can be generated
by the soft terms via their effects on the boundary conditions at
$\mu_{\rm sp}$.

Of course, the SUSY-scale threshold corrections are not the only way to
resolve the Yukawa matrix unification problem. Another well-known approach is
to relax the assumption about the minimal form~\eqref{yukGUT} of the
interactions that generate the MSSM Yukawa couplings at \mgut.
Modifications of the GUT field content aiming at solving this
problem were considered already in the 1970's for the
non-supersymmetric case, and are known as the Georgi-Jarlskog
mechanism\cite{Georgi:1979df}. The most exhaustively studied alterations
of the MSSM Yukawa coupling ratios at the GUT scale 
originate from assuming either various versions of the Georgi-Jarlskog
mechanism, or non-negligible effects from higher-dimensional operators in
the $SU(5)$ model Lagrangian. Different quark and lepton mass ratios
obtained by such manipulations are summarised, e.g., in
Refs.\cite{EmmanuelCosta:2003pu,Antusch:2009gu,Antusch:2013rxa}. 
Departures from the minimal assumption~\eqref{yukGUT} have been
considered also in the case of the bottom-tau
unification\cite{Monaco:2011wv}.

On the other hand, the low-energy boundary conditions turn out to give us more
freedom that one might have naively expected. The SM Yukawa couplings are
fixed by the mass measurements in the SM. However, their translation to the
MSSM renormalised couplings is affected by contributions involving the soft
terms. It has been observed that such threshold corrections at the
superpartner decoupling scale $\mu_{\rm sp}$ can significantly change or even
generate the light fermion masses\cite{Buchmuller:1982ye}. An application of
this mechanism in the context of grand unification was reported in
Ref.\cite{Hall:1985dx}. In particular, an adjustment of the lower-end boundary
values of $Y_s$ and $Y_{\mu}$ can lead to their unification, as illustrated in
Fig.~\ref{YukUniShift}.

However, in most of the contemporary phenomenological analyses, the
Yukawa coupling unification has been exhaustively studied only in the
third generation case. A quantitative study that obtained $Y_s(M_{\rm
GUT})=Y_{\mu}(M_{\rm GUT})$ within the renormalisable R-parity-conserving MSSM
was performed in Ref.\cite{DiazCruz:2000mn}. It considered only the threshold
corrections coming from gluino and higgsino loops, and pointed out that a
tension arises between the Yukawa unification and flavour observables. That
was to be expected, as the flavour off-diagonal soft terms were used
there to generate the Cabibbo angle, too. The analysis was later
expanded and simplified to the flavour-diagonal case in
Ref.\cite{Enkhbat:2009jt}. It contained examples of points in the MSSM
parameter space where the $SU(5)$ Yukawa unification was achieved for
$\tan\beta \leq 20$. In another publication\cite{Bajc:2013dea}, where the
leading MSSM threshold corrections were investigated, the problem of proton
decay was addressed by making the Higgs soft masses greater than 30$\,$TeV.

In the present work, as already stated above, we shall insist on the
minimal GUT Yukawa terms~\eqref{yukGUT}, and study whether the SUSY-scale
threshold effects can make the approximate condition~\eqref{yukunif} feasible
for all the three generations. The analysis will be restricted to the soft
terms not exceeding a few TeV, and we shall allow for the Cabibbo angle being
generated by the flavour-violating terms in $\mathbf{Y^u}$. Most importantly, the
up-to-date experimental constraints, including the Higgs boson mass, will be
taken into account.

%% file: chapters/2_1_2_Scenarios.tex
In this chapter, we define the two MSSM scenarios that we 
are going to investigate in the context of the Yukawa coupling
unification. The scenarios will differ in the specific assumptions
about soft terms at \mgut{}, on the top of the $SU(5)$ symmetry 
constraints. Even after imposing the latter constraints, the MSSM has a
large number of free parameters (${\mathcal O}(50)$ real ones), so some extra
assumptions need to be made for manageable scans over the parameter space.
%
%
%
%

\section{Large diagonal A terms}

The main phenomenological motivation behind our first scenario
is to achieve the Yukawa matrix unification and fulfil the
experimental constraints in the simplest possible manner,
adjusting as few parameters as possible. We start from the observation
that to independently influence the ratios
$Y^d_{ii}/Y^e_{ii}$ for all the three families, one has to
adjust at least three real parameters. The diagonal entries of the trilinear
$\mathbf{A^{de}}$ terms in the super-CKM basis can well serve this purpose, as
they have a strong influence on the relevant threshold corrections.

As it is common in the MSSM analyses, we will adjust the modulus of the
superpotential $\mu$-parameter to get a correct value of the VEV that gives
masses to the $W$- and $Z$-bosons. Moreover, to obtain a correct mass of
the lightest Higgs boson for given sparticle masses, one has to adjust
$A^u_{33}$ that governs the stop mixing\cite{Brummer:2012ns}, see
Sec.~\ref{mhsect}. Both the Higgs soft mass terms and
$\tan\beta=\tfrac{v_u}{v_d}$, which we employ to parameterise the Higgs
sector, are unconstrained by the $SU(5)$ unification conditions, and can serve
well other phenomenological purposes.

As far as the gaugino and the soft sfermion masses at the GUT scale are
concerned, we restrict ourselves to the common gaugino mass $M_{1/2}$ (which
is the simplest choice among relations that naturally arise in the framework
of SUSY $SU(5)$ GUTs), and a universal soft mass $m_0$ for all the sfermions
(but not the Higgs doublets). This choice reduces the number of free
parameters and makes the analysis transparent.  However, it is not necessary
for achieving the Yukawa matrix unification, so these assumptions could be
relaxed, if a need would arise.

In total, our large-diagonal-$A$-term scenario 
will be characterised in terms of 9 free parameters: $\tan\beta$,
$M_{1/2}$, $m_0$, $m_{h_u}$, $m_{h_d}$, $A^{de}_{11}$, $A^{de}_{22}$,
$A^{de}_{33}$, $A^{u}_{33}$. As far as the remaining $A$-terms in the
super-CKM basis are concerned, they are inessential for our problem, so
we shall set them to zero for simplicity. Of course, apart from the
above-mentioned free parameters, we need to take into account the measured SM
gauge couplings and fermion masses. The (adjusted) superpotential 
$\mu$-parameter will be assumed to be real, not to introduce flavour-blind sources
of CP-violation.\footnote{
Such effects are not excluded, but strongly constrained, e.g.,  by the neutron EDM.}
Choosing a negative sign of $\mu$ will appear to be preferable
(see~\refsec{TreshCorChap} below).

\section{General Flavour Violating MSSM}

Despite being consistent with the Yukawa matrix unification requirement, the
regions delivered by the first scenario will turn out to suffer from
the MSSM vacuum stability problem. Though this problem has a
phenomenologically acceptable solution, namely the long lifetime of a metastable
Universe, there is no reason for which it would need to be
the unification's necessary cost.

In principle, the soft masses and trilinear terms do not need to be 
flavour-blind. They could involve mixing terms between the squark
generations, as long as it does not lead to contradictions with
experimental data. Such a case is often referred to as the
\emph{General Flavour Violating} (GFV) MSSM.  If the flavour-violating
soft terms are sizeable, they could generate too large contributions
to processes known as the Flavour Changing Neutral Currents (FCNCs). We
shall discuss this issue in more detail in Sec.~\ref{flavPheno}.

The phenomenology of models with $SU(5)$ symmetry at the \gut\ scale and
flavour mixing in the squark mass matrices has been studied in various
contexts. Ref.\cite{Guasch:1999jp,Cao:2007dk,Fichet:2014vha} analysed possible
signatures of their spectra at the LHC. Ref.\cite{Herrmann:2011xe}
investigated properties of the dark matter (DM) candidate, while the
consequences for the Higgs mass, $B$-physics and electroweak (EW) observables
were discussed in
Refs.\cite{Heinemeyer:2004by,Cao:2006xb,AranaCatania:2011ak,Arana-Catania:2014ooa,Kowalska:2014opa}.

We shall start our analysis by defining a general set of SUSY
parameters at \mgut. A~priori, we do not know which GFV
parameters in the down-squark sector are indispensable to achieve the Yukawa
coupling unification, and which can be skipped. Therefore, 
initially, we allow all of them to assume non-zero values, and
perform a preliminary numerical scan, to identify those parameters
that are essential.  Next, we pass to our final scan,  described in 
Chapter~\ref{OffDiagChap}, where both the Yukawa
unification and the experimental constraints are tested.

We shall assume for simplicity that the soft SUSY-breaking parameters
are real, therefore neglecting the possibility of new SUSY sources of 
CP-violation. The GUT-scale $SU(5)$ boundary conditions for the 
soft masses are as follows:
\bea\label{softmass:su5}
\mllij=\mdrij\equiv \mdlij, \qquad \mqlij=\murij=\merij\equiv (m^2_{ue})_{ij}. 
\eea
We do not impose any additional conditions on the relative sizes of the
diagonal entries. The off-diagonal elements of the down-squark matrix are
required to satisfy the bound
$\left| {\color{black} (m^2_{dl})_{ij}/(m^2_{dl})_{33}}\right|\le 1$.
\par
We assume as well that
\be
(m^2_{ue})_{ij}=0,\;\;i\neq j.
\ee
Such a condition is not expected to cause any significant loss of generality
because relatively large off-diagonal elements of $\mqlij$ are generated
radiatively at the scale\footnote{
The scale \msusy\ is defined as the geometric average of masses of the 
two top-squark (\emph{stop}) mass eigenstates:
$\msusy=\sqrt{m_{\tilde{t}_1}m_{\tilde{t}_2}}$.}
\msusy\ due to the RG-running in the super-CKM basis. It will turn out
later that $(m^2_{\tilde q})_{23}(\msusy)>0$ and $(m^2_{\tilde q})_{13}(\msusy)<0$ are
the desired properties for the Yukawa unification. As we will see in
Chapter~\ref{TreshCorChap}, when those inequalities are satisfied, the
dominant gluino threshold correction turns out to be negative for the strange
quark, and positive for the down quark, as long as the trilinear term
$A^{de}_{33}$ is greater than zero. For similar reasons we
restrict our study to the case $(m^2_{dl})_{ij}>0$. At this point, we 
can introduce a short-hand notation
\be
m^{dl}_{ij}\equiv\sqrt{(m^2_{dl})_{ij}},\qquad m^{ue}_{ij}\equiv\sqrt{(m^2_{ue})_{ij}}.
\ee
The GUT-scale $SU(5)$ boundary conditions for the trilinear terms are, similarly as for the Yukawa matrices, given by
\bea\label{trili:su5}
A_{ij}^{d}=A_{ji}^{e}\equiv A_{ij}^{de}.
\eea
We constrain the relative magnitude of the diagonal entries by the corresponding Yukawa couplings
\begin{equation}
\frac{|A^f_{ii}|}{|A^f_{33}|} < \frac{Y^f_{ii}}{Y^f_{33}}.
\end{equation}
Doing so, we aim at relaxing the strong tension between the EW vacuum
stability condition and the Yukawa unification that has been
mentioned above in the context of large diagonal $A$-terms.  We also
impose that
\be
A^u_{ij}=0,\;\;i\neq j.
\ee
On the other hand, the off-diagonal entries in the down-sector trilinear
matrix are not constrained to scale proportionally to the corresponding Yukawa
matrix entries. They are only required to satisfy
$|(A^{de}_{ij})/(A^{de}_{33})|\le 0.5$.

Finally, as in the first scenario, we take a negative $\mu$, and 
assume that the gaugino mass parameters are universal at $M_{\rm GUT}$,
\be
M_1=M_2=M_3\equiv M_{1/2}.
\ee

Our usage of the squark mixing terms is a tactic to enhance and control the
threshold corrections to the down-type quark Yukawa couplings. Consistent SU(5)
symmetry conditions at \mgut{} force us though to include also a mixing between
the corresponding sleptons, as in~\refeq{softmass:su5}. The latter is in general
severely constrained by Lepton Flavour Violating (LVF) observables. Thus, we split
the analysis into two sub-scenarios:
\begin{enumerate}
\item $GFV_{23}$:\\
 First, we want to use only one additional parameter, the (strange squark)-(bottom 
 squark) mixing term, to unify the Yukawa couplings of the second and
 third generations. Therefore, the only non-zero off-diagonal entry of $m^{dl}$
 matrix at \mgut{} in this sub-scenario is $m^{dl}_{23}$.
\item $GFV_{123}$: \\
Then, we turn to a more general setup and address the complete SU(5) Yukawa
matrix unification, allowing all the off-diagonal entries of $m^{dl}$ matrix to
be non-zero at \mgut{}.
\end{enumerate}

%% file: chapters/2_2_0_intro.tex
In this section, we present and analyse explicit expressions
for the SUSY-scale threshold corrections to the Yukawa 
couplings. We investigate their form in the two
previously defined scenarios.  In the first scenario,
which depends on a lower number of free parameters, we shall plot
the GUT-scale Yukawa coupling ratios as functions of selected pairs
of the relevant soft parameters.
\SidePlotsTwoMM{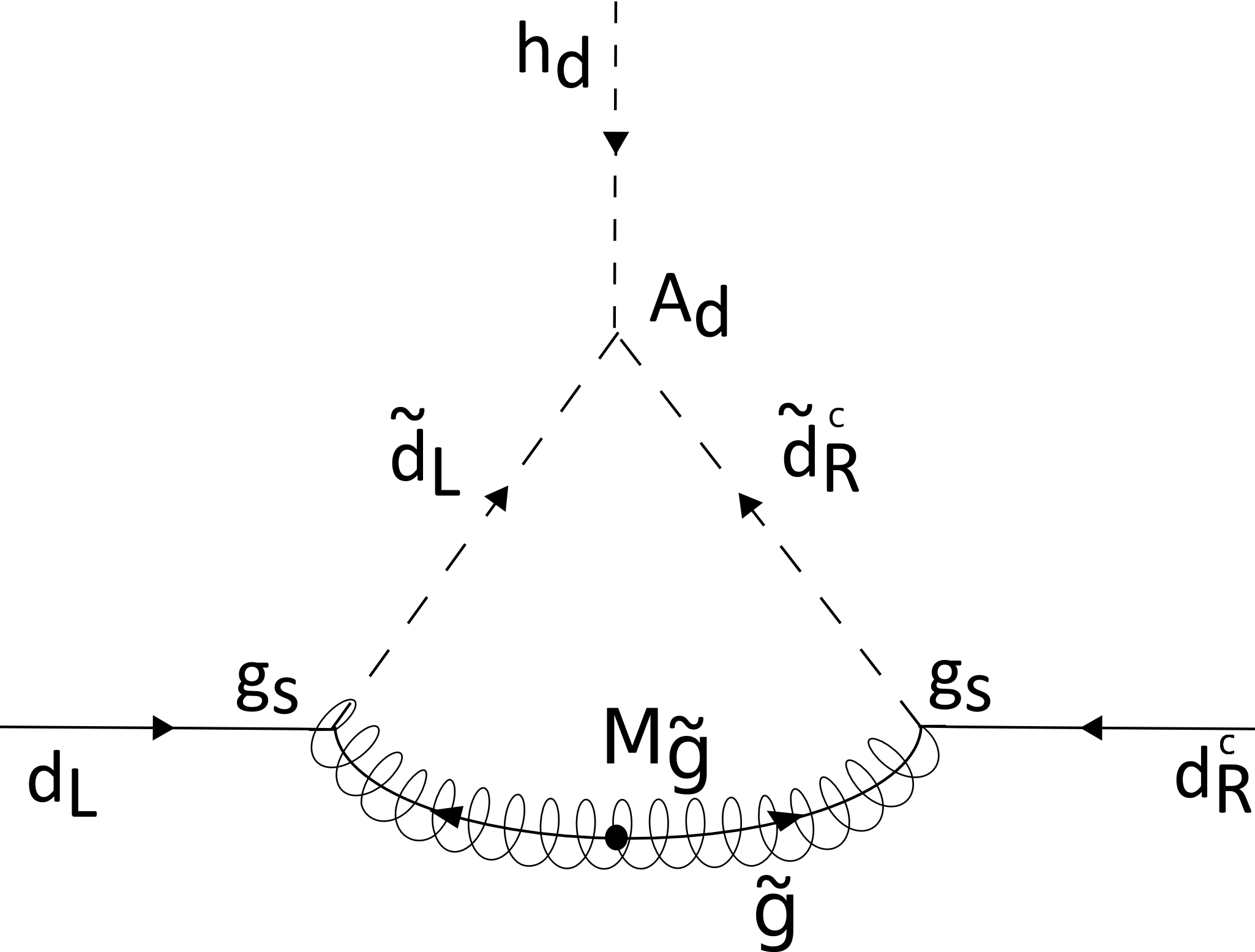}{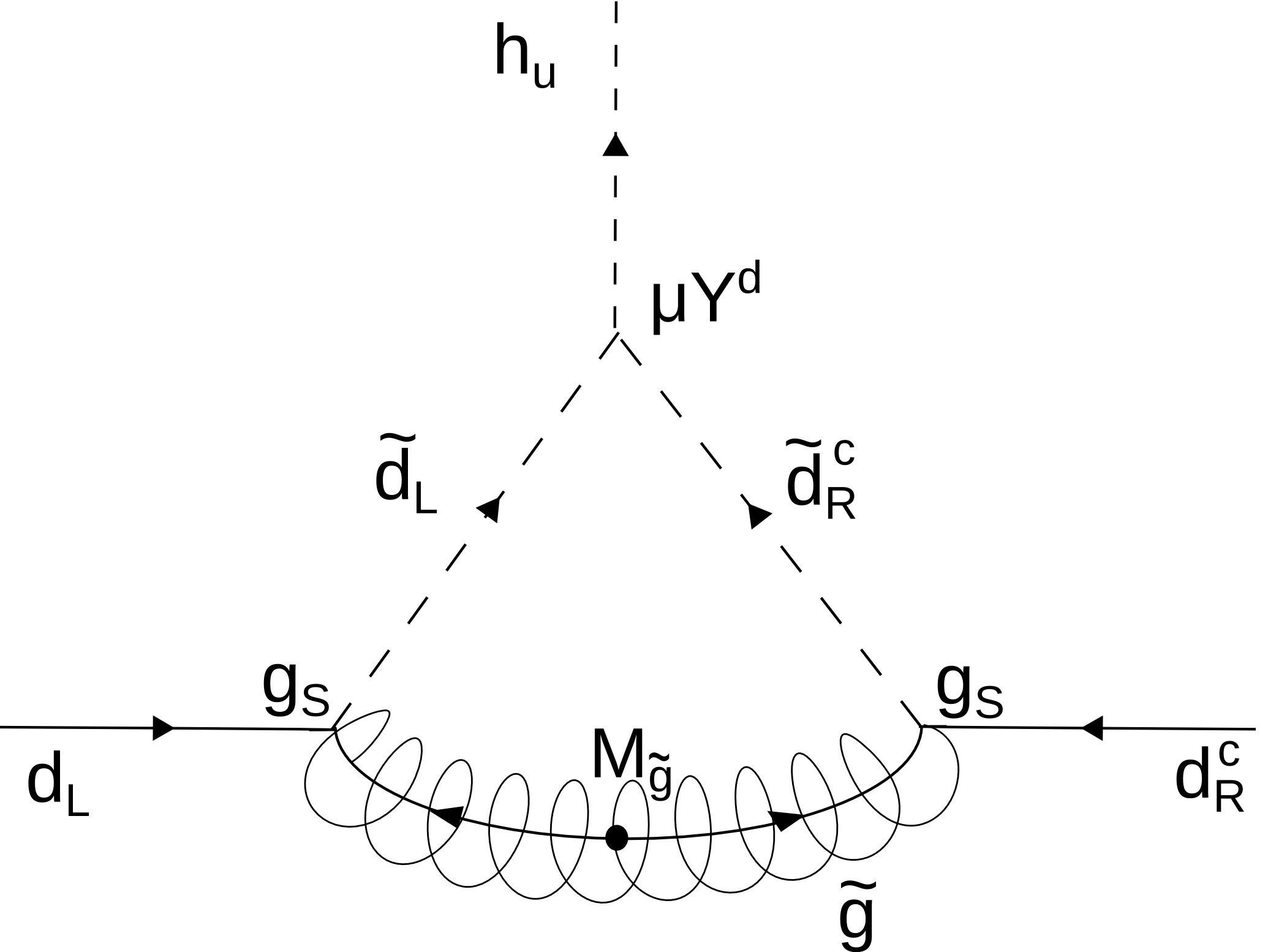}
{Diagrams describing the gluino-mediated one-loop threshold corrections to the Yukawa couplings at 
$\mu_{\rm sp}$. \label{diags}}
{0.26\textheight}
{Diagrams describing the gluino-mediated threshold corrections to the Yukawa couplings.}

The diagonal entries of the Yukawa couplings are constrained by measurements
of the quark and lepton masses that are performed at or below the electroweak
scale. Therefore, these entries are most easily first fixed within the
SM. One needs, though, to determine their renormalised values in the
MSSM.  This is done by evaluating threshold corrections $\Sigma_{ii}^f$
at the matching scale $\mu_{sp}$. Such corrections depend on values of the
soft supersymmetry-breaking terms. Examples of the relevant Feynman
diagrams are shown in Figs.~\ref{diags} and~\ref{diagsh}. 
For $f=u,d,e$ (and denoting $v_e\equiv v_d$), one can write
\be
v_f Y_{ii}^{f\,MSSM} = v_f Y_{ii}^{f\,SM}
-\Sigma_{ii}^f( \alpha_s, \mu, Y^{f'}_{jj}, A^{f'}_{jj},
                (m^2_{\tilde{f}'})_{jj},m_{\tilde{g}}),
\ee
where we have explicitly indicated dependence on those parameters that
matter for the numerically dominant effects in the flavour-diagonal case.
Once the threshold corrections are evaluated, the MSSM Yukawa
coupling values at \mgut{} are determined by solving their MSSM RGEs
(that do not depend on the soft parameters). Next, the Yukawa
unification for a given set of parameters can be tested.

The experimentally determined values of fermion masses can give a
qualitative feeling about the problems encountered in achieving the full
Yukawa matrix unification. As it is well known, the condition
$Y_b(M_{\rm GUT})=Y_\tau(M_{\rm GUT})$ can be satisfied without large
threshold corrections at $\mu_{\rm sp}$, at least for moderate $\tan\beta$. On
the other hand, achieving strict unification of the Yukawa couplings for the
remaining families ($Y_s(M_{\rm GUT})=Y_\mu(M_{\rm GUT})$ and $Y_d(M_{\rm
GUT})=Y_e(M_{\rm GUT})$) forces the threshold corrections to be of the same
order as the leading terms.  It does not contradict perturbativity of
the model because the considered Yukawa couplings are small enough,
$Y_{s,d} \ll 1$.  To satisfy the minimal $SU(5)$ boundary
conditions for the Yukawas, the MSSM tree-level strange-quark
mass needs to be larger than the SM one (see
Fig.~\ref{YukUniShift}), whereas the threshold correction effect
in the down-quark case should have an opposite sign.

The leading SUSY-scale threshold corrections to the Yukawa couplings
beyond the small \tanb\ limit have been calculated in
Ref.\cite{Crivellin:2011jt}.  It was shown that in the SUSY-decoupling limit,
the chirality-flipping parts of the renormalised quark (lepton) self energies
$\Sigma$ are linear functions of the Yukawa couplings, with a proportionality
factor $\epsilon$ and an additive term $\Sigma_{\cancel{Y}}$,
\be
m^{d(\ell)\,SM}_i-v_d Y^{d(\ell)MSSM}_{ii} \;=\;
\Sigma_{ii}^{d(\ell)\,LR} \;=\;
\Sigma_{ii\,{\cancel{Y}}}^{d(\ell)\,LR} \, + \,
\epsilon_i^{d(\ell)}\,v_u\,\,Y^{d(\ell)MSSM}_{ii} ~+~ O(\tfrac{v^2}{M_{SUSY}}), 
\label{eq:epsilon_b}
\ee
where $m^{d,SM}_i$ is the SM quark mass at $\mu_{\rm sp}$.
In this approximation, the relation can easily be inverted, and the corrected
MSSM Yukawa couplings in the super-CKM basis read
\be
\label{TCstructEq0}
Y^{d(\ell)MSSM}_{ii}=\frac{m^{d(\ell)\,SM}_i-\Sigma^{d(\ell)\,LR}_{ii\,{\cancel{Y}}}}{v_d (1+ \tan\beta \cdot\epsilon_{i}^{d(\ell)})}.
\ee

%% file: chapters/2_3_treshCor_SoftDiag.tex
\begin{figure}[t]
\centering
\includegraphics[width=0.45\textwidth]{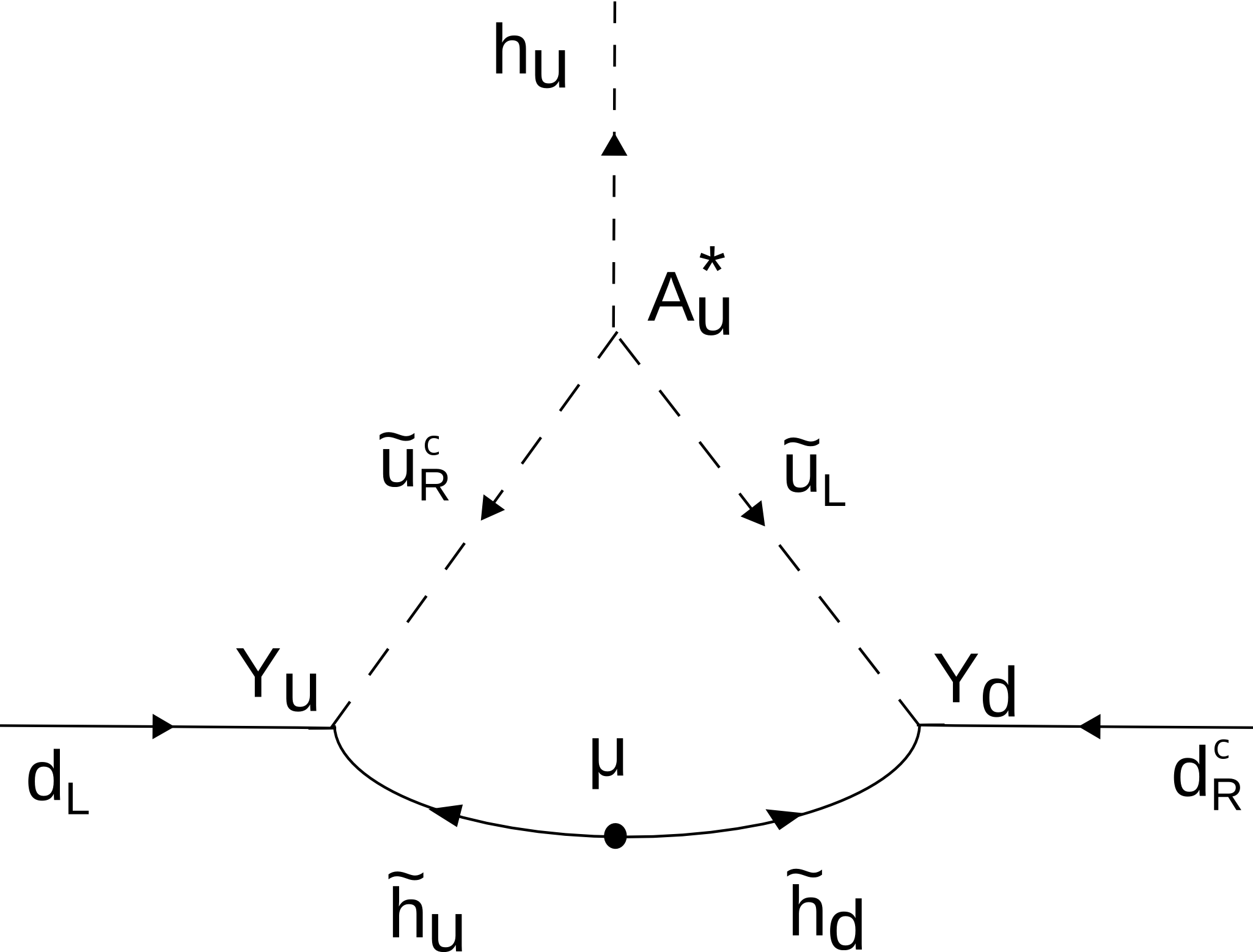}
\caption[A diagram contributing to the higgsino-mediated threshold corrections to the Yukawa couplings.]{
A diagram contributing to the higgsino-mediated threshold correction to the down-type quark Yukawa couplings at $\mu_{\rm sp}$.} \label{diagsh}
\end{figure}

What are the possible patterns of soft terms and Yukawa couplings at the
matching scale $\mu_{sp}$ in the first scenario? Given our choice of the
GUT-scale parameters, the only origin of flavour violation at this scale is
the Yukawa matrix $\mathbf{Y^u}$. As it affects the RGE for the remaining
parameters, neither $\mathbf{Y^d}$ nor the soft terms are going to remain
strictly flavour-diagonal below \mgut{}. Nevertheless, the corresponding
flavour violation is going to be given by the CKM matrix, and remain genuinely
small. Although such flavour violation is taken into account in our numerical
study, we shall neglect it for simplicity in the following discussion where
large corrections to the flavour-diagonal terms are of main interest. Within
this approximation, it is sufficient to consider only real diagonal Yukawa
matrices $\mathbf{Y^d} \equiv {\rm diag}(Y_d, Y_s, Y_b)$ and $\mathbf{Y^e}
\equiv {\rm diag}(Y_e, Y_\mu, Y_\tau)$ at all the renormalisation scales.

The threshold corrections $\Sigma^{d,LR}_{\cancel{Y}}$ to the down-type
 quark Yukawa couplings in Eq.~\ref{TCstructEq0} can be enhanced by either
 $\tan\beta$ or large values of the $A$-terms. For the first two
 generations, the most significant contribution comes from 
 loops with the gluino shown in Fig.~\ref{diags}. The 
 resulting correction in the flavour-diagonal case is given by
\be\label{thres:mfv}
(\Sigma^d_{ii})^{\tilde{g}}=\frac{2\alpha_{s}\mgluino v_d}{3\pi}
\left(A_{ii}^d-Y^d_{ii}\mu\tanb\right) 
C_0(\mgluino^{\!\!\! 2},m^2_{\tilde{q}^L_i},m^2_{\tilde{d}^R_i}),
\ee
where 
\begin{equation}
C_0 \left( {m_1^2 ,m_2^2 ,m_3^2 } \right)
= \dfrac{m_1^2 m_2^2 \ln \dfrac{m_1^2}{m_2^2} + m_2^2 m_3^2 \ln
  \dfrac{m_2^2}{m_3^2} + m_3^2 m_1^2 \ln \dfrac{m_3^2}{m_1^2}}{\left(
  m_1^2 - m_2^2 \right)\left( m_2^2 - m_3^2 \right) \left( m_3^2 -
  m_1^2 \right)}\,. 
\end{equation}
As we can see, two parameters play a major role here: the soft trilinear
coupling $A^d_{ii}$ and the superpotential $\mu$-parameter.
Interestingly, for the third family, the expression given in
Eq.~(\ref{thres:mfv}) often tends to cancel with the higgsino-mediated
contribution generated by the diagram in Fig.~\ref{diagsh}. This
fact makes the ratio $Y_b/Y_{\tau}$ relatively stable with respect to the
SUSY threshold corrections. On the contrary, for the first and second
generations, the gluino contribution is dominant, and can be
used to fix the ratios of the corresponding Yukawa couplings at $\mu_{\rm
sp}$. Such a possibility was considered in
Refs.\cite{DiazCruz:2000mn,Enkhbat:2009jt}.
\SidePlotsTwoMMx{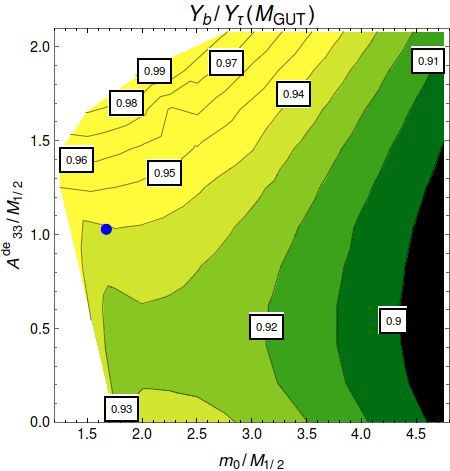}{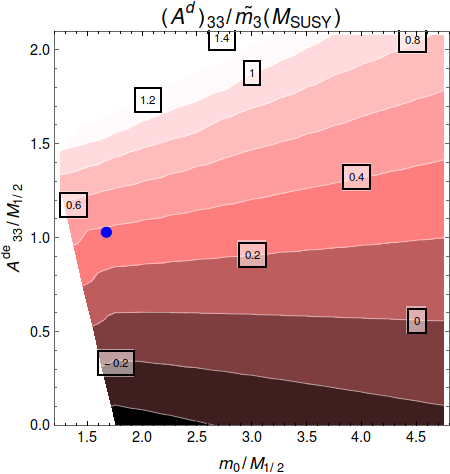} 
{Left: The ratio $Y_b/Y_{\tau}$
shown as a function of $A^{de}_{33}$ and $m_0$. Right: 
The corresponding values of $A^d_{33}/\tilde{m}_3$ at $M_{\rm
SUSY}$.  They are presented around point 3 from Tab.~\ref{Ex}
(marked by a blue dot). Both $A^{de}_{33}$ and $m_0$
are normalized to $M_{1/2}$ which equals to around 815 GeV at that
point. \label{m3A33}}{0.35\textheight}{$Y_b/Y_{\tau}$ at \mgut  \hspace{3pt}
 as a function of $A^{de}_{33}$ and $m_0$.}

Importantly, while $\mu$ affects corrections to all the Yukawa
couplings in a correlated manner, the diagonal $A$ terms can be used to
tune them independently for each family. Still, a large and negative
$\mu$ for a high \tanb{} is decisively helpful for unification of the second
family down-type quark and lepton Yukawa couplings.

In Figs.~\ref{m3A33}--\ref{m1A11}, we illustrate how the
ratios $Y^d_{ii}/Y^{e}_{ii}$ (at the GUT
scale) and $A^d_{ii}/\tilde{m}_i$ (at the SUSY scale) depend on the
most important GUT-scale soft parameters. Here, $\tilde{m}_i$ are
defined by
\begin{equation}
\tilde{m}_i=\sqrt{\tfrac{m_{\tilde{q}_{i}}^2+m_{\tilde{d}_i}^2
+m_{H_d}^2}{3}}.
\label{mtildeEq}
\end{equation}
Using the point no.~3 from Tab.~\ref{Ex} (Sec.~\ref{ExamplesSec}) as a
reference, we have varied only two parameters at a time, which gives some
estimates of the shape of the parameter correlations in the
vicinity of the considered point. In all these plots, we display
only the points which fulfil all the necessary
phenomenological requirements, in particular that the Higgs vacuum is a local
minimum of the scalar potential,\footnote{
No scalar tachyons appear in the spectrum, meaning that no $m^2$ terms are
negative.}
and that no Landau poles exist below \mgut{}. White regions in the
plots signal that either one of above conditions was not fulfilled, or that
the applied program, {\tt SOFTSUSY}\cite{Allanach:2001kg}, rejected the
point because its iterative algorithm had not converged.
\SidePlotsTwoMMx{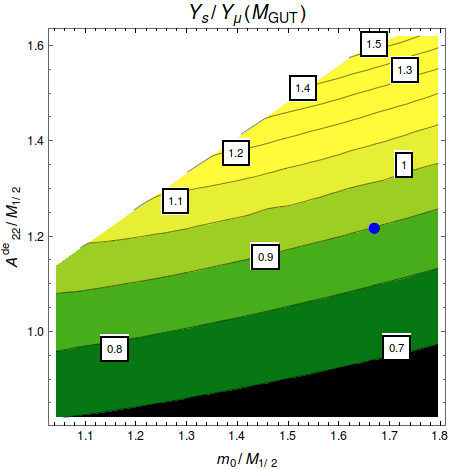}{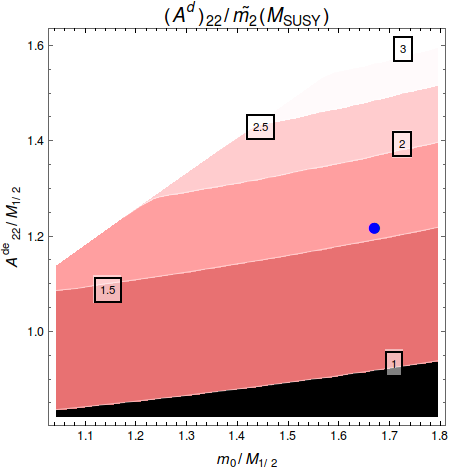}{Left: The ratio $Y_s/Y_{\mu}$
presented as a function of $A^{de}_{22}$ and $m_0$. Right: 
The corresponding values of $A^d_{22}/\tilde{m}_2$ at $M_{\rm
SUSY}$. They are shown around point 3 from Tab.~\ref{Ex}
(marked by a blue dot).
\label{m2A22}}
{0.35\textheight}{$Y_s/Y_{\mu}$ at \mgut \hspace{3pt}  as a function of $A^{de}_{22}$ and $m_0$.}
\SidePlotsTwoMMx{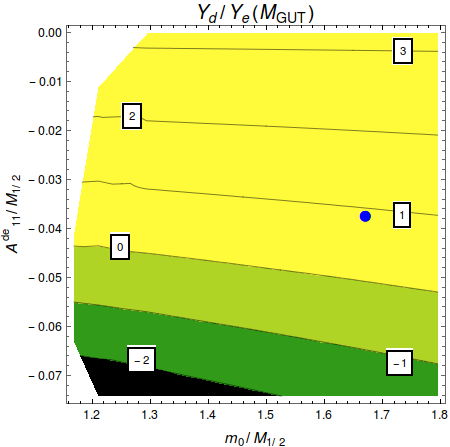}{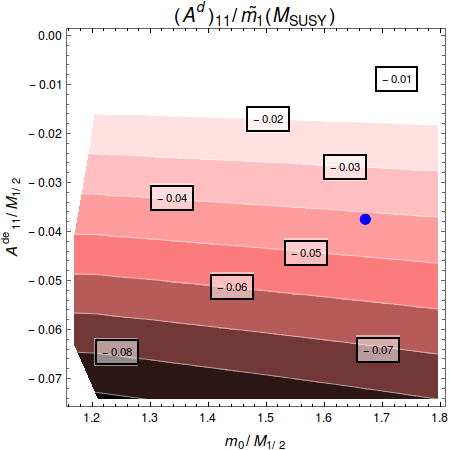}{Left: The ratio $Y_d/Y_{e}$ as
a function of $A^{de}_{11}$ and $m_0$.  Right: The corresponding
values of $A^d_{11}/\tilde{m}_1$ at $M_{\rm SUSY}$.  They are shown
around point 3 from Tab.~\ref{Ex} (marked by a blue dot).
\label{m1A11}}
{0.35\textheight}{$Y_d/Y_{e}$ at \mgut \hspace{3pt} as a function of $A^{de}_{11}$ and $m_0$.}

Beginning with the largest couplings, we notice that three parameters play a
crucial role in the case of bottom-tau unification: $A^{de}_{33}$, $\mu$ and
$m_0$ (which for given $M_{1/2}$ governs masses of the third family
sfermions). The non-universal sfermion masses, independent for each family,
could add some additional freedom to our model. Although they are not 
necessary for achieving the Yukawa unification, relaxation of the
universality could facilitate finding points with even higher $\tan\beta$ than
presented in the next section.

The obtained values of the ratio $Y_b/Y_{\tau}$ at \mgut{} are shown in
the left panel of Fig.~\ref{m3A33} as functions of $A^{de}_{33}$ and
$m_0$.  The equality of $Y_b$ and $Y_{\tau}$ at that scale might in
general require an adjustment of all the parameters, as the excluded
points tightly surround the allowed region.

In the second family case, the unification of $Y_s$ and
$Y_{\mu}$ is normally possible by a manipulation of just one parameter, namely
$A^{de}_{22}$, despite the fact that it influences both the Yukawa
couplings. For the first two families, $\mu$ has no noticeable
influence on the unification in the considered region because the higgsino
loop, being $\sim Y^u_{ii}$, gives a much smaller
contribution, due to $m_{u,c} \ll m_t$.

The ratio $Y_s/Y_{\mu}$ at \mgut{} is plotted in the left panel
of Fig.~\ref{m2A22} against $m_0$ and $A^{de}_{22}$. The plot shows
that a large value of $A^{de}_{22}$ is necessary to achieve the
unification.  The corresponding values of $A^d_{22}/\tilde{m}_2$ at $M_{\rm
SUSY}$ are shown in the right panel of Fig.~\ref{m2A22}. Such ratios
will be relevant in the context of our discussion of the vacuum metastability
in Sec.~\ref{vacSect}.

Analogous plots that describe unification of the down-quark and
electron Yukawa couplings are presented in Fig.~\ref{m1A11}.  It is the
simplest case because the necessary adjustment of the respective
$A$-term neither triggers phenomenological problems, nor influences any
parameters that are relevant for other families.

%% file: chapters/2_4_treshCor_OffDiag.tex
Now, after we have described the most important features of the
flavour-diagonal soft term case, let us explore the more general
framework of GFV MSSM. The dependence of threshold corrections on the GFV
parameters is non-trivial. The flavour-off-diagonal soft mass 
matrix elements enter~\refeq{eq:epsilon_b} through rotation matrices that
diagonalise the very sfermion mass matrices. On the other hand, the
off-diagonal trilinear couplings appear explicitly in the
expressions for the threshold corrections given in
Ref.\cite{Crivellin:2011jt}.  Such a treatment of the flavour-violating terms
is correct in the SUSY-decoupling limit, in which case the sfermion mass
matrix diagonalisation can be performed prior to considering the EWSB.

To gain some intuition about the functional dependence
of~\refeq{eq:epsilon_b} on various entries in the sfermion mass matrices, let
us consider again an example of threshold corrections to the self-energy of
down-type quarks.  A relatively simple analytic expression for
$(\Sigma^d_{22})^{\tilde{g}}$ can be obtained when the first-family couplings
are still assumed to be flavour-diagonal, while the second and third family
mixing is generated by $(m^2_{dl})_{23}(\mgut)$ alone.\footnote{
As we have already mentioned, not only $(m^2_{\tilde{d}})_{23}$ but also
$(m^2_{\tilde{q}})_{23}$ gets then generated at the SUSY scale, thanks to the
RG-evolution effects.}
In such a case, the dominant GFV contribution to the strange quark
self-energy is associated with a gluino loop. The
relevant diagrams are shown in Fig.~\ref{diagsOff}. 
They give\cite{Crivellin:2011jt}
\bea\label{sigma22}
(\Sigma^d_{22})^{\tilde{g}}&=&\frac{2 \alpha_s m_{\tilde{g}}v_{d}}{3\pi}
(A^d_{33}-Y_b\,\mu\tan\beta)\sum_{m,n=2,3}
C_0(\mgluino^{\!\!\!2},m^2_{\tilde{q}^L_m},m^2_{\tilde{d}^R_n})\times\\&&
\frac{(m^2_{\tilde{q}})_{23}}{\sqrt{\left[(m_{\tilde{q}}^2)_{22}-(m^2_{\tilde{q}})_{33}\right]^2+
4((m^2_{\tilde{q}})_{23})^2}}\frac{(m^2_{\tilde{d}})_{23}}{\sqrt{\left[(m^2_{\tilde{d}})_{22}-
(m^2_{\tilde{d}})_{33}\right]^2+4((m^2_{\tilde{d}})_{23})^2}},\nonumber
\eea
where it has been assumed that $(m^2_{\tilde{q}})_{23}$ and
$(m^2_{\tilde{d}})_{23}$ are real. It follows from~\refeq{sigma22} that
the chirality-conserving GFV interactions
$[(m^2_{\tilde{q}})_{23}\,\tilde{s}_L^* \tilde{b}_L + {\rm h.c.}]$
and
$[ (m^2_{\tilde{d}})_{23}\, \tilde{s}_R^* \tilde{b}_R + {\rm h.c.}]$ 
generate a threshold correction to \Ys\ of the order of $\Delta
\Ys\sim \alpha_s A_{33}^d/M_{\rm SUSY}$, which in general can be large enough
to enable a satisfactory Yukawa coupling unification for the second
family, even when $A_{22}^d$ is small.
\SidePlotsTwoMM{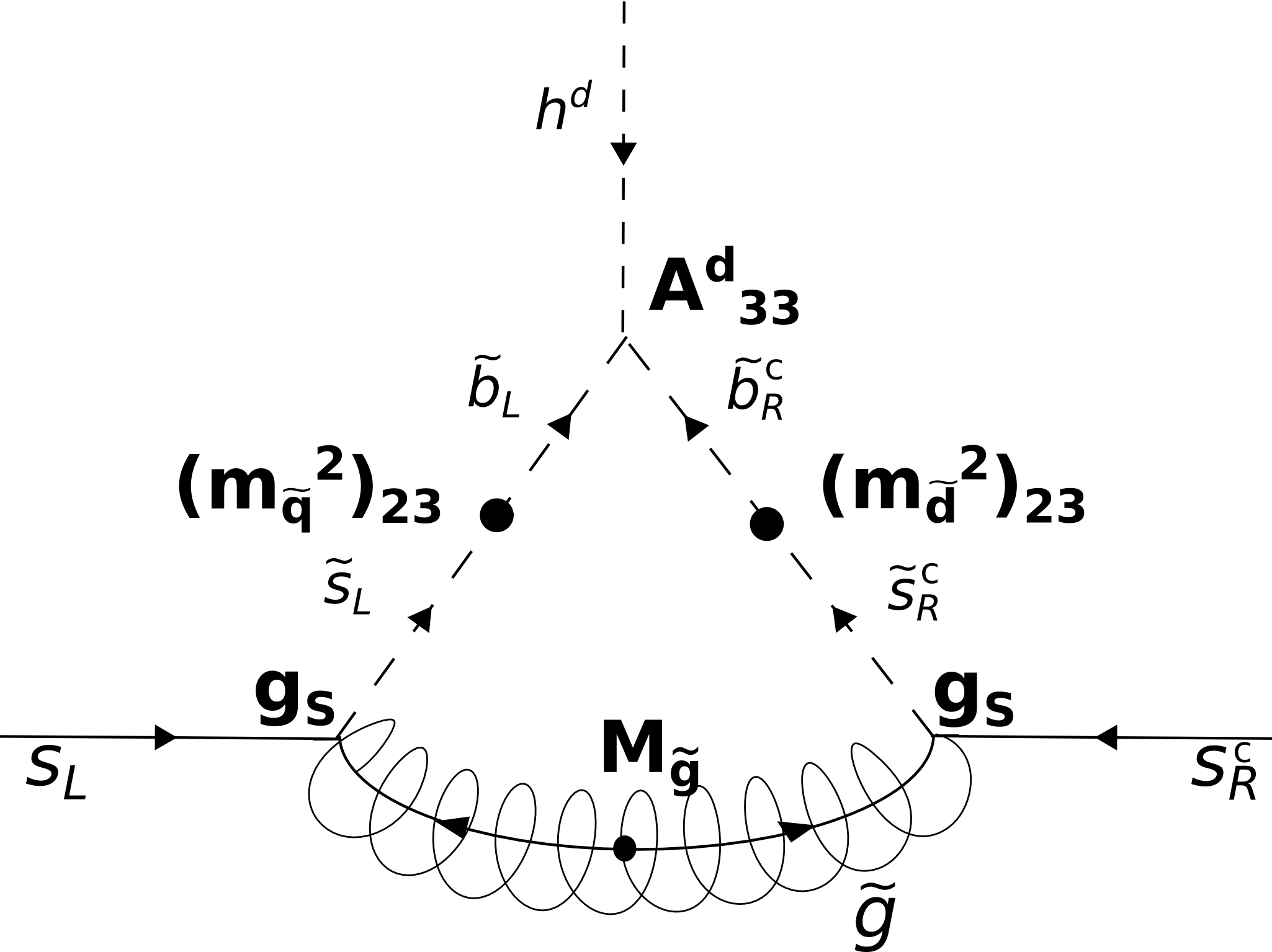}{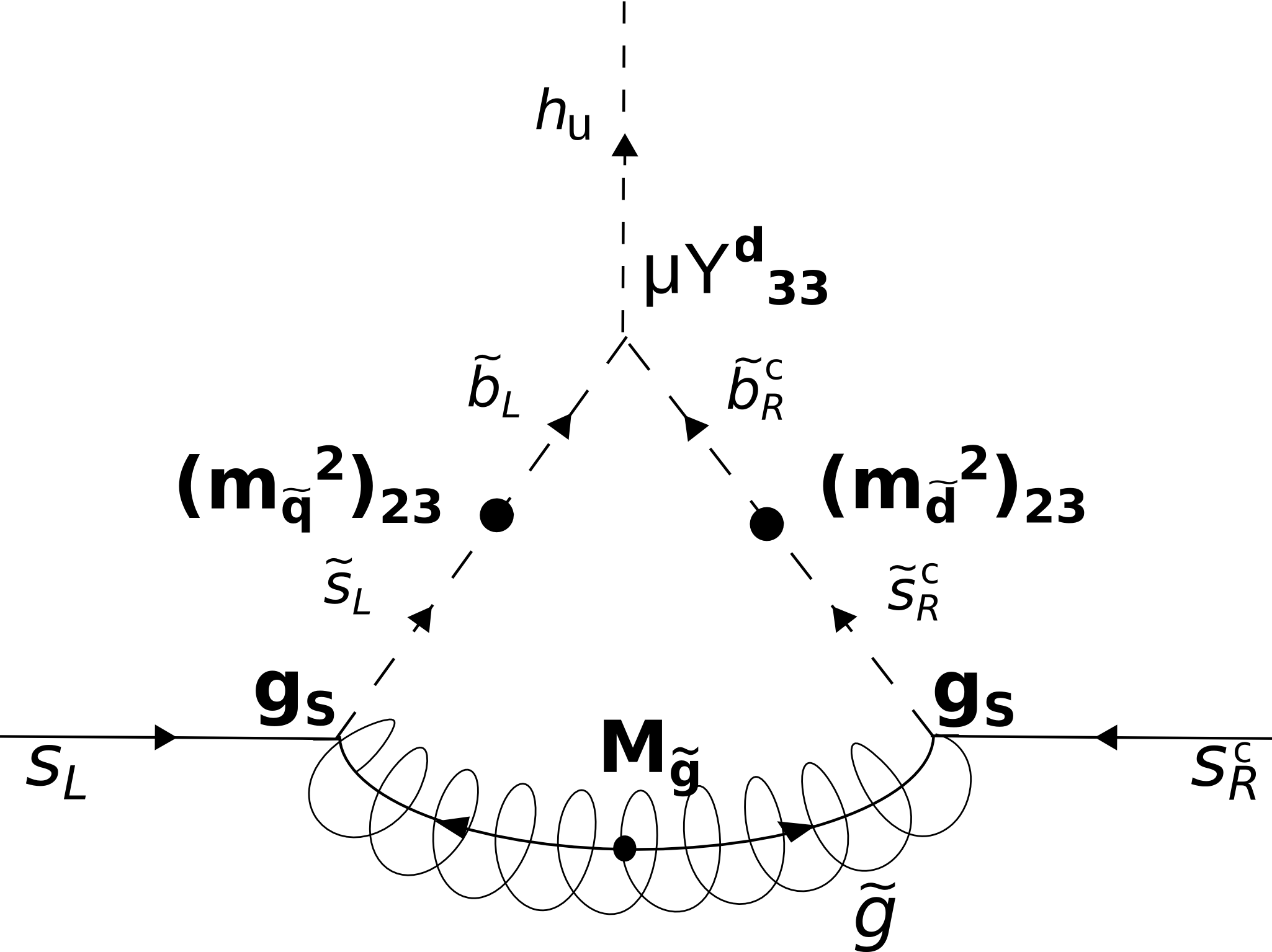}
{Diagrams contributing to the gluino-mediated threshold
correction to the strange-quark Yukawa coupling at $\mu_{\rm sp}$. It
arises when flavour mixing is present in the soft mass
matrix. \label{diagsOff}}
{0.26\textheight}
{Diagrams contributing to the gluino-mediated threshold correction with flavour mixing.}

The above discussion, however, should be treated only as a simplified
qualitative illustration. In a general case, other off-diagonal elements of
the squark mass matrix can significantly differ from zero, which makes 
the mixing among all the three generations important. To make sure that all
the relevant effects are properly taken into account, we shall
perform a complete numerical analysis of the GFV scenario in
Chapters~\ref{OffDiagChap} and~\ref{GFV123Num}.

%% file: chapters/3_1_Flavour.tex
This section is devoted to recalling basic properties of those Flavour
Changing Neutral Current (FCNC) processes that are going to serve as
experimental tests in the following chapters. The name FCNC refers to
processes where flavour is not conserved, but each open fermion line has
particles with the same electric charges on its ends. In the SM and MSSM,
there are no FCNCs at the tree level because all the flavour-changing vertices
involve a change in the electric charge, too.  Historically, this was a
postulate that allowed to explain the FCNC suppression via the so-called
Glashow-Iliopoulos-Maiani (GIM) mechanism\cite{GIMpaper}. This mechanism
implies that the FCNCs involving the first two generations of quarks alone are
suppressed not only by loop factors and small coupling constants but, in
addition, by tiny ratios of the light quark masses squared to the $W$-boson
mass squared.\footnote{
The same mechanism makes the FCNC processes in the lepton sector completely
negligible in the SM, due to tiny masses of the neutrinos.}
The GIM suppression by mass ratios is not in place for loops with the top
quark because $m_t > M_W$. However, other numerically small factors often
enter in this case, e.g., the CKM matrix element $V_{td}$.
 
The generic smallness of FCNC amplitudes in the SM renders them sensitive to
possible new physics effects, making all the beyond-SM theories with
tree-level FCNCs severely constrained. In the MSSM, no such tree-level effects
are present, but we encounter new loop contributions to the FCNC
amplitudes. We need to test the corresponding experimental constraints,
keeping in mind though that the SUSY FCNC effects tend to zero (decouple) when
all the superpartners become much heavier than the electroweak scale. This is
contrary to the Yukawa threshold corrections that may remain large also in the
heavy SUSY limit.

\subsection{Kaon mixing}

\begin{figure}[t]     
\includegraphics[width=\textwidth]{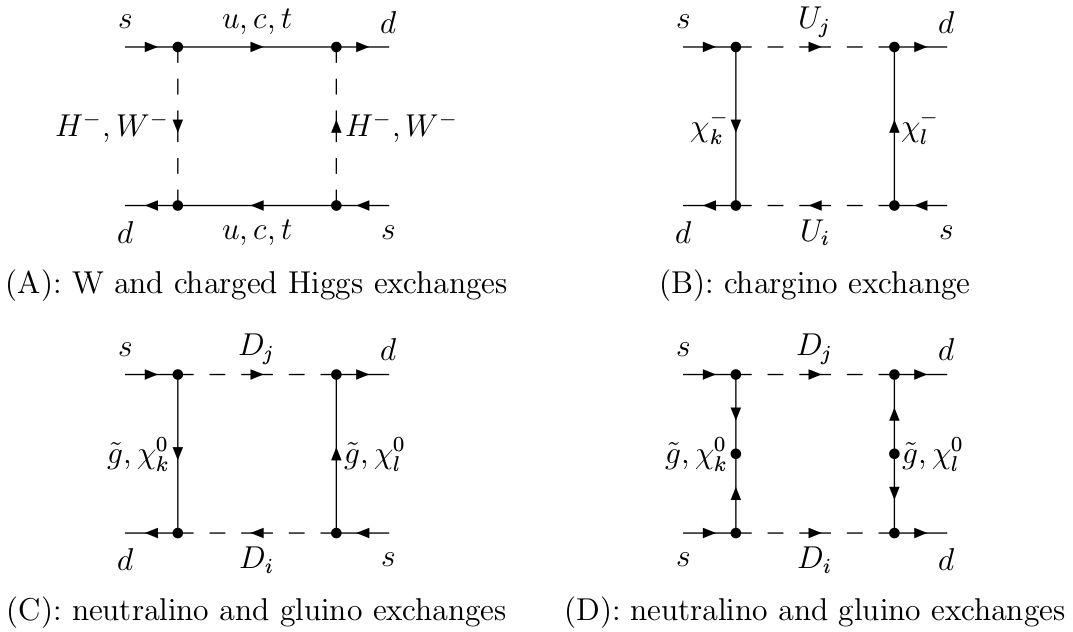}
   \caption[MSSM Feynman diagrams contributing to $\bar{K}^0 K^0$ mixing]{The MSSM
   Feynman diagrams contributing to $\bar{K}^0$-$K^0$ mixing. In this figure,
   $U_i$ and $D_i$ denote respectively the up- and down-squarks
   of any family. The drawings have been adopted from
   Ref.\cite{Misiak:1997ei}.\label{epsKdiags}}
    \end{figure}

 The lightest mesons containing an (anti)strange quark, called
 \emph{kaons}, are a significant object of study, both at present
 and in the past. The kaon system was the first 
 where CP-violation was observed\cite{CPkaonHist}. A
 mixing of the two neutral kaon flavour eigenstates 
 ($K^0\sim (d\bar s)$ and $\bar{K}^0 \sim (s\bar{d})$) 
 produces two mass eigenstates $K_S$ and $K_L$ that are detected
 in experiments. An important observable used to quantify CP-violation in
 the mixing phenomenon is the ratio $\epsilon_K$ of the decay amplitudes of
 $K_S$ and $K_L$ into two-pion states with vanishing total isospin:
 \begin{equation}
    \epsilon_K = \frac{\mathcal{A}[K_L \rightarrow (\pi\pi)_{I=0}]}{\mathcal{A}[K_S \rightarrow (\pi\pi)_{I=0}]}.
 \end{equation}
Here, $\ket{(\pi\pi)_{I=0}}=\tfrac{1}{\sqrt{3}}(\ket{\pi^+\pi^-}+\ket{\pi^-\pi^+}-\ket{\pi^0\pi^0})$.
One-loop diagrams contributing to $K^0$-$\bar{K}^0$ mixing in the MSSM are
 presented in Fig. ~\ref{epsKdiags}.

The current experimental result for $|\epsilon_K|$ which we adopt
for our analysis reads\cite{Agashe:2014kda}:
\begin{equation} \label{eps.exp}
|\epsilon_K^{\rm exp}|= (2.228\pm 0.011 )\times 10^{-3}.
\end{equation}
The present status of the SM calculations of $|\epsilon_K|$ is discussed
in Ref.\cite{Bailey:2015wta}. The theoretical uncertainty is much larger than
the experimental one. Until several years ago, it had originated mainly from
the non-perturbative parameter $B_K$ that is being determined using lattice
QCD simulations. Nowadays, the lattice QCD errors are subdominant, and the
main theory uncertainty in $|\epsilon_K|$ stems from poor convergence of
the QCD perturbation series for the double-charm contributions
in Fig.~\ref{epsKdiags}(A), as well as from the CKM matrix element
$|V_{cb}|$ which enters in the fourth power into the expression for
$|\epsilon_K|$. The value of $|V_{cb}|$ can be quite accurately extracted
either from the inclusive or exclusive semileptonic $B$-meson decay
data. However, a long-standing discrepancy between the two methods
persists. If the inclusive method for $|V_{cb}|$ is used, the SM result for
$|\epsilon_K|$ in Ref.\cite{Bailey:2015wta} agrees with Eq.~\eqref{eps.exp}
within the theory uncertainty of $\pm 2.3\times 10^{-4}$.  On the other hand,
a $3.4\sigma$ discrepancy is found (with $\sigma \sim 1.8\times 10^{-4}$) when
the exclusive method is followed. The discrepancy is going to decrease after
taking into account the very recent experimental update of the exclusive
semileptonic results at the EPS-HEP~2015 conference\cite{Belle.at.EPS.2015},
as it shifts the exclusive $|V_{cb}|$ towards the inclusive one.

Our numerical analysis here and in Ref.\cite{Iskrzynski:2014sba} relies on the
calculation of Ref.\cite{Brod:2011ty} where the SM result
\begin{equation} \label{eps.SM}
|\epsilon_K^{\rm SM}|= (1.81 \pm 0.28)\times 10^{-3}
\end{equation}
was obtained employing the 2010 weighted average of the exclusive and inclusive
$|V_{cb}|$, namely $|V_{cb}| = 0.0406(13)$. We do not make any use of the SM
central value from Eq.~\eqref{eps.SM}, but rather calculate $|\epsilon_K|$ in
the MSSM from the outset, using the code {\tt SUSY\_FLAVOR~v2.10}\cite{Crivellin:2012jv}.
As far as the SM parameters are concerned, we adopt the SM values and
uncertainties for all of them but the Wolfenstein parameters $\rho$ and $\eta$
in the CKM matrix description. In our numerical treatment, each point of the
MSSM scan corresponds to a particular value of $(\rho,\eta)$, and it is tested
against SUSY-sensitive loop observables that matter for determining the
allowed regions in the $(\rho,\eta)$ plane. At the same time, our initial
ranges for $\rho$ and $\eta$ are adopted from the so-called 'new physics fit'
by the UTFit collaboration\cite{utfit} where only tree-level observables are
included.

Once $|\epsilon_K|$ is calculated for a given point in the MSSM parameter
space, it needs to be tested against the experimental result~\eqref{eps.exp},
taking into account the theory uncertainty. To do so, we subtract the
Wolfenstein-parameter-induced error from the one in Eq.\eqref{eps.SM}, which
reduces the theory uncertainty to $\pm 1.7\times 10^{-4}$. The latter error is
treated as the theory one for particular values of $\rho$ and $\eta$ in our
MSSM scan.  To accept or reject a given point, we check its consistency with
Eq.~\eqref{eps.exp} at the desired level (see Chapter~\ref{OffDiagChap}),
after adding the theory error in quadrature to the experimental one. An
identical approach is followed for all the other flavour observables for which
the dependence on $\rho$ and $\eta$ (varied within their initial
tree-level-determined ranges) is non-negligible when compared to other
uncertainties.

\subsection{$B$-meson mixing}

Similarly to the neutral kaon mixing, an analogous phenomenon
occurs for the neutral $B$-mesons whose flavour contents are 
\be
B_d^0 \sim (d\bar b), \hspace{1cm} 
\bar{B}_d^0 \sim (b\bar d), \hspace{1cm} 
B_s^0 \sim (s\bar b), \hspace{1cm} 
\bar{B}_s^0 \sim (b\bar s). 
\ee
They are the lightest eigenstates of the strong interaction
Hamiltionian which possess these very flavour contents.

There are several observables in the neutral $B$-meson mixing which
we shall use for tests of our Yukawa unification scenarios. First, we
shall consider mass differences $\Delta M_{B_d}$ and $\Delta M_{B_s}$
between the corresponding mass eigenstates. Their experimentally determined
values are as follows\cite{Agashe:2014kda}:
\begin{align}
\Delta M_{B_s}^{\rm exp} &=(1.1691 \pm 0.0014) \times 10^{-11} \gev, \\ 
\Delta M_{B_d}^{\rm exp} &=(3.357 \pm 0.033) \times 10^{-13} \gev. 
\end{align}
Their theoretical estimates for fixed $\rho$ and $\eta$ receive the main
uncertainties\footnote{
Numerical estimates of all such uncertainties are collected in 
Tab.~\ref{tab:exp_constraints} of Chapter~\ref{OffDiagChap}.}
from overall factors that are determined in lattice QCD simulations --
see, e.g., Ref.\cite{Aoki:2013ldr}.  These uncertainties cancel to a large
extent in the ratio $\Delta M_{B_d}/\Delta M_{B_s}$.  In consequence, this
ratio has a smaller theory uncertainty, and will be considered as a
supplementary observable. Its experimental value reads\cite{Amhis:2014hma}
\begin{equation} 
 \left( \Delta M_{B_d}/\Delta M_{B_s}\right)^{\rm exp}
 =(2.87 \pm 0.02) \times 10^{-2}. 
\end{equation}

Another important observable is the CP-violating phase $\beta$ of the
$B_d^0$-$\bar{B}_d^0$ mixing amplitude. Measurements of CP-asymmetries in
$B^0_d \to J/\psi K_S$ and similar processes imply that\cite{Amhis:2014hma}
\be
\sin(2\beta)_{\rm exp} = 0.682 \pm 0.019.
\ee
The above result provides a valuable constraint in the GFV case
because the experimental error is small, and the theory uncertainty is even
smaller, in fact practically negligible at present.

\subsection{Rare B decays}

 Another set of precisely measured flavour observables is provided by
 rare decays of the $B$-mesons. Among them, we select
 three branching ratios to be investigated more closely when
 testing our scenarios:
 \begin{enumerate}
  \item $\overline{{\mathcal B}}_{s\mu} \equiv {\mathcal B}(B_s \to \mu^+ \mu^-)$ \\[1mm] 
  As for any FCNC process, contributions to this branching ratio arise only
  at one-loop level in the SM and MSSM. Two sample SM diagrams are presented in
  Fig.~\ref{bsmumudiagsSM}. Apart from loop factors, the branching ratio gets 
  additionally suppressed by $m^2_{\mu}/m^2_{B_{s}}$ where $m_{\mu}$ 
  is the muon mass. An analogous suppression occurs in all the other decays of the neutral
  $B$-mesons into lepton pairs.
  \begin{figure}[t]     
\includegraphics[width=\textwidth]{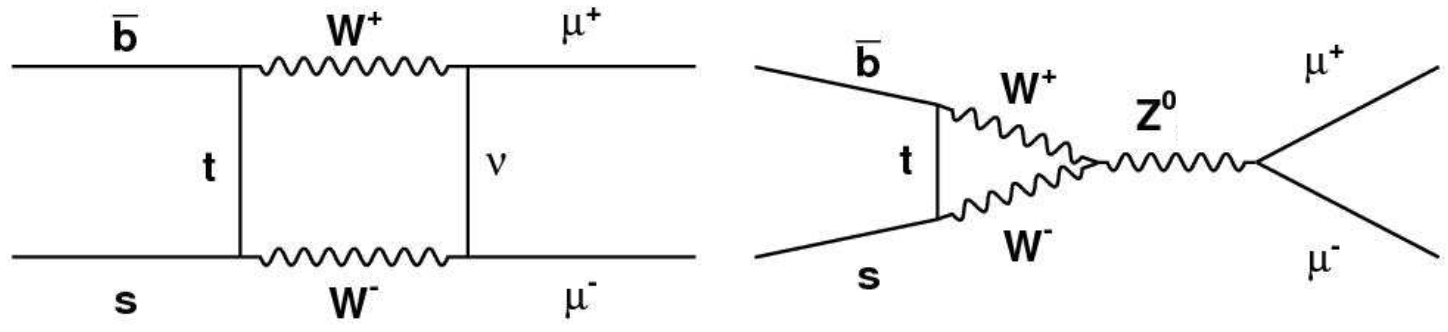}
   \caption[Examples of SM diagrams contributing to $\overline{{\mathcal B}}(B_s \to
   \mu^+ \mu^-)$.]{Examples of SM diagrams contributing to ${\mathcal B}(B_s
   \to \mu^+ \mu^-)$. The drawings have been adopted from
   Ref.\cite{Abazov:2013wjb}.\label{bsmumudiagsSM}}
    \end{figure}
    Current measurements by CMS and LHCb give the following weighted average\cite{CMS:2014xfa}:
\begin{equation} \label{bsmumu.exp}
 {\mathcal B}(B_s \to \mu^+ \mu^-)_{\rm exp}=\left(2.8^{+0.7}_{-0.6}\right) \times 10^{-9}, 
\end{equation}
whereas the SM prediction reads\cite{Bobeth:2013uxa}
\begin{equation} \label{bsmumu.SM}
{\mathcal B}(B_s \to \mu^+ \mu^-)_{\rm SM}=(3.65 \pm 0.23) \times 10^{-9}. 
\end{equation}
  \item $\overline{{\mathcal B}}_{d\mu} \equiv {\mathcal B}(B_d \to \mu^+ \mu^-)$ \\[1mm] 
  This is an analogous process involving the $B_d^0$ meson. The CMS and LHCb experiments 
  have published the following average of their branching ratio measurements\cite{CMS:2014xfa}:
 \begin{equation} \label{bdmumu.exp}
  \brbdmumu_{\rm exp}  =(3.9^{+1.6}_{-1.4}) \times 10^{-10}, 
  \end{equation}
while the SM calculation returns\cite{Bobeth:2013uxa}
 \begin{equation} \label{bdmumu.SM}
  \brbdmumu_{\rm SM} =(1.06 \pm 0.09)\times 10^{-10}. 
  \end{equation}
  \item ${\mathcal B}_\gamma \equiv {\mathcal B}(\bar B \to X_s \gamma)$ \\[1mm]
  This process is an inclusive decay of $\bar{B}^0$ ($\bar{d}b$) or
  $B^-$ ($\bar{u}b$) into a photon and charmless hadrons with non-zero 
  overall strangeness.\footnote{
  It means that \emph{none} of these hadrons is allowed to contain a valence $c$-quark,
  while the non-vanishing strangeness requirement refers to the total hadronic state
  $X_s$.}
  The leading SM contributions to this process are shown in Fig.~\ref{bsgdiagsSM}, whereas a 
  few examples of the MSSM contributions can be seen in Fig.~\ref{bsgdiagsMSSM}.
  \begin{figure}[t]
  \begin{center}     
  \includegraphics[width=0.7\textwidth]{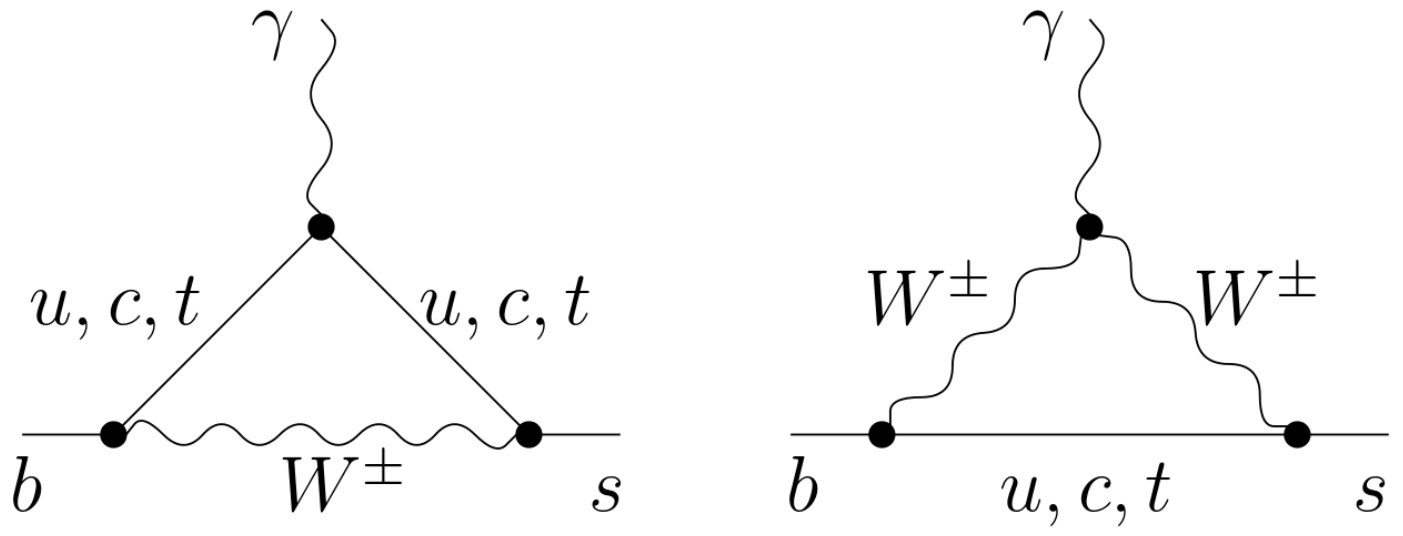}
  \end{center} 
  \caption[SM LO Feynman diagrams contributing to ${\mathcal B}(\bar B \to X_s
   \gamma)$]{Leading order diagrams contributing to ${\mathcal B}(\bar B \to X_s
   \gamma)$. The drawings have been adopted from Ref.\cite{Buras:2002er}.\label{bsgdiagsSM}}
    \end{figure}
    \begin{figure}[t]     
    \begin{center}
    \includegraphics[width=0.7\textwidth]{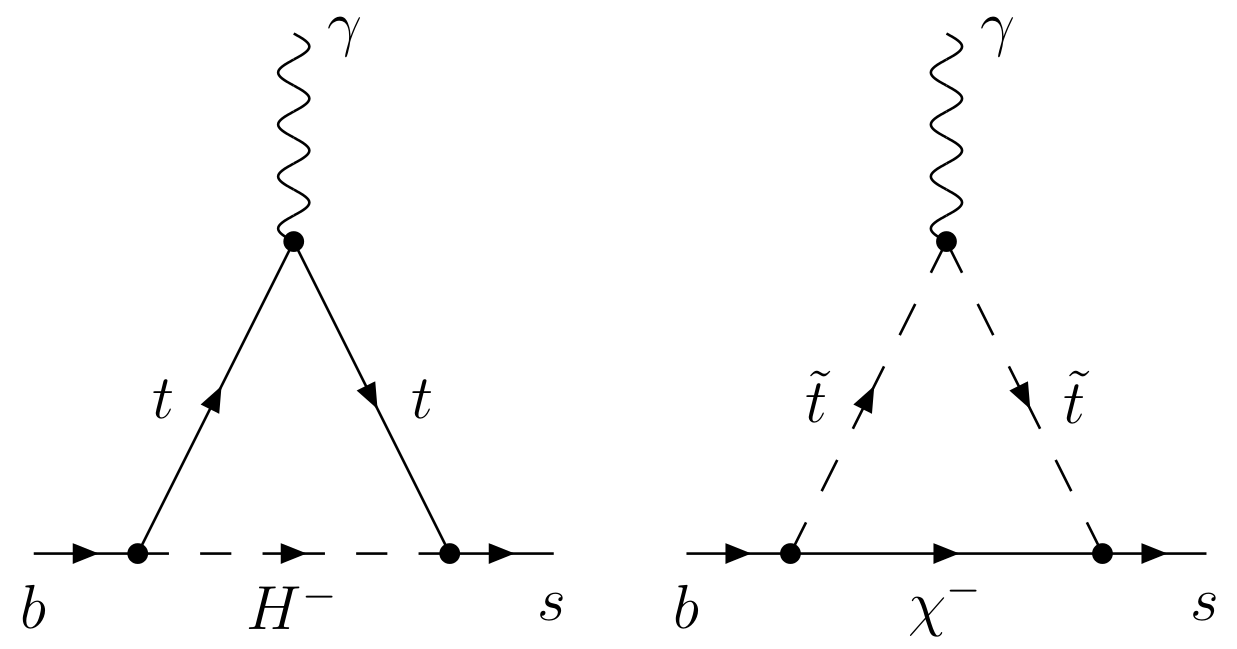}
    \end{center}
   \caption[Sample MSSM diagrams contributing to ${\mathcal B}(\bar B \to X_s
   \gamma)$]{Sample MSSM diagrams contributing to ${\mathcal B}(\bar B \to X_s
   \gamma)$. The drawings have been adopted from Ref.\cite{Misiak:1997ei}.\label{bsgdiagsMSSM}}
    \end{figure}
The inclusive branching ratio ${\mathcal B}(\bar B \to X_s \gamma)$ was
first measured in 1995 by the CLEO collaboration\cite{CLEObsg}.  The current experimental 
world average reads\cite{Amhis:2014hma}:
    \begin{equation} \label{bgamma.exp}
   {\mathcal B}(\bar B \to X_s \gamma)^{\rm exp}_{E_\gamma > 1.6\,{\rm GeV}}=
   (3.43 \pm 0.22) \times 10^{-4}.
    \end{equation}
It agrees very well with the SM prediction\cite{Misiak:2015xwa}
\begin{equation} \label{bgamma2015}
{\mathcal B}(\bar B \to X_s \gamma)^{\rm SM}_{E_\gamma > 1.6\,{\rm GeV}}=
(3.36  \pm 0.23) \times 10^{-4}.
\end{equation}
It is an untypical FCNC process, as the GIM mechanism in the SM does not
provide any extra suppression except for the loop factor and the
electromagnetic coupling.  In effect, the branching ratio is of the same order
as $\alpha_{\rm em}/(4\pi) \simeq 6\times 10^{-4}$. Such a ``big'' branching
ratio helps in performing accurate measurements. On the other hand, the
sensitivity to beyond-SM effects remains exceptional. To understand the latter
point one needs to recall that the leading contributions to the $B$-meson
decay originate from three-body partonic processes ($b \to c \ell \bar\nu$, $b
\to c d \bar u$, etc.), while the radiative decay is generated by a two body
transition $b \to s \gamma$ that comes with a CKM factor of practically the
same size. The phase-space enhancement of the two-body mode gets compensated
by the chirality suppression factor, namely by the fact that $b \to s \gamma$
proceeds via a dipole-type interaction that changes chirality of the
quarks. In the SM, the chirality shift results in an extra factor of $m_b/M_W$
in the amplitude. However, in the second MSSM diagram of
Fig.~\ref{bsgdiagsMSSM}, the factor of $m_b/M_W$ gets replaced by (roughly)
$m_b\tan\beta\, \sin\theta_{\chi}\, m_{\chi^-}/{\rm max}(m^2_{\chi^-},m^2_{\tilde t})$ 
where $\theta_{\chi}$ is the higgsino-wino mixing angle. Thus, the SUSY
contributions can get enhanced by $\tan\beta$, which acts against
the usual suppression by the superpartner masses. For heavy superpartners, one
additionally needs to take into account that the higgsino-wino mixing angle
tends to zero in the SUSY decoupling limit.

\end{enumerate}
  
\subsection{Lepton Flavour Violation}

Lepton Flavour Violating (LFV) processes belong to the observables that 
provide the most severe constraints on GUT-motivated models with new
sources of flavour mixing. In the SM, mixing between different
lepton generations comes only through higher-dimensional operators.\footnote{
By writing ``in the SM'' we understand working under the assumption that
  all the beyond-SM degrees of freedom are much heavier than the electroweak
  scale and could have been decoupled.}
One of such operators has been given in Eq.~\eqref{qnunu}. It is
responsible for generating the observed neutrino masses and mixings. These
tiny masses and sizeable mixing angles do imply existence of charged-lepton
LVF processes but with extremely low rates. If the neutrino mixings are
the only source of LFV then, for instance, the branching ratio of $\mu^+
\rightarrow e^+ \gamma$ decay is estimated at the level of
$10^{-54}$\cite{Bernstein:2013hba}. Thus practically no SM background is
present in using this process for testing new physics models with LFV
parameters.
\begin{figure}[t]
\includegraphics[width=\textwidth]{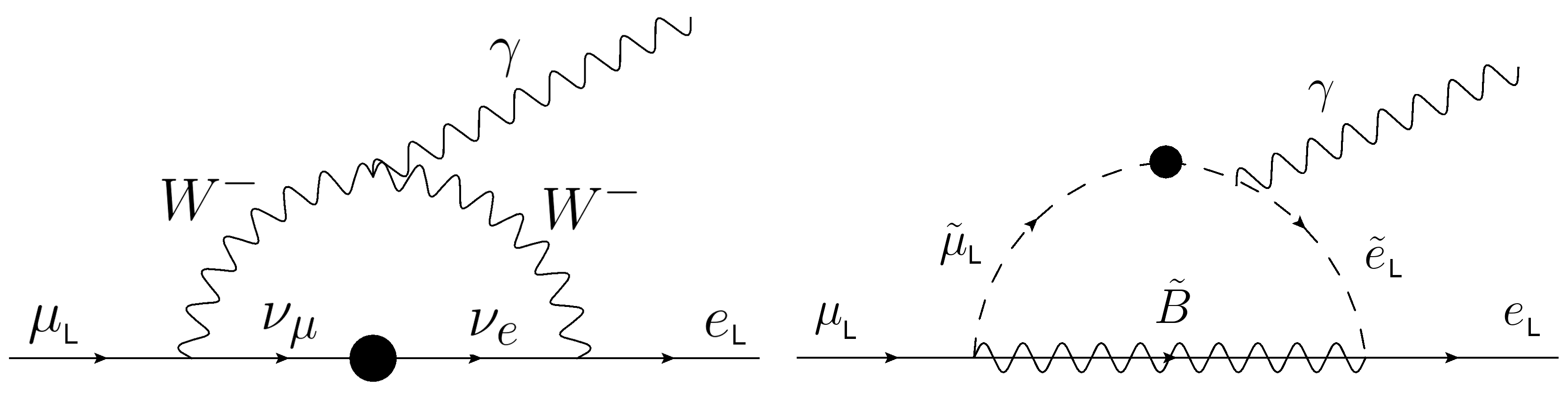} 
\caption[Sample SM and MSSM diagrams generating the $\mu^+ \rightarrow e^+
\gamma$ decay.]{Sample diagrams generating the $\mu^+ \rightarrow e^+ \gamma$
decay. The relevant flavour changing vertex is marked by a
dot. Left: decay via neutrino mixing. Right: A possible decay
mechanism in the GFV MSSM. \label{muegammaDiag}}
\end{figure}

Two experimental bounds on LFV are of special importance for our
scenario with flavour mixing in the soft mass terms:
\begin{enumerate}
\item ${\mathcal B}(\mu^+ \rightarrow e^+ \gamma$) \\
 Sample diagrams contributing to this process are presented in
 Fig.~\ref{muegammaDiag} for the SM (left) and GFV MSSM 
 (both). The first historical search was reported in
 Ref.\cite{Pontecorvo}, and the upper limit has been
 continuously improving over the following decades. The choice of $\mu^+$
 rather than $\mu^-$ has a technical motivation: $\mu^+$ does not get
 captured by nuclei, which helps to simplify the experimental background
 subtraction as compared to the $\mu^-$ decays. The present
 upper bound comes from the MEG experiment.  It
 reads\cite{Adam:2013mnn}
\begin{equation}
{\mathcal B}(\mu^+ \rightarrow e^+ \gamma) < 5.7  \times 10^{-13}.
\end{equation}
\item ${\mathcal B}(\tau^{\pm} \rightarrow \mu^{\pm} \gamma)$ \\ 

The mechanism of this decay is analogous to the previous one. The
current upper bound reads\cite{Aubert:2009ag}
\begin{equation}
{\mathcal B}(\tau^{\pm} \rightarrow \mu^{\pm} \gamma) < 4.4 \times 10^{-8}.
\end{equation}
\end{enumerate}

%% file: chapters/3_2_mh0.tex
In the years surrounding the official announcement of the Higgs boson
discovery by the LHC experiments ATLAS and CMS in 2012\cite{Aad:2012tfa,
Chatrchyan:2012ufa}, the main aim in most of the supersymmetric
phenomenology research was to correctly reproduce the observed
Higgs boson mass $m_{h^0}$. The combined measurements as of March 2015
yield\cite{Aad:2015zhl}
\begin{equation}
 m_{h^0}^{\rm exp} ~=~ 125.09 \pm 0.21({\rm stat.}) \pm 0.11({\rm syst.}) \gev. 
\end{equation}
As previously mentioned, it is not straightforward to obtain this number in a
simple supersymmetric extension of the SM. The MSSM requires certain adjustment
of parameters to provide a correct prediction. At the tree level, the mass of
the lightest neutral Higgs boson is lower than $M_Z$. Only the
higher-order contributions can raise it to the observed value. If the
pseudoscalar $A^0$ is much heavier than $M_Z$, the expression for $m_{h^0}^{2}$ 
including the leading one-loop corrections is given by\cite{Carena:1995bx}
%
%
%
\begin{equation}
 m_{h^0}^{2} \approx M_Z^2 \cos^2 2 \beta + \frac{3m_t^4 }{2 \pi^2 v^2} 
\left[ \log\frac{M_{\rm SUSY}^2}{m_t^2}
+ \frac{X_t^2}{ M_{\rm SUSY}^2} \left( 1 - \frac{X_t^2}{12M_{\rm SUSY}^2} \right) \right],
\end{equation}
with~ $v \simeq 246\,$GeV,~ 
$X_t = A^u_{33}/Y_t - \mu \cot \beta$~ 
and~ $\msusy =  \sqrt{m_{\tilde{t}_1}m_{\tilde{t}_1}}$.
The QCD-running couplings ($Y_t$, $A^u_{33}$) and masses ($m_t$,
$m_{\tilde{t}_i}$) in the above equations should be taken at the scale
\msusy{} to minimize higher-order effects. One can observe that to
maximise the Higgs mass, a high \tanb{} should be chosen, both stops should be
heavy, and the stop mixing encoded in the parameter $|X_t|$ should be 
large, i.e. $|X_t|$ should be close to $\msusy\sqrt{6}$. This is the
reason why for given soft mass terms it is the $A^u_{33}$ parameter
that can be used to adjust the Higgs mass.

An extant phenomenological trouble is the theoretical uncertainty of
the Higgs mass predictions. It has been estimated at the level
of $\pm 3 \gev$\cite{Allanach:2004rh} (i.e.\ much
larger than the current experimental error) on the basis of
differences between results obtained in various approximations by 
several of the available spectrum generators.

%% file: chapters/3_3_SUSY_LHC.tex
   After decades of searches for supersymmetry at various colliders, the LHC
   has provided the most recent constraints on particular MSSM parameters or
   specific configurations of them. The results collected at the center-of-mass energy of 7
   and 8$\,$TeV clearly indicate that the superpartners should have
   masses of at least $\mathcal{O}(\text{TeV})$. Each of the constraints is based on particular
   simplifying assumptions, which makes the overall picture extremely complex.
   However, the basic fact is that no search has provided any significant signal in
   favour of supersymmetry. The most stringent and model-independent
   constraints are observed for the gluino being the particle that should be
   most abundantly produced at the LHC. 
   
\begin{sidewaysfigure}
\begin{center}
\includegraphics[width=\textwidth]{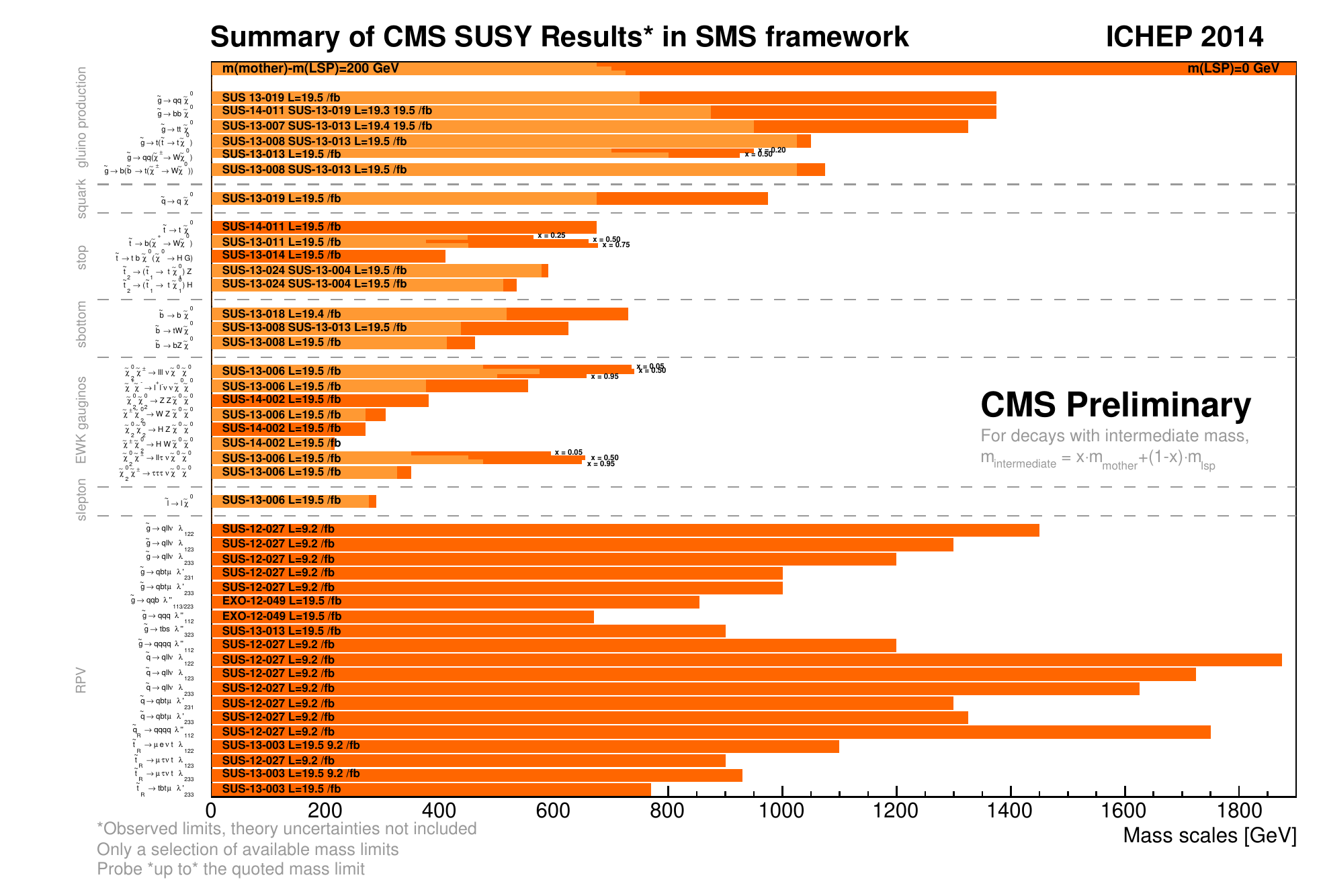}
\caption[CMS SUSY searches summary plot for the 8 TeV data.]{Plot summarising
the SUSY searches performed by the CMS collaboration with 8 TeV data. It shows
the best exclusion limits for masses of the mother particles when
$m_{\rm LSP} = 0$ (dark shades) and $m_{\rm mother} - m_{\rm LSP} = 200\,$GeV (light
shades). The indicated values are to be interpreted as upper bounds on the mass
limits. The plot has been adopted from Ref.\cite{CMSwebsite}. \label{CMSsummary}}
\end{center}
\end{sidewaysfigure}

\begin{sidewaysfigure}
\begin{center}
\includegraphics[width=\textwidth]{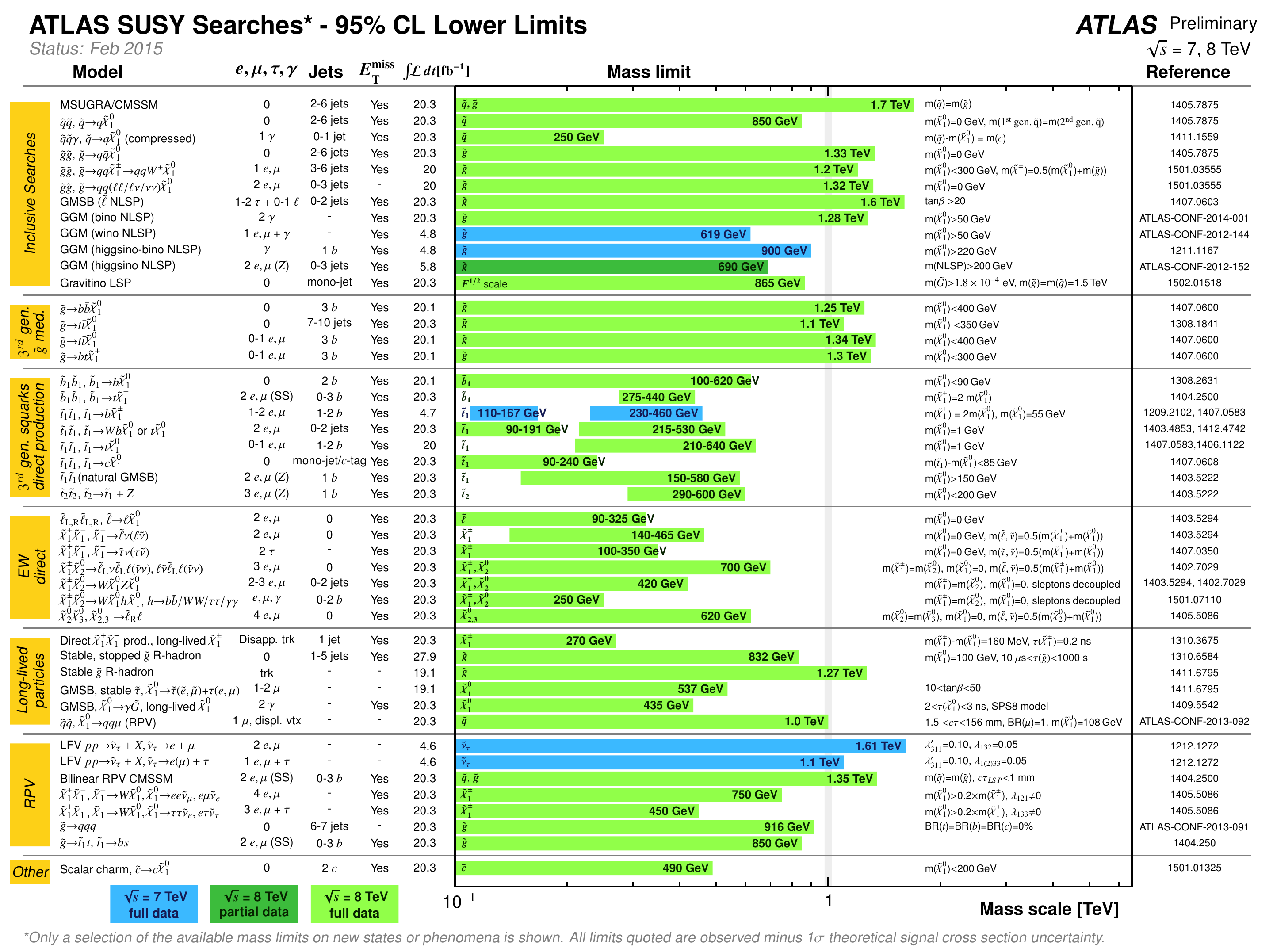}
\caption[ATLAS SUSY searches summary plot for the 8 TeV data.]{Plot
summarising the SUSY searches performed by the ATLAS collaboration with 8 TeV
data. It shows exclusion limits for masses of the sparticles under
the assumptions stated adjacent to each
bar. The plot has been adopted from Ref.\cite{ATLASwebsite}. \label{ATLASsummary}}
\end{center}
\end{sidewaysfigure}

   The summaries of searches at 8 TeV performed by the CMS and ATLAS
   collaborations are presented in Figs.~\ref{CMSsummary} and~\ref{ATLASsummary}.
   They provide a good illustration of the current frontier of direct SUSY
   searches.

%% file: chapters/3_4_vacStab.tex
To discuss another important requirement imposed on the MSSM,
let us briefly recall the issue of the EW vacuum stability. In the
Minkowski spacetime, a construction of the Hilbert space over which 
the operator fields act starts from a Lorentz invariant \emph{vacuum
state}.  In the local and perturbative QFT case, we begin with
determining the configuration of fundamental scalar fields
that minimises their classical Hamiltonian.  The positive definiteness
of the scalars' kinetic terms in the Hamiltonian ensures that 
only constant fields can be non-vanishing in the minimum. By minimising the
potential with respect to them, we find tree-level approximations for the
\emph{vacuum expectation values} (VEVs) of the scalar fields
$\phi(x)=v_{\phi}=\rm{const.}$ Next, builiding up a perturbative QFT
around the tree-level minimum, we find the loop-corrected effective potential
and, order-by-order, the loop-corrected VEVs.

The problem of finding a set of tree-level VEVs in the MSSM 
for arbitrary values of its parameters has not been solved so far due
to immense dimensionality of the field space. In practice, 
one only tests particular menaces at certain classes of directions
in this space, namely whether the potential is unbounded from below (UFB)
or has deeper minima than the standard Higgs one. In the latter case, the
VEVs in the global minimum would necessarily involve electrically charged
scalars or the sneutrinos (having $L\neq 0$), which would stand against the
experimental evidence. Such non-standard deeper minima are called
Charge- or Colour-Breaking (CCB) vacua, as the sneutrino
alone (without any charged scalar VEV) cannot produce a negative contribution
to the classical vacuum energy. The appearance of a CCB minimum or the
UFB configuration does not necessarily exclude the considered
set of the MSSM parameters. One can assume that the
cosmological history started with descending into a metastable
vacuum state (in our case -- the standard Higgs minimum). Then, as long as the
standard Higgs vacuum lives longer than the age of the Universe, the model would
not contradict any of our observations. There is a vast literature on testing
the vacuum stability conditions in the MSSM. Some of the most
dangerous field configurations have been considered in
Ref.\cite{Casas:1995pd}.

%% file: chapters/3_5_DarkMatter.tex
 In this section, we briefly sketch the main facts about Dark Matter (DM)
 which belongs to the main unsolved puzzles of contemporary fundamental
 science. Particle physics offers possible explanations to the problem 
 by means of introducing a beyond-SM stable particle. In
 this work, we find interesting correlations of the DM issue with our
 GFV scenario.
 
\begin{figure}[h]     
\includegraphics[width=\textwidth]{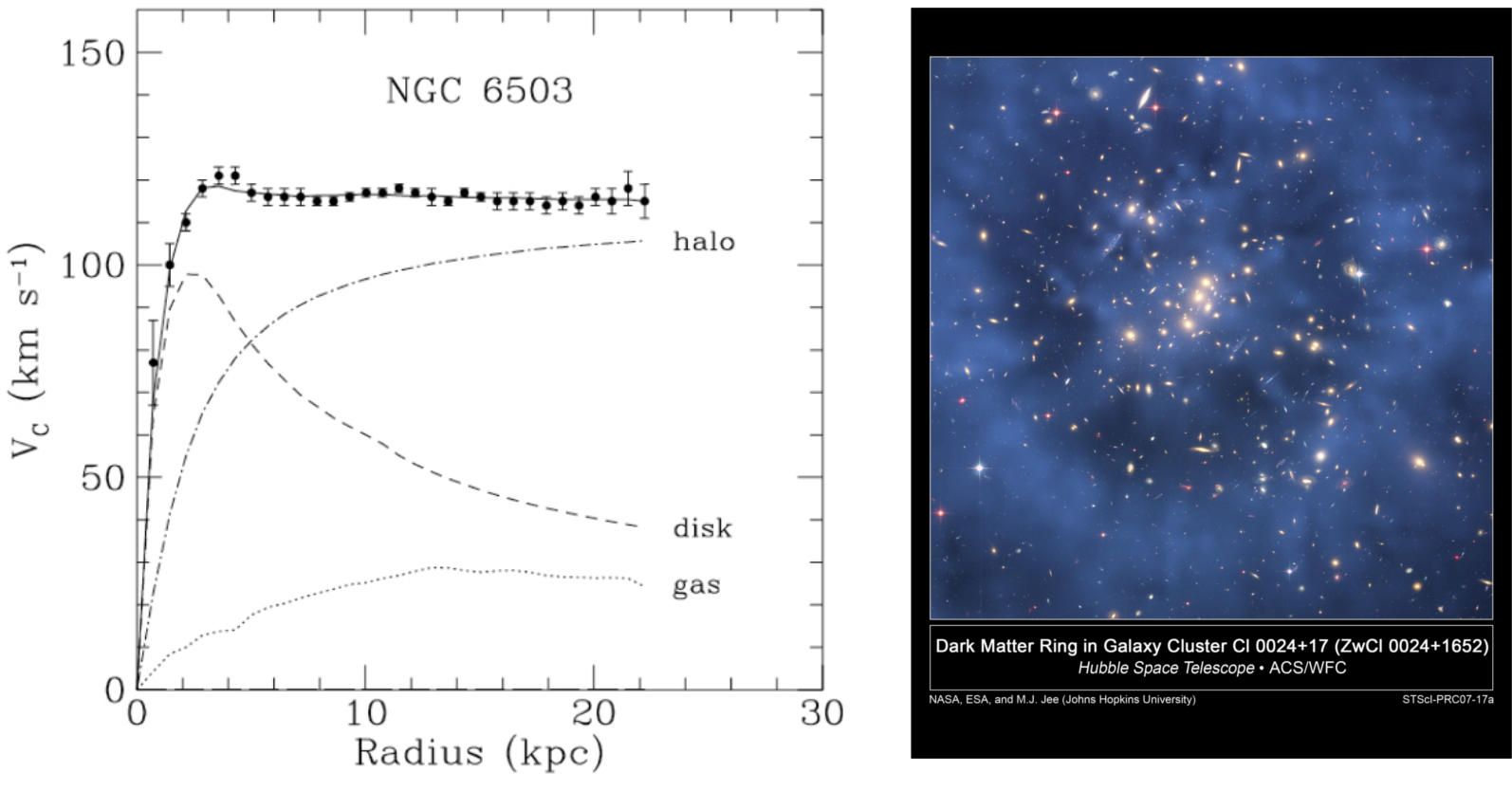}
   \caption[Evidence of dark matter - galaxy rotation curves and gravitational
   lensing.]{Examples of observational evidence for the dark matter existence.
   Left:
   the rotation curve of NGC 6503 galaxy -- the data points and a fit split into
   contributions from the light-emitting components (disk), gas, and the missing dark
   component (halo)\cite{rotCurvesIm}.
   Right:
   a blue map of the cluster Cl 0024+17 dark matter distribution superimposed
   on the Hubble Space Telescope image of the cluster. The map was inferred from
   the observed gravitational lensing (source:
   Ref.\cite{HubbleIm}).\label{ObsEvDM}}
   \end{figure}

 The existence of DM was postulated for the first time when astronomers could not explain velocities of stars in Milky Way relying only on masses of visible objects\cite{Oort}. The same problem arose also for the galaxies in the Coma cluster\cite{Zwicky}. By now,
 there is a vast amount of data consistently pointing that a yet-unknown
 type of gravitationally interacting matter exists (see e.g.,
 Ref.\cite{DMrev}). Two examples are shown in Fig.~\ref{ObsEvDM}.
   
Still, it is not known what kind of matter constitutes the DM. Many
astronomical-scale objects and all the SM particles are excluded as
viable candidates\cite{DMnotSM}. A range of proposals to explain the nature of
DM exists. A notable position among them have the ones that
assume that it consists of yet unknown elementary particles. This could have
an impact on our quantum theories of fundamental interactions, especially if the
DM candidate interacts not only gravitationally with the SM fields. An immensely popular
idea advertises new Weakly Interacting Massive Particles (WIMPs) that would be
consistent with present data coming from astronomy and particle physics, yet
could be described with the means of the familiar perturbative QFT
techniques. Moreover, a generic model involving such matter, $\Lambda_{\rm CDM}$,
is able to perfectly fit also the well-studied observational data that 
come from the Cosmic Microwave Background (CMB)\cite{Ade:2013zuv}.

One of the phenomenological side-effects of the $R$-parity assumption in the
MSSM is that it provides a viable DM candidate. The $R$-parity implies that the
Lightest Supersymmetric Particle (LSP) cannot decay to possibly lighter SM
particles and, being stable, it is bound to permeate the Universe. There still
remains, however, a quantitative test that is highly non-trivial to pass. This
crucial observable is the current DM density in the Universe.

The DM relic density can be calculated in the MSSM, as long as a
cosmological model is selected. The so-called \emph{standard
cosmological model} relies on the assumption that the Universe on sufficiently
large scales is homogeneous and isotropic. It can be then described by the
Friedman-Lema\^itre-Robertson-Walker (FLRW) metric
\begin{equation}
 ds^2= - dt^2 + a^2(t) d\Sigma^2,
\end{equation}
where $\Sigma$ is a metric on a 3-dimensional Euclidean manifold of uniform
curvature. In the polar coordinates $r, \theta, \phi$, we have
\begin{equation}
 d\Sigma^2=\frac{dr^2}{1-kr^2} + r^2 (d\theta^2 + \sin^2\theta d\phi^2),
\end{equation}
with $k\in\{-1,0,1\}$ describing the curvature type. The
dynamics of the Universe is captured in the term $a(t)$, and computable given
the universal energy density in the Universe $\rho(t)$. If the Universe
is spatially flat ($k=0$), the total energy density is said to
be \emph{critical} and obeys the following equation:
\begin{equation}
 \rho_{\rm crit}=\frac{3}{8 \pi G} \left( \frac{\dot{a}}{a} \right)^2,
\end{equation}
with $G$ being the Newton's gravitational constant ($6.70837 \times 10^{-39}
\gev^{-2}$).

The $\Lambda_{\rm CDM}$ model that currently provides the best fit to the CMB data
splits the energy density $\rho$ into three parts:
\begin{equation}
 \rho= \rho_b + \rho_{\chi} + \frac{\Lambda}{8 \pi G},
\end{equation}
where $\rho_b$ is the baryonic matter energy density (including also small 
contributions from electrons, photons and neutrinos), $\rho_{\chi}$ is that
of dark matter, and $\Lambda$ is the cosmological constant. The
energy density $\rho_\chi$ is usually expressed in units of the critical density via
\begin{equation}
 \Omega_{\chi} \equiv \frac{\rho_{\chi}}{\rho_{\rm crit}}.
\end{equation}
The $\Lambda_{\rm CDM}$ fit in Ref.\cite{Ade:2013zuv} returns the dark
matter relic abundance in terms of the parameter 
\begin{equation}
 \abundchi = 0.1199 \pm 0.0027	
\end{equation}
which additionally contains the dimensionless constant
 $h=0.673(12)$\cite{Agashe:2014kda} that parameterises
 the current expansion rate of the Universe 
\begin{equation}
  \frac{\dot{a}}{a}({\rm now}) = h\times 100\;{\textstyle \frac{{\rm km/s}}{{\rm Mpc}}}.
\end{equation}

The relic abundance of dark matter is often
the most stringent constraint in GUT-constrained SUSY scenarios. The lightest neutralino (see
\refsec{sec:MSSMspectr}) is a favoured candidate for the LSP. It
is well known that the properties of this DM candidate strongly depend on its
composition. If the lightest neutralino is almost purely a bino, the relic
density is generally too large.
 
In the cosmological past, the neutralinos could have been produced
thermally. Their present density depends on their annihilation rates to other
particles. Since in the bino case the neutralino relic abundance is too high,
its annihilation cross-section needs to be enhanced by a particular
mechanism. A good solution is a co-annihilation with the lightest sfermion, or
a resonance annihilation through one of the Higgs bosons. On the other hand, a
significant higgsino component of the neutralino opens a possibility of
efficient annihilation into gauge bosons. In fact, the annihilation
cross-section in such a case is usually too large, and it leads to the DM
underabundance.

Complementary to indirect estimates, a range of direct searches for WIMP DM
have been performed. Typically, an underground tank filled with a noble gas
(e.g., xenon), well isolated from other particle sources, is observed over a
long period of time in expectation for an interaction with a DM
particle. For a review of results and techniques applied in various
experiments, see, e.g., Ref\cite{Cushman:2013zza}. In this work,
we consider only the currently most stringent bounds on the
spin-independent neutralino-proton scattering cross section \sigsip{} from the
LUX experiment\cite{Akerib:2013tjd}.

%% file: chapters/4_Tools.tex
\section{Spectrum generators}

Let us briefly summarise the problem we face when quantitatively testing the
Yukawa matrix unification in the MSSM. In principle, we could
specify each point of the parameter space in terms of the gauge,
superpotential and soft couplings at \mgut{}. Next, we
could check whether it is acceptable using the low-energy SM 
determinations of the gauge and Yukawa couplings, the Higgs
boson mass, and its quartic coupling.  Finally, we could test all the other
observational constraints: the direct search limits, flavour and
electroweak observables, as well as the predicted dark matter relic density.

In practice, the problem is somewhat more complicated because some of the
input parameters need to be initially specified at the low-energy scale, to
avoid inefficient scans over the multi-dimensional parameter space. This
refers not only to the gauge and Yukawa couplings but also to the $\mu$ and
$B\mu$ parameters that are necessary for a proper EWSB.  For
every point in the parameter space, we employ an
iterative procedure that determines the complete MSSM mass spectrum, 
as well as values of all the couplings at \mgut{}.

As a first step to be made, one has to numerically solve the nonlinear MSSM
RGEs with boundary conditions subsequently specified at three
different scales:
\begin{enumerate}
 \item $\mu_{\rm sp}$~ where the SM gauge and Yukawa couplings are translated
 into the MSSM ones (with the threshold corrections 
 being calculated in an iterative manner, assuming no such corrections
 at the first step),
 \item $M_{\rm EWSB} \equiv M_{\rm SUSY} =\sqrt{m_{\tilde{t}_1}
 m_{\tilde{t}_2}}$~  where the minimisation of the MSSM scalar potential is
 performed. This scale is chosen with the aim to reduce the sensitivity
 of the EWSB conditions to higher-order perturbative corrections.
 \item \mgut{}~ where the soft terms are specified as the input parameters 
 that determine the superpartner spectra and, consequently, the threshold corrections
     for the next steps of the iteration.
\end{enumerate}
\begin{figure}[t]
\begin{center}
\includegraphics[width=0.7\textwidth]{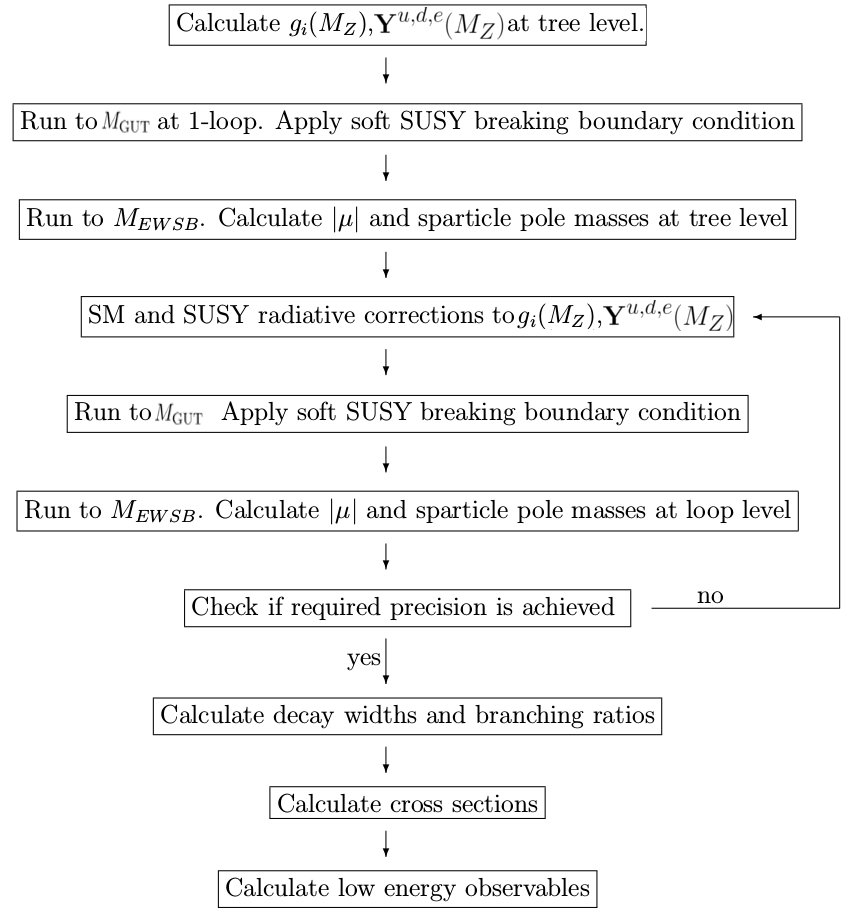}
\caption[Iterative procedure of solving RGE problem in MSSM.]{An iterative
algorithm finding a solution of the RGE problem in the MSSM. The
illustration has been adopted from the {\tt SPheno}
manual\cite{SPheno1}. \label{iterDiag}}
\end{center}
\end{figure}

Public programs, called \emph{spectrum generators}, implement this
iterative algorithm in a generically similar manner, but with
convention-dependent details. The procedure used by the code {\tt
SPheno}\cite{SPheno1} is sketched in Fig.~\ref{iterDiag}. The boundary with
the SM (i.e. the scale $\mu_{\rm sp}$) is currently set by most of the public
programs to be at $M_Z$.  Such a choice has considerable disadvantages, one of
which is excluding too many parameter points from the analysis. For instance,
some scalar fields may become formally tachyonic (having a
negative mass squared) only well below their proper
decoupling scale but above $M_Z$, which is still acceptable, though most
programs usually reject such points.

Another caveat is that there may exist acceptable points in the GUT-scale
parameter space to which such an iterative procedure can never converge.  In
other words, since the starting-point data are not specified in terms of
single-scale boundary conditions, the problem may have multiple
solutions. Examples of such situations have been presented in
Ref.\cite{Allanach:2013cda}. This fact is of phenomenological relevance, as it
affects the MSSM dark matter analyses performed with the standard
iterative codes\cite{Allanach:2013yua}. However, this problem hardly matters
for our present investigation, as we only demonstrate existence of regions in
the parameter space where the Yukawa unification constraint is satisfied.

\section{Numerical setup for the large $A$-term scenario}

For the purpose of the analysis of the large $A$-term scenario,
we have modified \linebreak {\tt SOFTSUSY~3.3.8}\cite{Allanach:2001kg} that
distinguishes itself among other spectrum generators by possessing a technical
documentation and clear structure. Following Ref.\cite{Crivellin:2011jt},
we modified this code by implementing threshold corrections to
the first and second family Yukawa couplings, as well as to the CKM matrix.
It was necessary because such corrections were missing in {\tt SOFTSUSY}
at the time when the work was performed. As a by-product, we found some
points to correct in both {\tt SOFTSUSY} and {\tt SPheno} which were then
implemented by the authors\cite{priv.corr}.
\begin{table}[htbp]
\centering
\renewcommand{\arraystretch}{1.5}
\begin{tabular}{| l | l | l | l | l | l | l | l |}
\hline
\input{tabs/SMinput_horizont.tab}
\end{tabular}
\caption[Standard Model parameters used in numerical calculations for the
large diagonal A-term scenario.]{Standard Model
parameters\cite{Beringer:1900zz} used in our numerical calculations for
the diagonal $A$-term scenario. The
light ($u$, $d$, $s$) quark masses are $\overline{\rm MS}$-renormalised
at 2$\,$GeV. \label{SMinput}}
\end{table}

Our input values of the SM parameters are collected in
Tab.~\ref{SMinput}. Flavour observables are calculated with the help of {\tt
SUSY\_FLAVOR~v2.10}\cite{Crivellin:2012jv}. This code evaluates the
renormalised MSSM Yukawa matrices and obtains the proper CKM matrix also
according to the prescriptions of Ref.\cite{Crivellin:2011jt}, taking the
previously determined soft parameters as input.

For the purpose of the final scan that delivered the regions consistent with
the Yukawa unification, we used the BayesFITSv3.2\cite{BayesFits} numerical
package that interfaces several publicly available codes. Except the
above-mentioned programs, it uses \texttt{\multinest\ v2.7}\cite{Feroz:2008xx}
which allows for fast and efficient Markov Chain Monte Carlo (MCMC) scanning
according to a pre-defined likelihood function.  For the $SU(5)$ boundary
condition in Eq.~(\ref{yukunif}), we assume a Gaussian likelihood distribution
\begin{equation}\label{TresholdGUT}
 \like_{\rm Yuk}=\sum_{i=1,2,3}\exp\left[-(1-Y^e_{ii}(\mgut)/Y^d_{ii}(\mgut))^2
/2\sigma_{\scriptscriptstyle\rm Yuk}^2\right],
\end{equation}
with $\sigma_{\scriptscriptstyle\rm Yuk}$ set to 5\% to allow for deviations
from the exact unification condition.

\section{Numerical tools for the GFV MSSM scenario}

To investigate the second scenario, we have employed the same package
BayesFITSv3.2. It was first developed in 
Ref.\cite{BayesFits}, and then modified to incorporate the
full GFV structure of the soft SUSY-breaking sector in
Ref.\cite{Kowalska:2014opa}. The package uses for sampling the
\texttt{\multinest~v2.7}\cite{Feroz:2008xx} code which allows for fast
and efficient scanning according to a predefined likelihood function using 
the MCMC algorithms. The likelihood corresponding to the $SU(5)$
boundary condition~\eqref{yukunif} is modelled as in Eq.~\eqref{TresholdGUT},
with the same allowed deviation from the exact unification condition,
$\sigma_{\scriptscriptstyle\rm Yuk} = 0.05$.
\begin{table}[t]\footnotesize
\centering
\renewcommand{\arraystretch}{1.4}
\begin{tabular}{| c | c | c | c | c | c | c | c |}
\hline
\multicolumn{2}{|c|}{\mtpole} & \multicolumn{2}{c|}{\mbmbmsbar} & 
\multicolumn{2}{c|}{\alphasmzms} & \multicolumn{2}{c|}{\alphaemmz} \\
\multicolumn{2}{|c|}{$173.34 \pm 0.76$ GeV} & \multicolumn{2}{c|}{$4.18 \pm 0.03$ GeV} & 
\multicolumn{2}{c|}{$0.1184 \pm 0.0007$} & \multicolumn{2}{c|}{$127.944 \pm 0.015$} \\
\hline
\mumbmsbar & \mdmbmsbar & \msmbmsbar & \mcmbmsbar & \mepole & \mmupole & \mtaupole & \mzpole \\
2.3 MeV & 4.8 MeV & 95 MeV & 1.275 GeV & 511 keV & 106 MeV & 1.777 GeV & 91.19 GeV \\
\hline
\multicolumn{2}{|c|}{$\bar{\rho}$} & \multicolumn{2}{c|}{$\bar{\eta}$} & 
\multicolumn{2}{c|}{$A$} & \multicolumn{2}{c|}{$\lambda$} \\
\multicolumn{2}{|c|}{$0.159 \pm 0.045$} & \multicolumn{2}{c|}{$0.363 \pm 0.049$} & 
\multicolumn{2}{c|}{$0.802 \pm 0.020$} & \multicolumn{2}{c|}{$0.22535 \pm 0.00065$} \\
\hline
\end{tabular}
\caption[Standard Model parameters used in numerical calculations for GFV
scenario.]{Standard Model parameters\cite{Agashe:2014kda,utfit} used in our
numerical calculations for the GFV scenario. The light ($u$, $d$, $s$) quark masses are
$\overline{\rm MS}$-renormalised at 2$\,\gev$.\label{SMinput.GFV}}
\end{table}

When our calculations for the first scenario were being performed,
another version of {\tt SPheno} was released. It offered us several
necessary features that had not been available before. In
particular, a more accurate treatment of the threshold corrections to the
Yukawa couplings was implemented. The mass spectra in our GFV analysis are
therefore calculated with {\tt SPheno v3.3.3}\cite{SPhenoTrzy} and checked
against results obtained with the modified {\tt SOFTSUSY~3.3.8} 
that has been mentioned earlier. Four SM parameters (\mtpole,
\mbmbmsbar, \alphaemmz\ and \alphasmzms) are treated as nuisance parameters,
and randomly drawn from a Gaussian distribution centered around their
experimentally measured values\cite{Agashe:2014kda}. The elements of the CKM
matrix in the Wolfenstein parameterisation ($\bar{\rho}$, $\bar{\eta}$, $A$,
$\lambda$) are scanned as well, with central values and errors given by the
UTfit Collaboration for the scenario allowing new physics effects in loop
observables\cite{utfit}. The other SM parameters which are passed as an input
to \spheno\ (\msmbmsbar, \mcmbmsbar, \mdmbmsbar, \mumbmsbar, \mtaupole,
\mmupole, \mepole, \mzpole) are fixed at their experimentally measured
values. Our SM input is collected in Table~\ref{SMinput.GFV}.

The relic density and the spin-independent neutralino-proton cross section
\sigsip\ have been calculated with the help of \texttt{\dsusy\
v5.0.6}\cite{Gondolo:2004sc}. For the EW precision constraints
\texttt{\feynhiggs\
v2.10.0}\cite{Hahn:2013ria,Frank:2006yh,Degrassi:2002fi,Heinemeyer:1998yj} has
been used. To include the exclusion limits from Higgs boson searches at LEP,
Tevatron, and LHC, as well as the \chisq\ contributions from the Higgs boson
signal rates from Tevatron and LHC, we applied \texttt{\higgsbounds\
v4.0.0}\cite{Bechtle:2008jh,Bechtle:2011sb,Bechtle:2013wla} interfaced with
\texttt{\higgssignals\ v1.0.0}\cite{Bechtle:2013xfa}. Like in the previous
scenario, all the flavour observables have been evaluated with the code
\texttt{\susyflav\ v2.10}\cite{Crivellin:2012jv}.

%% file: tabs/SMinput_horizont.tab
$\alpha_s^{\overline{MS}} (M_Z)$ & 
$\alpha_{em}^{-1} (M_Z)$ & 
\multicolumn{2}{|c|}{$G_F\;\left[{\rm GeV}^{-2}\right]$} & 
$M_Z^{\rm pole}$  & 
$m_{e}^{\rm pole}$ &
$m_{\mu}^{\rm pole}$ \\
0.1184 &  
127.944 & 
\multicolumn{2}{|c|}{$1.16638 \times 10^{-5}$} & 
91.19 GeV &   
511 keV &
106 MeV \\
\hline
$m_{\tau}^{\rm pole}$ &
$m_u^{\overline{MS}}$ & 
$m_d^{\overline{MS}}$ &
$m_s^{\overline{MS}}$ &  
$m_c^{\overline{MS}}(m_c)$  &   
$m_b^{\overline{MS}}(m_b)$  &   
$m_t^{\rm pole}$ \\
1.777$\,$GeV &
2.3 MeV &  
4.8 MeV &
95 MeV  & 
 1.275 GeV & 
4.18$\,$GeV & 
173.5$\,$GeV \\
\hline

%% file: chapters/5_1_RegionsSoftDiag.tex
In this chapter, following our article in Ref.\cite{Iskrzynski:2014zla},
we describe sample regions in the MSSM parameter space that are consistent with
the unification condition $\mathbf{Y}^d \simeq
\mathbf{Y}^{e\,T}$ at \mgut{} in the flavour-diagonal $A$-term scenario. Due
to the large dimensionality of the considered parameter space, our
description of these regions will amount to showing a few of
their projections onto surfaces spanned by the most relevant 
parameters, as well as to presenting selected benchmark points.

In Figs.~\ref{scatPlotTb}--\ref{scatPlotTb2} and in Tab.~\ref{Ex}, we
present sample parameter-space regions and benchmark points where a proper
Yukawa matrix unification has been achieved in our setup. In selecting these
regions and points, we aimed at fulfilling the unification constraints and
reproducing the lightest Higgs particle mass (up to the theoretical
uncertainty of 3 GeV) for a broad range of $\tan\beta$. We have chosen the
sparticle masses so that the gluino is heavy enough to have evaded the current
bounds, but could possibly be detected in the second LHC phase.

Plots in Figs.~\ref{scatPlotTb}--\ref{scatPlotTb2} show points investigated in
our MCMC scans performed for three $\tan\beta$ intervals: [5,20], [15,30], and
[30,45]. Different colours are used to indicate successful Yukawa matrix
unification, either for all the three families or for some of them only. We
observe that Yukawa unification for all the three generations can be achieved
for a wide range of $\tan\beta$. Generically, larger values of the $A$-terms
are necessary for larger $\tan\beta$ because the down-type quark Yukawa couplings
(and thus the required threshold corrections) scale proportionally to
$\tan\beta$. For this reason, finding acceptable points for larger values of
$\tan\beta$ in each random scan required collecting much more statistics.
\SidePlotsTwoW{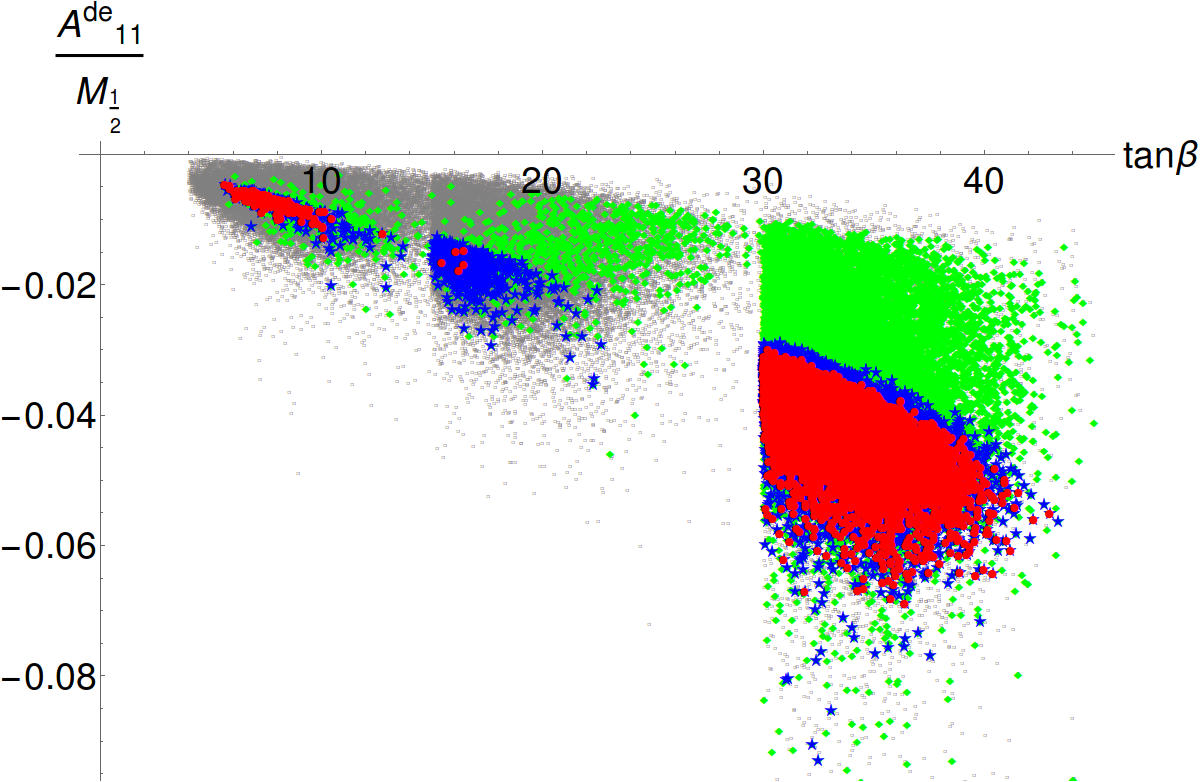}{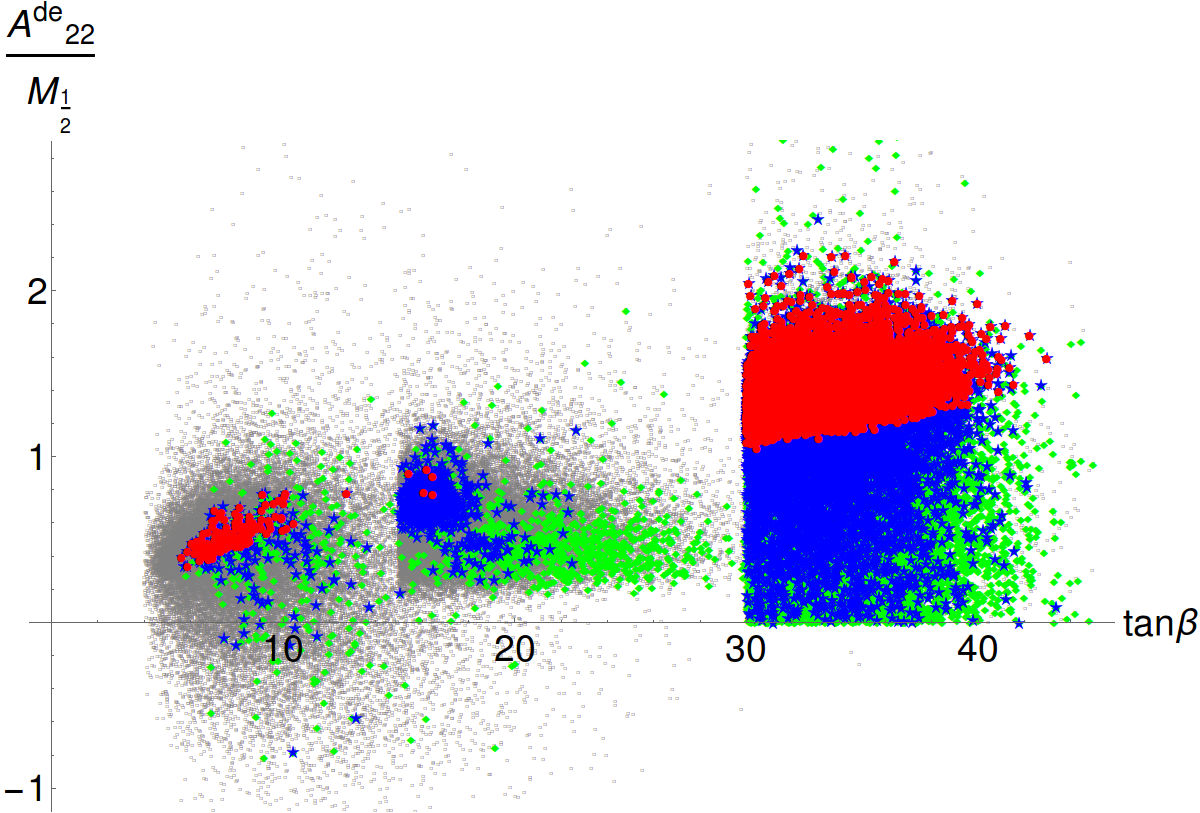}{
(a): points gathered in three of our MCMC scans (grey), shown in the
$\tan\beta \times (A^{de}_{11}/M_{1/2})$ plane.  For some of them,
the respective Yukawa couplings get unified within a 10\% bound and the Higgs
boson mass prediction lies in the interval $[122.5, 128.5]$ GeV: green
diamonds mark the $b$--$\tau$ unification, blue stars fulfil also the $d$--$e$
one, while red circles include also the $s$--$\mu$ one (i.e. the full Yukawa
matrices get unified).
(b): the same data projected onto the $\tan\beta \times
(A^{de}_{22}/M_{1/2})$ plane.  

For efficiency reasons in covering a broad $\tanb$ range, the scans were made in disjoint $\tanb$ intervals. The algorithm was finding points consistent with all the requirements quicker for lower $\tanb$, what is visible in the plot.
\label{scatPlotTb}}{
Points consistent with SU(5) Yukawa matrix unification in the $\tan\beta
\times (A^{de}_{11}/M_{1/2})$ and $\tan\beta \times
(A^{de}_{22}/M_{1/2})$ planes.}
\SidePlotsTwoW{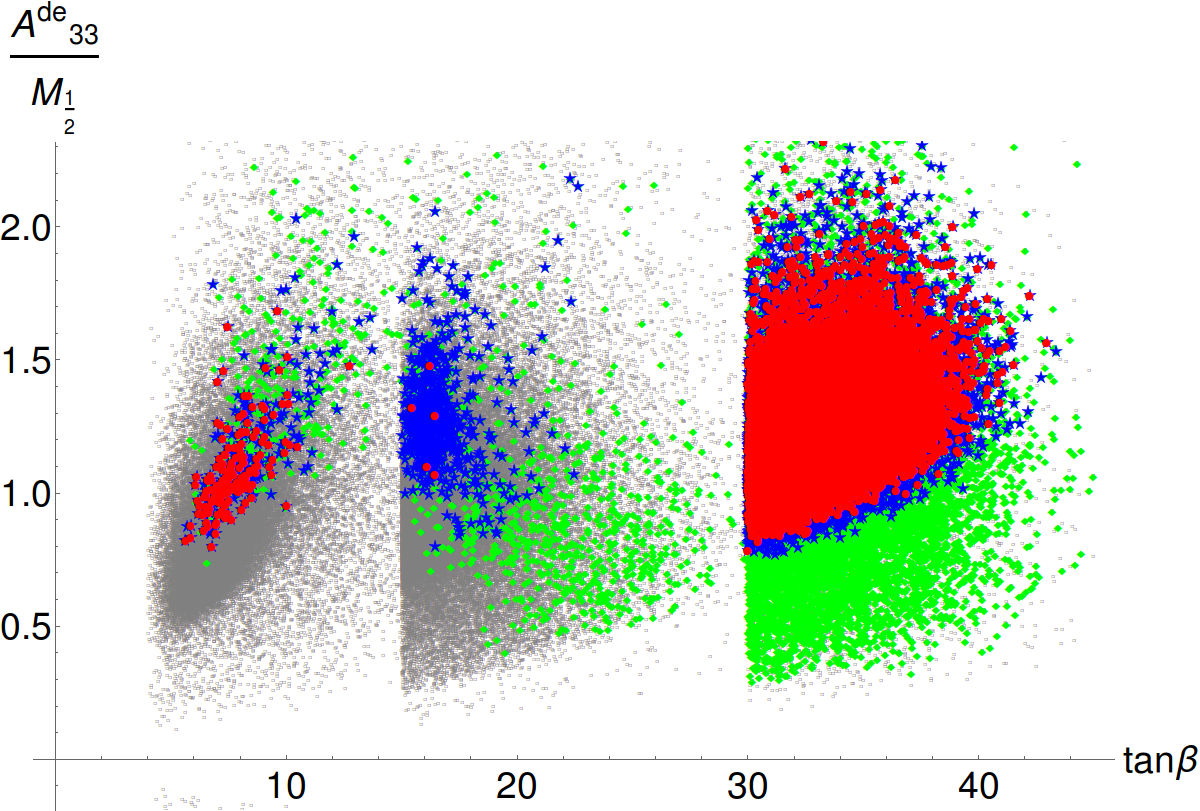}{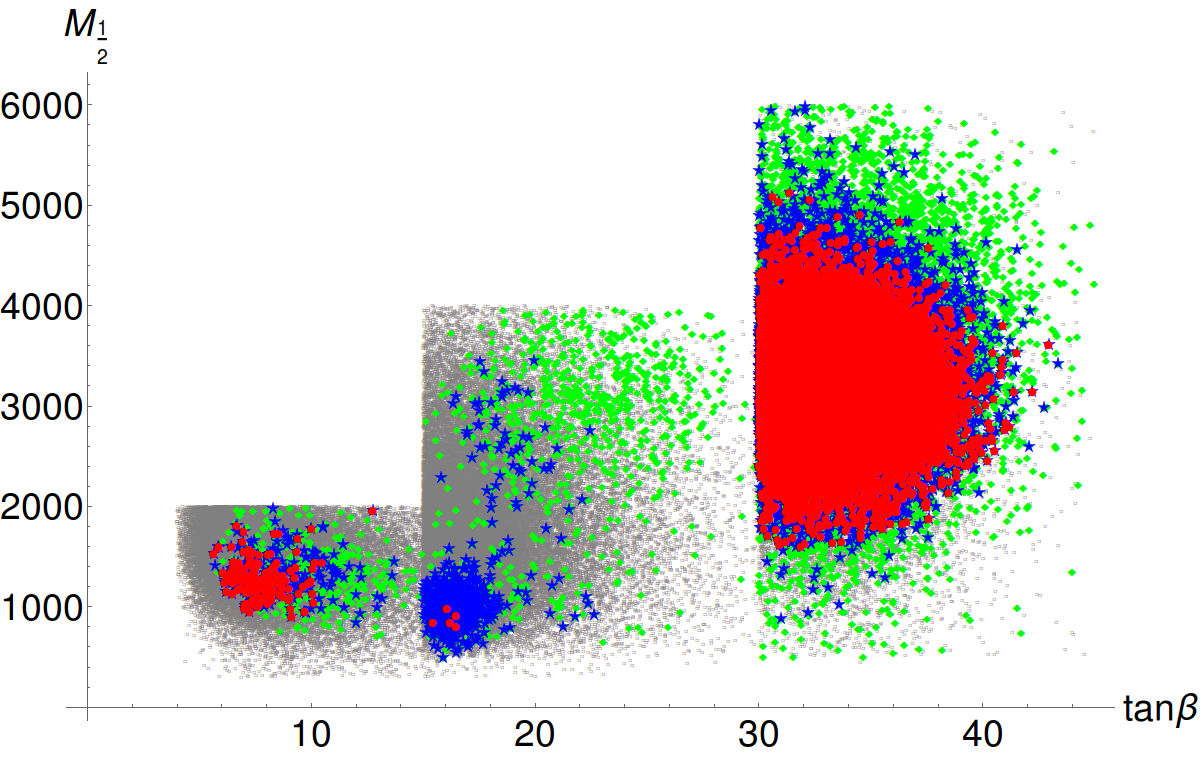}{
The same data as in Fig.~\ref{scatPlotTb} projected onto the $\tan\beta \times
(A^{de}_{33}/M_{1/2})$ (a) and $\tan\beta \times M_{1/2}$ 
(b) planes. \label{scatPlotTb2}}{
Points consistent with SU(5) Yukawa matrix unification in the $\tan\beta
\times (A^{de}_{33}/M_{1/2})$ and $\tan\beta \times M_{1/2}$ planes.}

Tabs.~\ref{Ex}-\ref{Outinos} contain information on the input parameters and
particle spectra in four sample points with a proper Yukawa matrix
unification. We also present the ratios $m_{h_d}/m_0$ and 
$m_{h_u}/m_0$ at the GUT scale that quantify departures from the scalar
mass universality for each of the points.  Tab.~\ref{OutEx} shows the
corresponding SUSY-scale threshold corrections
\begin{equation} \label{eq:dY}
\delta Y_{x} \equiv \frac{v_d Y_{x}^{MSSM}-m^{\rm SM}_x}{m^{\rm SM}_x},
\hspace{1cm} x = d,s,b,e,\mu,\tau,
\end{equation}
as well as the GUT-scale ratios $Y^d_{ii}/Y^e_{ii}$ which
parameterise the unification quality. The observed small deviations 
of the latter ratios from unity determine sizes of the necessary 
GUT-scale threshold corrections.  
\begin{table}[t]
\centering
\renewcommand{\arraystretch}{1.5}
\begin{tabular}{| c | c | c | c | c | c | c | c | c | c |c | c | c | c | }
\hline
 & $\tan\beta$ & $\tfrac{M_{1/2}}{{\rm GeV}}$ &  $\tfrac{m_0}{M_{1/2}}$ & 
$\tfrac{m_{h_u}}{m_0}$  & $\tfrac{m_{h_d}}{m_0}$  & $\tfrac{A^{de}_{11}}{M_{1/2}}$ &
$\tfrac{A^{de}_{22}}{M_{1/2}}$ & $\tfrac{A^{de}_{33}}{M_{1/2}}$ 
& $\tfrac{A^{u}_{33}}{M_{1/2}}$ 
 \\
\hline
\input{tabs/SamplePoints_pars.tab}
\hline
\end{tabular}
\caption[Examples of points with a successful Yukawa unification in the large
diagonal A terms scenario.]{Examples of points with a successful Yukawa
unification.  They are given by their defining sets of MSSM parameters:
$\tan\beta$, common gaugino mass $M_{1/2}$, common sfermion mass $m_0$, soft
masses of Higgs doublets $m_{h_u}$ and $m_{h_d}$, and soft trilinear couplings
$A_{ii} (\mgut)$. \label{Ex}}
\end{table}
\begin{table}[t]
\centering
\renewcommand{\arraystretch}{1.5}
\begin{tabular}{| c | c | c | c | c | c | c |  }
\hline
& $\delta Y_d$ & $\delta Y_s$ & $\delta Y_b$ & $\tfrac{Y_d}{Y_e}$ & $\tfrac{Y_s}{Y_{\mu}}$ & $\tfrac{Y_b}{Y_{\tau}}$  \\
\hline
\input{tabs/SamplePoints_outs.tab}
\hline
\end{tabular}
\caption[Values of the threshold corrections and other characteristics
of the points.]{Values of the threshold corrections and other characteristics
of the points from Tab.~\ref{Ex} (see the text). \label{OutEx}}
\end{table}
\begin{table}[t]
\centering
\renewcommand{\arraystretch}{1.5}
\begin{tabular}{| c | c | c | c | c | c | c | c | c | c | c |}
\hline
& $m_{\tilde{s}_L}$ & $m_{\tilde{s}_R}$ & $m_{\tilde{\mu}_L}$ & $m_{\tilde{\mu}_R}$
& $m_{\tilde{t}_1}$ & $m_{\tilde{t}_2}$ & $m_{\tilde{b}_1}$ & $m_{\tilde{b}_2}$ & $m_{\tilde{\tau}_1}$ & $m_{\tilde{\tau}_2}$  \\
\hline
\input{tabs/SamplePoints_sfermions.tab}
\hline
\end{tabular}
\caption[Masses of selected sfermions (in GeV).]{Masses of selected sfermions
(in GeV) corresponding to the points from Tab.~\ref{Ex}. In the case of the
second generation, where the left-right mixing is negligible, mass eigenstates
are labeled according to their largest interaction eigenstate component.
\label{OutSf}} \end{table}
\begin{table}[t]
\centering
\renewcommand{\arraystretch}{1.5}
\begin{tabular}{| c | c | c | c | c | c | c | c | c | c |}
\hline
& $m_{\tilde{g}}$ & $m_{\chi^0_1}$ & $m_{\chi^0_2}$ & $m_{\chi^0_3}$
& $m_{\chi^0_4}$ & $m_{\chi^{\pm}_1}$ & $m_{\chi^{\pm}_2}$  & $m_{A_0}$ & $\mu$ \\
\hline
\input{tabs/SamplePoints_inos.tab}
\hline
\end{tabular}
\caption[Masses of the gluino, neutralinos, charginos, pseudoscalar $A_0$ and
the value of $\mu$ parameter (in GeV).]{Masses of the gluino, neutralinos,
charginos, pseudoscalar $A_0$ and the value of $\mu$ parameter (in GeV)
corresponding to the points from Tab.~\ref{Ex}. \label{Outinos} }
\end{table}

%% file: tabs/SamplePoints_pars.tab
1 & 7.47 & 1032.27 &  2.28 &  0.49 & 2.35 & -0.008 &  0.63 &  1.63 & -2.13 \\
2 & 16.4 & 815.081 &  1.72 &  0.37 & 2.46 & -0.017 &  0.89 &  1.30 & -1.23 \\
3 & 30.3 & 2733.63 &  1.67 &  0.19 & 1.79 & -0.037 &  1.22 &  1.04 &  1.96 \\
4 & 40.4 & 2663.45 &  1.72 &  0.26 & 2.58 & -0.055 &  1.53 &  1.65 &  1.56 \\

%% file: tabs/SamplePoints_outs.tab
1 & -0.6 & 2.4 & 0.19 &  1.10 &  0.94 &  0.97    \\
2 & -0.66 & 2.3 & 0.089 &  0.95 &  0.91 &  0.90    \\
3 & -0.69 & 2.1 & 0.085 &  0.92 &  0.90 &  0.94    \\
4 & -0.66 & 2.1 & 0.13 &  1.01 &  0.92 &  1.08    \\

%% file: tabs/SamplePoints_sfermions.tab
1 &     2031. &     1856. &     1433. &     1433. &     1328. &     1649.  &     1603.  &     1652.  &     1130.  &      1185.  \\
2 &     3049. &     2958. &     2579. &     2579. &     1542. &     2085. &     2068.  &      2154.  &     1916. &       2011.  \\
3 &  5999. &    4944.  &      3752.  &      3752.  &     5596 &    4933 &    5620  &       5698 &  3937 &      4180.  \\
4 &    5700.  &     4290.  &      3714.  &     3714.  &       4902  &      4951 &      4904.  &   3984.  &      2658. &       2658.  \\

%% file: tabs/SamplePoints_inos.tab
1 &  2350. &   453. &   865. &  2688. &  2688. &   865. &  2689. &  5192. & -2698. \\
2 &  1858. &   350. &   670. &  1626. &  1626. &   670. &  1628. &  2981. & -1626. \\
3 &  5719. &  1223. &  2269. &  4529. &  4530. &  2270. &  4531. &  5523. & -4521. \\
4 &  5545. &  1189. &  2205. &  3690. &  3691. &  2205. &  3692. &  6958. & -3686. \\

%% file: chapters/5_2_FlavourSoftDiag.tex
In this section, we discuss the impact of large $A$-terms on a few selected flavour
observables.  The MSSM scenario we consider does not include any sources of
flavour- and CP- violation at $M_{GUT}$ other than the CKM matrix. Therefore,
flavour off-diagonal entries of the soft terms remain small, as they arise
solely from the RGE running.\footnote{ {\tt SOFTSUSY 3.3.8} assumes that all
the MSSM parameters are real, i.e. it neglects the CP-violating phases. A
separate numerical evaluation of the soft term imaginary parts has been
performed with the help of {\tt SPheno 3.3.3}\cite{SPheno1,SPhenoTrzy}. No
observable impact on CP-violating observables has been found for the MSSM
parameter space points discussed in the previous section.}

In the following, we shall illustrate how the flavour observables change when
the $A$-terms grow from 0 to $150\%$ of the value that is necessary for the
Yukawa unification $\mathbf{Y}^d(M_{\rm GUT})= \mathbf{Y}^{e\,T}(M_{\rm GUT})$
to take place. Among the observables calculable with the help of {\tt
SUSY\_FLAVOR~v2.10}, only three turn out to be significantly altered:
\begin{center}
${\mathcal B}_\gamma \equiv {\mathcal B}(\bar B \to X_s \gamma)$,~~~
    $\overline{{\mathcal B}}_{s\mu} \equiv {\mathcal B}(B_s \to \mu^+ \mu^-)$~~ and~~
    $\overline{{\mathcal B}}_{d\mu} \equiv {\mathcal B}(B_d \to \mu^+ \mu^-)$.
\end{center}
Moreover, the only $A$-term component they noticeably depend on is
$A^{de}_{33}$. Another important parameter to which these observables
are sensitive is $\tan\beta$.

In Fig.~\ref{bsgammaPlot}, we show the dependence of
$\delta{\mathcal B}_\gamma \equiv ({\mathcal B}_\gamma^{\rm MSSM}-{\mathcal
B}_\gamma^{\rm SM})/{\mathcal B}_\gamma^{\rm SM}$
on $A^{de}_{33}$ and $\tan\beta$. For each example listed in
Tab.~\ref{Ex} (and also for 17 other examples), we have plotted
$\delta{\mathcal B}_\gamma$ keeping all the parameters but 
$A^{de}_{33}$ fixed.  As one can see, SUSY contributions in our examples can
enhance ${\mathcal B}_\gamma$ by up to $25\%$ with respect to the SM
prediction. Moreover, up to $10\%$ relative differences are observed between
points with vanishing and maximal $A^{de}_{33}$. These conclusions
are practically insensitive to whether the calculation is performed with
the very recently updated SM prediction that was quoted in Eq.~\eqref{bgamma2015} 
or rather with the previous one
\be \label{bgamma2006}
{\mathcal B}_\gamma^{\rm SM, 2006} = (3.15 \pm 0.23)\times 10^{-4}~\mbox{\cite{Misiak:2006zs}}.
\ee
The two predictions differ mainly due to an additive correction to the $b
\to s\gamma$ amplitude from diagrams with the charm quark loops, which are the
same in the SM and MSSM.

Our actual results displayed in Fig.~\ref{bsgammaPlot} were obtained in
Ref.\cite{Iskrzynski:2014zla} with the use of Eq.~\eqref{bgamma2006}. This
fact has little effect on the plotted data points, but strongly affects the
position of the horizontal lines marking the (unchanged) experimental
world average~\eqref{bgamma.exp}. The solid horizontal lines corresponds to
the central value of this average, while the dashed lines describe the
$1\sigma$ range that is obtained after adding the experimental and theoretical
uncertainties in quadrature.  With the new SM result of Eq.~\eqref{bgamma2015},
all the horizontal lines would shift downwards by around $0.07$, with the
distances among them remaining unchanged. This would move around half of the
displayed points outside the $1\sigma$ range. However, most of them would
remain in the $2\sigma$ range.  We can thus see that the observed
significant variation of ${\mathcal B}_\gamma$ could lead to a possible 
exclusion of some of our points once the uncertainties get reduced.
\SidePlotsTwo{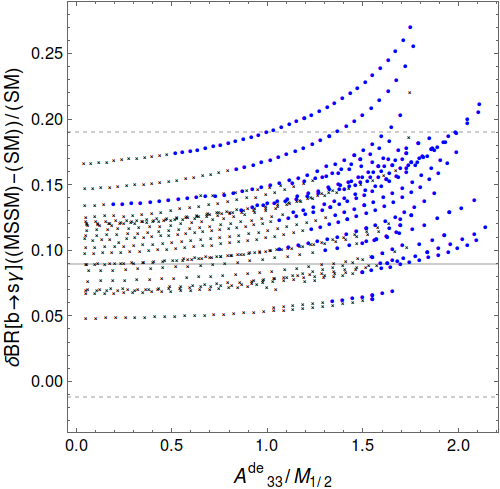}{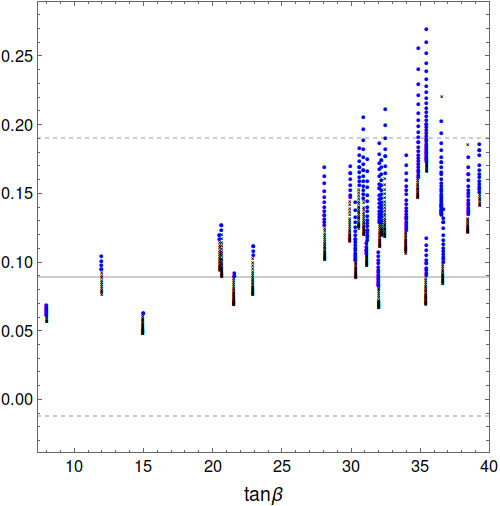}
{Dependence of $\delta{\mathcal B}_\gamma$ on
$A^{de}_{33}$ and $\tan\beta$. Points fulfilling $\mathbf{Y}^d(M_{\rm
GUT})=\mathbf{Y}^{e\, T}(M_{\rm GUT})$ are marked in blue. 
The $1\sigma$ experimental error band is represented by horizontal lines. \label{bsgammaPlot}}
{0.35\textheight}{Dependence of $\delta{\mathcal B}_\gamma$ on
$A^{de}_{33}$ and $\tan\beta$.}
\SidePlotsTwo{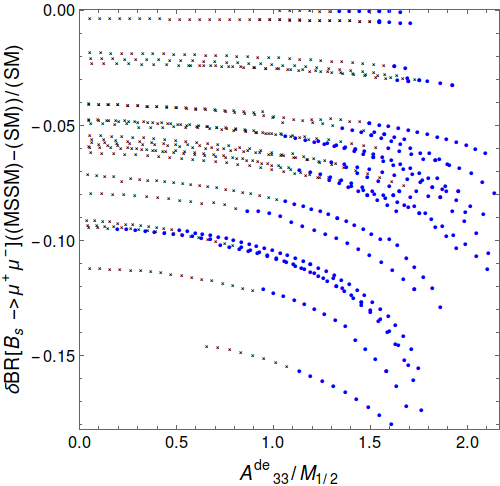}{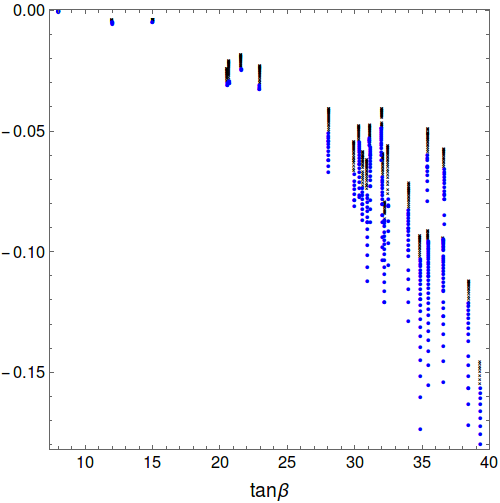}
{Dependence of $\delta\overline{\mathcal B}_{s\mu}$ on
$A^{de}_{33}$ and $\tan\beta$. Points fulfilling
$\mathbf{Y}^d(M_{\rm GUT})=\mathbf{Y}^{e\, T}(M_{\rm GUT})$ are marked in
blue. The results for $\delta\overline{\mathcal B}_{d\mu}$ are practically
identical.
\label{bsmumuPlot}}
{0.35\textheight}{Dependence of $\delta\overline{\mathcal B}_{s\mu}$ on
$A^{de}_{33}$ and $\tan\beta$.}

Analogous plots for 
$\delta\overline{\mathcal B}_{s\mu} \equiv (\overline{\mathcal B}_{s\mu}^{\rm MSSM}-
\overline{\mathcal B}_{s\mu}^{\rm SM})/\overline{\mathcal B}_{s\mu}^{\rm SM}$
are shown in Fig.~\ref{bsmumuPlot}. All our sample results for
$\overline{\mathcal B}_{s\mu}^{\rm MSSM}$
fall within the $1\sigma$ band above the experimental average~\eqref{bsmumu.exp},
and the branching ratio can be smaller by about $15\%$ compared to the SM
prediction~\eqref{bsmumu.SM}.

As far as $\overline{{\mathcal B}}_{d\mu}$ is concerned, it undergoes an almost identical
alteration with respect to the SM. However, it remains in perfect
agreement with the present experimental result~\eqref{bdmumu.exp}
within its large uncertainties. The experimental sensitivity would need to be
improved by more than an order of magnitude to distinguish between the SM 
prediction~\eqref{bdmumu.SM} and the corresponding MSSM results for our sample points.

The three considered decays share the crucial property of being sensitive to
supersymmetric contributions even if no sources of flavour violation beyond
the CKM matrix are present. It follows from the fact that they are all
chirally suppressed in the SM.

%% file: chapters/5_3_EWSB_SoftDiag.tex
As described in \refsec{EWSBstab}, the MSSM contains a large number of scalar
fields. In a proper analysis of the electroweak symmetry breaking, one would
need to prove that only the neutral Higgs fields acquire non-zero values in
the global minimum of the MSSM scalar potential. However, it is well known
that there exist large regions in the MSSM parameter space where other, deeper
minima arise. At such minima, also sfermions develop non-vanishing VEVs.

In particular, along the direction in the MSSM scalar field space where 
%
$ 
 | H_1|=|\tilde{s}_L|=|\tilde{s}_R|,
$
%
a deeper, CCB minimum arises when $A_{s}(M_{\rm
SUSY})$ is large. Actually, all our examples in Tabs.~\ref{Ex}-\ref{OutEx}
strongly violate the stability condition\cite{Casas:1995pd}
   \begin{equation}\label{roughstab}
    \frac{A_{ii}}{Y_{ii} \widetilde{m}_i} < O(1),
   \end{equation}
  with $\widetilde{m}_i$ defined by Eq.~(\ref{mtildeEq}). Instead, we have
$ A_{s}/(Y_{s} \widetilde{m}_2)~\sim~ 10^2 $
at the scale $M_{\rm SUSY}$.
    
However, the usual Higgs vacuum does not need to be absolutely stable.  The
standard viability condition is that the its lifetime must be longer than the
age of the Universe. According to Ref.\cite{Borzumati:1999sp}, such a
condition is fulfilled when
%
$A_{s}/\widetilde{m}_2<1.75\,$.
%
This requirement turns out to be satisfied in all our examples of Yukawa
unification. One can verify this by inspecting Tab.~\ref{AEx} where the ratios
$A^d_{ii}/\widetilde{m}_i$ have been presented for all the three generations.
\begin{table}[t]
\centering
\renewcommand{\arraystretch}{1.5}
\begin{tabular}{| c | c | c | c |  }
\hline
& $\tfrac{A_d}{\widetilde{m}_1}$ &
$\tfrac{A_s}{\widetilde{m}_2}$ & $\tfrac{A_b}{\widetilde{m}_3}$ \\
\hline
\input{tabs/SamplePoints_aMtilde.tab}
\hline
\end{tabular}
\caption[Values of the ratios $A^d_{ii}/\widetilde{m}_i$ at the scale
$M_{SUSY}$.]{Values of the ratios $A^d_{ii}/\widetilde{m}_i$ at the scale $M_{SUSY}$
for the points from Tab.~\ref{Ex}. \label{AEx}}
\end{table}

%% file: tabs/SamplePoints_aMtilde.tab
1 & -0.0064 & 0.49 & 1.26 \\
2 & -0.016 & 0.89 & 1.01 \\
3 & -0.041 & 1.56 & 0.38 \\
4 & -0.048 & 1.54 & 0.18 \\

%% file: chapters/6_00_introOffDiag.tex
In this chapter, we describe the phenomenological aspects of the
$GFV_{23}$ scenario. We do it step by step, describing how the imposed
constraints affect the parameter space of the model. We start by presenting
the region in which the bottom-tau and strange-muon Yukawa unification is
fulfilled. Further, points giving the correct DM relic density are presented
(Sec.~\ref{sec:dm}). Next, it is shown that reproducing the measured
Higgs mass and results from flavour physics is not in any tension with the
unification condition (Sec.~\ref{sec:flav}). Finally, we find that having a
proper DM candidate means that the $GFV_{23}$ parameter space points with
SU(5) Yukawa matrix unification that have been found by our scan
turn out to be testable at the LHC with $\sqrt{s}
\simeq 14\,$TeV (Sec.~\ref{sec:lhc}). In Tab.~\ref{tab:exp_constraints},
all the experimental constraints discussed in this chapter are 
collected.
\begin{table}[t]
\begin{center}
\begin{tabular}{|l|l|l|l|l|l|}
\hline
Measurement & Mean or range & Error [~exp.,~th.] & Reference\\
\hline
\abundchi      & $0.1199$ 	& [$0.0027$,~$10\%$]		& \cite{Ade:2013zuv}\\
LUX (2013) & See Sec.~3 of\cite{Kowalska:2014hza} 	& See Sec.~3 of\cite{Kowalska:2014hza} & \cite{Akerib:2013tjd}\\ 
\mhl\ (by CMS) & $125.7\gev$ & [$0.4$, $3.0$] \gev & \cite{CMS:yva} \\
\sinsqeff 			& $0.23155$     & [$0.00012$, $0.00015$] &  \cite{Agashe:2014kda}\\
$M_W$                     	& $80.385\gev$  & [$0.015$, $0.015$] \gev &  \cite{Agashe:2014kda}\\
\brbxsgamma $\times 10^{4}$ & $3.43$   	& [$0.22$, $0.23$] & \cite{Amhis:2014hma} \\ 
$\brbsmumu\times 10^9$	  & $2.8$ & [$0.7$, $0.23$] & \cite{CMS:2014xfa}\\ 
$\brbdmumu\times 10^{10}$ & $3.9$ & [$1.6$, $0.2$]  & \cite{CMS:2014xfa}\\ 
$\Delta M_{B_s}\times 10^{11}$ & $1.1691\gev$ & [$0.0014$, $0.1580$] \gev & \cite{Agashe:2014kda}\\
$\Delta M_{B_d}\times 10^{13}$ & $3.357\gev$ & [$0.033$, $0.340$] \gev & \cite{Agashe:2014kda} \\
$\Delta M_{B_d}/\Delta M_{B_s}\times 10^{2}$ & $2.87$ & [$0.02$, $0.14$] & \cite{Amhis:2014hma}\\ 
$\sin(2\beta)_{\rm exp}$ & $0.682$ & [$0.019$, $0.003$] & \cite{Agashe:2014kda}\\ 
\brbutaunu $\times 10^{4}$  & $1.14$ & [$0.27$, $0.07$] & \cite{Agashe:2014kda}\\ 
${\mathcal B}(K^+ \to \pi^+ \nu \bar{\nu})\times 10^{10}$  &$1.73$  
            & [$1.15, 0.04$] & \cite{Agashe:2014kda}\\ 
$|d_n|\times 10^{26}$  &  $< 2.9$ $e$ cm & [--, $30\%$] & \cite{Baker:2006ts} \\
$|\epsilon_K|\times 10^3$  & $2.228$ & [$0.011$, $0.17$] &  \cite{Agashe:2014kda}  \\
\brmuegamma $\times  10^{-13}$ & $< 5.7$ & [--,0] & \cite{Adam:2013mnn} \\
\brmueee $\times  10^{-12}$ & $< 1.0$ & [--,0] & \cite{mueeeArt} \\
\brtmg $\times  10^{-8}$ & $< 4.4$ & [--,0] & \cite{Aubert:2009ag} \\
\brtaummm $\times  10^{-8}$ & $< 2.1$ & [--,0] & \cite{taummmArt} \\
LHC \eight\ & \refsec{sec:lhc} & \refsec{sec:lhc} & \cite{CMS-PAS-SUS-13-018,Aad:2014wea,Aad:2014vma}\\
\hline
\end{tabular}
\caption{
Experimental constraints applied in the analysis.} 
\label{tab:exp_constraints}
\end{center}
\end{table}

We present the results of our final MCMC scan performed with the
software described in Chapter~\ref{ToolsChap}. The principal aim of the scan
was to verify whether the questionable strange-muon Yukawa coupling
unification can be achieved alongside with well tested bottom-tau case
when all phenomenological constraints are satisfied. In particular, we
focus on fulfilling the flavour physics bounds and the phenomenological
patterns that emerge when the observed DM relic density is 
correctly reproduced. Although deliberately avoiding a CCB
scalar potential minimum triggered by the squark fields of the second
generation, we did not guide the scan in such a way that the CCB bounds from
electroweak vacuum stability on the third family couplings were automatically
satisfied. The latter issue, already covered in many other works, is discussed
in Sec.~\ref{OffDiagEWSBsec}. The ranges of input variables defining our
search are summarized in Tab.~\ref{rangesGFV23}.
\begin{table}[t]
\centering
\renewcommand{\arraystretch}{1.1}
\begin{tabular}{|c|c|}
\hline 
Parameter &  Scanning Range\\
\hline 
$M_{1/2}$	& [$100$, $4000$] \gev\ \\
\mhu            & [$100$, $8000$] \gev\ \\
\mhd            & [$100$, $8000$] \gev\ \\
\tanb	        & [$3$, $35$] \\
\signmu		& $-1$ \\
\hline 
$A_{33}^{de}$   & [$-5000$, $5000$] \gev\ \\
$A_{33}^{u}$    & [$-9000$, $9000$] \gev\ \\
$A^{de}_{11}/A^{de}_{33}$   & [$-0.00028$, $0.00028$] \\
$A^{de}_{22}/A^{de}_{33}$  & [$-0.065$, $0.065$] \\
$A^{u}_{22}/A^{u}_{33}$    & [$-0.005$, $0.005$] \\
\hline 
$m_{ii}^{dl},\; i=1,2,3$   & [$100$, $7000$] \gev\ \\
$m^{dl}_{23}/m^{dl}_{33}$  & [$0$, $1$] \\
$m_{ii}^{ue},\; i=1,2,3$   & [$100$, $7000$] \gev\ \\
\hline 
\end{tabular}
\caption[Ranges of the input SUSY parameters used in the final $GFV_{23}$
scan.]{Ranges of the input SUSY parameters used in the final $GFV_{23}$
scan. The omitted soft SUSY-breaking parameters at the GUT scale have been set
to zero, as discussed in Chapter~\ref{TwoScenarios}.}
\label{rangesGFV23}
\end{table}

%% file: chapters/6_1_Regions_OffDiag.tex
\begin{figure}[h]
\centering
\subfloat[]{
\includegraphics[width=0.48\textwidth]{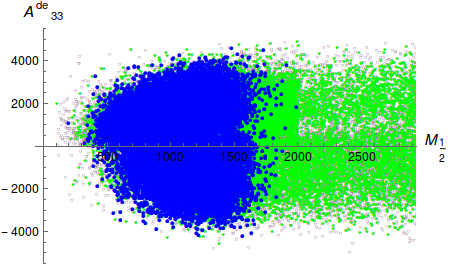}}
\subfloat[]{
\includegraphics[width=0.48\textwidth]{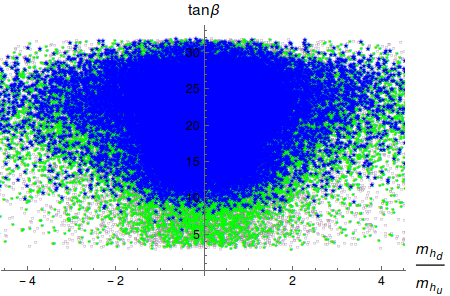}}\\
\subfloat[]{
\includegraphics[width=0.48\textwidth]{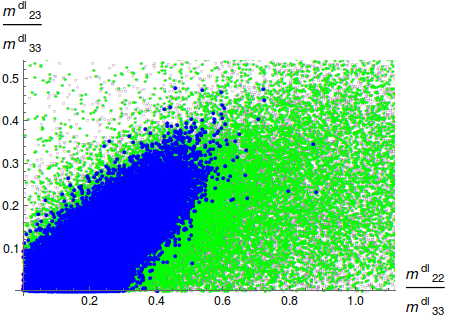}}
\caption[$GFV_{23}$ points in the planes
($\mhalf$, $A^d_{33}$), ($\mhd/\mhu$, $\tanb$), and
($m^{dl}_{23}/m^{dl}_{33}$,
$m^{dl}_{22}/m^{dl}_{33}$) ]{
Scatter plot of the $GFV_{23}$ points in the planes: 
(a) ($\mhalf$, $A^d_{33}$), 
(b) ($\mhd/\mhu$, $\tanb$), 
(c) ($m^{dl}_{23}/m^{dl}_{33}$, $m^{dl}_{22}/m^{dl}_{33}$). 
Colour code: 
grey rectangles -- all the points for which the spectrum was returned; 
green diamonds -- points satisfying the Yukawa unification condition for the third generation at $2\sigma$; 
blue stars -- points for which both the heavier generations are unified at $2\sigma$.}
\label{m23ad22}
\end{figure}

The performed scan has returned the MSSM spectrum for 121986 
parameter-space points, among which 34758 yield the Yukawa couplings of
leptons and down-quarks at \mgut{} equal within 10\% for the second and third
generations. \reffig{m23ad22} presents distributions of the points
collected by our scan in the planes
($\mhalf$, $A^d_{33}$) (a), 
($\mhd/\mhu$, $\tanb$) (b), and
($m^{dl}_{23}/m^{dl}_{33}$, $m^{dl}_{22}/m^{dl}_{33}$) (c). 
All the points for which the spectrum was returned by {\tt SPheno} are depicted as
grey rectangles, whereas those that satisfy the Yukawa unification condition
for the third generation at $2\sigma$ ($0.9<Y_b/Y_{\tau}<1.1$) as green
diamonds. Finally, blue stars correspond to those points for which both
heavier generations are unified at $2\sigma$. The latter points constitute around 28\% of
the points collected in the scan.

Several observations can now be made. First of all, it is known that a
satisfactory unification of the third family Yukawa couplings is possible
in the MFV $SU(5)$ with universal scalar masses for moderate values
of \tanb. This is confirmed by \reffig{m23ad22} where green points can easily
be found for vanishing flavour-violation in the GUT-scale soft parameters,
and for values of $m^{dl}_{22}/m^{dl}_{33}$ close to 1. One
can also observe that the values of \mhalf\ are limited for the points
with a successful second- and third-family Yukawa unification.
The ratio $\mhd/\mhu$ seems unconstrained by the
unification requirement, whereas moderate values of \tanb{} are
slightly preferred.

Secondly, the scan includes values of $A^d_{33}$ that are large as compared to
the superpartner masses. It is of relevance for the minimisation of the
scalar potential, and will be discussed in \refsec{OffDiagEWSBsec}.

\begin{figure}[h]
\centering
\subfloat[]{
\includegraphics[width=0.48\textwidth]{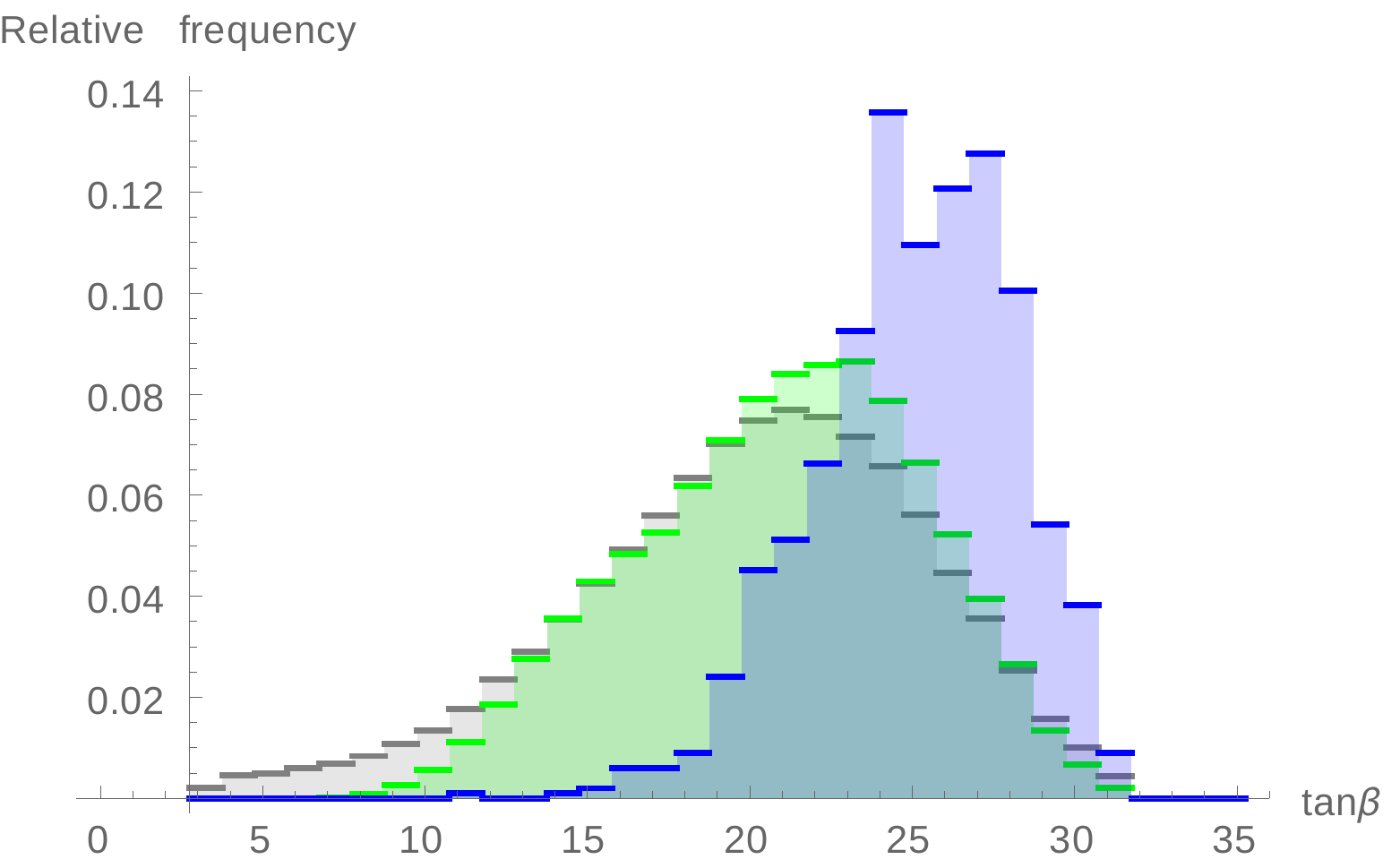}}
\subfloat[]{
\includegraphics[width=0.48\textwidth]{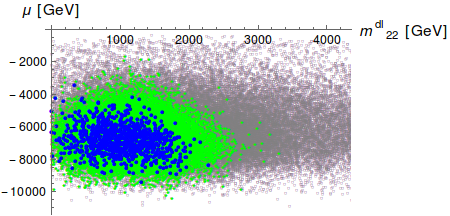}}
\caption[Relative distributions of points gathered by the $GFV_{23}$ scan in
\tanb, and a scatter plot showing these points in the plane
($m^{dl}_{22}$,$\mu$).]{(a) Relative frequencies of \tanb{} values among
the points of our $GFV_{23}$ scan; (b) Scatter plot showing these
points in the plane ($m^{dl}_{22}$,$\mu$). Colour code: grey -- all the points
for which a mass spectrum was returned; green -- points with the bottom-tau and
strange-muon Yukawa unification; blue -- points that in addition are
characterised by $m^{dl}_{23}/m^{dl}_{33} < 0.005$.}
\label{noMixProps}
\end{figure}

Finally, the blue points, signalling that the strange-muon Yukawa unification
has taken place, appear also for almost vanishing values of the soft-mass
element $m^{dl}_{23}$. The scan has collected 999 points (about 0.8\% of all
points) with $m^{dl}_{23}/m^{dl}_{33} < 0.005$ for which the bottom-tau and strange-muon
Yukawa unification takes place within 10\%. As already discussed in the
context of Eq.~\eqref{thres:mfv}, large and negative values of $\mu$ might
decisively contribute to the Yukawa unification issue for
sizeable values of \tanb{} and relatively light sfermions 
(as compared to $\sqrt{|\mu m_{\tilde g}|}$). Such a situation corresponds to the 
blue points in Fig.~\ref{noMixProps}.

%% file: chapters/6_2_1_DM_OffDiag.tex
\begin{figure}[t]
\centering
\subfloat[]{
\includegraphics[width=0.41\textwidth]{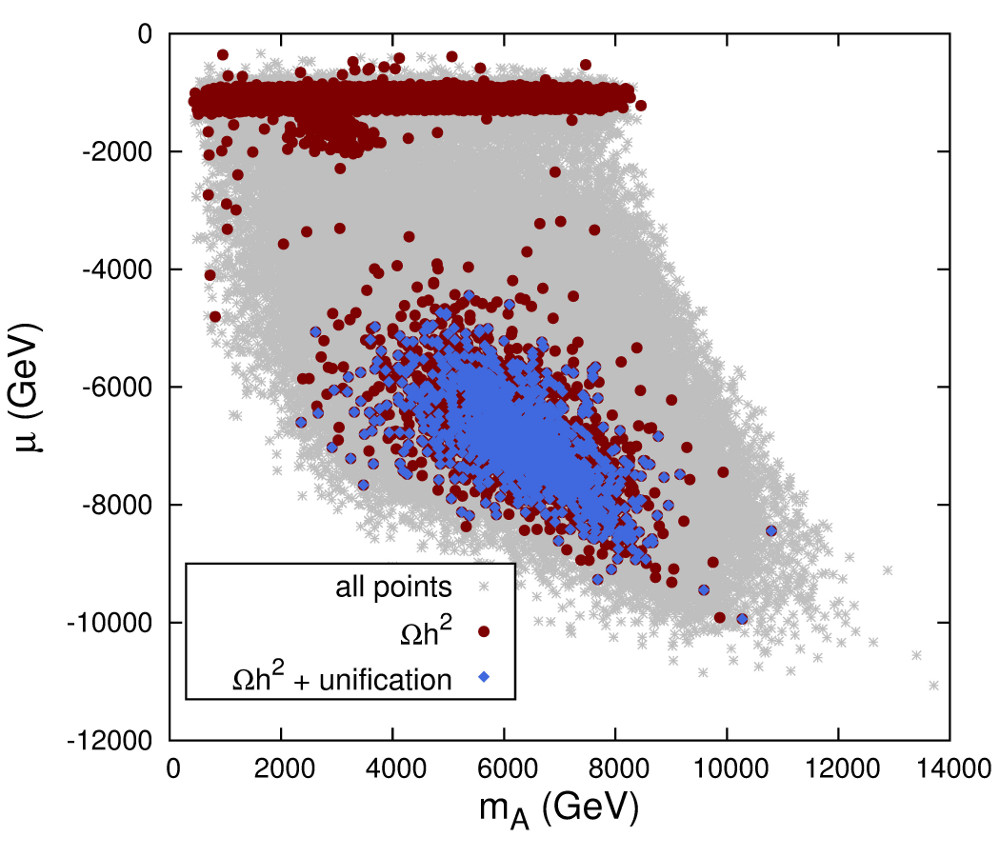}}
\subfloat[]{
\includegraphics[width=0.41\textwidth]{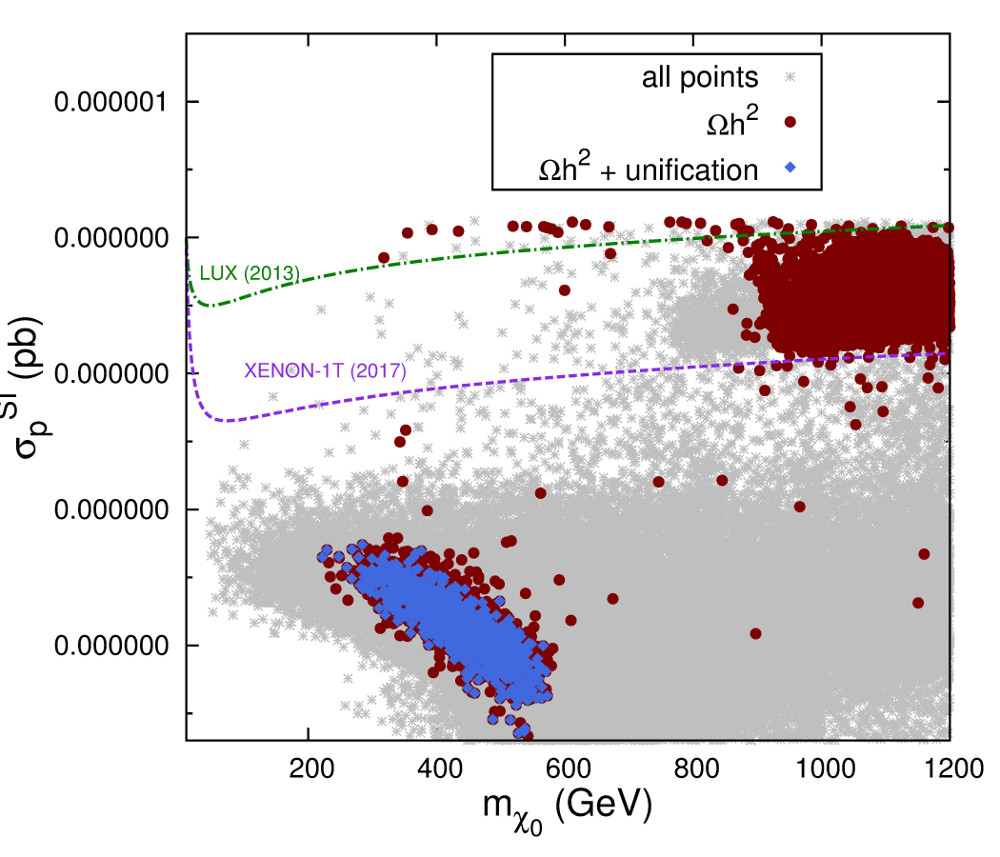}}\\
\subfloat[]{
\includegraphics[width=0.41\textwidth]{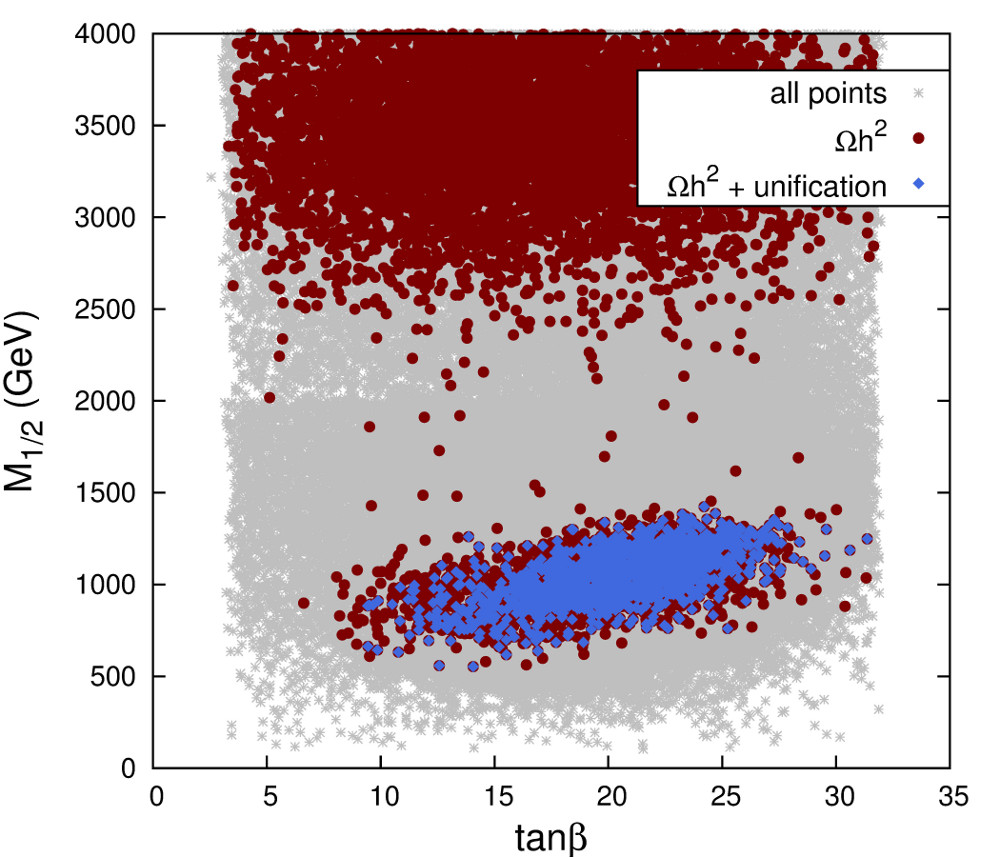}}
\caption[Scatter plot of the $GFV_{23}$ points in the planes ($\ma$,
$\mu$), ($m_{\neutone}$, \sigsip) and (\tanb, \mhalf)]{Scatter plot of the
$GFV_{23}$ points in the planes ($\ma$, $\mu$) (a), ($m_{\neutone}$,
\sigsip) (b) and (\tanb, \mhalf) (c). Grey stars: all the points collected by the
scan; brown dots: points satisfying the DM relic density constraint at
$3\sigma$; blue diamonds: points for which additionally the Yukawa unification
holds. The meaning of dashed lines is described in the text.}
\label{pheno:CDM_111}
\end{figure}

The DM relic density is influenced by multiple parameters of the model. For
the GFV scenario it requires a very specific mass spectrum, which leads to
interesting phenomenological conclusions.

In \reffig{pheno:CDM_111}, we show distributions of points found by our
scanning procedure in the planes ($\ma$, $\mu$) (a), ($m_{\neutone}$, \sigsip)
(b) and (\tanb, \mhalf) (c). All the points collected by our scan are drawn as
grey stars, while those that satisfy at $3\sigma$ the observational bound on
the DM relic density appear as brown dots. Blue diamonds correspond to those
scenarios for which also the Yukawa coupling unification holds. The green
dashed line indicates the $90\%$ \cl\ exclusion bound on the \sigsip\ based on
the 85-day measurement by the LUX collaboration\cite{Akerib:2013tjd}.  The
purple dashed line is a projection of XENON1T
sensitivity~\cite{Aprile:2012zx}. By comparing the panels (a) and (b) of
\reffig{pheno:CDM_111}, one can see that in the region where Yukawa unification
is achieved, the higgsino mass $\mu$ is much larger than the LSP mass,
which means that the LSP is bino-like. It corresponds to a relatively low
spin-independent proton-neutralino cross-section. In other words, the
condition of Yukawa coupling unification strongly disfavours purely or partly
higgsino-like neutralino as the LSP. This is due to the fact that the $\mu$
parameter value associated with this region is quite large in magnitude
($|\mu| \gsim 5\tev$) to enhance the $\mu$-dependent contribution in
Eq.~(\ref{sigma22}) that aids unification of the second family Yukawa
couplings. For this reason, only a bino-like neutralino was found by the
scan, which is an important phenomenological feature of our GFV
Yukawa unification scenario.

A unique mechanism that makes the effective increase of the DM annihilation
cross-section possible in this case is the neutralino co-annihilation with the
lightest sneutrino. The pseudoscalar is too heavy to allow a resonant
\neutone\ annihilation (as can be read from the panel (a) of
\reffig{pheno:CDM_111}), while the masses of coloured sfermions in the
GUT-constrained unification scenarios are always larger than those of the
sleptons. It is due to a renormalisation effect, as their RGE running is
strongly driven by the gluino. Such a property of the spectrum, however, has
important consequences for experimental testability. In
\reffig{pheno:CDM_111}(b) the dashed lines indicate the present reach of LUX
and the expected sensitivity for XENON1T experiment. The region favoured by
the relic density constraint in the Yukawa unification scenario remains far
beyond the reach for any of them. On the other hand, a light 
bino-like \neutone\ and a sneutrino having a similar
mass can be tested by the LHC 14 TeV, as will be discussed in
\refsec{sec:lhc}.

The value of \tanb\ favoured by the DM measurement is also strongly limited,
$\tanb\in [4, 15]$. This is a characteristic feature of the co-annihilation
mechanism, as the value of \tanb\ influences the sneutrino mass. It becomes
somewhat heavier for larger \tanb, and the efficiency of neutralino-sneutrino
co-annihilation drops. What is interesting, however, is that such a limited
value of \tanb\ has important consequences for flavour physics observables, as
will be discussed in \refsec{sec:flav}.

\begin{figure}[t]
\centering 
\subfloat[]{
\includegraphics[width=0.49\textwidth]{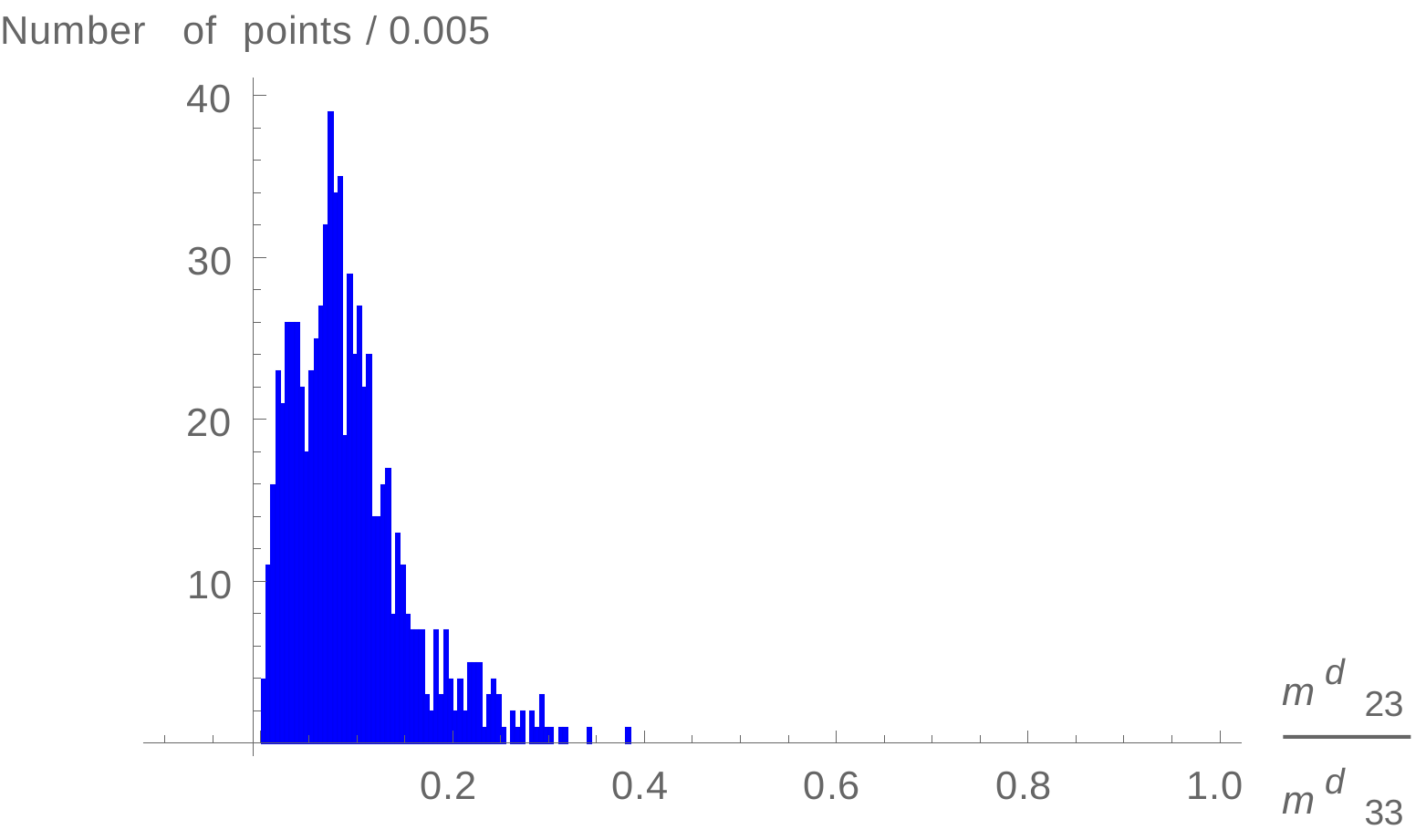}}
\subfloat[]{
\includegraphics[width=0.49\textwidth]{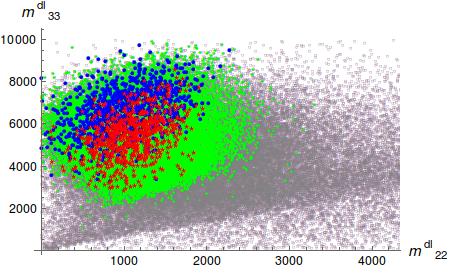}}
\caption[Properties of points that successfully fit DM relic
density.]{Characteristics of points satisfying the DM relic density
constraint in the $GFV_{23}$ scenario:
(a) Histogram showing the distribution (as a function
$m^{dl}_{23}/m^{dl}_{33}$) of points that satisfy the Yukawa
unification and the DM constraints.
(b) Scatter plot in the plane ($m^{dl}_{22}$,$m^{dl}_{33}$). Grey
rectangles -- all the points for which the MSSM spectrum was
returned; green diamonds -- points satisfying the bottom-tau and
strange-muon Yukawa unification; red dots -- points that fit (in
addition) the observed DM relic density; blue stars -- those of the
Yukawa-unified (green) points that satisfy $m^{d}_{23}/m^{d}_{33} <
0.005$.}  \label{DMnoMix} 
\end{figure}

Fig.~\ref{DMnoMix} is devoted to illustrating the dependence of our results on
the sfermion-mixing parameter $m^{dl}_{23}/m^{dl}_{33}$. It can be seen
in the left panel of this plot that the DM and Yukawa unification requirements
restrict this parameter to remain below around 0.2 for most of the points. 
As mentioned at the end of the previous section, the $GFV_{23}$ Yukawa
unification constraint alone could be satisfied also for very small 
$m^{dl}_{23}/m^{dl}_{33}$ (smaller than $0.005$) thanks to the
flavour-diagonal $Y_s\mu\tan\beta$ term in Eq.~\eqref{thres:mfv}. However, the
DM constraints reject all but four such points. The corresponding
distributions in the $(m^{dl}_{22}/m^{dl}_{33})$ plane are shown in the right
panel of Fig.~\ref{DMnoMix}.

%% file: chapters/6_2_2_Flavour_OffDiag.tex
\begin{figure}[t]
\vspace*{-8mm}
\centering
\subfloat[]{
\includegraphics[width=0.4\textwidth]{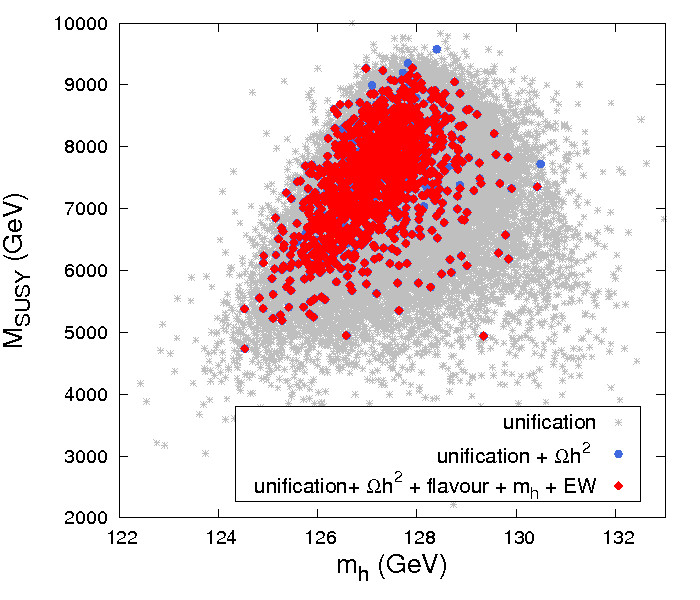}}
\subfloat[]{
\includegraphics[width=0.4\textwidth]{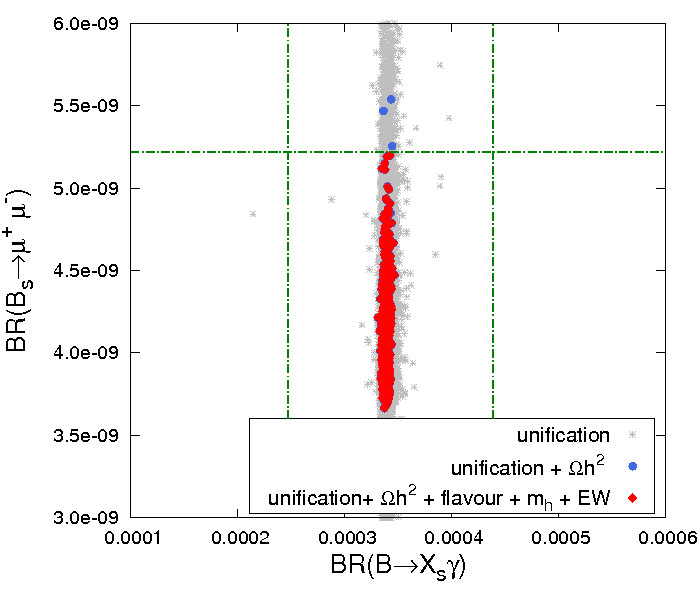}}\\
\subfloat[]{
\includegraphics[width=0.41\textwidth]{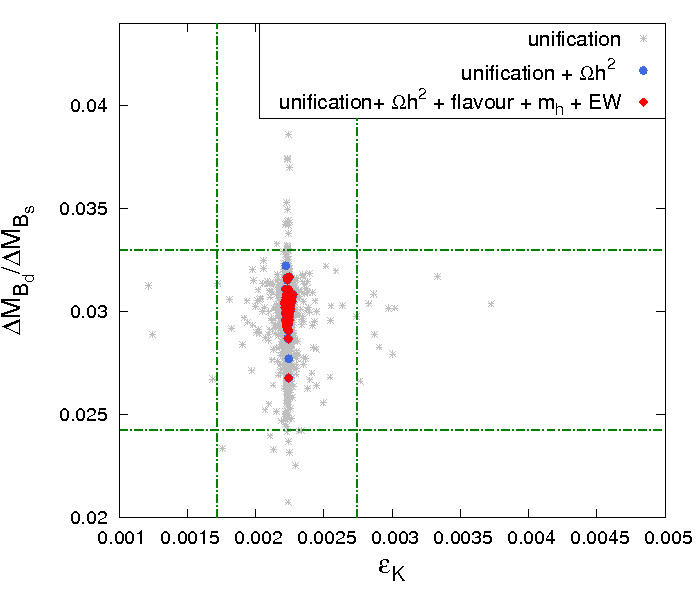}}
\caption[Scatter plot of the $GFV_{23}$ points in the planes (\mhl,
\msusy), (\brbxsgamma, \brbsmumu) and ($\epsilon_K$, $\Delta M_{B_d}/\Delta
M_{B_s}$)]{Scatter plot of the $GFV_{23}$ points in the planes (\mhl,
\msusy) (a), (\brbxsgamma, \brbsmumu) (b), and ($\epsilon_K$, $\Delta
M_{B_d}/\Delta M_{B_s}$) (c). Grey stars: all the points that allow the Yukawa
coupling unification; blue dots: points satisfying additionally the DM relic
density constraint at $3\sigma$; red diamonds: points with good Yukawa
coupling unification and all the constraints listed in
Table~\ref{tab:exp_constraints} satisfied at $3\sigma$ (except the
LHC). Dashed lines correspond to $3\sigma$ experimental limits on the
corresponding observables.}
\label{flavour}
\end{figure}

In \reffig{flavour}, we present distributions of points for several relevant
observables: (\mhl, \msusy) (a), (\brbxsgamma, \brbsmumu) (b) and
($\epsilon_K$, $\Delta M_{B_d}/\Delta M_{B_s}$) (c). Grey stars indicate all
the points for which it is possible to achieve the Yukawa coupling unification
for the second and third generations. Points that satisfy the relic density
constraint at $3\sigma$ are shown as blue dots, while red diamonds correspond
to those cases where additionally all the other constraints listed in
Table~\ref{tab:exp_constraints} are met at $3\sigma$ (except for the LHC bounds
from direct SUSY searches that will be discussed in \refsec{sec:lhc}).

The Higgs boson mass dependence on the GFV parameters has been discussed in
Ref.\cite{Cao:2006xb,AranaCatania:2011ak,Arana-Catania:2014ooa,Kowalska:2014opa}. It
was shown that while \mhl\ can be enhanced by non-zero $(2,3)$ entries of the
trilinear down-squark matrix, its dependence on the off-diagonal soft-mass
elements is negligible. Therefore, in the scenario considered in this study,
the only parameters relevant for the Higgs physics remain $A^u_{33}$ and
\msusy. That is confirmed by the panel (a) of \reffig{flavour} where no
tension between the correct value of the Higgs boson mass and the Yukawa
unification constraint (driven by the large GFV parameter $m^{dl}_{23}$) is
observed. The EW observables are not affected either, because the dominant GFV
contribution to $M_W$ and $\sinsqeff$ would be controlled by the element
$m^{ue}_{23}$\cite{Heinemeyer:2004by}.

On the other hand, the presence of sizeable off-diagonal entries in the squark
mass matrices might lead to disastrously high SUSY contributions to FCNC
processes. It turns out, however, that in the considered scenario most of the
flavour constraints are quite easily satisfied for the points that have
survived imposing the DM experimental limit. This is mainly due to the fact
that the coloured sfermions are relatively heavy, while \tanb\ needs to be low
in our setup, in order to allow an efficient neutralino-sneutrino
co-annihilation, as discussed in Sec.\ref{sec:dm}. In consequence, SUSY
contributions to the FCNC processes are suppressed.

\begin{figure}[t]
\centering
\subfloat[]{
\includegraphics[width=0.4\textwidth]{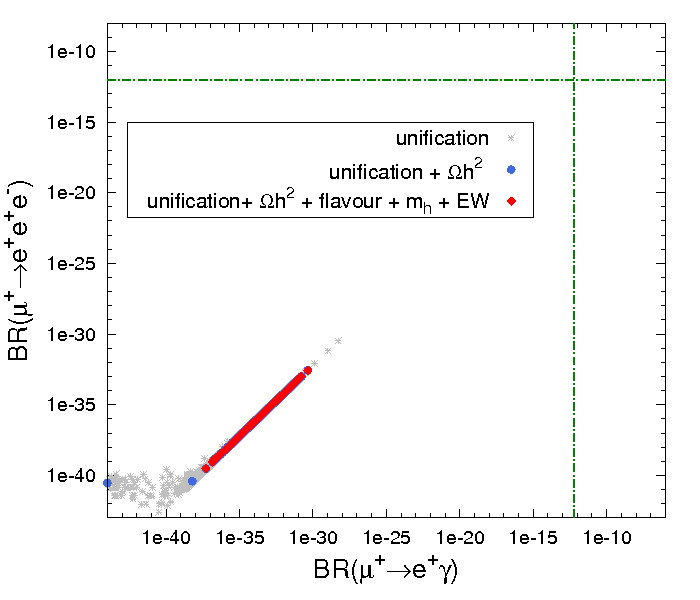}}
\subfloat[]{
\includegraphics[width=0.4\textwidth]{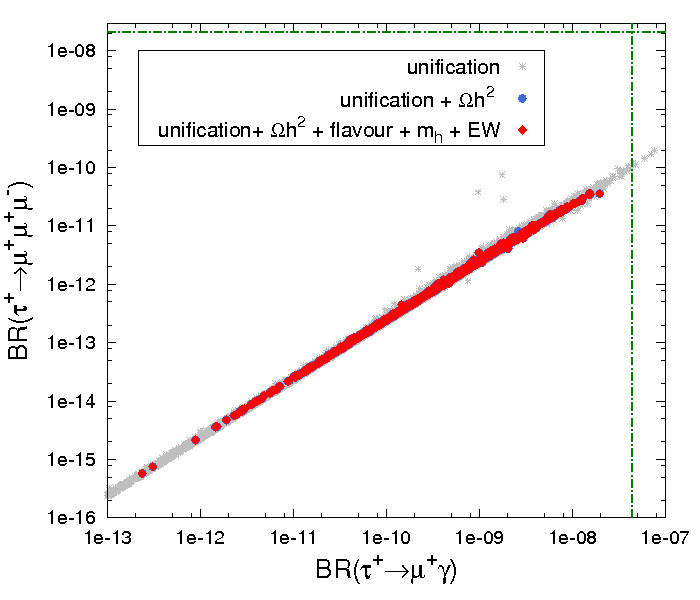}}
\caption[$GFV_{23}$ points in the planes (\brmuegamma, \brmueee), and
(\brtmg, \brtaummm)]{Scatter plot of the $GFV_{23}$ points in the planes
(\brmuegamma, \brmueee) (a), and (\brtmg, \brtaummm) (b). grey stars: all the
points that allow the Yukawa coupling unification; blue dots: points
satisfying additionally the DM relic density constraint at $3\sigma$; red
diamonds: points with good Yukawa coupling unification and all the constraints
listed in Table~\ref{tab:exp_constraints} satisfied at $3\sigma$ (except the
LHC). Dashed lines correspond to the upper experimental limits on the
corresponding observables.}
\label{lfvPlots}
\end{figure}

A potential threat to the GFV scenario is posed by the LFV observables
that severely constrain any non-zero flavour mixing of
sleptons\cite{Arana-Catania:2013nha}. When $m^{dl}_{23}$ is the only
non-zero flavour-violating soft term at \mgut{}, the current constraints on
the relevant processes are still satisfied without a need to tune any
parameters, as can be seen in Fig.~\ref{lfvPlots}.

In summary, the flavour observables can easily be accommodated into the $GFV_{23}$ scenario,
and no tension with either the bottom-tau or strange-muon Yukawa unification
is noticed.

%% file: chapters/6_2_3_LHC_OffDiag.tex
\begin{figure}[t]
\centering
\includegraphics[width=0.65\textwidth]{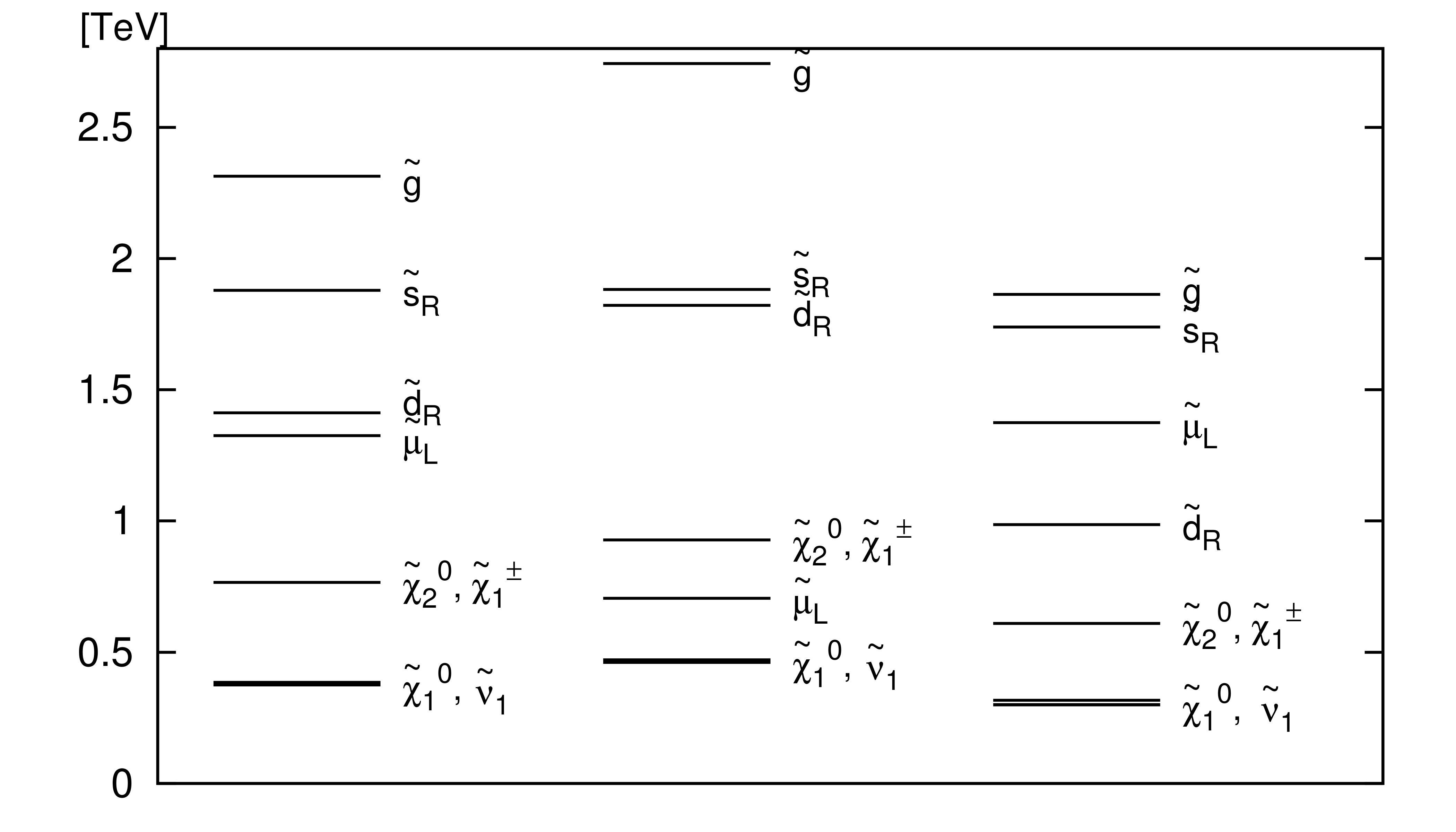}
\caption[Characteristic spectra.]{Examples of spectra characteristic for the $GFV_{23}$ Yukawa unification scenario.}
\label{spectrum}
\end{figure}

All the points from our scan that fulfil the bottom-tau and
strange-muon Yukawa unification requirement and satisfy the
experimental constraints from Table~\ref{tab:exp_constraints} share the
same pattern of the light part of the spectrum.  Three examples are shown in
\reffig{spectrum}.  The Next-to-Lightest SUSY particle (NLSP) is the
lightest sneutrino, while one charged slepton, neutralino \neuttwo\ and
chargino \charone\ are slightly heavier. The presence of light sleptons in the
spectrum is very important, as it leads to a characteristic 3-lepton signature
at the LHC. The next particle on the mass ladder is the lightest down-type
squark which is followed by the gluino. All the other coloured particles, the
remaining sleptons and heavy Higgses are much heavier and effectively
decoupled.

The strongest limit on the gluino mass comes from the ATLAS no-lepton,
2-6~jets plus Missing Transverse Energy (MET) inclusive
search\cite{Aad:2014wea}. It provides a stringent 95\% \cl\ exclusion
bound $m_{\gluino} \gtrsim 1400\gev$ when the neutralino LSP is
lighter than $300\gev$. The strongest bound on the lightest sbottom mass comes
from the CMS no-lepton, 2 jets and MET
search\cite{CMS-PAS-SUS-13-018}, while in the electroweakino sector the most
stringent experimental exclusion limits are obtained using the 3-lepton plus
MET CMS search\cite{Aad:2014vma}.

However, one needs to keep in mind that the bounds provided by the
experimental collaborations are interpreted in the Simplified Model Scenarios
(SMS) that make strong assumptions about the hierarchy of the spectrum and the
decay branching ratios. Usually it is assumed that there is only one light
SUSY particle apart from the neutralino LSP, and only one decay channel of the
NLSP with the branching ratio set to 100\% is considered. In a more general
case, however, the presence of other light particles in the spectrum may alter
the decay chain, and the assumption concerning the branching ratio may not
hold either. In such a case, the exclusion limits for the SMS should be
treated with care, and the actual limits are expected to be weaker.
\begin{figure}[t]
\centering
\subfloat[]{
\includegraphics[width=0.41\textwidth]{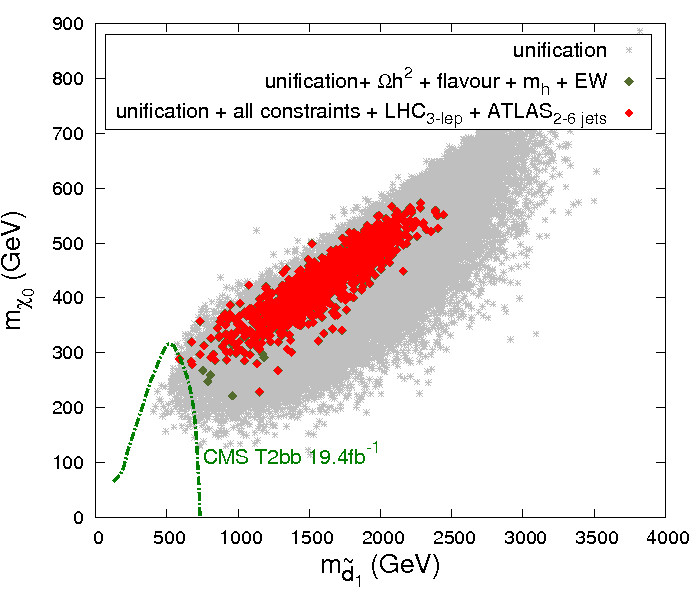}}
\subfloat[]{
\includegraphics[width=0.41\textwidth]{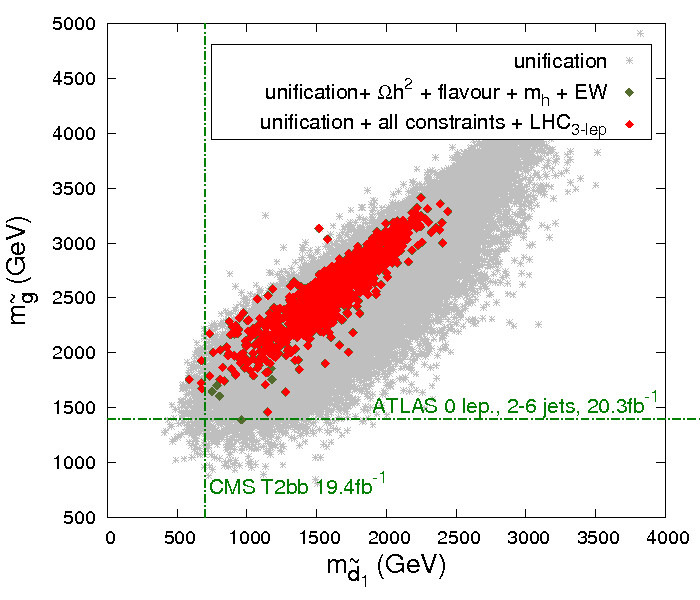}}
\caption[$GFV_{23}$ points in the planes ($m_{\tilde{d}_1}$,
$m_{\neutone}$), and ($m_{\tilde{d}_1}$, $m_{\gluino}$)]{Scatter plot of the
$GFV_{23}$ points in the planes ($m_{\tilde{d}_1}$, $m_{\neutone}$) (a),
and ($m_{\tilde{d}_1}$, $m_{\gluino}$) (b). grey stars: all the points that
allow the Yukawa coupling unification; dark green dots: points with good
Yukawa coupling unification and all the constraints listed in
Table~\ref{tab:exp_constraints} (except the LHC) satisfied at $3\sigma$; red
diamonds: points additionally surviving the CMS 3-lepton search at
$3\sigma$. Dashed lines correspond to 95\% \cl\ limits provided by other LHC
SUSY searches discussed in the text.}
\label{lhc}
\end{figure}

In our analysis\cite{Iskrzynski:2014sba}, we have used the experimental
exclusion bound for the gluino mass\cite{Aad:2014wea} at face value because
this search is inclusive, and therefore tests any gluino-driven multijet
signature, irrespectively of a particular decay chain. We have also decided to
use a direct 95\% \cl\ limit from the CMS sbottom production
search\cite{CMS-PAS-SUS-13-018}. For the SMS T2bb, it reads $m_{\tilde{b}_1}
\gtrsim 700\gev$ for $m_{\neutone}\simeq 150\gev$, and $m_{\tilde{b}_1}
\gtrsim 640\gev$ for $m_{\neutone}\simeq 250\gev$.  In our scenario, the
sbottom decay corresponds exactly to the SMS T2bb, i.e. ${\mathcal
B}(\tilde{b}\to b\neutone)=100\%$.  We neglect here a possibility that the
actual limit can be weakened due to a significant mixing between the sbottoms
and other down-type squarks, and we leave a detailed analysis of the GFV
effects in such a case for future studies. We will see, however, that this
simplifying assumption is justified by the fact that the limits derived from
Ref.\cite{CMS-PAS-SUS-13-018} are not the dominant ones.

On the other hand, interpretation of the CMS 3-lepton search strongly depends
on hierarchy in the considered spectrum, as well as on actual branching ratios
for neutralino and chargino decays. Therefore, in order to correctly quantify
the effect of this search in the GFV scenario, a full
reinterpretation of the experimental analysis was performed, using the
tools developed first in Ref.\cite{Fowlie:2012im}, and modified to recast the
limits from SMS in Ref.\cite{Kowalska:2013ica}. For the purpose of the present
study\cite{Iskrzynski:2014sba} the previously
implemented\cite{CMS-PAS-SUS-12-022} CMS 3-lepton search was updated to
include the full set of data with integrated luminosity of
$19.5\invfb$\cite{Aad:2014vma}.

In \reffig{lhc}, we present a distribution of the model points in the
($m_{\tilde{d}_1}$, $m_{\neutone}$) plane (a), and in the ($m_{\tilde{d}_1}$,
$m_{\gluino}$) plane (b). All the points for which the Yukawa coupling
unification is possible are shown as grey stars, and those that additionally
satisfy at $3\sigma$ the experimental constraints listed in
Table~\ref{tab:exp_constraints} (except the LHC) as dark green dots. Finally,
red diamonds depict the points that survive (at $3\sigma$) the CMS 3-lepton
plus MET search. Dashed lines correspond to the 95\% \cl\ exclusion bounds
from the CMS and ATLAS multijet searches described above, and should be
interpreted as lower bounds on the sbottom and gluino masses. One can see that
already at the LHC \eight, the 3-lepton search provides a very strong
constraint on the Yukawa unification scenario, much stronger than the limits
from Refs.\cite{CMS-PAS-SUS-13-018} and\cite{Aad:2014wea}. It is due to the
presence of one generation of light sleptons in the spectrum, although the
efficiency of the search is weakened with respect to the corresponding SMS,
for which the interpretations are provided in the experimental
analysis\cite{Aad:2014vma}. Such spectra, however, will be fully tested by the
LHC \four.

%% file: chapters/6_2_4_EWSB_Offdiag.tex

\begin{figure}[t]
\centering
\subfloat[]{
\includegraphics[width=0.48\textwidth]{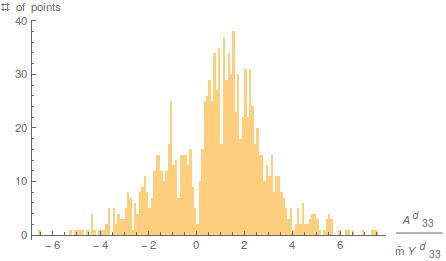}}
\subfloat[]{
\includegraphics[width=0.48\textwidth]{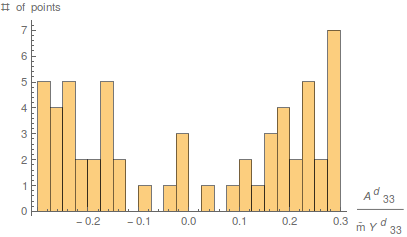}}
\caption[Histograms of points gathered by the $GFV_{23}$ scan as a
function of $A^d_{33}/(Y^d_{33} \widetilde{m}_3)$.]{Histograms of points
gathered by the $GFV_{23}$ scan as a function of $A^d_{33}/(Y^d_{33}
\widetilde{m}_3)$ at the scale $M_{SUSY}$: (a) all the points gathered
by the scan; (b) enlarged vicinity of zero.}
\label{fig:EWSB_GFV_23}
\end{figure}

As already mentioned at the beginning of this chapter, our $GFV_{23}$
scan allows significant values of $A^{de}_{33}/M_{1/2}$ at \mgut. As
in the case of the large $A$-term scenario, some of the points might
therefore be characterised by a global CCB minimum of the MSSM scalar
potential. Indeed, a large value of $A^{d}_{33} (M_{SUSY})$ makes
tuning the threshold corrections to Yukawa couplings easier, and most
of the points that are consistent with bottom-tau and strange-muon Yukawa
coupling unification show $A^d_{33}/(Y^d_{33}\widetilde{m}_3) \sim {\mathcal O}(1)$
at the scale $M_{SUSY}$, with $\widetilde{m}_i$ defined in
Eq.~\eqref{mtildeEq}.  For some points however, this ratio is close to
zero, as can be seen in Fig.~\ref{fig:EWSB_GFV_23}.

Therefore, we do not see it necessary that the $GFV_{23}$ 
scenario leads to a metastable vacuum, as the relevant factor
$A^d_{33}/(Y^d_{33} \widetilde{m}_3)$ at $M_{SUSY}$ can be fitted to lie an
order of magnitude below the coarse bound~\ref{roughstab}. However, this is
only a very weak statement, and a more detailed numerical analysis and new
scans will be necessary in the future to investigate this issue
properly, with inclusion of at least the (already complicated) tree-level
stability bounds\cite{Casas:1995pd}.

%% file: chapters/7_0_introGFV123.tex
In this chapter, we investigate the full $3\times 3$ Yukawa matrix 
$SU(5)$ unification within the GFV MSSM. We show the numerical results
obtained within the $GFV_{123}$ scenario that was introduced in
Chapter~\ref{TwoScenarios}. The scans have returned regions consistent with
the unification of lepton and down-type quark Yukawa couplings of all 
the families, and fulfilling all the imposed experimental
constraints except those coming from the LFV observables.

%% file: chapters/7_1_Regions_OffDiag_123.tex
To find regions consistent with the Yukawa matrix unification, we perform a
scan in the parameter space of the $GFV_{123}$ scenario within the bounds
presented in Table~\ref{tab:priors$SU(5)$}.
\begin{table}[t]
\centering
\renewcommand{\arraystretch}{1.1}
\begin{tabular}{|c|c|}
\hline 
Parameter &  Scanning Range\\
\hline 
$M_{1/2}$	& [$100$, $4000$] \gev\ \\
\mhu            & [$100$, $8000$] \gev\ \\
\mhd            & [$100$, $8000$] \gev\ \\
\tanb	        & [$3$, $45$] \\
\signmu		& $-1$ \\
\hline 
$A_{33}^{de}$   & [$0$, $5000$] \gev\ \\
$A_{33}^{u}$    & [$-9000$, $9000$] \gev\ \\
$A^{de}_{11}/A^{de}_{33}$   & [$-0.00028$, $0.00028$] \\
$A^{de}_{22}/A^{de}_{33}$  & [$-0.065$, $0.065$] \\
$A^{u}_{22}/A^{u}_{33}$    & [$-0.005$, $0.005$] \\
$A^{de}_{ij}/A^{de}_{33},\; i\neq j$    & [$-0.5$, $0.5$] \\
\hline 
$m_{ii}^{dl},\; i=1,2,3$   & [$100$, $7000$] \gev\ \\
$m^{dl}_{23}/m^{dl}_{33}$  & [$0$, $1$] \\
$m^{dl}_{13}/m^{dl}_{33}$  & [$0$, $1$] \\
$m^{dl}_{12}/m^{dl}_{33}$  & [$0$, $1$] \\
$m_{ii}^{ue},\; i=1,2,3$   & [$100$, $7000$] \gev\ \\
\hline 
\end{tabular}
\caption[Ranges of the input SUSY parameters used in the initial $GFV_{123}$
scan.]{Ranges of the input SUSY parameters used in our \emph{initial} $GFV_{123}$
scan. The omitted soft SUSY-breaking parameters at the GUT scale ($A^{u}_{11}$
as well as $A^{u}_{ij}$ and $m^{ue}_{ij}$ for $i\neq j$) have been set to
zero.}  \label{tab:priors$SU(5)$}
\end{table}

\begin{figure}[t]
\centering
\subfloat[]{
\includegraphics[width=0.41\textwidth]{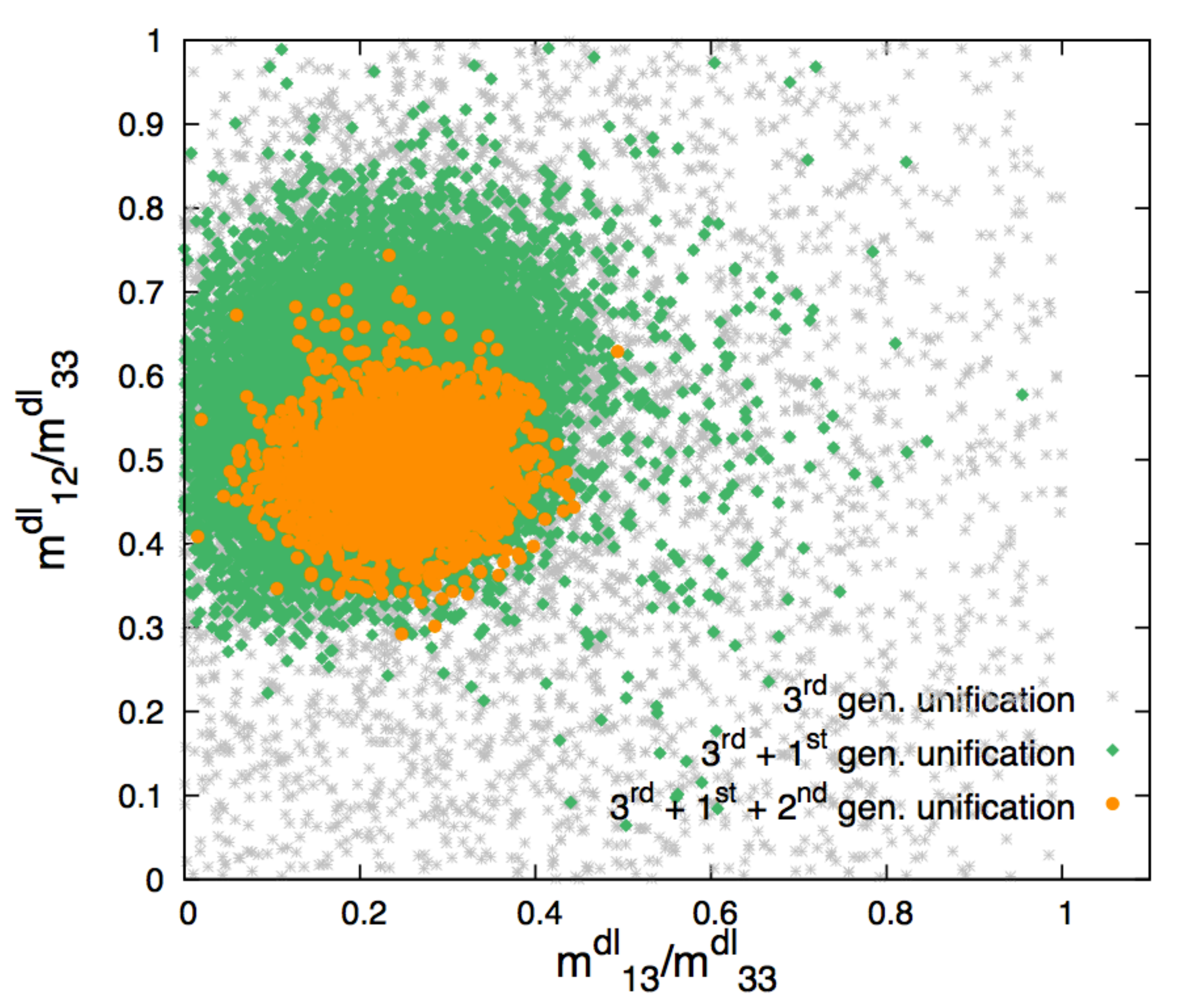}}
\subfloat[]{
\includegraphics[width=0.41\textwidth]{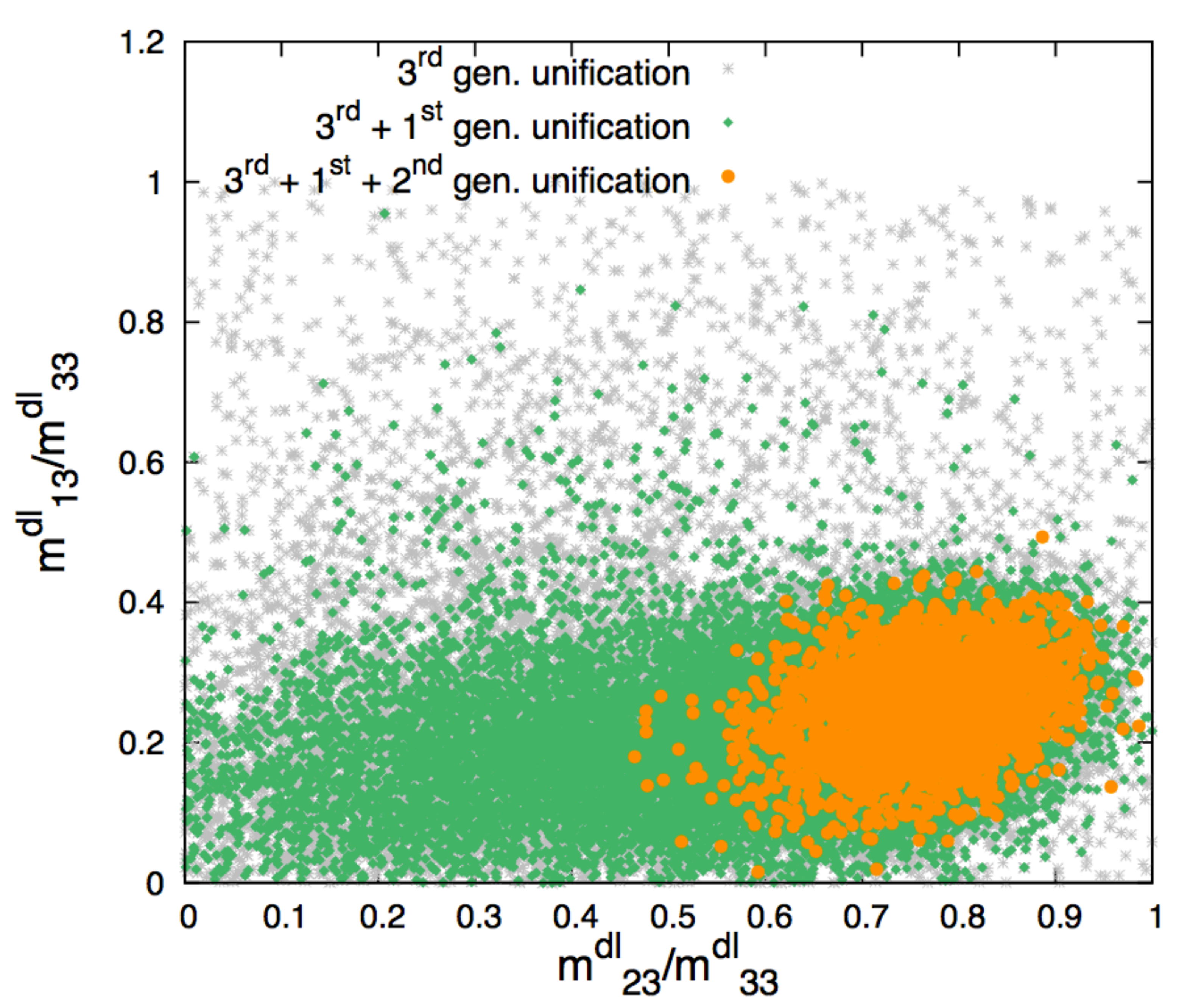}}
\caption[Scatter plot of the $GFV_{123}$ points in the planes
($m^{dl}_{13}/m^{dl}_{33}$,
$m^{dl}_{12}/m^{dl}_{33}$) and
($m^{dl}_{23}/m^{dl}_{33}$,
$m^{dl}_{13}/m^{dl}_{33}$)]{Scatter plot of the $GFV_{123}$ points in the planes
($m^{dl}_{13}/m^{dl}_{33}$,
$m^{dl}_{12}/m^{dl}_{33}$) (a), and
($m^{dl}_{23}/m^{dl}_{33}$,
$m^{dl}_{13}/m^{dl}_{33}$) (b). grey stars: all the points
satisfying the Yukawa unification condition for the third generation at
$2\sigma$; green diamonds: points additionally requiring $2\sigma$ unification
of the first family; orange dots: points for which all the three generations are
unified at $2\sigma$.}
\label{m12m13m23}
\end{figure}

\begin{figure}[t]
\centering
\subfloat[]{
\includegraphics[width=0.41\textwidth]{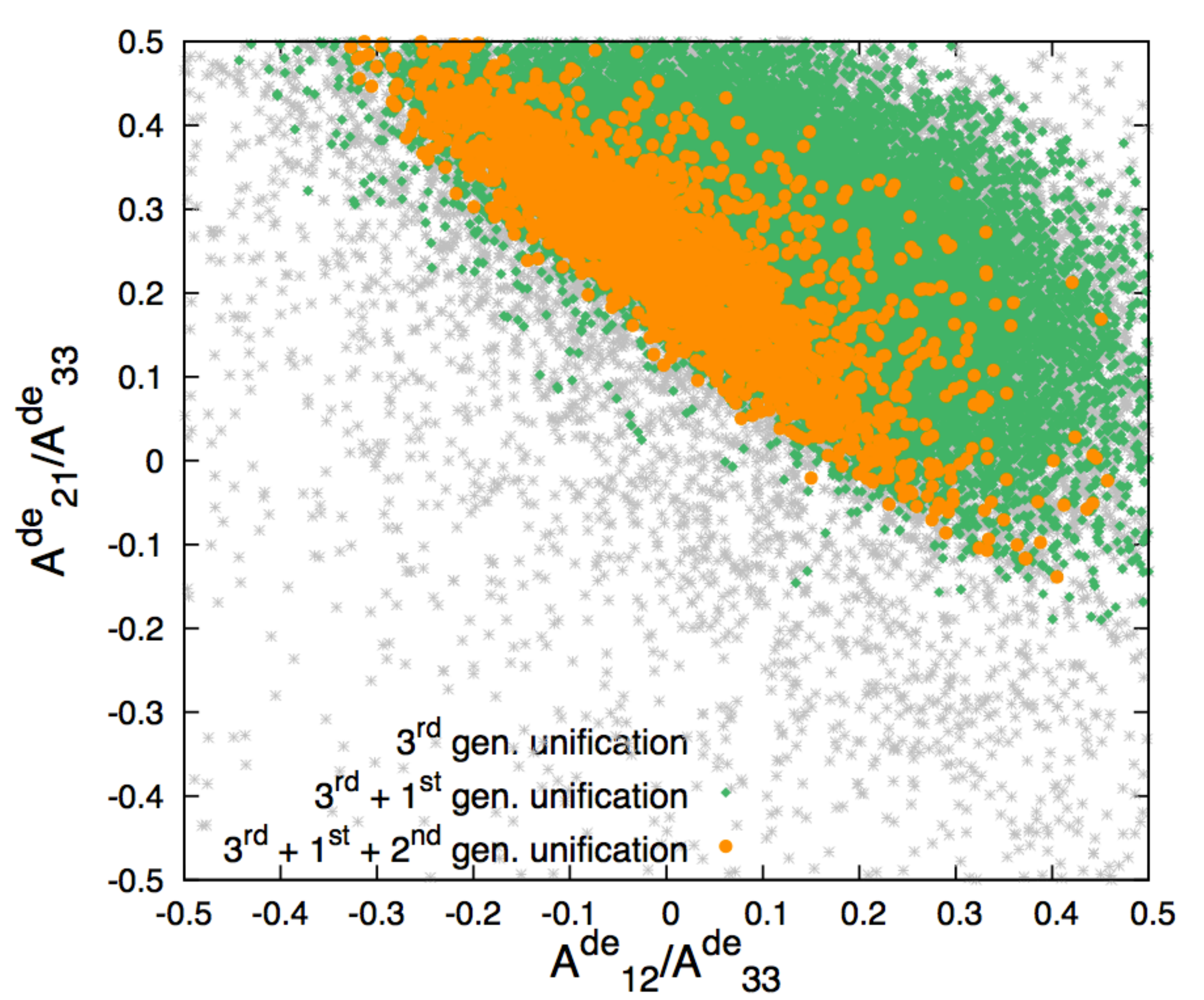}}
\subfloat[]{
\includegraphics[width=0.41\textwidth]{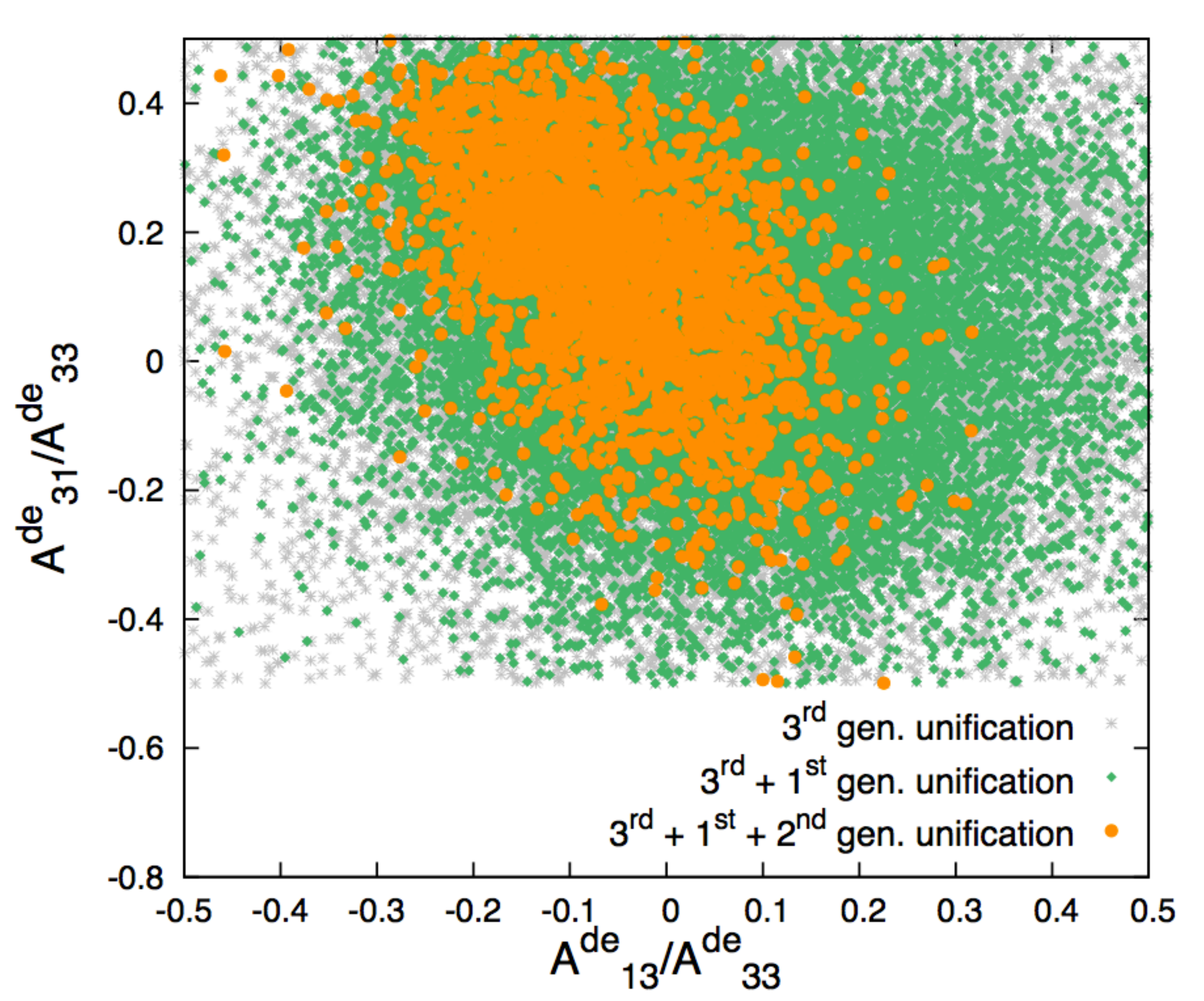}}\\
\subfloat[]{
\includegraphics[width=0.41\textwidth]{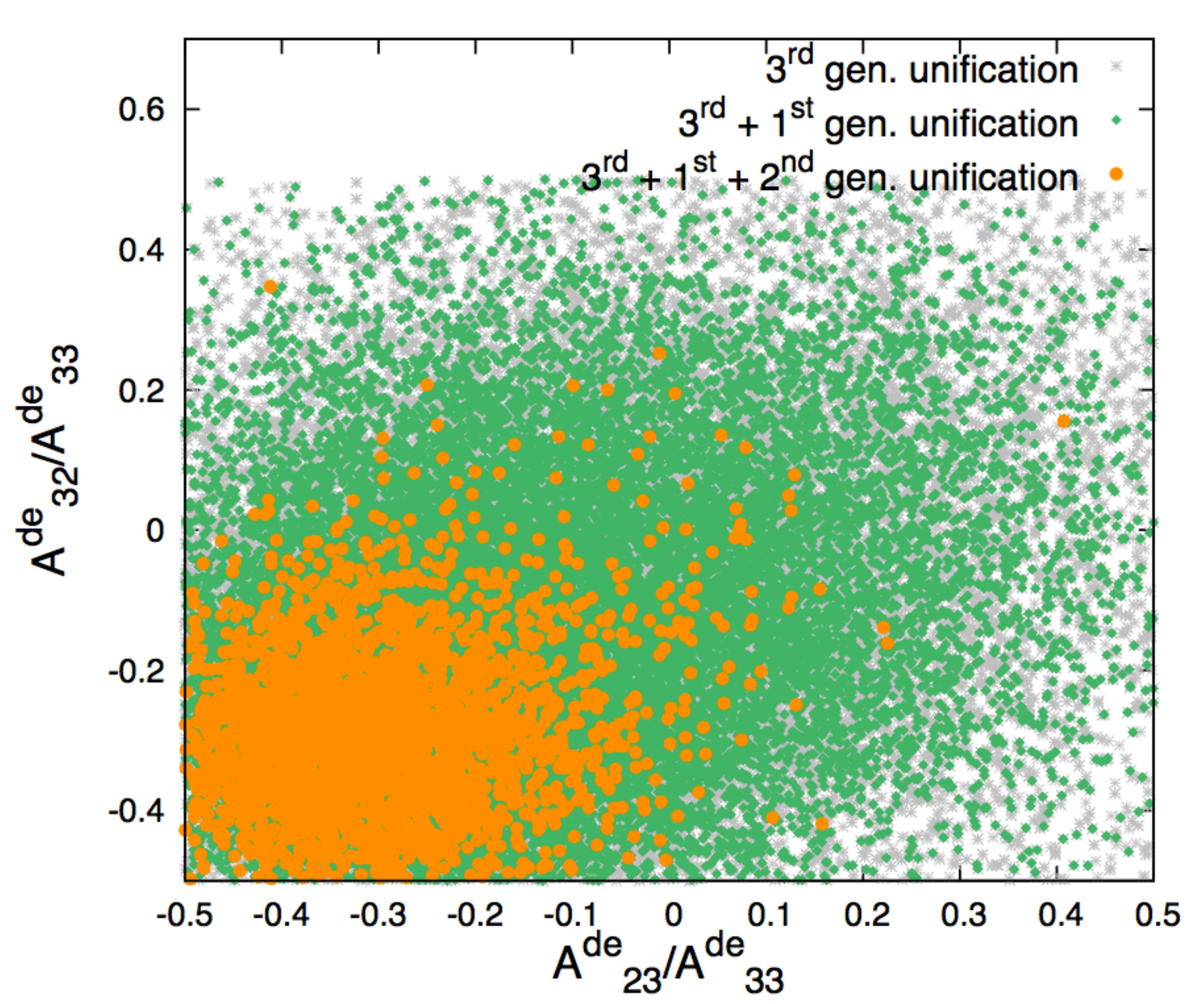}}
\caption[Scatter plost of the $GFV_{123}$ points in the planes
($A^{de}_{12}/A^{de}_{33}$,
$A^{de}_{21}/A^{de}_{33}$),
($A^{de}_{13}/A^{de}_{33}$,
$A^{de}_{31}/A^{de}_{33}$) and
($A^{de}_{23}/A^{de}_{33}$,
$A^{de}_{32}/A^{de}_{33}$)]{Scatter plot of the $GFV_{123}$ points in the planes
($A^{de}_{12}/A^{de}_{33}$,
$A^{de}_{21}/A^{de}_{33}$) (a),
($A^{de}_{13}/A^{de}_{33}$,
$A^{de}_{31}/A^{de}_{33}$) (b), and
($A^{de}_{23}/A^{de}_{33}$,
$A^{de}_{32}/A^{de}_{33}$) (c). The colour code is the same
as in \reffig{m12m13m23}.}
\label{t12t13t23}
\end{figure}

In \reffig{m12m13m23}, we present distributions of the points we collected in
our scanning procedure in the planes
($m^{dl}_{13}/m^{dl}_{33}$,
$m^{dl}_{12}/m^{dl}_{33}$) (a) and
($m^{dl}_{23}/m^{dl}_{33}$,
$m^{dl}_{13}/m^{dl}_{33}$) (b). 
All the points that satisfy the Yukawa unification condition for the third
generation at $2\sigma$ ($0.9<Y_b/Y_{\tau}<1.1$) are depicted as grey stars
(they account for $49\%$ of all the points), while those that additionally
fulfil $2\sigma$ unification of the first family as green diamonds ($29\%$ of
all the points). Finally, orange dots correspond to those points for which all
three generations are unified at $2\sigma$ ($1.7\%$ of all the points
collected by the scan). In \reffig{t12t13t23}, similar distributions are shown
for the flavour-violating entries of the trilinear down-sector matrix, in the
planes corresponding to
($A^{de}_{12}/A^{de}_{33}$,
$A^{de}_{21}/A^{de}_{33}$) (a),
($A^{de}_{13}/A^{de}_{33}$,
$A^{de}_{31}/A^{de}_{33}$) (b), and
($A^{de}_{23}/A^{de}_{33}$,
$A^{de}_{32}/A^{de}_{33}$) (c).

Exactly as in the case of the $GFV_{23}$ scenario, both \reffig{m12m13m23} and
\reffig{t12t13t23} confirm, that a satisfactory unification of the third
family Yukawa couplings can be easily achieved in the Minimal Flavour
Violating $SU(5)$ for moderate values of \tanb.

The functional form of the threshold correction in Eq.~(\ref{sigma22}) might
suggest that non-zero soft-mass elements $m^{dl}_{23}$ and $m^{dl}_{13}$ 
together with the RG-generated $(m^2_{\tilde{q}})_{23}$ and $(m^2_{\tilde{q}})_{13}$ 
are sufficient to allow the Yukawa unification in both the second and first
family cases. Such a simplistic picture, however, is not true, as can be seen
from the panel (a) of \reffig{m12m13m23} where large $m^{dl}_{12}$ is clearly
favoured. To understand what happens, let us note that the GFV corrections
$(\Sigma^d_{22})^{\tilde{g}}$ and $(\Sigma^d_{11})^{\tilde{g}}$ (obtained from
Eq.~(\ref{sigma22}) by replacing indices ``2'' with ``1'') are determined by
overlapping sets of parameters, in particular \mhalf\ and $A_{33}^{de}$. On
the other hand, sizes of those corrections, as required by the Yukawa coupling
unification, differ by two orders of magnitude. Let us now assume that
$(\Sigma^d_{11})^{\tilde{g}}$ is fixed by the unification condition for the
first family. Thus \mhalf\ and $A_{33}^{de}$, already constrained by
unification of the third family, are even more limited. With such a choice of
parameters, however, the correction $(\Sigma^d_{22})^{\tilde{g}}$ is still too
small to allow unification of the second family, and needs to be further
enhanced by another contribution. Such a contribution comes from a diagram
like the one shown in Fig.~\ref{diagsh}, but with the trilinear term
$A_{21}^{de}$ in the vertex, and $m^{dl}_{12}$ mixing in the right-handed
sector. However, a similar diagram also exists for the first family, and the
corresponding contribution should be added to the one driven by
$m^{dl}_{13}$. That explains why all the five parameters $m^{dl}_{12}$,
$m^{dl}_{13}$, $m^{dl}_{23}$, $A_{12}^{de}$ and $A_{21}^{de}$ must be adjusted
simultaneously. Note also that $A_{12/21}^{de}$ can be kept relatively low, as
this contribution is always enhanced by a large value of $m^{dl}_{12}$.
\begin{figure}[ht]
\centering
\subfloat[]{
\includegraphics[width=0.41\textwidth]{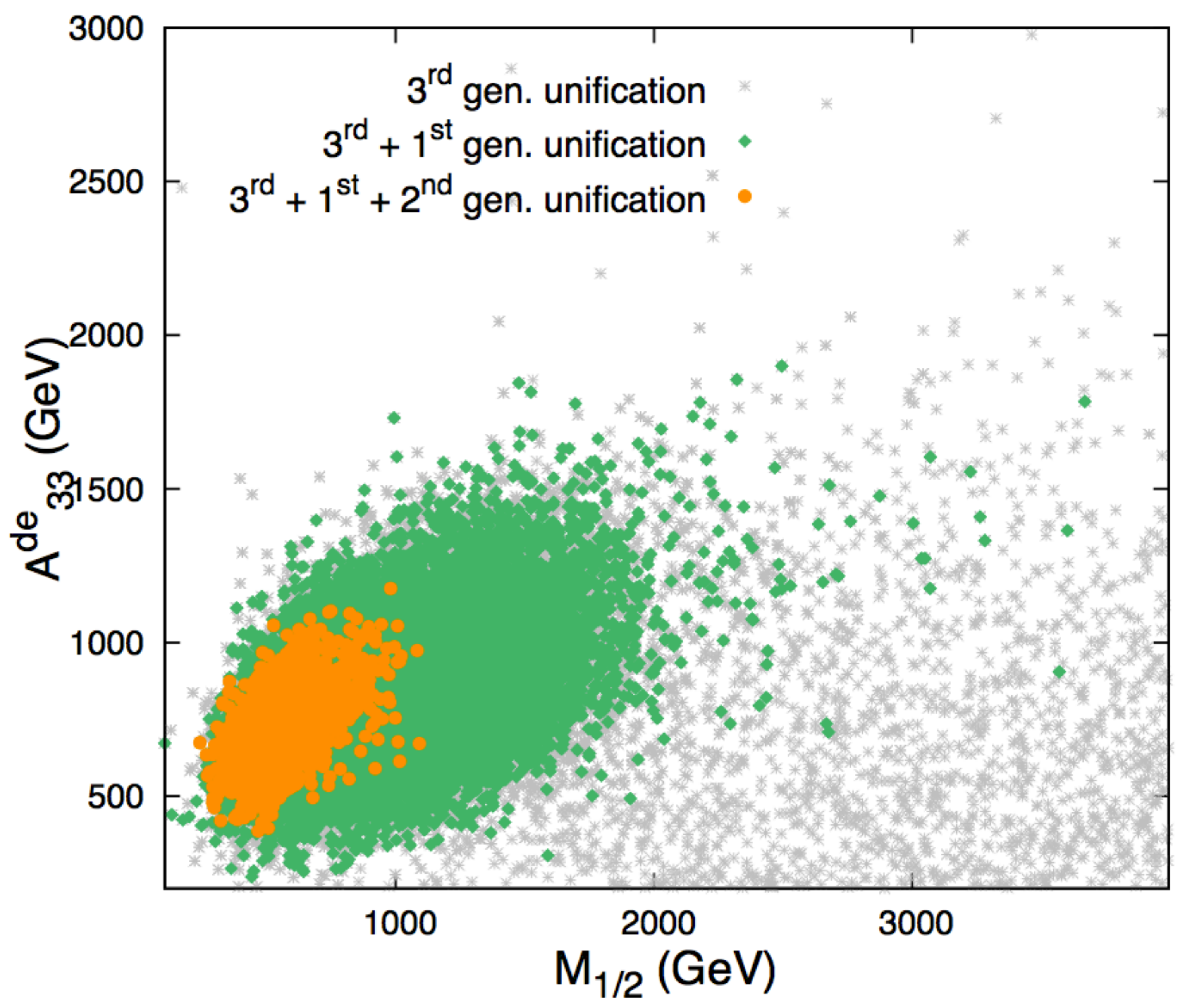}}
\subfloat[]{
\includegraphics[width=0.41\textwidth]{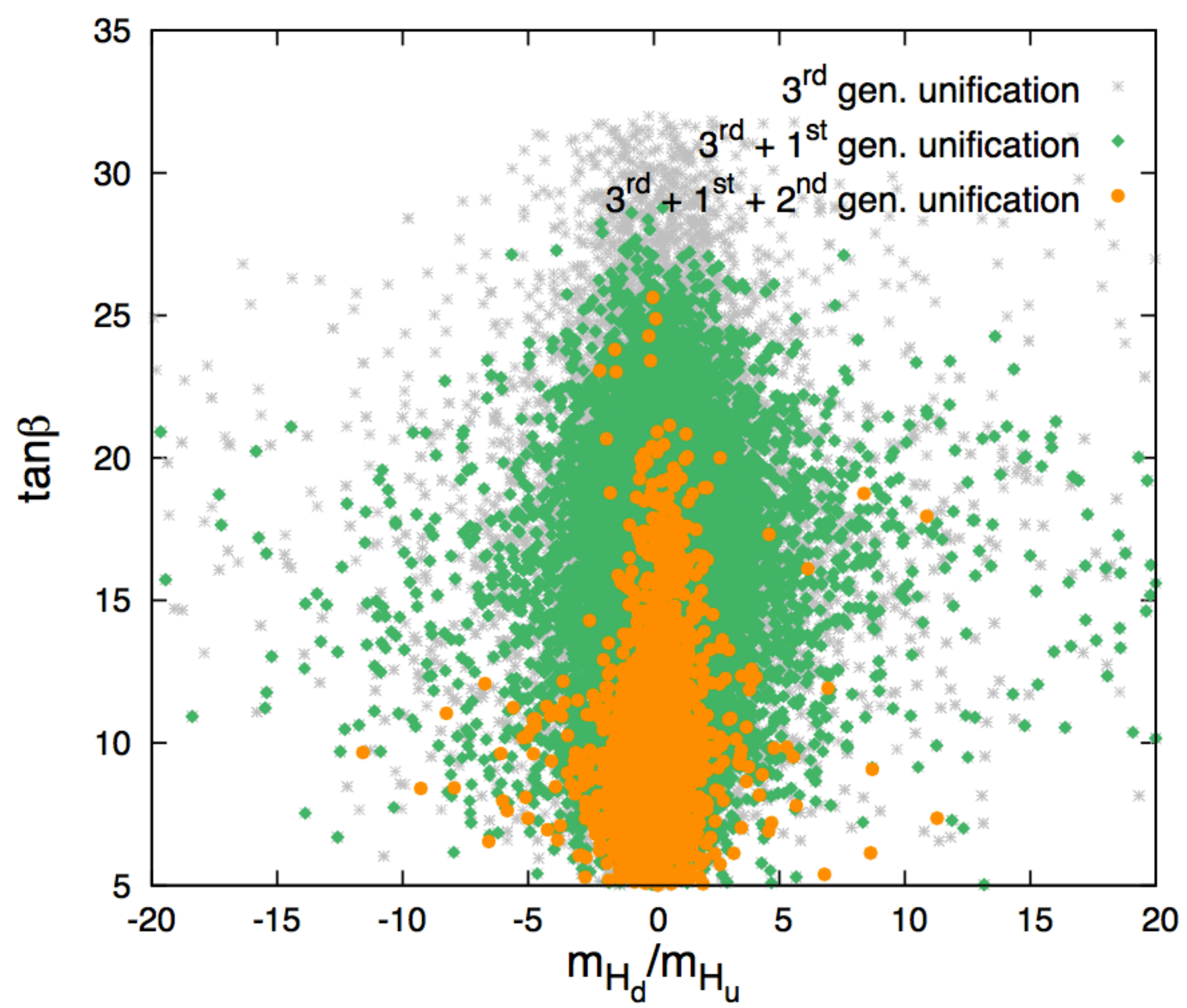}}\\
\subfloat[]{
\includegraphics[width=0.41\textwidth]{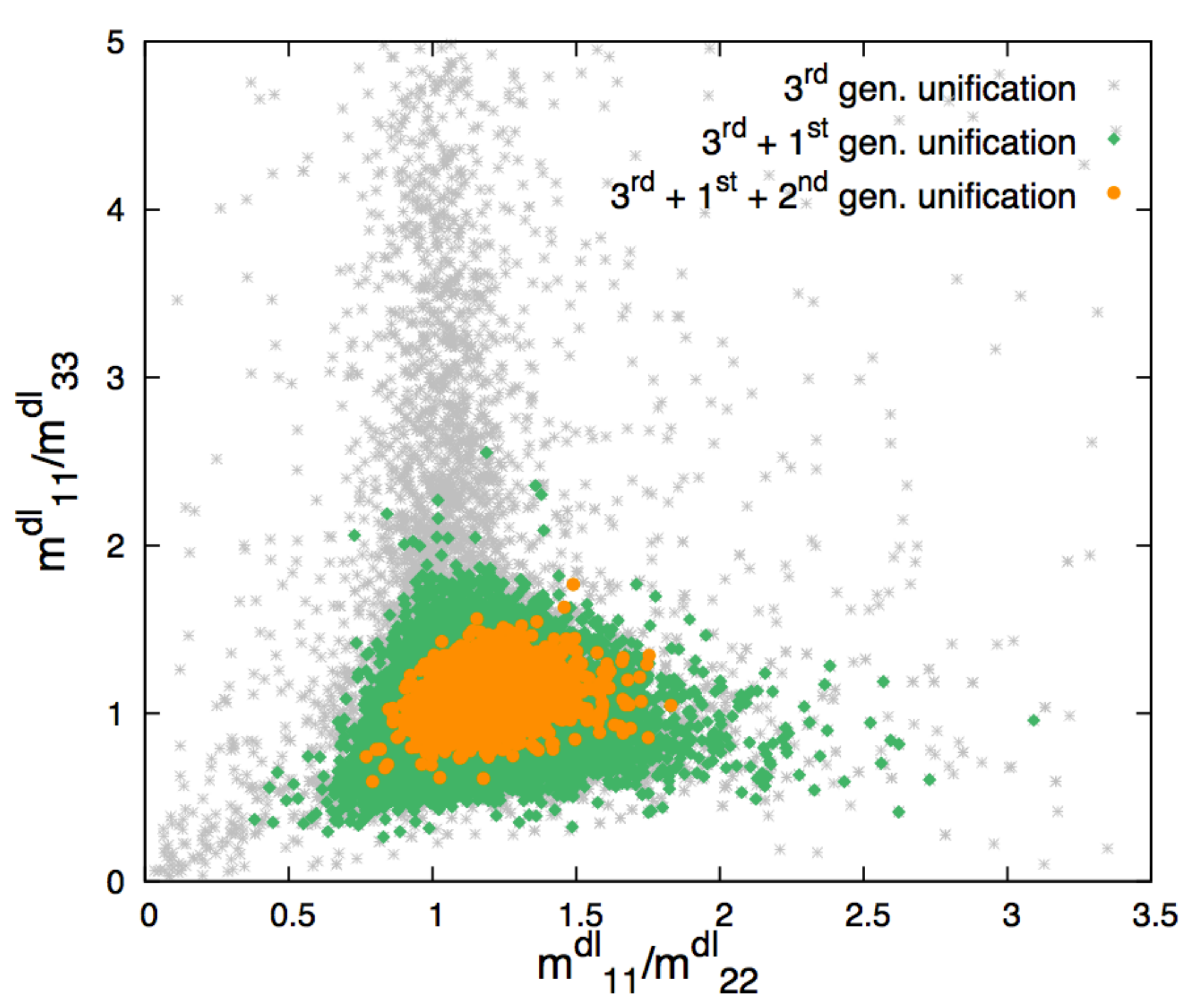}}
\caption[$GFV_{123}$ points in the planes
($\mhalf$, $A^d_{33}$), ($\mhd/\mhu$, $\tanb$), and
($m^{dl}_{11}/m^{dl}_{22}$,
$m^{dl}_{11}/m^{dl}_{33}$) ]{Scatter plot of the $GFV_{123}$ points in the planes
($\mhalf$, $A^d_{33}$) (a), 
($\mhd/\mhu$, $\tanb$) (b), and
($m^{dl}_{11}/m^{dl}_{22}$,
$m^{dl}_{11}/m^{dl}_{33}$) (c). 
The colour code is the same as in \reffig{m12m13m23}.}
\label{m12a33tanb}
\end{figure}

Flavour-violating parameters are not the only ones constrained by the Yukawa
unification condition. In \reffig{m12a33tanb}, we present distributions of
points in the planes ($\mhalf$, $A^{de}_{33}$) (a), ($\mhd/\mhu$, $\tanb$)
(b), and ($m^{dl}_{11}/m^{dl}_{22}$, $m^{dl}_{11}/m^{dl}_{33}$) (c). The
colour code is the same as in \reffig{m12m13m23}. One can observe that values
of both \mhalf\ and $A^d_{33}$ need to be very limited in order to facilitate
unification in the first and second family cases, as they directly enter
Eq.~(\ref{sigma22}). The ratio $\mhd/\mhu$ in the range $[0,2]$ allows the
unification of the second family for larger values of \tanb, namely
$\tanb\in [15,25]$. Finally, large mass splittings between the diagonal entries
of the down-squark mass matrix are disfavoured because they would lead to a
strong suppression of SUSY threshold corrections, as can be deduced from
Eq.~(\ref{sigma22}).

We conclude this section with summarising the allowed ranges of the non-zero
GFV parameters that characterise the $SU(5)$ GUT scenario with the full Yukawa
coupling unification:
\bea
0.5<m^{dl}_{23}/m^{dl}_{33}<1,&\quad 0 < m^{dl}_{13}/m^{dl}_{33} < 0.5,&\quad 0.3 < m^{dl}_{12}/m^{dl}_{33} < 0.7, \nonumber
\eea
\bea\label{eq2}
0 <A^{d}_{12}/A^{d}_{33} < 0.2, &\quad 0 <A^{d}_{21}/A^{d}_{33} < 0.2.
\eea

%% file: chapters/7_2_phenoIntro.tex
In order to identify points satisfying both the GUT-scale Yukawa
unification requirements and the experimental constraints, we use
the tools described in Chapter~\ref{ToolsChap}, and scan the parameter space
specified in Table~\ref{tab:priorsYuk}.  The scanning
ranges are chosen to contain the region consistent with full $3\times 3$
Yukawa matrix unification, as discussed in \refsec{sec:unif}. Thus, we
restrict our search to $M_{1/2}$ lower than 1 TeV, $\tanb<25$ and
$A^{de}_{33}$ in the range of [$400$, $1100$] \gev. The off-diagonal terms
$A^{de}_{12}/A^{de}_{33}$, $A^{de}_{21}/A^{de}_{33}$,
$m^{dl}_{23}/m^{dl}_{33}$, $m^{dl}_{13}/m^{dl}_{33}$,
$m^{dl}_{12}/m^{dl}_{33}$ are limited according to \refeq{eq2}. The remaining
parameters are within the initial scanning ranges (cf. Table
~\ref{tab:priors$SU(5)$}), as they do not affect the Yukawa unification. The
experimental constraints considered in the analysis are listed in
Table~\ref{tab:exp_constraints}.
\begin{table}[t]
\centering
\renewcommand{\arraystretch}{1.1}
\begin{tabular}{|c|c|}
\hline 
Parameter &  Scanning Range\\
\hline 
$M_{1/2}$	& [$200$, $1100$] \gev\ \\
\mhu, \mhd            & [$100$, $8000$] \gev\ \\
\tanb	        & [$3$, $25$] \\
\signmu		& $-1$ \\
\hline 
$A_{33}^{de}$   & [$400$, $1100$] \gev\ \\
$A_{33}^{u}$    & [$-9000$, $9000$] \gev\ \\
$A^{de}_{11}/A^{de}_{33}$   & [$-0.00028$, $0.00028$] \\
$A^{de}_{22}/A^{de}_{33}$  & [$-0.065$, $0.065$] \\
$A^{u}_{22}/A^{u}_{33}$    & [$-0.005$, $0.005$] \\
$A^{de}_{12}/A^{de}_{33}$,\; $A^{de}_{21}/A^{de}_{33}$ & [$-0.2$, $0.2$] \\
\hline 
$m_{ii}^{dl},\; i=1,2,3$   & [$100$, $7000$] \gev\ \\
$m^{dl}_{23}/m^{dl}_{33}$  & [$0.5$, $1.0$] \\
$m^{dl}_{13}/m^{dl}_{33}$  & [$0.0$, $0.5$] \\
$m^{dl}_{12}/m^{dl}_{33}$  & [$0.3$, $0.7$] \\
$m_{ii}^{ue},\; i=1,2,3$   & [$100$, $7000$] \gev\ \\
\hline 
\end{tabular}
\caption[Ranges of the input SUSY parameters in the final scan for the $GFV_{123}$
scenario.]{Ranges of the input SUSY parameters in our \emph{final} scan for the $GFV_{123}$
scenario. The parameters that are not explicitly listed in the table (namely
$A^{u}_{11}$, $A^{u}_{ij}$ and $m^{ue}_{ij}$ for $i\neq j$, $A^{de}_{23/32}$,
$A^{de}_{13/31}$) have been set to zero.}
\label{tab:priorsYuk}
\end{table}

%% file: chapters/7_3_LFV.tex
\begin{figure}[h!]
\centering 
\includegraphics[width=0.7\textwidth]{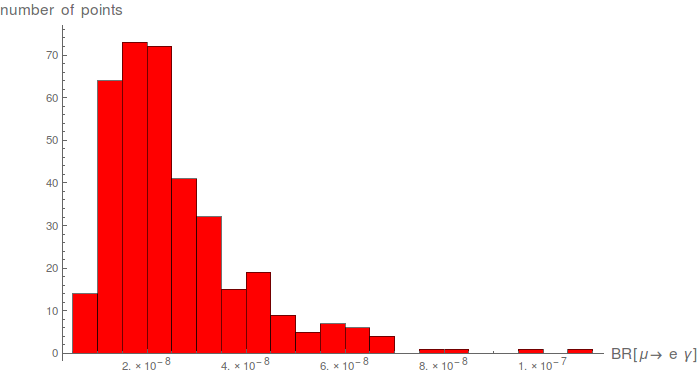}
\caption[\brmuegamma{} calculated for points from $GFV_{123}$ scan.]{Histogram
showing \brmuegamma{} calculated for points satisfying other phenomenological
constraints in $GFV_{123}$ scenario. It exceeds the experimental upper bound
by five orders of magnitude. They were found by the scan specified by the
ranges presented in Table~\ref{tab:priorsYuk}.\label{mueGammaHist}}
\end{figure}

In the $GFV_{123}$ scenario, the muon-electron conversion observables might be
strongly enhanced. All of them are influenced by the relatively large values of
$m^{dl}_{13}$, $m^{dl}_{12}$, $A^{de}_{12}$, $A^{de}_{21}$, $A^{de}_{13}$,
$A^{de}_{31}$ that are required for the electron-(down quark) Yukawa
unification. The 90\% \cl\ upper bound on \brmuegamma{} reported in
Ref.\cite{Adam:2013mnn} equals $5.7\times 10^{-13}$. All the points of our
scan that satisfy the full $3 \times 3$ Yukawa matrix unification requirement exceed this bound by
five orders of magnitude, as shown in \reffig{mueGammaHist}.

It is theoretically possible to evade this constraint while unifying 
the electron and down-quark Yukawa couplings by raising the overall
scale of the superpartner masses. However, this could not be
achieved with the applied numerical tools that assume
$\mu_{sp}=M_Z$. All our scans consistently found values of 
$M_{1/2}$ to be around 1 TeV at \mgut. The fixed condition
$\mu_{sp}=M_Z$ is one of the weaknesses of the spectrum generators that were
available and widely tested at the moment when our research work at
the Yukawa unification problem started.

%% file: chapters/7_4_EWSB_OffDiag.tex
Let us for the moment assume that if the superpartners were heavier, we
could satisfy both the full $3\times 3$ Yukawa matrix unification and
all the experimental conditions listed in
Table~\ref{tab:exp_constraints}. There is still an important
non-decoupling effect of phenomenological importance, namely the
already discussed vacuum (meta)stability problem. Here, we study 
this issue for the points found in our TeV-scale $GFV_{123}$
scenario.

As mentioned in \refsec{sec:unif}, non-zero elements $A^{de}_{12/21}$ are
required to achieve the Yukawa coupling unification for both the first and
second family. However, off-diagonal entries of the trilinear couplings (as
well as the diagonal ones) are strongly constrained by the requirement of EW
vacuum stability. When the flavour-violating entries are too large, a
CCB minimum may appear in the MSSM scalar potential,
and it may become deeper than the standard EW one. The potential may also
become unbounded from below
(UFB)\cite{Frere:1983ag,AlvarezGaume:1983gj,Derendinger:1983bz,Kounnas:1983td,Casas:1995pd,
Casas:1996de}. An important feature of all such constraints is that, unlike
the FCNC ones, they do not become weaker when the scale \msusy\ is increased.

\begin{figure}[t]
\centering
\subfloat[]{
\includegraphics[width=0.41\textwidth]{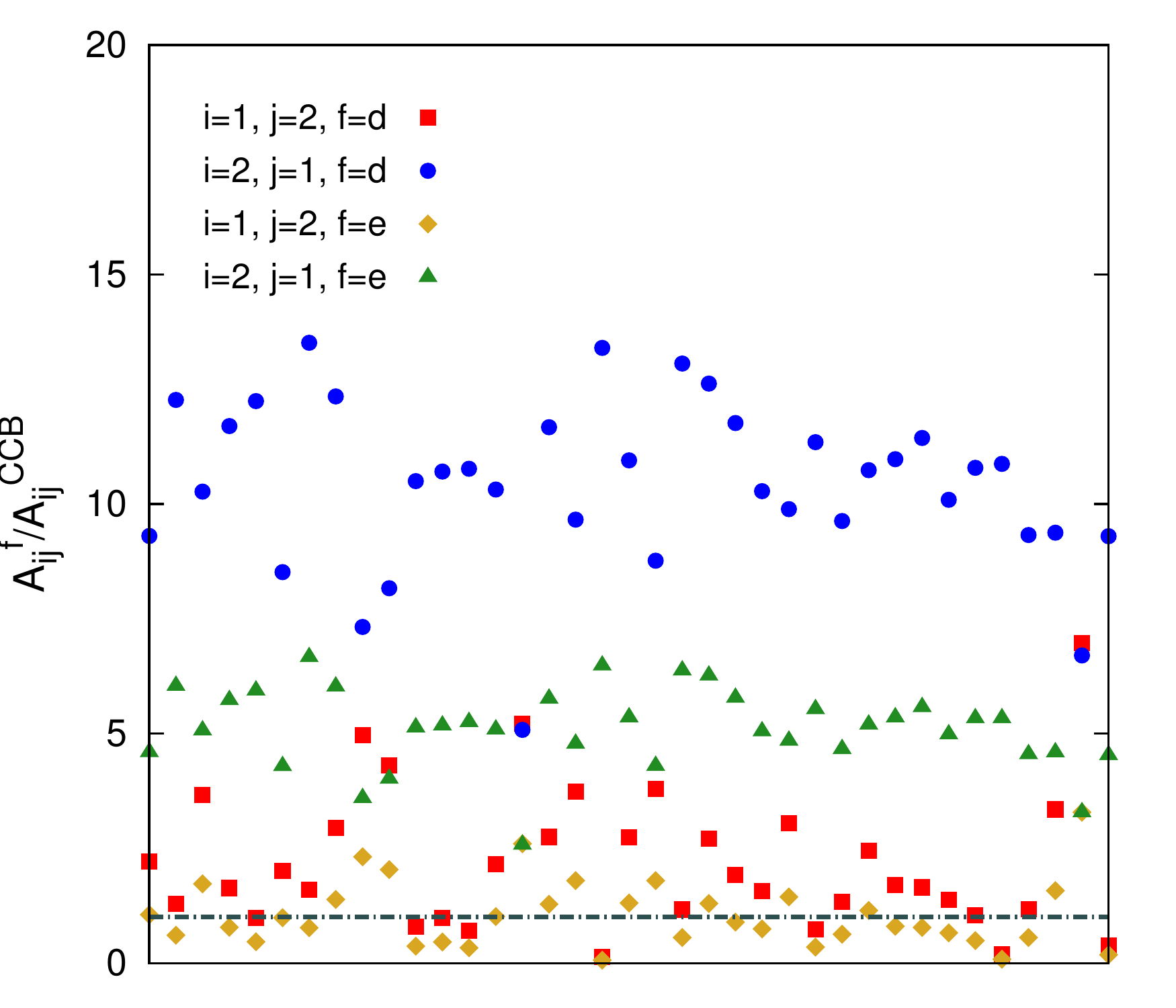}}
\subfloat[]{
\includegraphics[width=0.41\textwidth]{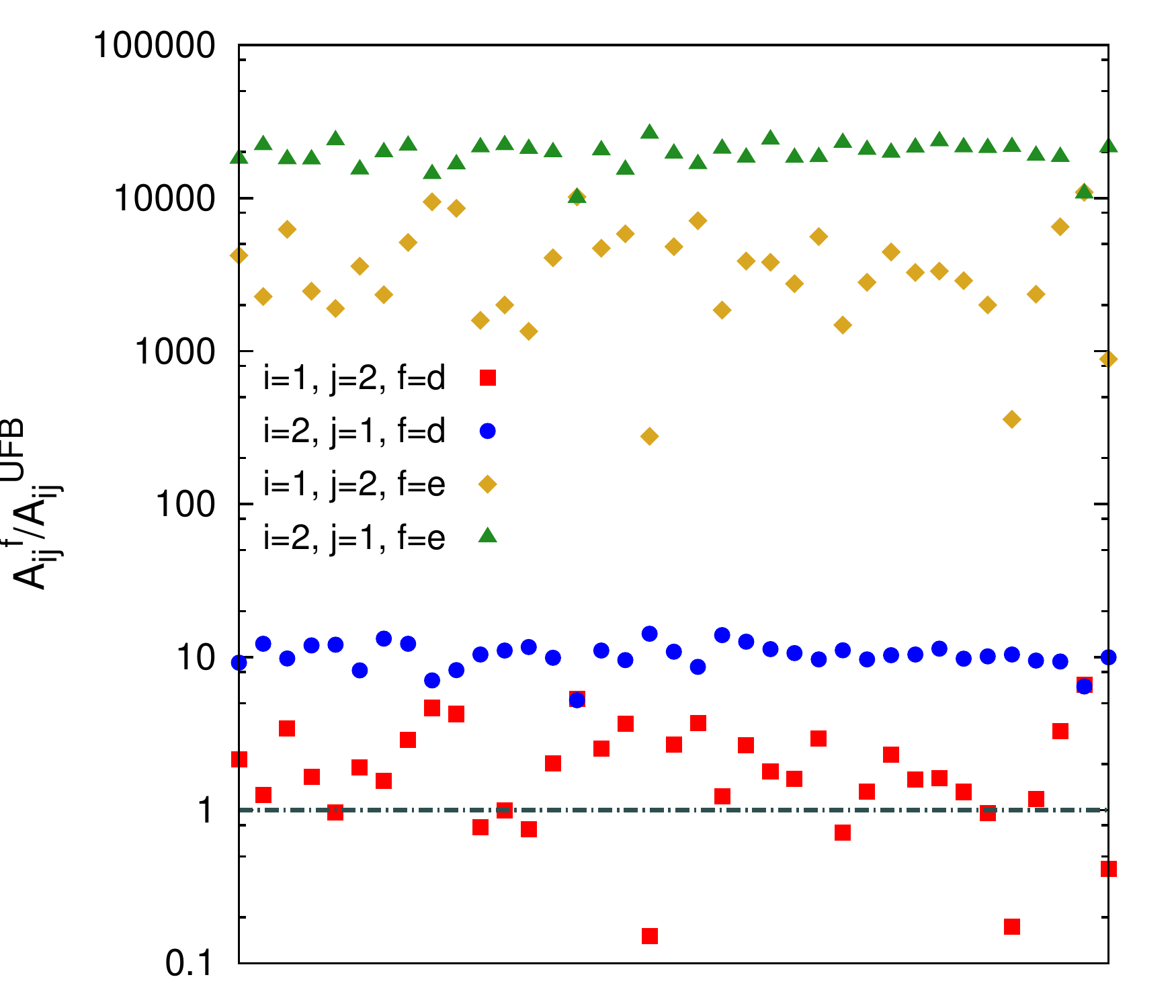}}
\caption[EW vacuum CCB (a) and UFB (b) stability bounds on the elements
$A^{d}_{12/21}$ and $A^{e}_{12/21}$.]{EW vacuum CCB (a) and UFB (b) stability bounds on the elements
$A^{d}_{12/21}$ and $A^{e}_{12/21}$. Dashed line indicates the upper limit on
the allowed size of the off-diagonal trilinear terms.}
\label{vacuum}
\end{figure}

In the down-squark sector, tree-level formulae for the CCB bounds are given
by\cite{Casas:1996de}
\bea\label{ccb}
(v_d/\sqrt{2})A^{d}_{ij}&\leq& 
m^d_{k}[(m^2_{\tilde{q}})_{ii}+(m^2_{\tilde{d}})_{jj}+\mhd^2+\mu^2]^{1/2},\quad k
=\textrm{Max}(i,j).
\eea
The limits on $A^{e}_{ij}$ have an analogous form, up to replacing the matrix
$m^2_{\tilde{d}}$ by $m^2_{\tilde{e}}$. Similarly, the UFB bounds
read\cite{Casas:1996de}
\bea\label{ufb}
(v_d/\sqrt{2})A^{d}_{ij}&\leq& 
m^d_{k}[(m^2_{\tilde{q}})_{ii}+(m^2_{\tilde{d}})_{jj}+(\mll^2)_{ii}+(m^2_{\tilde{e}})_{jj}]^{1/2},\nonumber\\
(v_d/\sqrt{2})A^{e}_{ij}&\leq& \sqrt{3}m^l_{k},\quad k=\textrm{Max}(i,j).
\eea
In \reffig{vacuum}, we show to what extent the CCB and UFB limits are
satisfied for the points that allow the Yukawa coupling unification and
satisfy at $3\sigma$ all the experimental constraints listed in
Table~\ref{tab:exp_constraints}. The dashed lines indicate upper limits on the
allowed size of the off-diagonal trilinear terms. One can see that the CCB
stability bounds are violated by around an order of magnitude. The
situation is even worse in the case of the UFB bounds where the size of
the elements $A^e_{ij}$ are around four orders of magnitude larger than
it is allowed by the stability constraint. It results from the fact
that the UFB limit on $A^e_{ij}$ is of the order of the muon mass. Therefore,
we conclude that the EW MSSM vacuum is not stable in the $GFV_{123}$ Yukawa
unification scenario.

On the other hand, even if a CCB minimum appears, it does not imply that a
considered model is not valid, as long as the standard EW vacuum lives longer
than the age of the universe. Moreover, in such a case, the UFB bound becomes
irrelevant because the probability of a tunnelling process along the CCB
direction is much higher\cite{Park:2010wf}. To derive metastability bounds,
the bounce action for a given scalar potential should be calculated
numerically, which is beyond the scope of our present work. To evaluate the impact of
metastability on the validity of the unification scenario, we will use instead
the results of the analysis performed in Ref.\cite{Park:2010wf}. The derived
metastability bounds do not depend on the Yukawa couplings, and therefore are
in general much less stringent than the CCB ones. The effect is the strongest
for the $A_{12/21}$ elements, in which case the CCB limit can be weakened by
three to four orders of magnitude, depending on a particular choice of the
model parameters. We can therefore conclude that the $GFV_{123}$ scenario
leads to an unstable, but a long-lived vacuum.

It should be stressed, though, that the tension between the Yukawa unification
condition and the EW vacuum stability is less severe in $GFV_{123}$ scenario
than in the case of Yukawa unification through large diagonal $A$-terms where
the CCB bounds were violated by two orders of magnitude~\refsec{vacSect}. On
the other hand, theoretical calculations of the stability conditions are still
marred with many uncertainties. This leaves a possibility that future
improvements might further reduce (or even eliminate) the tension between the
vacuum stability and the Yukawa unification.

%% file: chapters/8_conclusions.tex
\section{Summary}

This work provided evidence in support of the following statement: 
\par
\emph{There exist regions in the R-parity conserving MSSM parameter space for
    which the unification of the down-type quark and lepton Yukawa matrices takes
    place, while the predicted values of flavour, electroweak and other
    collider observables are consistent with experimental constraints.}
    
We pointed out three distinct MSSM scenarios that encompass points
consistent with the bottom-tau and strange-muon Yukawa unification, and for
which all the investigated experimental constraints are
satisfied. These constraints include the mass and decay rates of the
lightest Higgs boson, EW precision tests, flavour observables, limits
on the spin-independent proton-neutralino scattering cross-section, as
well as the 8\tev\ LHC exclusion bounds from the direct SUSY searches.  
Our MSSM scenarios are not exclusive, and should be viewed as
complementary strategies of tuning the MSSM threshold corrections to
the Yukawa couplings. At \mgut{}, they can be characterised by the
following parameters expressed in the super-CKM basis:
\begin{enumerate}
 \item large diagonal entries of the trilinear soft terms $A^{de}_{22}$ and $A^{de}_{33}$; 
 \item large and negative $\mu$, 
       low value of the ratio $m^{dl}_{22}/m^{dl}_{33}$, 
       moderate or high \tanb; 
 \item non-zero off-diagonal soft mass element $m^{dl}_{23}$.
\end{enumerate}

The second of the above options was analysed as a special case of 
our $GFV_{23}$ scenario that allows a non-zero $m^{dl}_{23}$ element. Both
the second and the third approach assume diagonal $A$-terms that are
limited in magnitude by the corresponding Yukawa couplings. We provided
evidence that the relic density of the neutralino dark matter can be
reproduced within the second and third scenario. Points found for both of them
are characterised by a large magnitude of the $\mu$ term, which increases
the well-known fine-tuning problem of the MSSM.

In the first scenario with large $A$-terms, the MSSM scalar
potential possesses a charge- and colour-breaking (CCB) global
minimum. However, the decay time of the standard electroweak vacuum in this
case is longer than the age of the Universe, so this scenario is not
phenomenologically excluded.  In the $GFV_{23}$ scenario, we
assumed that $\tfrac{|A^f_{ii}|}{|A^f_{33}|} < \tfrac{Y^f_{ii}}{Y^f_{33}}$ to
avoid this problem.

We showed that the electron-(down quark) Yukawa unification can take place 
for any of the following types of parameter configurations in the
super-CKM basis:
\begin{enumerate}
 \item[\it (i)] a large diagonal trilinear term $A^{de}_{11}$;
 \item[\it (ii)] non-zero off-diagonal soft terms 
                    $m^{dl}_{13}$, $m^{dl}_{12}$, $A^{de}_{12}$, $A^{de}_{21}$, $A^{de}_{13}$, $A^{de}_{31}$.
\end{enumerate}
The latter scenario, however, is excluded by the LFV observables, as
all the points consistent with the electron-(down quark)
unification that were found by our scan predict values of \brmuegamma{} that
exceed the experimental 90\% \cl\ upper limit by at least five orders of
magnitude. On the other hand, in the case {\it (i),} the large
$A^{de}_{11}$ makes the MSSM vacuum metastable but long-lived,
similarly to the case of large $A^{de}_{22}$ in the second generation
case.

\section{Open questions and discussion}

In our analysis, we have assumed validity of the MSSM up to the GUT scale,
    i.e.\ we ignored the likely fact that the right-handed neutrinos do not
    decouple at \mgut{} but rather at an intermediate scale. Given the
    existing bounds on the neutrino masses, the size of this scale is
    correlated with assumed values of the Yukawa couplings in the matrix
    $\mathbf{Y^\nu}$ that is responsible for generating the dimension-five operator in
    Eq.\eqref{qnunu}. In the $SU(5)$ context (contrary to the $SO(10)$ one),
    the size of $\mathbf{Y^\nu}$ is unrelated to $\mathbf{Y^u}$ and
    $\mathbf{Y^{de}}$. If the elements of $\mathbf{Y^\nu}$ are numerically small, 
    the intermediate scale is much lower than \mgut, but the RG-evolution of all
    the other MSSM parameters is hardly affected due to the very smallness of
    $\mathbf{Y^\nu}$. In such a situation our approach is valid to a very
    good approximation. On the other hand, if some of the couplings in 
    $\mathbf{Y^\nu}$ are large, a new analysis would be necessary. In
    fact, potential effects of $\mathbf{Y^\nu}$ on the Yukawa
    unification have been already discussed in the literature -- see, e.g.,
    Ref.\cite{Gomez:2009dr}.

If the MSSM were to provide an explanation for the exact $SU(5)$ Yukawa
matrix unification at \mgut{}, the case of the heavier two families could be
resolved by any of the strategies 1-3 mentioned in the previous section. The
electron-(down quark) Yukawa unification was shown to be consistent
with experiment only in the large $A$-term case. Nevertheless, the
$GFV_{123}$ scenario could be modified, so that the LFV bounds are not
violated. One possibility would be to raise the superpartner 
masses. Another option is to abandon the $SU(5)$ condition of
equality of the squark and slepton soft terms, which might be
possible if the soft terms are affected by the $SU(5)$-breaking scalar
fields. This way, the off-diagonal elements of $m^{d}_{ij}$ could
still be used to adjust the light SM fermion mass ratios, while 
the off-diagonal $m^{l}_{ij}$ might remain low 
enough to satisfy the LFV constraints.

Another natural question is whether it might be possible to unify 
the down-quark and electron Yukawa couplings in a scenario where
neither large diagonal $A$-terms nor flavour violation in the soft
terms are allowed. In such a case, the only source of possible SUSY-scale
threshold corrections are the (\tanb)-enhanced contributions proportional to the
$\mu$ parameter~\eqref{thres:mfv}. In this case, however, the threshold
corrections to the down and strange quark Yukawa
couplings would have the same sign, while opposite signs are
phenomenologically required. Therefore, it is unlikely that one could
unify both of them in such a setup.

The observed DM relic density was correctly reproduced in the GFV scenarios,
showing that the \ymu{} is not at tension with \abund{}. Although
the DM constraint was not taken into account in our large $A$-term
scenario\cite{Iskrzynski:2014zla}, we do not expect it could change our
qualitative conclusions. A possible price to pay might be abandoning 
our assumption of the scalar soft mass universality at the GUT scale.

One might wonder whether including higher-order corrections to our
analysis could affect the conclusions. We used one-loop 
formulae for the threshold corrections, which was consistent
with employing the two-loop MSSM RGEs. However, in the case of
high \tanb, some of the two-loop threshold corrections might get
enhanced\cite{Crivellin:2012zz}. Although including such corrections would
require an extended numerical analysis, we do not expect our final qualitative
conclusions could be affected.

The vacuum metastability issue was investigated at tree level. A
one-loop calculation might be a valid supplement to the
analysis. In fact, there exists public software which
could be employed for such a calculation\cite{Camargo-Molina:2013qva}. 
In any case, again, our qualitative conclusions are unlikely to be modified.

Some issues related to the $SU(5)$ GUT, like the proton lifetime, 
have remained unaddressed in this work. On the other hand, 
several ways of saving the proton from decaying too fast in the $SU(5)$
framework have been proposed in the literature, for example by employing
higher-dimensional operators\cite{EmmanuelCosta:2003pu}.

In the present study we aimed at checking whether large
GUT-scale threshold corrections or departures from the minimal
$SU(5)$ boundary conditions to the Yukawa couplings can be avoided
in the R-parity conserving MSSM. We did not aspire, however, to construct a
full and self-consistent model valid above the GUT-scale. In reality, the
Yukawa unification problem, in the absence of the Georgi-Jarlskog mechanism,
might receive a combined solution from both the SUSY-scale and GUT-scale
threshold effects. Our study should be viewed as a qualitative test to what
extent the GUT-scale corrections are absolutely necessary. We hope that it
contributes in a constructive manner to the long-lasting search for a simple
GUT model that works.

%% file: chapters/Acronyms.tex
\label{acronymsRef}
\vspace*{-4mm}
The following acronyms are used throughout the text: \\[4mm]
CCB = Charge- and Colour-Breaking \\[2mm]
CKM = Cabibbo-Kobayashi-Maskawa \\[2mm]
CMB = Cosmic Microwave Background \\[2mm]
DM = Dark Matter \\[2mm]
EW = electroweak \\[2mm]
EWSB = Electroweak Symmetry Breaking \\[2mm]
GFV = General Flavour Violating \\[2mm]
GIM = Glashow-Iliopoulos-Maiani \\[2mm]
GUT = Grand Unified Theory \\[2mm]
LHC = Large Hadron Collider \\[2mm]
LSP = Lightest Supersymmetric Particle \\[2mm]
MCMC = Markov Chain Monte Carlo \\[2mm]
MET = Missing Transverse Energy \\[2mm]
MFV = Minimal Flavour Violating \\[2mm]
MSSM = Minimal Supersymmetric Standard Model \\[2mm]
NLSP = Next-to-Lightest Supersymmetric Particle \\[2mm]
QFT = Quantum Field Theory\\[2mm]
RGE = Renormalisation Group Equations \\[2mm]
SM = Standard Model \\[2mm]
SMS = Simplified Model Scenarios \\[2mm]
SUSY = supersymmetry \\[2mm]
UFB = Unbounded From Below \\[2mm]
VEV = Vacuum Expectation Value \\[2mm]
WIMP = Weakly Interacting Massive Particle \\[4mm]
soft terms/couplings = soft supersymmetry breaking terms/couplings \\[2mm]
$A$-terms = trilinear soft Higgs-squark-squark couplings (matrices in the flavour space) \\[2mm]
%